\documentclass[12pt, letterpaper, oneside,openright,titlepage]{book}

\usepackage{amsmath,  amsfonts, amssymb}
\usepackage{graphicx}
\usepackage[usenames, dvipsnames]{color}
\usepackage{caption}
\usepackage{subcaption}
\usepackage{tikz}
\usepackage{geometry}
\usepackage[abs]{overpic}
\usepackage[toc,page]{appendix}
\usepackage{varioref}
\usepackage{url}
\usepackage{titlesec}
\usepackage{float}
\usepackage[normalem]{ulem}
\usepackage{fancyhdr}

\usepackage{tocloft}

\newcommand{\dprime}{{\prime\prime}}
\newcommand{\tprime}{{\prime\prime\prime}}

\renewcommand{\thechapter}{\arabic{chapter}}
\titleformat{\chapter}[display]
  {\bfseries\Huge\sffamily}
  {\filcenter\Huge\thechapter}
  {1ex}
  {\filcenter}

\renewcommand{\thesection}{\arabic{chapter}.\arabic{section}}
\titleformat{\section}[hang]
  {\bfseries\Large\sffamily}
  {\Large\thesection}
  {0.5cm}
  {\titlerule*\filright}

\begin{document}

\newpage
\pagenumbering{roman}
\thispagestyle{empty}
{
\centering 
\Large{ \bf {Numerical Studies on Correlations in

 Dynamics and Localization of 

Two  Interacting  Particles in Lattices}}

\vspace{1cm}
by 

\vspace{1cm}

\large{Tirthaprasad Chattaraj}
\vspace{0.5cm}

M. Sc., INDIAN INSTITUTE OF TECHNOLOGY KANPUR, 2012

\vspace{1.5cm}

A THESIS SUBMITTED IN PARTIAL FULFILLMENT OF 

THE REQUIREMENTS FOR THE DEGREE OF

\vspace{1cm}

DOCTOR OF PHILOSOPHY

\vspace{1.3cm}
in

\vspace{0.5cm}
The Faculty of Graduate and Postdoctoral Studies

\vspace{0.5cm}

(CHEMISTRY)

\vspace{0.5cm}

THE UNIVERSITY OF BRITISH COLUMBIA

(Vancouver)

\vspace{0.5cm}
September 2018
\vspace{0.5cm}

\copyright ~ Tirthaprasad Chattaraj, 2018

}

\newpage

\noindent  The following individuals certify that they have read, and recommended to the Faculty of Graduate and Postdoctoral Studies for acceptance, a dissertation entitled:

\vspace{1cm}
\noindent \uline{\textsc{Numerical Studies on Correlations in}\hfill}

 \noindent \uline{\textsc{Dynamics and Localization of Two  Interacting Particles in Lattices}\hfill}
\vspace{1cm}

\noindent submitted \hspace{5cm} in partial fulfillment of the requirement for

\noindent by  \hspace{1.5cm} \underline{Tirthaprasad Chattaraj} 

\vspace{0.5cm}
\noindent the 

\noindent degree of~~~ \uline{\textsc{Doctor of Philosophy}\hfill} 

\vspace{0.5cm}
\noindent in ~~~~~~~~~~~~\uline{\textsc{Chemistry}\hfill}

\vspace{3.5cm}
\noindent \textbf{Examining Committee:}

\vspace{0.5cm}
\noindent \uline{\textsc{Professor Roman V. Krems, Chemistry}\hfill}

\noindent Supervisor

\vspace{0.5cm}
\noindent \uline{\textsc{Professor Grenfell N. Patey,  Chemistry}\hfill}

\noindent Supervisory Committee Member

\vspace{0.5cm}
\noindent \uline{\textsc{Professor Mark Thachuk,  Chemistry}\hfill}

\noindent Supervisory Committee Member

\vspace{0.5cm}
\noindent \uline{\textsc{Professor Yan Alexander Wang,  Chemistry}\hfill}

\noindent University Examiner

\vspace{0.5cm}
\noindent \uline{\textsc{Professor Andrew MacFarlane,  Chemistry}\hfill}

\noindent University Examiner

\newpage

\textbf{\Large{Abstract}}

\addcontentsline{toc}{section}{Abstract}

\vspace{2cm}

Two interacting particles in lattices, in the absence of dissipation, can not distinguish between attractive or repulsive interaction when the range of their tunnelling is limited to nearest neighbor sites. However, we find that, in the case of long-range tunnelling, the particles exhibit different dynamics for different types of interactions of the same strength. The nature of dynamical correlations between particles also becomes significantly different. For weak interactions, particles develop a character in correlation which is in between that of antiwalking and cowalking when the tunnelling is long-range. For strong interactions, particles cowalk independently of their statistics. A few recent experiments have demonstrated such effects of interactions on quantum walk of photons, atoms and spin excitations on various lattice platforms.

In disordered lattices the effect of coherent backscattering makes particles localize to their initial position. We find  that a weak repulsive interaction reduces localization and a strong interaction enhances localization. We also calculate the correlations between the particles in the disordered 1D and 2D systems. The effect of long-range tunnelling on localization of particles in disordered 1D systems has been explored.

For large ordered or disordered lattices, computation of localization parameters becomes difficult. In these cases, an efficient recursive algorithm is used to calculate Green's functions exactly. We extend such algorithm to disordered systems in  both one and two dimensions. We also illustrate that this recursive algorithm maps directly to some graph structures  like binary trees. We perform calculations for quantum walk of interacting particles on such graphs. The method is also used to calculate the properties of interacting particles on lattices with gauge fields. For disordered 2D lattices, we introduce and test approximations which produce accurate results and make the calculations more efficient. We examine the localization parameters for a broad range of interaction and disorder strengths and try to find differences among parameters within the range.


\newpage

\textbf{\Large{Lay Summary}}
\addcontentsline{toc}{section}{Lay Summary}

\vspace{1cm}
\noindent \textit{While many of us tend to think of owning a rule after finding one, it is the nature that rules. From mathematics to sociology, some of us only are fortunate or worked hard to see these rules first. - folklore}
\vspace{1cm}

In nature there are two types of particles: bosons and fermions. The bosons have unique behaviour of togetherness while the fermions want to keep a distance between them. Such particles in lattices can be described as hopping from one site to other sites and when more than one particles occupy a site then they interact with their hopping modified. Depending on different interaction strengths, their behaviour might be totally different in both ordered and disordered lattices. For two such interacting particles, which is the focus of this thesis, range of hopping affects their dynamics differently for attractive or repulsive interactions. In finite disordered lattices, both of one and two dimensions, these particles get localized. This thesis illustrates how the correlations between the particles in disordered systems change depending on the interaction strength.

\newpage
\textbf{\Large{Preface}}

\addcontentsline{toc}{section}{Preface}

\vspace{1cm}
The work described in this thesis has been published or in preparation for publication as mentioned in the following.

\vspace{1cm}
\noindent  \textbf{Publications on thesis work: }
\vspace{1cm}

\noindent Chattaraj, T. and Krems, R. V.  (August 1, 2016), Effects of long-range hopping and interactions on quantum walks in ordered and disordered lattices, \emph{Phys. Rev. A} 94: 023601 

\noindent \url{arxiv.org/pdf/1605.04349.pdf}
\vspace{1cm}

\noindent Chattaraj, T. (2018), Recursive computation of Green’s functions for interacting particles in disordered lattices and binary trees.

\noindent \url{arxiv.org/pdf/1808.04898.pdf}
\vspace{1cm}

\noindent Chattaraj, T. (2018), Localization parameters for two interacting particles in disordered two-dimensional lattices.

\noindent \url{arxiv.org/pdf/1808.06141.pdf}
\vspace{1cm}

\noindent Chattaraj, T. (2018), Spectral weights of doublon in interacting Hofstadter model.

\noindent \url{arxiv.org/pdf/1808.10112.pdf}

\newpage
\cleardoublepage
\addcontentsline{toc}{section}{Table of Contents}
\tableofcontents

\cleardoublepage
\addcontentsline{toc}{section}{List of Figures}
\listoffigures


\newpage
\textbf{\Large{Acknowledgements}}
    
\addcontentsline{toc}{section}{Acknowledgements}

\vspace{3cm}
    I acknowledge the support that I have received from my supervisor since the beginning of my doctoral study. I acknowledge those persons who have worked tirelessly for many many years to build the computational facilities that we enjoy today. Without such computational power, this work would be impossible to accomplish. I acknowledge my friends in lab whose presence and active participation made my work enjoyable and insightful. Finally I acknowledge my family who have always stayed on my side throughout the entire journey.

\newpage
\addcontentsline{toc}{section}{Dedication}

\thispagestyle{plain}
\begin{flushright}

\null\vspace{\stretch{1}} 

\emph{amar ma ke}

\vspace{\stretch{5}}\null

 \end{flushright}

\newpage

\newpage
\pagenumbering{arabic}
\chapter{Introduction}

The interference of quantum objects has been found to give rise to many phenomena that cannot be understood classically. One of such phenomena was discovered during middle of the last century by R. Hanbury-Brown and R. Twiss \cite{hbt} who observed that detection of two photons by two detectors was correlated. In a similar experiment from the 1980s performed by C. Hong, Z. Ou and L. Mandel \cite{hom} it was found that two identical photons, when interfered and guided toward two separate detectors,  tend to appear together. Such correlations are now understood to be caused by fundamental statistics of particles. While bosons show bunching behaviour in their correlations, fermions anti-bunch. In the presence of interactions between particles, however, these effects are known to become significantly different. Today one can perform similar experiments  not only with photons but also with atoms as well as with electronic or spin excitations. This has been made possible by the experiments on trapping atoms in external fields \cite{chu}.

In the 1990s it was proposed that random walk of quantum particles \cite{aharonov} in lattices can be used for quantum information purposes \cite{aharonovd, ambainis}, which inspired numerous experiments \cite{karski, schreibera, schreibera1, peruzzo, schreibera2, zahringer} studying wavepacket dynamics in various atomic and optical systems. Algorithms for fast spatial search \cite{ashwin, childs1} are particularly promising. However, in these algorithms one has to optimize the search speed and search probability on a graph. Ballistic propagation of wavepackets on lattices, when applied for search \cite{shenvi} in databases, has promised a speedup of $\sqrt{N}$ ($N =$ database size), over classical search algorithms. Further generalization to multiparticle quantum walk \cite{childs2} has been shown to be effective not only for quantum information transfer but also for understanding isomorphism of graphs \cite{coppersmith1, coppersmith2, wang}. Quantum walk in the presence of an impurity has been proposed for the preparation of entangled states \cite{sougoto}. The effect of quantum interference on transport of excitations has been shown to be present even in biological systems such as photosynthetic light harvesting complexes \cite{mohan, engel}.

In the case of more than one particles, quantum statistics affect quantum walk in lattices \cite{qin, bordone}. For bosonic and fermionic particles, the nature of multiparticle quantum walk is very different. Bosons exhibit bunching correlations while fermions and hard-core bosons show anti-bunching correlations in the absence of any interaction. These correlations can be used to determine the character of particles from the studies of their quantum walk when no such information is available otherwise. However, to study such phenomena one needs the most advanced atomic and optical systems where not only single particle resolution \cite{bakr, sherson} has been achieved but two-particle correlations \cite{folling, greiner1, fukuhara, preiss} can also be experimentally measured.

Photons have been at the forefront of understanding the effects of statistics and quantum interference for a long time. Lahini et al. \cite{lahini1, lahini2} performed experiments on both one particle and two particle photonic quantum walk on waveguide lattices. Recently, Greiner et al. have shown that such experiments  can also be performed with atoms \cite{preiss} in optical lattices. Bloch et al. have implemented such schemes for spin excitations \cite{fukuhara} in optical lattices.  Quantum walks in disordered systems have also been of interest since the work of Anderson \cite{anderson} explaining the role of disorder in low-dimensional systems leading to exponential localization of non-interacting particles. All these effects can be studied in optical lattice systems within a range of experimentally accessible parameters. Anderson localization has been recently illustrated experimentally for $^{87}$Rb atoms in optical speckle lattices \cite{billy}. In photonic wave guides, it was found that even in disordered systems, localized photons still remain correlated \cite{lahini2}. 

For more than one particle, the interaction between particles affects both the dynamics and localization. A huge amount of study has been done since the 1980s to understand these effects just for two particles \cite{shepelyansky, imry, pichard,  schreiberm1, ortuno, flach}. The two-particle interactions were shown to reduce localization in disordered lattices. A few of them predicted that interactions may enhance localization \cite{pichard, schreiberm1}. These are some of the fundamentally important investigations which can only now be explored and understood. Our work makes an effort to understand these effects not only in one dimensional systems but also in two dimensional systems. The effect of range of tunnelling on localization of interacting particles in 1D systems has been calculated in this thesis. The effects of interactions on two particle correlations in disordered 1D and 2D lattices have also been explored in this thesis.

In the case of atoms in optical lattices \cite{guidoni}, tunnelling beyond nearest neighbor site is not significant because of the length scales of these lattices. However, for dipolar molecules, tunnelling of excitations to sites far apart can be observed. These tunnelling parameters can also be tuned by an external field \cite{ping}. Preparation of such molecules in optical lattices has been achieved \cite{deborah, jun, volz, ospelkaus} very recently. However, the experiments to achieve a higher filling fraction still remain to be developed. Dipolar molecules in optical lattices are expected \cite{sowinski} to have various separate phases. Quantum random walk using Rydberg atoms has also been proposed \cite{cote} for long-range tunnelling models.

 In theoretical investigations at the most simple and fundamental level, the Hubbard model \cite{hubbard} plays a central role. It is very useful for understanding the effects of interaction between particles and their dynamics. The extended Hubbard model, where particles interact and tunnel in lattices beyond their nearest neighbors, has also become a highly investigated research topic since a few years ago. In this thesis, I try to understand the interplay of such long-range interactions and tunnelings in both ordered and disordered lattices. One of the most interesting findings of this thesis is the effect of such long-range tunnelling on dynamical correlations of two particles in 1D lattices.

There has also been a lot of interest recently in simulating particles on 2D lattices under synthetic gauge fields. In the case of atoms, these  fields are created by a periodic shaking of lattice potentials \cite{spielman1, spielman2}. We attempt to understand the implications of such gauge fields on quantum walk of interacting particles.

The methods that have been used for the calculations presented in this thesis are mostly based on full diagonalization of hamiltonians and recursive calculations of Green's functions. The recursion method used in this thesis is an extension of previous work in our group \cite{ping-thesis}. This method makes the calculations significantly more efficient compared to full diagonalization and also allows for performing calculations with a much larger basis size. We introduce new boundary conditions that make the method exact and approximations that make it more efficient while maintaining accuracy. The underlined mathematics of the method has been also elucidated in calculations for interacting particles on graphs such as binary trees in this thesis.

There are many experimental systems relevant for the research presented here. Our research is most relevant for but not limited to cold atom \cite{greiner-thesis} and trapped-ion \cite{rajibul-thesis} systems. (See Appendix F for a brief introduction to optical lattices.) The results are applicable wherever the model systems can be mapped to the approximate physics of the systems under investigation. Two particle correlations, in essence, describe the fundamental physics of many interacting particles.  For the cold atom systems, manipulation of interactions between particles has been pursued from the 1980s \cite{feshbach,chin, volz1}. See Appendices B and C for the discussion of such controls.   The range of tunnelling in lattices can be controlled within a broad range of tunability using the ideas of M{\"o}lmer and Sorensen \cite{molmer-sorensen}. Section 1.1 describes how such long-range tunnelling of excitations can be achieved in lattices. More details can be found in Appendix E. Phonons also play an important part in controlling interactions between electrons and excitations, as described in Appendix D.  Excitonic systems (see Appendix G) can also exhibit the interactions, for which the fundamental physics is expected to be very similar to that which has been described in this thesis.

\section{Thesis overview}

This thesis deals mainly with two aspects of two interacting particles. One is long-range hopping of particles in lattices and the interplay of such hopping with the effect of interactions between the particles.  Section 1.3 describes how long-range hopping can be engineered in most advanced atomic and optical systems. The other is the behaviour of interacting particles in the presence of impurities. The model that is used in this thesis to understand the effects of interactions between the particles in both ordered and disordered lattices is mainly an extension of the Hubbard model. The origin of the terms in the model is explained in Section 1.2.

The chapters are organized to describe the main results of the work that has been performed over the years. Chapter 2 describes the effects of long-range hopping of particles in lattices in presence of the interplay with interactions between the particles. Chapter 3 presents a numerical approach to extend the size of calculations for bigger lattices. Chapter 4 contains results that have been calculated from the dynamics of two interacting particles in disordered systems. The effects of the interaction on localization of particles in finite disordered lattices is the focus of that chapter. Finally Chapter 5 summarizes the conclusions that can be drawn from the work of the whole thesis.

\newpage
\section{Hubbard model}
In this section we introduce the notation that will be used extensively throughout the thesis. For  particles in lattices, one can describe them as hopping from site to site in the lattice and interacting with each other. This simple physical model is not only intuitive but also provides the basic understanding for other models.

The Hubbard model makes the notations more simplified than in 1st quantized form. Although it is very easy to write down, it is very difficult to solve  for more than one particle. The model has a nearest neighbor hopping term and onsite energies. In the presence of interactions, most effective models add the onsite two-body interaction term. Terms that describe tunnelling and interactions  beyond nearest neighbors are also added in many models. 

The starting hamiltonian for particles on lattices consists of the kinetic energy term and various potential energy terms deriving from electron-electron interaction, electron-ion interaction and ion-ion interactions

\begin{equation}
\mathcal{H} = \mathcal{H}_{kin} + V_{e-ion} + V_{e-e} + V_{ion-ion}
\end{equation}
where the first two terms are the one-particle terms

\begin{equation}
\mathcal{H}_0 = \mathcal{H}_{kin}  + V_{e-ion} =\sum_{i=1}^{N_e} \left[ \frac{p_{i}^2}{2m} + \sum_{n=1}^{N} V(\mathbf{r_i}, \mathbf{R}_n) \right],
\end{equation}
where $i$ is the index for the electrons and $n$ for the nuclei.

For a basis one can start with the atomic wavefunctions localized on each site

\begin{equation}
\mathcal{H}_{atom} \vert \psi \rangle = \mathcal{E}_{\psi} \vert \psi \rangle,    ~~~\mbox{where}~~~ \langle \psi^\prime\vert \psi \rangle = \delta_{\psi \psi^\prime}.
\end{equation}
One can take these solutions as those for the electron belonging to the nucleus at site $n$
\begin{equation}
\mathcal{H}_{atom}^n \vert \psi n \rangle = \mathcal{E}_{\psi n} \vert  \psi n \rangle,    ~~~\mbox{where}~~~ \langle \psi^\prime n^\prime\vert \psi  n\rangle = \delta_{\psi \psi^\prime}\delta_{n^\prime n},
\end{equation}
where we have applied the tight binding approximation for the overlap integral. 

One can also conveniently write these terms using creation (annihilation) operators

\begin{equation}
\vert  \psi n \rangle  = a_{\psi n}^\dagger \vert 0 \rangle 
\end{equation}
where the operators follow (anti)commutation relations as described in Appendix A for (fermions)bosons
\begin{align}
\left[ a_{\psi^\prime n^\prime}, a_{\psi n}^\dagger \right]_{\pm} &= \delta_{\psi \psi^\prime}\delta_{n^\prime n} \nonumber\\
  \left[ a_{\psi^\prime n^\prime}^\dagger, a_{\psi n}^\dagger \right]_{\pm}  &= 0 \nonumber\\
\left[ a_{\psi^\prime n^\prime}, a_{\psi n} \right]_{\pm} & = 0
\end{align}

The one-particle hamiltonian term then becomes

\begin{equation}
\mathcal{H}_0 = \sum_{\psi^\prime n^\prime, \psi n } \mathcal{E}_{\psi^\prime n^\prime,  \psi n } a_{\psi^\prime n^\prime}^\dagger a_{\psi n}
\end{equation}
where 
\begin{align}
 \mathcal{E}_{\psi^\prime n^\prime,  \psi n } &= \int d\mathbf{r} ~ \langle \psi^\prime n^\prime\vert \mathbf{r}\rangle \langle \mathbf{r}\vert\mathcal{H}_0\vert\mathbf{r}\rangle \langle \mathbf{r} \vert\psi  n\rangle \nonumber \\
&= \int d\mathbf{r}  ~\langle \psi^\prime n^\prime\vert \mathbf{r}\rangle   \left[ \frac{p^2}{2m} + \sum_{n^{\dprime}=1}^{N} V(\mathbf{r}, \mathbf{R}_{n^{\dprime}}) \right]    \langle \mathbf{r} \vert\psi  n\rangle \nonumber \\
&= \int d\mathbf{r} ~\phi_{\psi^\prime n^\prime} (\mathbf{r})   \left[\mathcal{E}_{\psi n} + \sum_{n^{\dprime} = n} V(\mathbf{r}, \mathbf{R}_{n^{\dprime}}) \right]    \phi_{\psi n}(\mathbf{r}) \nonumber \\
&= \left[\mathcal{E}_{\psi}\delta_{\psi\psi'}\delta_{n,n'} + \int d\mathbf{r}~ \phi_{\psi} (\mathbf{r})   \sum_{n^{\dprime} = n} V(\mathbf{r}, \mathbf{R}_{n^{\dprime}}) ~  \phi_{\psi }(\mathbf{r}) \nonumber  \right] \nonumber \\
& \hspace{5cm}+  \int d\mathbf{r}~ \phi_{\psi^\prime } (\mathbf{r})   \sum_{n^{\dprime} = n} V(\mathbf{r}, \mathbf{R}_{n^{\dprime}} - \mathbf{R}_n) ~  \phi_{\psi}(\mathbf{r}) \nonumber \\
 \mathcal{E}_{\psi^\prime n^\prime,  \psi n } &= \left[ \tilde{\mathcal{E}_{\psi}}\delta_{\psi\psi'}\delta_{n,n'} \right]  + \mathcal{W}_{\psi^\prime, \psi}\delta_{n,n'}
\end{align}

The one-particle hamiltonian can then be written as

\begin{equation}\label{H_0}
\mathcal{H}_0 =  \sum_{\psi} \tilde{\mathcal{E}_{\psi}} a_{\psi}^\dagger a_{\psi} +  \sum_{\psi\neq \psi^\prime}\mathcal{W}_{\psi^\prime, \psi} a_{\psi^\prime}^\dagger a_{\psi},
\end{equation}
Here the site indices are removed for simplicity. The terms on the right hand side of Eq. \ref{H_0} are the energies of the states and the excitation gap between the states on each site. For ideal two level systems one can disregard these terms as they contribute to a constant term. It also turns out that the divergent terms (for the case of Coulomb interactions) in $V_{ion-ion}$ and $V_{e-ion}$ cancel each other so that we get a stable system. The two body part of $V_{e-e}$ can now be written as an addition of  multiple parts

\begin{equation}
V_{e-e} = V_{00} + V_{11} + V_{01}^\prime + V_{01}^{\dprime}     ~~~~~~~ \psi \in \{0, 1\},
\end{equation}
where

\begin{align}
V_{00} &= \frac{1}{2} \sum_{n\neq n^\prime} V^{\{ \}} a_{0 n}^\dagger a_{0n^\prime}^\dagger a_{0n^\prime} a_{0n} \nonumber \\
V_{11} &= \frac{1}{2} \sum_{n\neq n^\prime} V^{\{ \}} a_{1 n}^\dagger a_{1n^\prime}^\dagger a_{1n^\prime} a_{1n} \nonumber \\
V_{01}^\prime &= \frac{1}{2} \sum_{n, n^\prime} V^{\{ \}} a_{1 n}^\dagger a_{0n^\prime}^\dagger a_{0n^\prime} a_{1n} \nonumber \\
V_{01}^{\dprime} &= \frac{1}{2} \sum_{n, n^\prime} V^{\{ \}} a_{1 n}^\dagger a_{0n^\prime}^\dagger a_{1n^\prime} a_{0n}.
\end{align}
Here $V^{\{ \}}$ contains all the relevant indices for the interaction terms. The first three terms here are also the interaction energies between the states of the same and different energy levels located at different sites. The last term is  responsible for transfer of the states between sites. Writing an excitation (or quasiparticle) as $q_{n}^\dagger = a_{1n}^\dagger a_{0n}$, one can find 

\begin{equation}
V_{01}^{\dprime} =  \sum_n \frac{1}{2} V^{\{ \}} q_n^\dagger q_n +  \sum_{n\neq n^\prime} \frac{1}{2} V^{\{ \}} q_n^\dagger q_{n^\prime}.
\end{equation}

We can simplify all these forms by writing the final Hubbard hamiltonian consisting of the onsite excitation energy term and the inter-site hopping terms limited to nearest neighbors only:

\begin{equation}
H_{\text{Hubbard}}=  \sum_n \varepsilon_n q_n^\dagger q_n +  \sum_{\langle n n^\prime\rangle} t_{n n^\prime} q_n^\dagger q_{n^\prime}.
\end{equation}

The rest of the thesis builds on this form with interaction terms such as $V_{11}$ added and calculates properties of two particles. The simplification of the physical system to such a model after the elimination of many details makes the calculations significantly easier to implement.

\newpage
\section{Engineering range of coupling in lattices}
In most physical systems of interest either nearest neighbor hopping or hopping extended to few nearest neighbors are observed. However, using modern optical methods, it is possible to engineer the hopping ranges. In this section we discuss a method where phonons can be used effectively to control the coupling between  particles at different sites of a chain.

Following from appendix  E,  the effective hamiltonian after including the spatial variance of the optical field leads  to the following equation:

\begin{align}
\mathcal{H} &=  - \frac{\Omega}{2}  \left[ e^{ \imath ( \delta t - \mathbf{k \cdot r})} \sigma^- + e^{ -\imath (  \delta  t - \mathbf{k \cdot r})} \sigma^+ \right].
\end{align}
In  the presence of phonon modes at some frequency $\omega_p$, we can write  the position of the mode with the time dependency of field operators included as following:

\begin{equation}
X(t) = X_0 \left(a e^{-\imath \omega_p t} + a^\dagger e^{\imath \omega_p t}  \right),
\end{equation}
where $X_0 = 1/\sqrt{2M\omega_p}$ ($M$ is the mass of atoms). One can make  $\delta$ equal to that of the phonon frequency $\omega_p$. When $\delta > 0$, it can excite the phonon mode by one quantum. 

Writing the spatial dependence of the laser in phonon modes of certain frequency $\omega_p$, we find the  following, assuming the momentum of the laser mode along the motional mode:

\begin{align}
\mathcal{H} &=  - \frac{\Omega}{2}  \left[ e^{ \imath ( \delta t - kX)} \sigma^- + e^{ -\imath ( \delta  t - kX)} \sigma^+ \right]  \nonumber \\
&=  - \frac{\Omega}{2}  \left[ e^{ \imath \left( \omega_p t - k X_0 \left(a e^{-\imath \omega_p t} + a^\dagger e^{\imath \omega_p t}  \right) \right)} \sigma^- + e^{ -\imath \left( \omega_p  t - kX_0 \left(a e^{-\imath \omega_p t} + a^\dagger e^{\imath \omega_p t}  \right) \right)} \sigma^+ \right]  \nonumber \\
&=  - \frac{\Omega}{2}  \left[ [1 - \imath k X_0 \left(a e^{-\imath \omega_p t} + a^\dagger e^{\imath \omega_p t}  \right)] e^{ \imath \omega_p t} \sigma^- + [1 - \imath kX_0 \left(a e^{\imath \omega_p t} + a^\dagger e^{-\imath \omega_p t}  \right) ] e^{- \imath  \omega_p  t  } \sigma^+ \right], \nonumber \\
\mathcal{H}_{+} & \simeq  \frac{\imath kX_0\Omega}{2} \left[a\sigma^- + a^\dagger \sigma^+ \right] + \cdot \cdot \cdot
\end{align}
where in the last step the \textit{rotating wave approximation} has been applied. Similarly, for the negative detuning ($\delta < 0$), one can find 
\begin{align}
\mathcal{H}_{-} & \simeq  \frac{\imath kX_0\Omega}{2} \left[a^\dagger\sigma^- + a \sigma^+ \right].
\end{align}

These methods are very effective in cooling down the vibrational modes to its ground states. One can raise the electronic states higher while going down in phonon numbers using $\pi-$pulses, then decouple from phonon states while returning to the ground electronic state and repeat the processes.

In the case of two particles at sites $i$ and $j$ in a lattice, phonon modes can be used to effectively couple ($J_{ij}$) them irrespective of the range of distance between the particles.

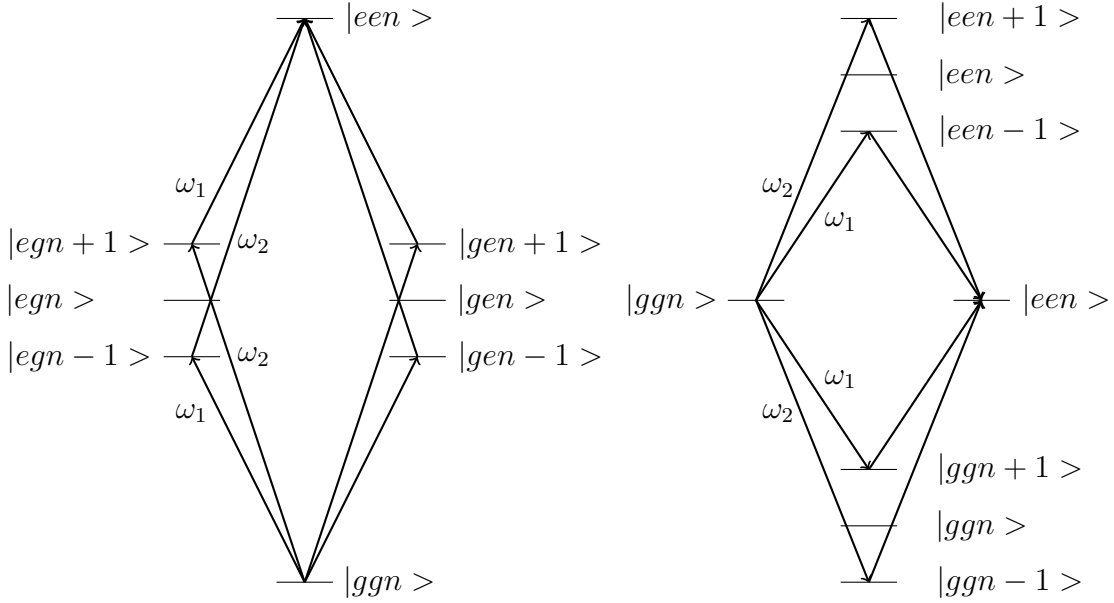
\begin{figure}[H]
\centering
\hspace{0.0cm}
\begin{tikzpicture}[scale=1.5]
\draw (5.0,0) --(5.5,0);
\node at (6.0,0) {$|ggn>$};
\draw (5.0,5) --(5.5,5);
\node at (6.0,5) {$|een>$};
\draw (4.0,2.0) --(4.5,2.0);
\node at (3.25,2.0) {$|egn-1>$};
\draw (4.0,2.5) --(4.5,2.5);
\node at (3.0,2.5) {$|egn>$};
\draw (4.0,3.0) --(4.5,3.0);
\node at (3.25,3.0) {$|egn+1>$};
\draw (6.0,2.0) --(6.5,2.0);
\node at (7.25,2.0) {$|gen-1>$};
\draw (6.0,2.5) --(6.5,2.5);
\node at (7.0,2.5) {$|gen>$};
\draw (6.0,3.0) --(6.5,3.0);
\node at (7.25,3.0) {$|gen+1>$};
\draw [->,thick] (5.25,0) -- (4.25,2.0);
\draw [->,thick] (5.25,0) -- (4.25,3.0);
\draw [->,thick] (5.25,0) -- (6.25,2.0);
\draw [->,thick] (5.25,0) -- (6.25,3.0);
\draw [<-,thick] (5.25,5) -- (4.25,2.0);
\draw [<-,thick] (5.25,5) -- (4.25,3.0);
\draw [<-,thick] (5.25,5) -- (6.25,2.0);
\draw [<-,thick] (5.25,5) -- (6.25,3.0);
\node at (4.25,1.5) {$\omega_1$};
\node at (4.25,3.5) {$\omega_1$};
\node at (4.8,2.0) {$\omega_2$};
\node at (4.8,3.0) {$\omega_2$};
\draw (10.0,0) --(10.5,0);
\node at (11.5,0.0) {$|ggn-1>$};
\draw (10.0,0.5) --(10.5,0.5);
\node at (11.25,0.5) {$|ggn>$};
\draw (10.0,1.0) --(10.5,1.0);
\node at (11.5,1.0) {$|ggn+1>$};
\draw (10.0,5) --(10.5,5);
\node at (11.5,5.0) {$|een+1>$};
\draw (10.0,4.5) --(10.5,4.5);
\node at (11.25,4.5) {$|een>$};
\draw (10.0,4.0) --(10.5,4.0);
\node at (11.5,4.0) {$|een-1>$};
\draw (9.0,2.5) --(9.5,2.5);
\node at (8.5,2.5) {$|ggn>$};
\draw (11.0,2.5) --(11.5,2.5);
\node at (12.0,2.5) {$|een>$};
\draw [->,thick] (9.25,2.5) -- (10.25,0.0);
\draw [->,thick] (9.25,2.5) -- (10.25,1.0);
\draw [->,thick] (9.25,2.5)  -- (10.25,4.0);
\draw [->,thick] (9.25,2.5)  -- (10.25,5.0);
\draw [<-,thick] (11.25,2.5)  -- (10.25,0.0);
\draw [<-,thick] (11.25,2.5) -- (10.25,1.0);
\draw [<-,thick] (11.25,2.5) -- (10.25,4.0);
\draw [<-,thick] (11.25,2.5) -- (10.25,5.0);
\node at (10.0,1.8) {$\omega_1$};
\node at (10.0,3.2) {$\omega_1$};
\node at (9.45,1.5) {$\omega_2$};
\node at (9.45,3.5) {$\omega_2$};
\end{tikzpicture}
\vspace{0.5cm}
\caption[M{\"o}lmer-Sorensen scheme]{The M{\"o}lmer-Sorensen scheme. Left: both spins excited. Right: spins exchange excitations with effectively the same coupling parameter as in the left. Adapted from Reference \cite{molmer-sorensen}.}
\label{molmer-sorensen}
\end{figure}
The  scheme is known after M{\"o}lmer-Sorensen \cite{molmer-sorensen}. The laser frequency can be detuned at the M{\"o}lmer-Sorensen detuning $\mu = \delta - \omega_p$. Now, both types of laser detuning can be applied to a system of two particles ($\delta < 0$ and $\delta > 0$). When both particles are in the ground state with $n$ phonons in state $\vert g, g, n\rangle$, any of them can absorb the negatively detuned photon and undergo the transition to the excited state while the phonon number goes to $n-1$. Thus the state $\vert g, e, n - 1\rangle$ or  $\vert e, g, n - 1\rangle$ is reached in this process. Alternatively, any state can absorb a positively detuned photon and go to the $\vert g, e, n + 1\rangle$ or  $\vert e, g, n + 1\rangle$ state. Both of these states can absorb just the oppositely detuned photon than the first time to reach the $\vert e, e, n\rangle$ state. The intermediate states can be made negligible in the whole transition by similar procedures as described in Appendix E, where we virtually made the excited state contributing little in the dynamics of two states but now with the M{\"o}lmer-Sorensen detuning ($\delta < \omega_p$) and the coupling $g$  replaced by $\eta \Omega \sqrt{n}$. Here $\eta = kX_0$, and the $\sqrt{n} (\sqrt{n+1} )$  factor comes from the phonon annihilation (creation) operation. 
The amplitude of the transition $\vert g, g, n\rangle \rightarrow \vert g, e, n-1\rangle \rightarrow \vert e, e, n\rangle$ is

\begin{align}
\Omega_{21-+} =- \frac{\eta \Omega \sqrt{n} \eta \Omega \sqrt{n}}{4\mu}.
\end{align}

The amplitude of the transition  $\vert g, g, n\rangle \rightarrow \vert g, e, n+1\rangle \rightarrow \vert e, e, n\rangle$ is

\begin{align}
\Omega_{21+-} = \frac{\eta \Omega \sqrt{n+1} \eta \Omega \sqrt{n+1}}{4\mu}.
\end{align}
So the  amplitude for the transition through exciting the first particle without any significant transition into intermediate states with different phonon numbers is given by

\begin{align}
\Omega_{21} = \Omega_{21-+} + \Omega_{21+-} = \frac{(\eta \Omega)^2}{4\mu}.
\end{align}

The contribution from the other two paths, where the first particle changes the state first $\vert g, g, n\rangle \rightarrow \vert e, g, n-1\rangle \rightarrow \vert e, e, n\rangle$ and 
$\vert g, g, n\rangle \rightarrow \vert e, g, n+1\rangle \rightarrow \vert e, e, n\rangle$, adds to total amplitude

\begin{align}
\Omega_{12}  = \frac{(\eta \Omega)^2}{4\mu},
\end{align}

\begin{align}
\Omega_{tot}  = \Omega_{12} + \Omega_{21} = \frac{(\eta \Omega)^2}{2\mu}.
\end{align}

Additional  detuning (with respect to the phonon frequency) of the photon (with respect to the energy gap of two states of the particles)  $\mu = \delta - \omega_p$ and $\mu = \omega_p + \delta$ gives the full coupling between the two particles

\begin{align}
J_{ij}  &=\frac{(\eta \Omega)^2}{2(\delta - \omega_p)} - \frac{(\eta \Omega)^2}{2(\delta + \omega_p)} \nonumber \\
&= \frac{(\eta \Omega)^2 \omega_p}{\delta^2 - \omega_p^2}.
\end{align}

This method has recently been used (tuning $\delta$ at different sites) to engineer coupling between spins in a 1D chain with variable ranges, where one can achieve a regular power law form of the coupling with respect to the distance between the spins \cite{jurcevic}.

\begin{equation}
J_{ij} = \frac{J}{\vert i - j \vert^\alpha}  ~~~~~~~~~~~~~ \text{with} ~~~~~~ 0 \le \alpha \le 3.
\end{equation}
In Chapter 2 of this thesis, we will consider hamiltonians for two particles with such long-range hopping in lattices.

\newpage
\chapter{Correlations in Dynamics of interacting particles}

In this chapter we mainly discuss the effects of long-range hopping on correlations of two interacting particles. The statistics of particles play a crucial role in determining correlations between the particles. This role of statistics is fundamental and has been described  in the introductory quantum mechanics books \cite{griffith}. The bunching of bosons and anti-bunching of fermions has been known to be a result of their fundamental statistics. However the role of interaction in determining the dynamics has been studied only recently \cite{lahini1, lahini2, winkler}.  A few recent experiments have explored these effects with photonic wave guides, trapped ion and trapped atom systems. The presence of repulsively bound pairs has been observed in cold atomic systems in the absence of dissipation of energy \cite{winkler}. Such systems for two particles can be modelled effectively by the Hubbard hamiltonian with a conserved number of particles and total energy. The similarity between the attractive and  repulsive interactions is also well known for these models with nearest neighbor hopping. In the case of long-range hopping, however, an asymmetry in the effect of the attractive and repulsive interactions is observed in our study. It is described  in the later part of this chapter. 

In the following section we describe a few important results that were obtained from exact diagonalization of the full hamiltonian of a 1D system of two particles. These studies were motivated by the experimental and theoretical studies of  quantum  walk on lattices \cite{lahini1, lahini2, winkler} and the studies on the effect of the long-range hopping on  eigenstates of the particles in 1D lattices \cite{levitov, malyshev, malysheva}. The existence of the bound pairs in the presence of both repulsive and attractive interactions was established by these studies. However, the effect of long-range nature of hopping on such bound pairs was not fully understood. Our calculations try to elucidate this effect.

\section{Two particle systems}

The case of  two distinguishable particles can be described by the composite wavefunction of the two particles 

\begin{equation}
\psi (x_1, x_2) = \psi_1 (x_1) \psi_2 (x_2).
\end{equation}
However, if the particles are indistinguishable,  the wavefunctions have to be symmetrized (anti-symmetrized) for the bosonic (fermionic) particles

\begin{equation}
\psi (x_1, x_2)_\pm = \frac{1}{\sqrt{2}}\left[ \psi_1 (x_1) \psi_2 (x_2) \pm \psi_2 (x_1) \psi_1 (x_2) \right].
\end{equation}
The effect of this symmetrization (anti-symmetrization) can be observed in the expectation value of the square of the relative distance

\begin{equation}
\Delta = \langle(x_1 -x_2)^2\rangle = \langle x_1^2\rangle + \langle x_2^2\rangle - 2\langle x_1 x_2\rangle,
\end{equation}
which for the distinguishable particles is

\begin{equation}
\Delta  = \langle x_1^2\rangle + \langle x_2^2\rangle - 2\langle x_1 \rangle\langle x_2\rangle,
\end{equation}
and for the indistinguishable particles is

\begin{equation}
\Delta  = \langle x_1^2\rangle + \langle x_2^2\rangle - 2\langle x_1 \rangle\langle x_2\rangle  \mp  2\vert \langle x_{12}\rangle\vert^2,
\end{equation}
where $x_{12}$ is an interference term \cite{griffith}. This interference effect (due to the fundamental statistics)  makes two bosons bunch together, while it results in anti-bunching for two fermions. 

This effect of distinguishability can be easily seen when two particle dynamics is simulated in an ideal 1D lattice with nearest neighbor hopping. One can simulate such dynamics under the effect of the Hubbard hamiltonian with onsite interactions for two bosons. There have been such studies of distinguishability with photons \cite{pathak}. 

The hamiltonian of two bosonic particles in ideal lattices can be written as the following simplified form 
\begin{equation}
\mathcal{H} =   \sum_{\langle n m\rangle} t a_n^\dagger a_{m} +  \sum_n U a_n^\dagger  a_n^\dagger a_n a_n 
\end{equation}
where $n$ and $m$ are the lattice site indices, $t$ is the hopping amplitude between two sites and $U$ is the onsite interaction energy.

The joint density distribution ($\varrho_{n, m}$) then can be calculated  from eigenfunctions ($|\lambda\rangle$) and eigenenergies ($E_\lambda$) of the hamiltonian

   \begin{equation}\label{joint-density}
       \varrho_{n, m}(\tau) = |\langle n, m | \sum_{\lambda} e^{-i E_{\lambda} \tau} |\lambda\rangle \langle \lambda|n',  m'\rangle|^2.
    \end{equation}
The initially occupied sites are denoted as $n', m'$ and the evolution time as $\tau$. This joint density describes the correlation between the two particles

   \begin{equation}\label{correlation1}
     C_{n, m}(\tau) =  \varrho_{n, m} (\tau)  = \langle a_n^\dagger a_m^\dagger a_m a_n \rangle (\tau).
    \end{equation}

One can also define the correlations as in Eq. \ref{correlation2}
   \begin{equation}\label{correlation2}
     C_{n, m}   = \langle a_n^\dagger a_m^\dagger a_m a_n \rangle - \langle a_n^\dagger  a_n \rangle\langle  a_m^\dagger a_m \rangle.
    \end{equation}
 However, our interest is in comparing the effects of interactions and the  last term in previous equation is independent of any interactions. This term only act as some constant additive which can be neglected for further simplification.
The total density distribution can be calculated from the joint density distribution as

   \begin{equation}\label{density}
     \rho (n,\tau)  =  \frac{1}{2} \sum_{m\neq n}  \varrho (n, m, \tau) =  \langle a_n^\dagger a_n \rangle (\tau).
    \end{equation}

\begin{figure}[H]
\centering
\includegraphics[width=0.85\textwidth]{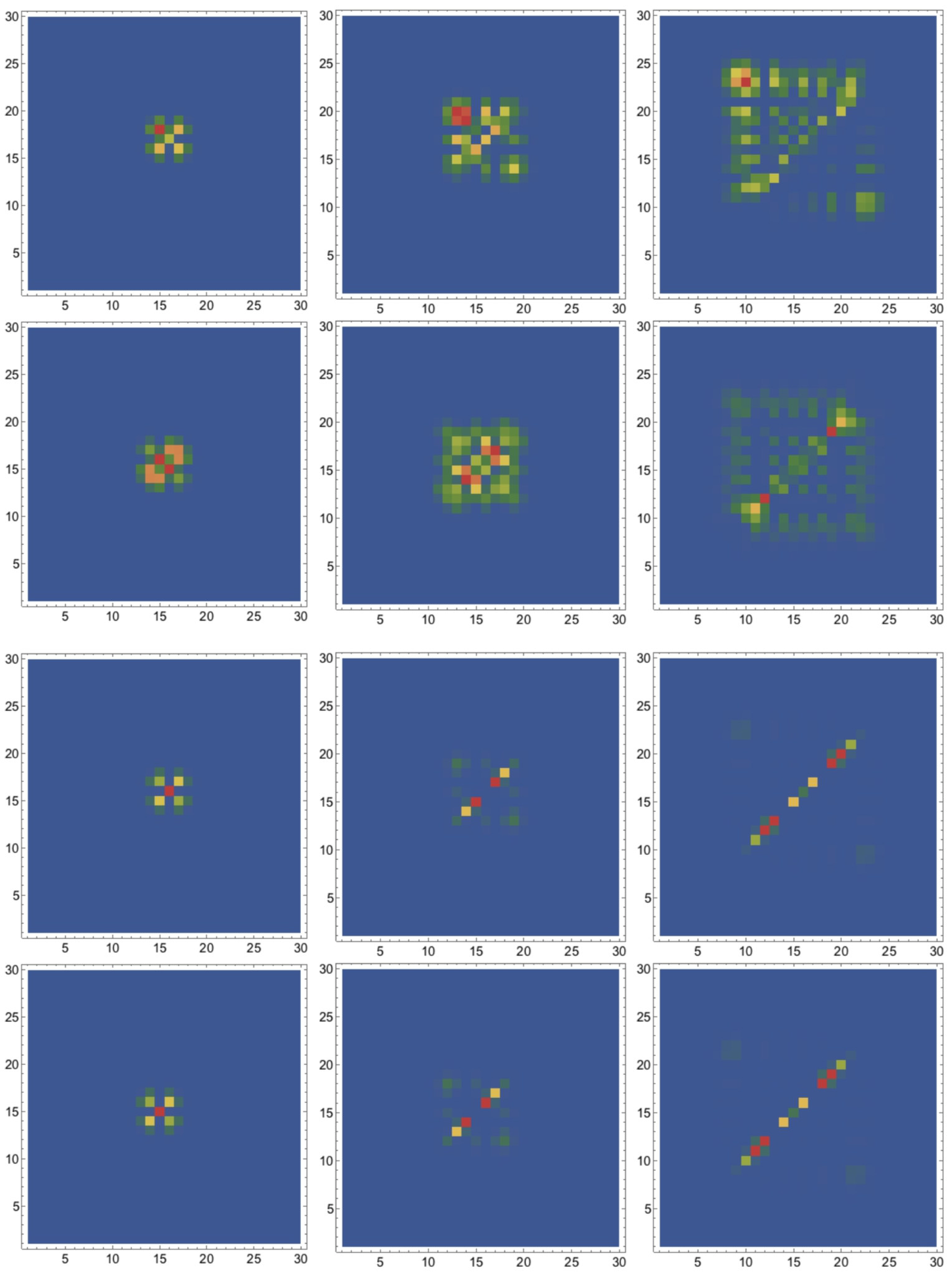}
\caption[Correlations in two (in)distinguishable bosonic quantum walks]{Joint densities (Eq. \ref{joint-density}) in quantum walk for distinguishable bosons (rows 1 and 3)) and indistinguishable bosons (rows 2 and 4). $U/t=2$ for all cases.  Time = 1$/t$, 2$/t$ and 4$/t$ respectively for columns 1, 2 and 3. Particles start from adjacent sites  in cases of rows 1 and 2, while from same site in cases of rows 3 and 4. Two axes are site indices for two particles. Color scheme- red, yellow, green, blue show lower joint density in that order.}
\label{distinguishable-correlation}
\end{figure}

Figure \ref{distinguishable-correlation} shows a simulation for quantum walk of two distinguishable and indistinguishable bosons  on an ideal lattice in the presence of the interaction ($U=2$). The effect can be clearly seen in terms of the correlation elements which include four creation and annihilation operators and also in the density terms which include only two creation or annihilation operators.  The correlations which describe the joint probablities of finding two particles on the same site or nearest neighbor sites can be termed as cowalking correlations which describe the effect of bunching. The joint densities which describe the particles moving in the opposite direction  are termed as antiwalking correlations and describe anti-bunching.

As can be seen from the correlations of two particles in Fig. \ref{distinguishable-correlation}, the bosonic particles tend to bunch together and cowalk. However, fermions and hardcore bosons tend to anti-walk as we will see in  later sections. When the particles start the quantum walk from adjacent sites, the correlation dynamics is different as for the indistinguishable particles the correlations are symmetric. However, when they start from the same lattice site,  the distinguishability has no effect and no difference in quantum walk can be observed.

\newpage

\begin{figure}[H]
\centering
\includegraphics[width=0.88\textwidth]{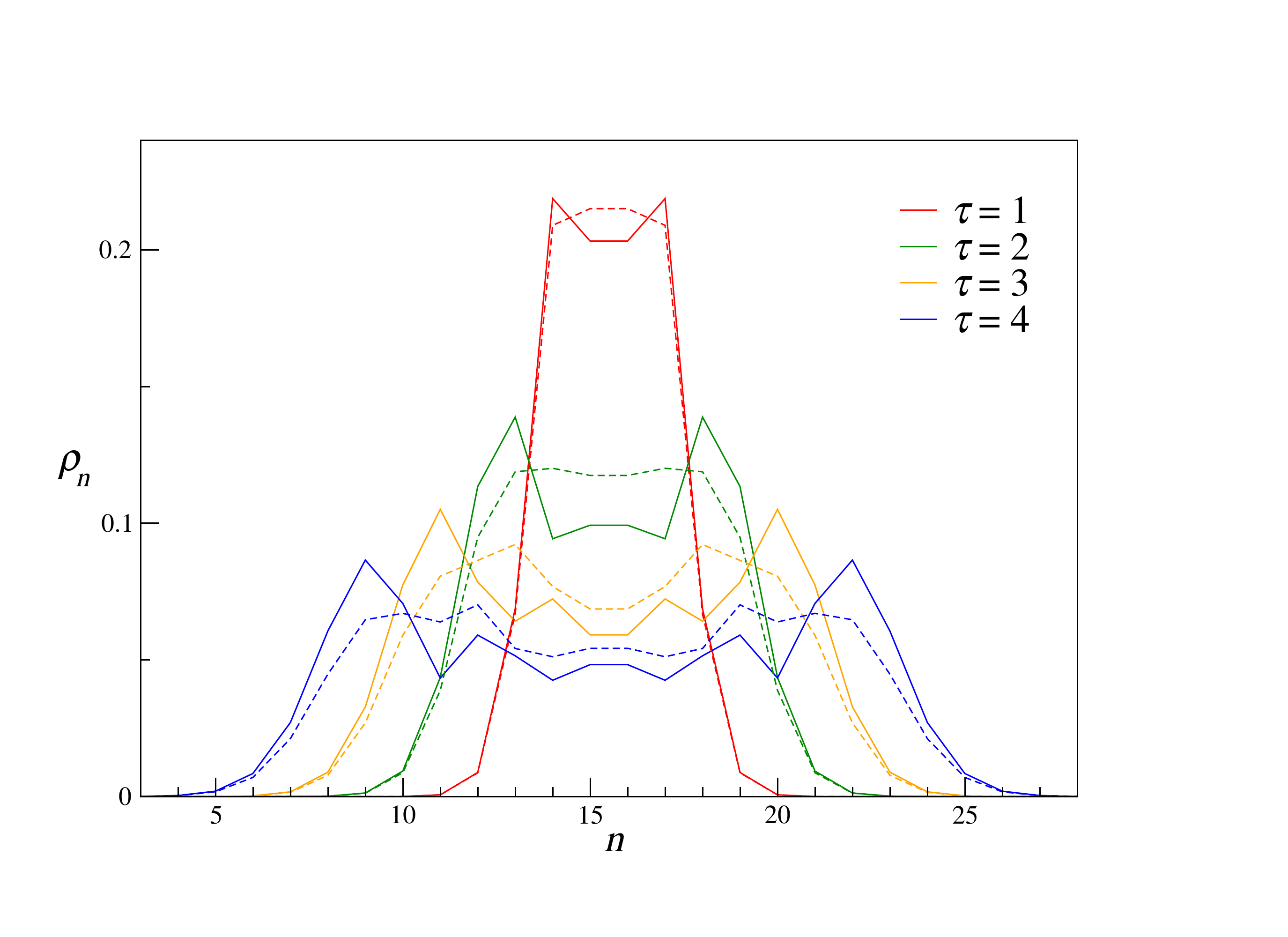}
\caption[Density distributions in two (in)distinguishable bosonic quantum walk]{Density distributions (Eq. \ref{density}) in quantum walk of two distinguishable and indistinguishable bosonic particles. The solid lines correspond to distinguishable bosons while dashed lines to indistinguishable ones. Time $\tau$ is measured in the units of the inverse of the hopping integral $t$. The interaction $U=2$. }
\label{distinguishable-density}
\end{figure}

This difference due to the distinguishability can be observed also from the simulated density terms on a lattice. From Fig. \ref{distinguishable-density} it can be seen that the bunching of indistinguishable bosons (dashed lines) tends to interfere constructively in between the dynamical wavepacket peaks compared to the case of the distinguishable ones (solid lines).  For strongly interacting particles, this difference in densities is expected to become small as they  form a bound state which will behave very similar to a single composite particle for both cases.

\section{Two particle states}
For a Hamiltonian \ref{2pH}  with onsite interaction term, the two particle states and energies can be analytically derived following the description of Valiente and Petrosyan \cite{valiente1}. The theoretical description can also be followed from the discussion of Hecker Denschlag and Daley \cite{hecker} or Piil and Molmer \cite{molmer1}.

\begin{equation}\label{2pH}
H =   \sum_{m} t \left(a_{m+1}^\dagger a_{m} + a_m^\dagger a_{m+1} \right)+  \sum_{m} \frac{U}{2} a_m^\dagger  a_m^\dagger a_m a_m.
\end{equation}

In absence of interaction, this Hamiltonian moves a single particle from position state $\vert x_m\rangle$ to $\vert x_{m\pm1}\rangle$ and the wavefunction and energy can be obtained from the following single particle Schrodinger equation \ref{sse}
\begin{equation}\label{sse}
t\left[\psi(x_{m-1} + \psi(x_{m+1})\right] = E^{(1)} \psi (x_m),
\end{equation}
where $\psi (x_m)$ is coefficient for position state $\vert x_m\rangle$ in the full wavefunction. Taking a plane wave solution $\psi_{q} (x_m) = \exp(\imath q m)$, provides the energy

\begin{equation}
E_q^{(1)} = 2t \cos{q}.
\end{equation}

In presence of interaction, the two particle Schrodinger equation takes the following form

\begin{eqnarray}\label{2pse}
t\left[\psi(x_{m-1}, y_{m'}) + \psi(x_{m+1}, y_{m'}) +\psi(x_{m}, y_{m'+1}) + \psi(x_{m}, y_{m'-1})\right] \nonumber \\
 + U\psi(x_{m}, y_{m'})\delta_{m,m'} = E^{(2)} \psi(x_{m}, y_{m'}),
\end{eqnarray}
where $x_m$ and $y_{m'}$ are the coordinates of two particles at sites $m$ and $m'$ respectively. This equation can be simplified in terms of centre of mass $R = \frac{1}{2} (x + y)$ and relative $r = (x - y)$ coordinates. The wavefunction in momentum basis then become 

\begin{equation}
\psi(x, y) = e^{\imath K R} \psi_{K} (r),
\end{equation}
and Eq. \ref{2pse} simplifies to
\begin{eqnarray}\label{2psek}
t_K \left[\psi_K(m-1) + \psi_K(m+1)\right] + U\psi_K(m)\delta_{r,0} = E_K^{(2)} \psi(m)
\end{eqnarray}
with $t_K = 2t \cos(\frac{K}{2})$ yields 

\begin{equation}\label{2pek}
 E_{K,k}^{(2), U=0}  = 4t\cos(\frac{K}{2}) \cos(k)
\end{equation}
upon plane wave basis $\psi_{K,k} (m) = \exp(\pm\imath k m)$.

A solution to the interacting problem can be approached from substitution of $E_{K,k}^{(2),U=0}$ into Eq. \ref{2psek} with $\psi_{K,k}(0) = C$. Given the symmetry $\psi_{K,k}(r) = \psi_{K,k}(-r)$ for bosonic particles, this yields

\begin{equation}
\psi_{K,k}(r) = C \left[ \cos(kr) + \frac{U}{2t_K} \csc (k) \sin(k|r|) \right].
\end{equation}

For $|K| = \pi$ ($t_K = 0$), Eq. \ref{2psek} is simply $U = E^{(2)}_{\pi}$ and $\psi_{\pi} = \delta_{r0}$. For $K \in (-\pi, \pi)$, an ansatz $\psi_{K}(r) = C \alpha_K^{|r|}$ can be used. With this ansatz, Eq. \ref{2psek} provides

\begin{equation}\label{ansatz1}
2t_K \alpha_K + U = E^{(2)}_K,
\end{equation}
 
\begin{equation} \label{ansatz2}
t_K \frac{\alpha_K^{|m+1|} + \alpha_K^{(m-1)}}{\alpha_K^{|m|}} = E^{(2)}_K,
\end{equation}
Following which the solution for $\alpha_K$ is found

\begin{equation}
\alpha_K = \frac{U}{2t_K} \pm \sqrt{1+ \left(\frac{U}{2t_K}\right)^2},
\end{equation}
with the wavefunction (normalized) and energy taking the following form

\begin{equation}
 E^{(2)}_K = \sqrt{U^2 + 4t_K^2}, 
\end{equation}

\begin{equation}
\psi_K(m) = \frac{\sqrt{\vert \frac{U}{2t_K} \vert}}{\sqrt[4]{1+ \left(\frac{U}{2t_K}\right)^2}} \left(  \sqrt{1+ \left(\frac{U}{2t_K}\right)^2} - \vert \frac{U}{2t_K} \vert  \right).
\end{equation}

\section{Role of interaction and bound state}

In the presence of strong interaction of both attractive and repulsive type, two particles co-walk irrespective of their statistics. The presence of  bound states is responsible for such behaviour. It is best explained in the momentum space for ideal lattices. The real space hamiltonian can be written as

\begin{equation}
\mathcal{H} =  \mathcal{T} + \mathcal{V} =  \sum_{ mn} t a_n^\dagger a_{m} +  \sum_{mn} V_{mn} a_m^\dagger  a_n^\dagger a_n a_m.
\end{equation}
Both in the absence or presence of the interaction between  two particles,  the momentum dependent eigenenergies of the hamiltonian can be obtained by the Fourier transform. For the case of 1D lattices, one obtains the following expressions which can then be numerically diagonalized to find the eigenenergies:

\begin{equation}
\langle k_1', k_2'\vert \mathcal{H} \vert k_1, k_2\rangle =\frac{1}{N^2} \sum_{m',n',m,n} \langle m', n'\vert \mathcal{T} + \mathcal{V} \vert m, n\rangle 
e^{-\imath(-k_1'm' -k_2'n' + k_1m + k_2n)}.
\end{equation}
The hopping part can be simplified further

\begin{align}
\langle k_1', k_2'\vert \mathcal{T} \vert k_1, k_2\rangle &=\frac{1}{N^2} \sum_{m',n',m,n} \langle m', n'\vert t_{m^{\dprime}n^{\dprime}}q_{m^{\dprime}}^\dagger q_{n^{\dprime}}\vert m, n\rangle 
e^{-\imath(-k_1'm' -k_2'n' + k_1m + k_2n)} \nonumber \\
&= \frac{1}{N^2}  \sum_{m',n',m,n} \left[t_{mm'}\delta_{nn'} + t_{mn'}\delta_{nm'} +t_{m'n}\delta_{mn'} +t_{nn'}\delta_{mm'}   \right]
\nonumber \\
&\hspace{7cm} e^{-\imath(-k_1'm' -k_2'n' + k_1m + k_2n)} \nonumber \\
&= 2 \sum_{m-n} t_{m-n} \left[  e^{\imath k_1 (m-n)} + e^{\imath k_2 (m-n)} \right]\delta_{k_1k_1'}\delta_{k_2k_2'}.
\end{align}
Similarly,

\begin{align}
\langle k_1', k_2'\vert \mathcal{V} \vert k_1, k_2\rangle &=\frac{1}{N^2} \sum_{m',n',m,n} \langle m', n'\vert  \sum_{m"n"} V_{m"n"} q_{m"}^\dagger  q_{n"}^\dagger q_{n"} q_{m"} \vert m, n\rangle \nonumber \\
& \hspace{7cm}  e^{-\imath(-k_1'm' -k_2'n' + k_1m + k_2n)} \nonumber \\
&= \frac{1}{N} \sum_{m-n} V_{m-n} \left[ e^{\imath (k_1 -k_1')(m-n)} + e^{\imath (k_2 -k_2')(m-n)} \right] \delta_{k_1+k_2,k_1'+k_2'}
\end{align}

\begin{figure}[H]
\centering
\includegraphics[width=0.65\textwidth]{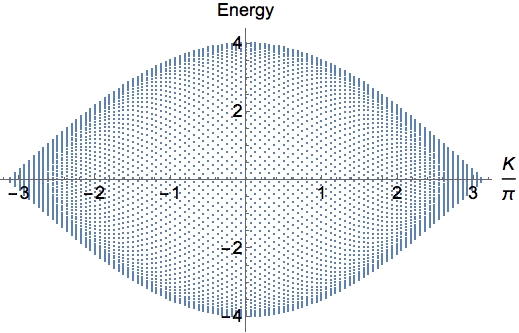}
\caption[Lattice spectrum for non-interacting tight-binding model]{Lattice spectrum for non-interacting tight-binding model. The spectrum is calculated for 50 lattice sites. Energy is calculated in the units of the hopping integral. Each dot for a fixed $K$ denote separate $(k_1,k_2)$ combinations with $K = k_1 + k_2$ where $-\pi < k_1, k_2 \le \pi$.}
\label{kspectrum-tb-0}
\end{figure}

These simplified equations can be used in diagonalization to obtain the  eigenenergies for the two particles in momentum basis or $K$-space. Figure \ref{kspectrum-tb-0} shows the eigenenergies in $K$-space for the non-interacting particles with only nearest neighbor hopping, which is the case in the tight binding model

\begin{equation}
\mathcal{H}  =  \sum_{\langle mn\rangle} t a_n^\dagger a_{m} +  \sum_{\langle mn\rangle} V a_m^\dagger  a_n^\dagger a_n a_m .
\end{equation}

For interacting particles, the bound states separate from the continuum beyond a critical interaction strength between the particles. The wavefunction and energy of the bound states have been derived analytically before   \cite{vektaris, fumika-thesis, ping-thesis, winkler, valiente1, valiente3, lieb-wu}. The energy of the bound states ($E^b$) for sufficiently strong interactions can be solved in any dimension. It takes the following form in 1D. For very strong interaction the dispersion of the bound states become flat. The two bound particles, for very strong interaction, can be represented as a single composite particle with modified hopping, which is of the order $t^2/V$

\begin{equation}
E^b(K) \simeq  V + \frac{4t^2\cos^2(K)}{V}.
\end{equation}

\begin{figure}[H]
\centering
\includegraphics[width=1.0\textwidth]{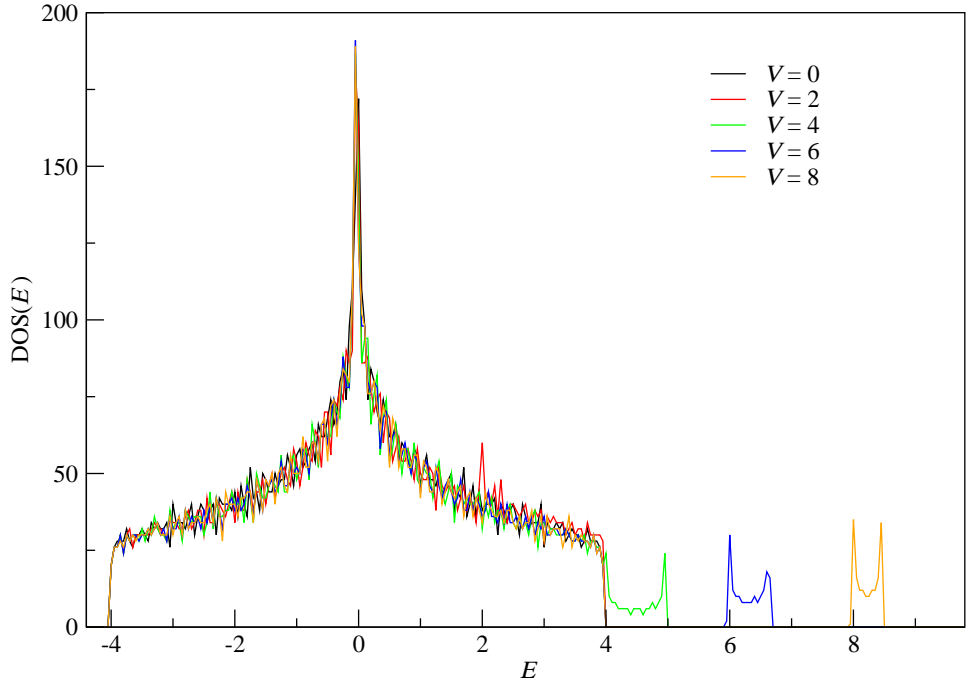}
\caption[Density of states in 1D tight-binding model]{Density of states in 1D tight-binding model. The exact diagonalization is done on a lattice of 86 sites. The bound state separates from the continuum beyond the critical interaction strength $V=4$.}
\label{dos-tb-1D}
\end{figure}

As shown in Fig. \ref{dos-tb-1D}, the bound states in 1D separate from the continuum at the interaction strength $V=4$. The states responsible for co-walking, in the non-interacting case, lie around middle of the continuum. With interaction strength increased, these  states move away from the centre within the continuum band. At the critical interaction strength $V=\pm4$, these states separate from the continuum as the bound states. This will be illustrated better in next chapter.  At very strong interactions, the energy of the bound state becomes that of the interaction strength.  The continuum states, however, remain very much unaffected by the interaction between the two particles.

\section{Recent experiments}
To elucidate this effect of binding on the dynamics and the correlations, several experiments have been performed on various lattice systems. Peruzzo et al \cite{peruzzo} studied quantum walk of two identical photons in an array of 21 coupled waveguides on a silicon oxynitride quantum photonic platform. In their setup, the photons were made distinguishable by a temporal delay larger than the coherence time when they arrived on the photonic lattice. When the photons arrived at the same time on the lattices side by side, a correlation of cowalking was measured. Lahini et al \cite{lahini1} performed similar experiments on waveguide lattices \cite{christodoulides} but  with photons arriving on two adjacent channels of waveguides in random phases. When averaged over many measurements, they also found cowalking photon correlations.   In waveguide lattices, there have also been experiments \cite{sebabrata} to understand the tunnelling properties of bound particles. Recent developments  \cite{hartmann, firstenberg, gorshkov}  have made it possible to control the interaction between two photons in such systems. 

In another study from Bloch et al \cite{fukuhara}, such two particle correlations  were measured between two spin excitations in a magnetic spin chain of $ ^{87}$Rb atoms. From a two dimensional quantum degenerate gas of $^{87}$Rb atoms, multiple one dimensional chains/tubes were first formed. Two spins in adjacent sites at the centre of these chains were then excited before freezing the dynamics by increasing the confining potential. Measurements were made with single site resolution \cite{sherson}, after removing excess atoms except from the desired state. These measurements accounted only for the tubes or lattices with two spin atoms left. Joint measurements of the two spins then revealed the bosonic cowalking character of the quantum dynamics of two correlated spins. In such studies the onsite interaction energy has been modified with remarkable control.

The most recent study on the two-particle quantum walk in cold-atom systems has been performed by Greiner et al \cite{preiss}.  In their study, the $^{87}$Rb atoms themselves are measured as quantum walkers. In the experiment, a similar prescription is followed. Preparation of 1D chains from a 2D degenerate gas by confining the gas in one direction with an optical lattice beam followed by narrow confining beams to retain only two atoms side by side when lattice depth was decreased to remove all other atoms. After the initial state preparation, the lattice depth was then again increased for the dynamics to take place under controlled parameters. A joint measurement is then performed with the single site resolution \cite{bakr}.

These studies have experimentally verified the effect of interaction on correlations of two-particle quantum walk. However, in all such studies, only nearest neighbor hopping was predominant. The case of long-range hopping has so far not been studied experimentally for two particle quantum walk. An interplay between the long-range hopping with the long-range interaction is now predicted by our study to make the dynamics different for different types of interactions.

\section{Effects of long-range hopping and interaction}

The dynamics of the interacting particles in the case of nearest neighbor hopping and interaction is independent of the sign of the interaction \cite{tirtha}. Both attractive and repulsive interactions have the same effect on the correlations and dynamical behaviour in the quantum walk. However, this dynamical symmetry with respect to the sign of the interaction is no longer the case when the particles can hop to sites at long-range beyond nearest neighbors in the lattice. For a single particle, the distribution of  eigenenergies on both sides of the zero energy line in $K$-space is symmetric (cosine) for the case of nearest neighbor hopping. For two particles, this symmetry remains in the absence of interaction as shown in Fig. \ref{kspectrum-tb-0}. For long-range hopping, this symmetry breaks even for a single particle. In addition to this asymmetry, the presence of interaction produces the bound state which can be controlled by tuning the strength of interaction. 

We simulate the dynamics of the correlations for the simplest long-range hopping case, which is taken as the isotropic power law decay of the hopping integral with respect to the distance between the sites. We find that not only the dynamics become different for the different signs of the interactions, but the nature of the correlations also becomes significantly different from that of the nearest neighbor models. 

For the simulations we consider two hardcore bosons intended to map Frenkel excitons (composite electron-hole pairs, see Appendix B), which do not change sign when exchanged, but cannot occupy same sites under the effect of following hamiltonian:

    \begin{equation}\label{H2p1d}
     \mathcal{H} =   \sum_{nm}   t _{nm}  a_n^\dagger  a_m + \sum_{nm}  V_{nm}  a_n^\dagger a_m^\dagger a_m a_n.
    \end{equation}  
The hardcore bosons follow mixed statistics
    \begin{equation}
       a_n a_m^\dagger =\delta_{nm} + \left( 1 - 2\delta_{nm}\right)  a_m^\dagger a_n,
    \end{equation}   
 where   $ a_n^{(\dagger)}$ anihiliates(creates) a particle at site $n$. Calculations are done for both the nearest-neighbor and long-range interaction and tunneling, which decay isotropically as an inverse power of distance. Both short and long range of the tunneling and interaction are considered:
    \begin{align}
           t_{nm} & =  \frac{t}{|n-m|^\alpha},  \hspace{2cm} ( \alpha = 1,  3)  \\
           V_{nm} & =  \frac{V}{|n-m|^\beta}.   \hspace{2cm}  (\beta = 1,  3)
    \end{align} 
We  define the interaction as attractive (\textit{t}/\textit{V}\textless0) or repulsive (\textit{t}/\textit{V}\textgreater0)  by the sign of for the ratio of the interaction and hopping.

 The initial state is indexed to one of such vectors, which is symmetrized
    \begin{equation}
     \Psi   ( 0) =   | n' m' \rangle \equiv \frac{1}{\sqrt{2}} \left( n_1' m_2' + n_2' m_1'  \right)
    \end{equation}   
 and the time evolution of this state is calculated using the eigenenergies $E_{\lambda}$ and eigenstates $|\lambda\rangle$ of the full hamiltonian,
     \begin{equation}
     \Psi   ( t) = \sum_{\lambda}  \exp(-\frac{\it i \normalfont E_{\lambda} t}{\hbar}) \vert \lambda \rangle \langle \lambda \vert \Psi(0)\rangle.
    \end{equation}

The wavefunctions and energies in can be analytically derived following Eq. \ref{2pek}, 

\begin{equation}
E_K^{(2), V=0}(k) = \sum_{d > 0} t_d \cos\left(\frac{Kd}{2}\right) \cos(kd)
\end{equation}
where $d = |n-m|$. However, the non-local character of the hopping may render mean field analysis inaccurate. The states can be analytically derived following Eqs. \ref{ansatz1} and \ref{ansatz2}, with $\alpha_K$ and $E_K^{(2)}$ as unknowns.

The pair correlations (or joint probabilities) are calculated directly from the coefficients of the two particle basis vectors
     \begin{equation}\label{pair_corr}
      C_{nm} = \langle a_n^\dagger a_m^\dagger a_m a_n \rangle.
     \end{equation}

For different combinations of  $\alpha$ and $\beta$, we observe a few features of the quantum walk for hardcore bosons. Two fermions would also have similar features but it was found  \cite{qin} that the correlations in momentum space would be different between two hardcore bosons and two fermions. In real space, both hardcore bosons and fermions have been observed to have the same correlations.

\begin{figure}[H]
\centering
\includegraphics[width=1.0\textwidth]{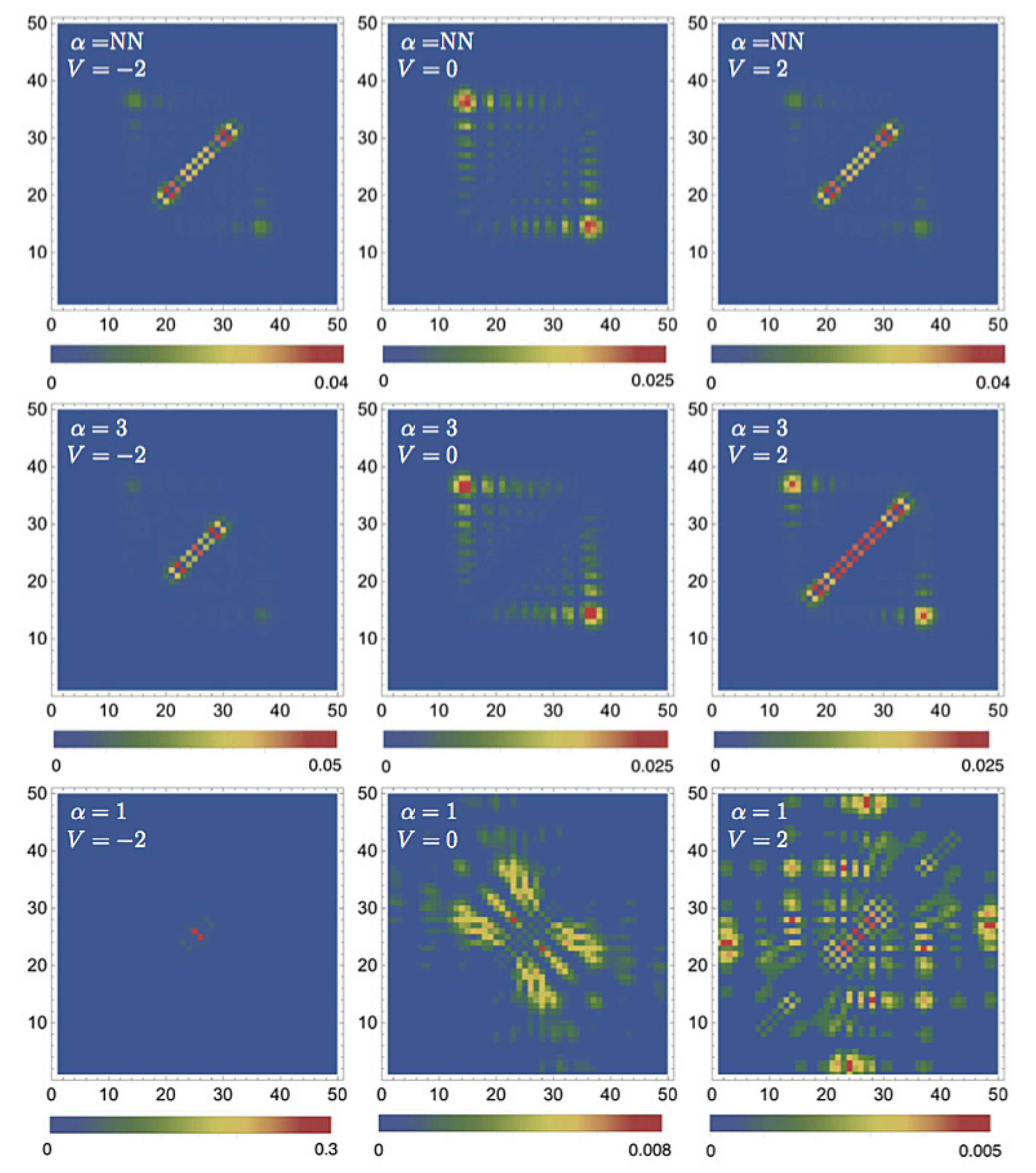}
\caption[Correlation dynamics of two hardcore bosons]{Correlation dynamics (pair correlations at time $2\pi/t$, Eq. \ref{pair_corr}) of two hardcore bosons with different range of hopping with long-range interaction ($\beta = 1$). For hopping limited to only nearest neigbors (NN), the dynamics is symmetric with respect to the sign of the interaction. For long-range hopping ($\alpha = 3$ and $\alpha = 1$), the particles spread faster for repulsive interactions. }
\label{long-range-interaction}
\end{figure}

\begin{figure}[H]
\centering
\includegraphics[width=1.0\textwidth]{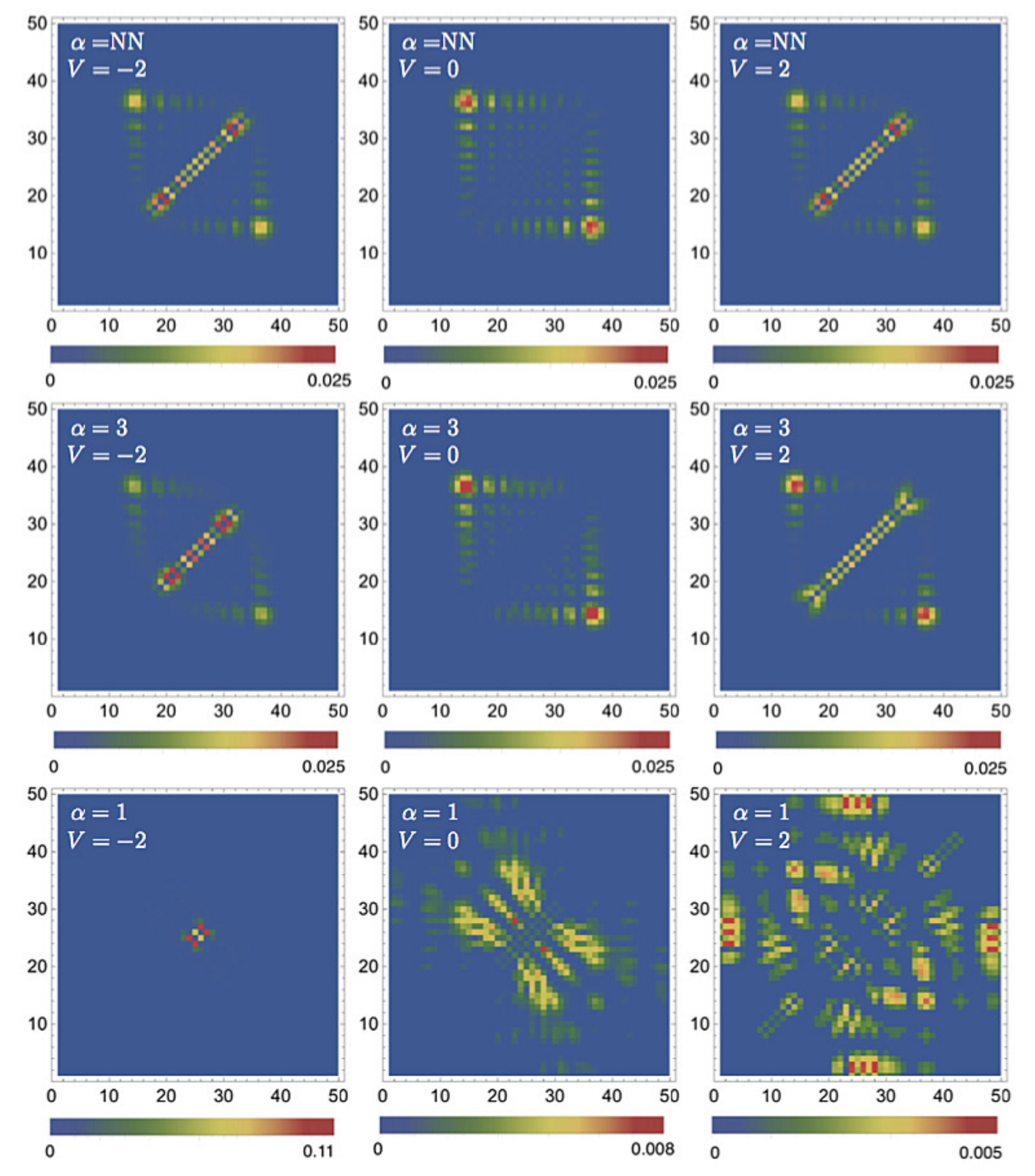}
\caption[Correlation dynamics of two hardcore bosons]{Correlation dynamics ((pair correlations at time $2\pi/t$, Eq. \ref{pair_corr}) ) of two hardcore bosons with different range of hopping with short-range interaction ($\beta = 3$). For hopping limited to only nearest neigbors (NN), the dynamics is symmetric with respect to the sign of the interaction. For long-range hopping ($\alpha = 3$ and $\alpha = 1$), the particles spread faster for repulsive interactions. }
\label{short-range-interaction}
\end{figure}

We find that, when the hopping is long-range, the dynamics for repulsive interactions are faster and for attractive interactions they are slower for the same magnitude of the interaction strength, as displayed in Figs. \ref{long-range-interaction} and \ref{short-range-interaction}. The dispersion of the continuum states below zero energy becomes flatter for the case of long-range hopping. These states contribute to make the dynamics slower for the attractive case. The dispersion of continuum states above zero energy becomes steep, which contributes to the faster dynamics in the repulsive case in the presence of long-range hopping. 

For short-range nearest neighbor hopping there is no asymmetry in dynamics with respect to the sign of the interaction. The expected anti-walking character is observed without any interaction. For sufficiently strong interactions the particles become bound and show cowalking character in the dynamics. However, in the case of the long-range hopping, the correlations are no longer only of cowalking or antiwalking types. The correlations develop a character in between that of cowalking and antiwalking, where one particle stays at the initial position, while the other particle extends to the boundaries.

\begin{figure}[H]
\centering
\includegraphics[width=0.65\textwidth]{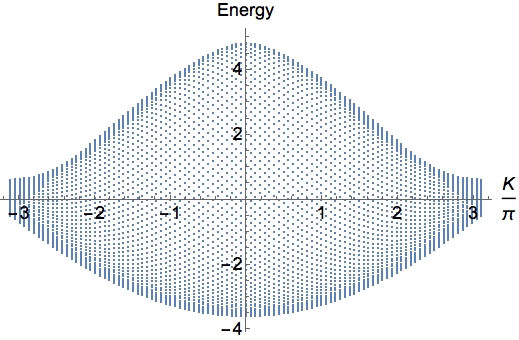}
\caption[Lattice spectrum for non-interacting coulombic hopping]{Lattice spectrum for non-interacting particles with Coulombic hopping. The spectrum is calculated for 50 lattice sites. Energy is calculated in the units of hopping integral. Each dot for a fixed $K$ denotes separate $(k_1,k_2)$ combinations with $K = k_1 + k_2$ where $-\pi < k_1, k_2 \le \pi$.}
\label{kspectrum-cb-0}
\end{figure}

\begin{figure}[H]
\centering
\includegraphics[width=0.65\textwidth]{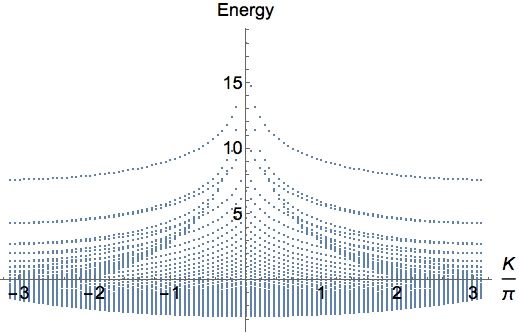}
\caption[Lattice spectrum for non-interacting dipolar hopping]{Lattice spectrum for non-interacting particles with dipolar hopping. The spectrum is calculated for 50 lattice sites. Energy is calculated in the units of hopping integral. Each dot for a fixed $K$ denotes separate $(k_1,k_2)$ combinations with $K = k_1 + k_2$ where $-\pi < k_1, k_2 \le \pi$.}
\label{kspectrum-dp-0}
\end{figure}

These effects can be understood when the lattice spectrum similar to the tight binding model is calculated. As can be seen from Figs. \ref{kspectrum-cb-0} and \ref{kspectrum-dp-0}, the dispersions become asymmetric in the case of the long-range hopping. In presence of interaction, one state moves out of the continuum states, which is termed as bound state.  This bound state separates from the continuum with lesser strength of interaction for the attractive case and requires higher strength of interaction to make it move out of the continuum in the repulsive case for the long-range hoping cases, as the dispersions become asymmetric. One can utilize this phenomenon by simply changing the sign of the interaction between the two particles to control their quantum walk on a lattice. How this can be done is explained in Appendix C.

When  Figs. \ref{long-range-interaction} and \ref{short-range-interaction} are compared, one can observe that the effect of long-range hopping is much more dominant than that of long-range interaction. In the limit of infinitely large lattices, the upper and lower bounds for the lattice dispersion can be calculated from the values of Riemann zeta functions and are (in the units of $t$) equal to $(+4.80, - 3.60), (+6.58, - 3.30), (+\infty, - 2.80),$ respectively, for $\alpha = 3, 2,$ and 1.

All calculations mentioned in  this chapter were performed by the method of full diagonalization which limits the size of the lattice that can be considered. In the next chapter we discuss how similar calculations can be performed for far larger lattice systems. This can be performed by exploiting the properties of the model hamiltonians. These hamiltonian matrices are generally sparse. This sparsity can be used to develop a method based on recursion (similar in essence to the famous Lanczos method) which will allow one to calculate desired properties from these matrices in an efficient and accurate manner. However, as we will see, one will then require to perform the same iterative calculations many times for each selection of energy within the full band of the dispersion.

\section{Phase transition}
A qualitative argument can be made on the effect of the asymmetry in spectrum with respect to the sign of the interaction in presence of long-range hopping on the superfluid to Mott insulator (MI) phase diagram. For a Hamiltonian in Eq. \ref{phase},
\begin{equation}\label{phase}
H = \sum_m \mu a_m^\dagger a_m + \sum_m t\left( a_m^\dagger a_{m+1} + a_{m+1}^\dagger a_m \right) + \sum_m U a_m^\dagger a_m^\dagger a_m a_m
\end{equation}
the diagram \cite{bloch} shows transition to Mott insulator state when the particles get bound. For repulsively interacting particles in presence of long-range hopping, the Mott insulator phase is expected to be smaller as transition to the bound state now requires higher energy and smaller $t/U$. The Mott insulator state can even be absent for the $\alpha = 1$ in 1D, when interaction is repulsive. For the attractive cases, the Mott insulator region is expected to grow larger, as binding becomes easier in presence of long-range hopping. 

\begin{figure}[h]
\centering
\hspace{0.0cm}
\begin{tikzpicture}[scale=2.3]
\draw [ ->]  (0,0) -- (5.3,0);
\draw [ ->]  (0,0) -- (0,3.3);
\draw (0,0) .. controls (4.0,0.3) and (4.0,0.7) .. (0,1);
\draw [dotted] (0,0) .. controls (3.5,0.3) and (3.5,0.7) .. (0,1);
\draw [dashed] (0,0) .. controls (4.5,0.3) and (4.5,0.7) .. (0,1);
\draw (0,1) .. controls (3.0,1.3) and (3.0,1.7) .. (0,2);
\draw [dotted] (0,1) .. controls (2.5,1.3) and (2.5,1.7) .. (0,2);
\draw [dashed] (0,1) .. controls (3.5,1.3) and (3.5,1.7) .. (0,2);
\draw (0,2) .. controls (2.0,2.3) and (2.0,2.7) .. (0,3);
\draw [dotted] (0,2) .. controls (1.5,2.3) and (1.5,2.7) .. (0,3);
\draw [dashed] (0,2) .. controls (2.5,2.3) and (2.5,2.7) .. (0,3);
\node at (-0.1,0) {\large $0$};
\node at (-0.1,1) {\large $1$};
\node at (-0.1,2) {\large $2$};
\node at (-0.1,3) {\large $3$};
\node at (2.55,-0.2) {\large $t/U$};
\node at (-0.25,1.55) {\large $\mu/U$};
\node at (0.5,0.5) {\large MI};
\node at (0.5,1.5) {\large MI};
\node at (0.5,2.5) {\large MI};
\node at (3.5,1.5) {\large Superfluid};
\end{tikzpicture}
\vspace{0.5cm}
\caption[Superfluid - Mott insulator phase transition]{Qualitative phase diagram for transition between Mott insulator and superfluid state when hopping is long range and interaction is repulsive (dotted lines) or attractive (dashed lines).}
\label{Hoflat}
\end{figure}
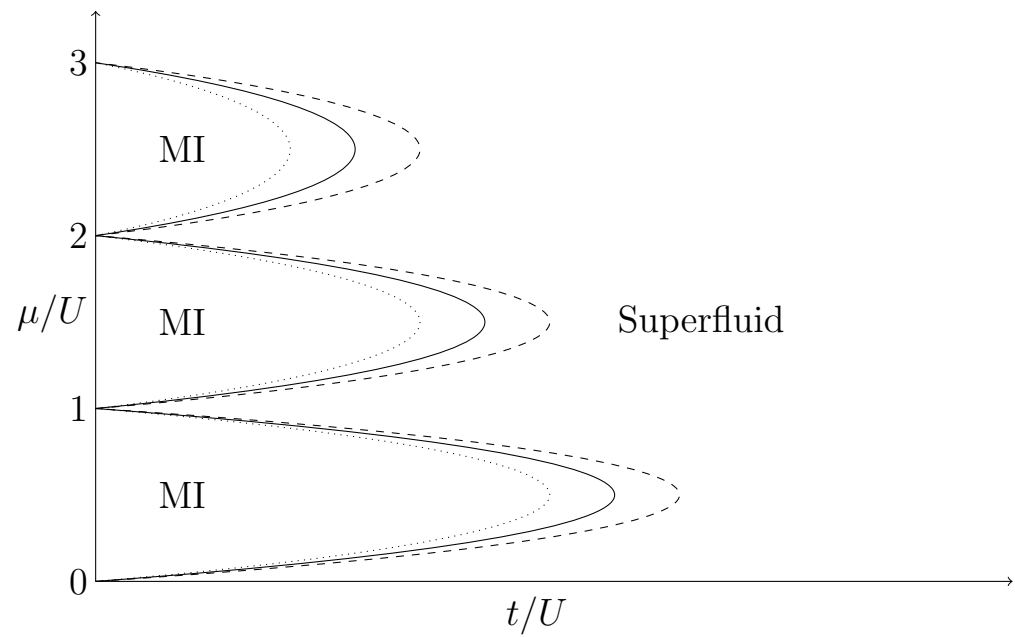

\section{Conclusion}

In this chapter, fundamental physics of the most simple model systems  has been found to be very rich in structure. Such simple model systems consisting of only two particles show different types of correlations in dynamics in the presence of interaction when their tunnelling in lattices is long-range. However, method of full diagonalization, which was applied to obtain results of this chapter,  limits the size of the lattices that can be considered. A recursive algorithm to compute Green's functions can be used as described in next Chapter to obtain values related to properties of interest of larger lattice systems.

\newpage
\chapter{Green's Functions of Interacting Particles}

Solving coupled differential equations for each lattice site in quantum random walks becomes extremely difficult for a large system size. Full diagonalization has so far been employed in most studies to understand the properties of random walk with both  short-range and long-range hopping cases.  However there exists a method of recursion which can be used to calculate Green's functions of interacting particles in fairly large lattice systems \cite{ping-thesis, berciu1, berciu2, berciu3}. Both the one-particle and two-particle Green's functions can be calculated exactly by this method for any ordered or disordered systems, as will be described in this chapter. In the case of the tight binding model, the  hamiltonian is readily solved by a continued fraction method. This continued fraction method was applied  by Haydock et al \cite{haydock} and Morita \cite{morita} who developed such algorithms for calculations of the density of states \cite{economou, licciardello} in ideal 3D lattices of various kinds (fcc, bcc, sc) for non-interacting particles.  In the case of the disordered 1D systems, Thouless et al. \cite{thouless} computed Green's functions iteratively to find the effect of different onsite energy distributions on conductivity. In our case, we adapt the recursive formulation to real space for finite lattices, both ordered and disordered, of both one and two dimensions.

 The two-particle correlations are known to play an important role in the properties of many lattice systems  \cite{sawatzky, gutzwiller, cini, kanamori, nolting}. One must account for the two-particle correlations to understand such systems. Recently, an efficient formulation in momentum space for ideal lattices  was developed to calculate few-particle Green's functions, which also elucidated  the effect of the interaction on few-particle bound complexes \cite{berciu1, berciu2, berciu3}.

A method, where such few-particle Green's functions can be efficiently calculated in disordered systems, was under development in our group \cite{ping-thesis}. In this thesis, it is extended to 2D systems and shown to be exactly mappable to some arbitrary graphs (e.g. binary trees). We here illustrate  the method  with the use of recursive Green's functions. We limit ourselves to discussing the method for the two-particle Green's functions in lattices and trees with nearest neighbor hopping. However, the method can be easily generalized to the cases of longer-range hopping and a larger number of particles.  

For fairly large lattices, this recursive method is very useful. For calculations of properties related to two-particle correlations  in 1D lattices, one can go beyond one thousand lattice sites thus eliminating finite size effects. For two particles in 2D lattices, around two thousand lattice sites can be considered.  Using this recursion, we  perform calculations for two particles in binary trees consisting of up to 9 generations. There is also a possibility to improve upon this and make the calculations even more efficient. 

The calculation of the density of states of various systems from the real space Green's functions and the spectral profile of the two-particle bound state is also efficient irrespective of the strength of the interaction between the particles. The dynamics of the interacting particles and their correlations are also shown to be  calculated efficiently once the important Green's elements are found. However, to do calculations for large 2D lattices, approximations have to be applied, as the basis size becomes very large even for systems with as few as twenty sites per dimension. We introduce such approximation and their usefulness in the later part of this chapter, which is mostly relevant to disordered systems. We also perform some preliminary calculations for the two-particle Green's functions in 2D lattices with complex hopping parameters, intended to simulate the effects of gauge fields.

The algorithm is explained in the next section. Later sections will present a few of the calculations  of dynamics and properties such as the density of states and the spectral weight of the two interacting particles in 1D and 2D lattices and in binary trees. 

\newpage
\section{Method of recursion}

We start with the most simple and extensively studied case of two particles in a one dimensional lattice. The lattice can be perfect or disordered. Each case can be simulated very efficiently using the recursive Green's function method in real space. 

The Green's function for some hamiltonian $H$ is defined as following:

     \begin{equation}\label{idtt}
              G(\omega) = \frac{1}{\omega - H} 
       \end{equation} 
    where $\omega = E + \imath\eta$ is a complex number with $\eta$ a very small positive real number and  $  G(m,n,\omega) = \langle m n | G(\omega) | m' n' \rangle $  is a time-independent propagator from two particles occupying sites $m'$, $n'$ to sites $m$, $n$ in the 1D lattice. We omit the indices $m'$, $n'$  wherever unnecessary for brevity from now on. 

 For a hamiltonian of the form of Eq. \ref{ham}, where $\epsilon_{m}$ is the onsite energy,  $t_{mn}$ is the hopping element moving the particle from site $m$ to site $n$ and $V_{mn}$ is the interaction between particles at sites $m$ and $n$,
  \begin{equation}\label{ham}
H = \sum_{m} \epsilon_{m} a_{m}^\dagger a_m  + \sum_{\langle mn\rangle} t_{mn} a_{m}^\dagger a_n + \sum_{\langle mn\rangle} V_{mn} a_{m}^\dagger a_{n}^\dagger a_n a_m,
 \end{equation} 
 the following type of recurrence relations will emerge. Here, the vectors $\langle m n |$ from left and $| m' n'\rangle$ from right are applied to the identity $(\omega - H) G(\omega) = 1$ from Eq. \ref{idtt} to find the relations for functions like $G(m, n, \omega)$ sorted on the left hand side of Eq. \ref{grn} and their related Green's functions on the right hand side.

\begin{equation*}
\begin{aligned}
  .   .   &=    .   . \\
(\omega - \epsilon_{m-1} -\epsilon_{n+1} - V_{m-1 n+1})G(m-1,n+1,\omega) &=   \delta_{m-1,m'}\delta_{n+1,n'} + \delta_{m-1,n'}\delta_{n+1,m'}   \\
& ~~~ -  t_{m-2, m-1}  G(m-2, n+1, \omega) \\
& ~~~ -  t_{m, m-1}  G(m, n+1, \omega) \\
& ~~~ - t_{n, n+1}  G(m-1, n, \omega) \\
& ~~~ - t_{n+2, n}  G(m-1, m+2, \omega)  \\
\end{aligned}
\end{equation*}
\begin{equation}\label{grn}
\begin{aligned}
(\omega - \epsilon_m -\epsilon_n - V_{mn})G(m,n,\omega) &=   \delta_{m,m'}\delta_{n,n'} + \delta_{m,n'}\delta_{n,m'} \\
 & ~~~ -  t_{m-1, m}  G(m-1, n, \omega) \\
& ~~~ - t_{m+1, m}  G(m+1, n, \omega)\\
& ~~~ - t_{n-1, n}  G(m, n-1, \omega) \\
& ~~~ - t_{n+1, n}  G(m, n+1, \omega)  \\
(\omega - \epsilon_{m+1} -\epsilon_{n-1} - V_{m+1 n-1})G(m+1,n-1,\omega) &=   \delta_{m+1,m'}\delta_{n-1,n'} + \delta_{m+1,n'}\delta_{n-1,m'}  \\
 & ~~~ -  t_{m, m+1}  G(m, n-1, \omega) \\
& ~~~ -  t_{m+2, m+1}  G(m+2, n-1, \omega)\\
 & ~~~ - t_{n-2, n-1}  G(m+1, n-2, \omega)\\
& ~~~ - t_{n, n-1}  G(m+1, n, \omega)  \\
 .   .   &=   .   . 
\end{aligned}
\end{equation}
Here, only nearest neighbor hopping and interaction is considered. Once all $G(m,n,\omega)$ are found, the dynamics can be easily computed by the Fourier transformation of the Green's function amplitudes from the energy domain to the time domain
   \begin{equation}\label{dyn}
     G(m,n,t)  = \sum_{\omega} e^{-\imath \omega t}  G(m, n, E + \imath\eta).
    \end{equation}

The spectral weights of eigenstates for any initial state or wave packet of the two particles at sites $m'$ and $n'$, can be computed from a single Green's element

   \begin{equation}\label{spec}
     \text{A} (m', n', E) = \frac{-1}{\pi}  \text{Im}[G(m', n', E + \imath\eta)].
    \end{equation}

The density of states (DOS) of the lattice systems up to a scaling factor can also be computed from all such single Green's elements

   \begin{equation}\label{dos}
     \text{DOS} (E)  = \sum_{m', n'} \text{A} (m', n', E).
    \end{equation}
If there is  translational symmetry present in the system, then only a few initial states with increasing relative distance $(|m' - n'|)$ might prove sufficient for convergence. Transport properties calculated from Green's elements such as $G(m'\pm1, n', \omega)$ or $G(m'\pm1, n'\pm1, \omega))$ might also be of key interest.

Now, the recursive functions are formulated in the form of Eq. \ref{grn} consisting of vectors in a chain. One needs to first find some good quantum numbers and group Green's elements according to such numbers. We find $R$=$m+n$ for the Green's functions of the form $G(m, n, \omega)$ in real space is such a number, as the hamiltonian does not connect functions with same  $R$ directly, as can be checked from Eq. \ref{grn}. We sort all such functions in a single vector $\mathcal{G}_{R}$, as in Eq. \ref{gv}. One can also notice that $\mathcal{G}_{R}$ is only connected to $\mathcal{G}_{R-1}$ and $\mathcal{G}_{R+1}$ by the hamiltonian, as in the Eq. \ref{vrec}

 \begin{equation}\label{gv}
    \mathcal{G}_{R =  m+n}(\omega) = \begin{pmatrix}  . \\ . \\ G(m-1, n+1, \omega) \\ G(m, n, \omega) \\ G(m+1, n-1, \omega) \\ . \\ .  \end{pmatrix}
    \end{equation}

   \begin{equation}\label{vrec}
     \mathcal{G}_{R} = \alpha_R \mathcal{G}_{R-1} + \beta_R \mathcal{G}_{R+1} + \bf C
    \end{equation}
where $\bf C = 0$ ( or $\neq \bf 0$) when $R\neq$ $m' + n'$ (or $= m' + n' = R'$).

These vectors form a one dimensional chain in terms of their connection to only the nearest neighbor vectors and each of their elements can be solved exactly by the following prescription. This particular form also appears in many other areas of quantum physics and therefore a similar method in principle can be constructed.

On the left and right boundary of the chain, the following equations hold for systems with open boundary condition

   \begin{equation}\label{vlr}
     \mathcal{G}_{0} =  \beta_0 \mathcal{G}_{1}   \text{     and     } \mathcal{G}_{L} = \alpha_L \mathcal{G}_{L-1},
    \end{equation}
where $0$ and $L$ are the minimum and maximum index possible for $R$.

If we can simplify Eq. \ref{vrec} as in Eq. \ref{vlr}, then all the calculations will become a recursion of vectors:

   \begin{equation}\label{vr}
    \mathcal{G}_{R} = \mathcal{A}_R \mathcal{G}_{R-1}   \text{     and     }  \mathcal{G}_{R} = \mathcal{B}_R \mathcal{G}_{R+1},   \text{             if  } R \neq R'.
    \end{equation}
We find $\mathcal{A}_0 = \beta_0$ and  $\mathcal{B}_L = \alpha_L$. These are our open boundary conditions. Now, substituting Eq. \ref{vr} to  Eq. \ref{vrec}  for $R < R'$ and $R > R'$, we find the following equations respectively:

   \begin{equation}\label{rl}
\begin{aligned}
 \mathcal{G}_{R} &= \alpha_R \mathcal{G}_{R-1} + \beta_R \mathcal{G}_{R+1}  \\
     \mathcal{B}_{R}\mathcal{G}_{R+1} &= \alpha_R \mathcal{B}_{R-1}\mathcal{G}_{R} + \beta_R \mathcal{G}_{R+1} \\
     \mathcal{B}_{R}\mathcal{G}_{R+1} &= \alpha_R \mathcal{B}_{R-1} \mathcal{B}_{R}\mathcal{G}_{R+1} + \beta_R \mathcal{G}_{R+1} \\
  [1 -  \alpha_R \mathcal{B}_{R-1}]  \mathcal{B}_{R} &=  \beta_R \\
 \mathcal{B}_{R} &=  [1 -  \alpha_R \mathcal{B}_{R-1}]^{-1} \beta_R 
\end{aligned}
    \end{equation}

\vspace{0.4cm}

   \begin{equation}\label{rr}
\begin{aligned}
  \mathcal{G}_{R} &=  \alpha_R \mathcal{G}_{R-1} + \beta_R \mathcal{G}_{R+1}  \\
     \mathcal{A}_{R}\mathcal{G}_{R-1} &=  \alpha_R \mathcal{G}_{R-1} + \beta_R \mathcal{A}_{R+1} \mathcal{G}_{R} \\
       \mathcal{A}_{R}\mathcal{G}_{R-1} &= \alpha_R \mathcal{G}_{R-1} + \beta_R \mathcal{A}_{R+1} \mathcal{A}_{R} \mathcal{G}_{R-1} \\
  [1 -  \beta_R \mathcal{A}_{R+1}]  \mathcal{A}_{R} &=  \alpha_R \\
  \mathcal{A}_{R} &=  [1 -  \beta_R \mathcal{A}_{R+1}]^{-1} \alpha_R 
\end{aligned}
    \end{equation}

One can compute these $\mathcal{A}_{R}$ and $\mathcal{B}_{R}$ matrices recursively starting from Eq. \ref{vlr} before one reaches $R=R'$ from both sides of the chain. At $R= R'$, applying $\mathcal{A}_{R'+1}$ and $\mathcal{B}_{R'-1}$ to Eq. \ref{vrec}, one finds the following:

   \begin{equation}\label{vr'}
   \begin{aligned}
\mathcal{G}_{R'} &=\alpha_{R'} \mathcal{G}_{R'-1} + \beta_{R'} \mathcal{G}_{R'+1} + \bf C, \\
 \mathcal{G}_{R'} &=\alpha_{R'} \mathcal{B}_{R'-1} \mathcal{G}_{R'} + \beta_{R'} \mathcal{A}_{R'+1}\mathcal{G}_{R'} + \bf C, \\
    \mathcal{G}_{R'} &=[1 - \alpha_{R'}\mathcal{B}_{R'-1} -\beta_{R'}\mathcal{A}_{R'+1}]^{-1} \bf C.
   \end{aligned}
    \end{equation}

Once $\mathcal{G}_{R'}$ is found, all other  $\mathcal{G}_{R}$ can be found by Eq. \ref{vr}, hence solving the problem of finding all the Green's elements for a given $m'$ and $n'$ for a single $\omega$ without diagonalization. To find all the eigenenergies, one has to scan over a range of $\omega$ that can be used in Eq. \ref{dyn} to compute dynamics. The calculations for each $\omega$ are distinct from each other and can be parallelized. The value of $\eta$ has to be chosen arbitrarily. This choice can be benchmarked by comparing a few sample calculations with full diagonalization.

\section{Two interacting particles in 1D}
\begin{figure}[H]
\centering
\includegraphics[width = 0.99\textwidth]{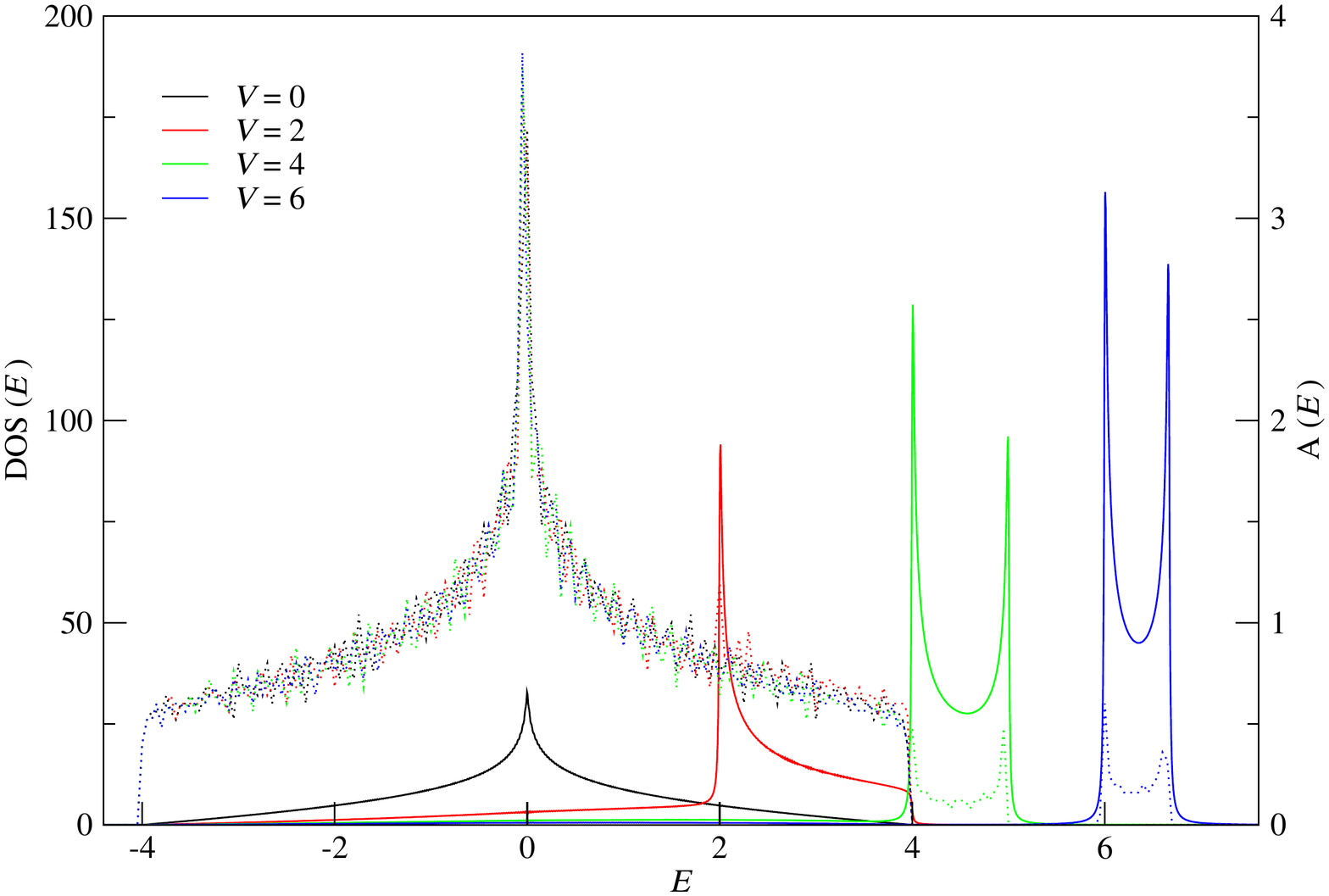}
\llap{\shortstack{%
        \includegraphics[scale=.15]{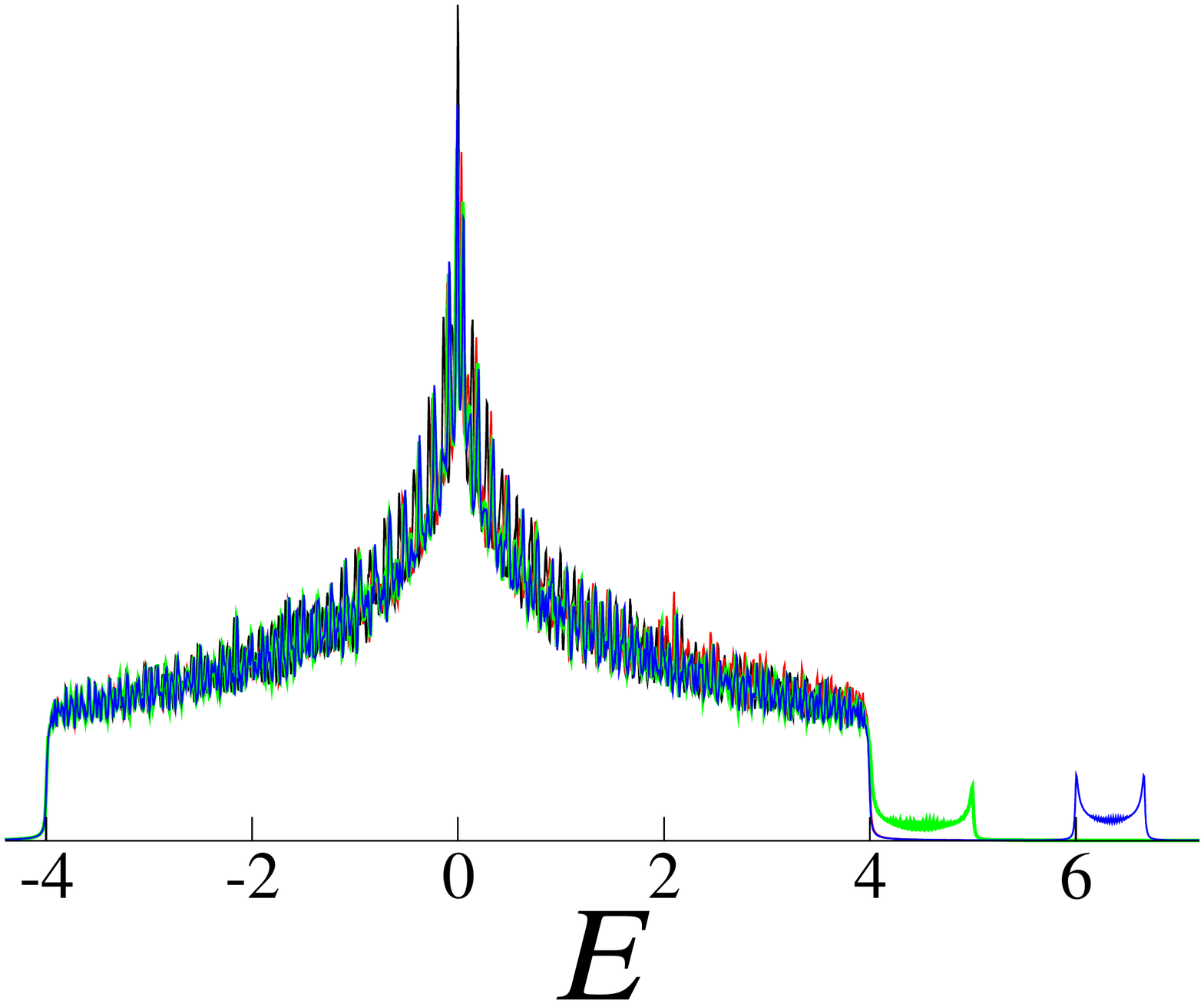}\\
        \rule{0ex}{2.54in}
      }
  \rule{1.85in}{0ex}}  
\caption[Spectral weight of bound state in 1D]{Density of states (DOS) calculated for a 1D ideal lattice of two interacting partcles from full diagonalization are plotted as dotted lines. The spectral weights for two particles initially prepared at one lattice site apart calculated using Eq. \ref{spec} is shown in solid lines. The inset shows the calculation of density of states using Eq. \ref{dos}.}
\label{1D-dos-sw}
\end{figure}

As has been remarked in the previous section, this recursive algorithm enables us to do calculations for much larger system sizes than what can be performed using full diagonalization procedures. These calculations are also very efficient. We can calculate dynamics in ideal or disordered lattices. We can calculate the density of states and spectral weight for any initial wave packet preparation.  In Fig. \ref{1D-dos-sw} we calculate $G(m', n', E)$ for $|m' - n'|=1$ for a lattice size of 500 sites. The particles were taken as hard-core bosons and the hamiltonian as in Eq. \ref {ham1D}

\begin{equation}\label{ham1D}
H =  \sum_{\langle mn\rangle} t a_{m}^\dagger a_n +  \sum_{\langle mn\rangle} V a_{m}^\dagger a_{n}^\dagger a_n a_m \end{equation}

These calculations reveal the weight of the bound state even inside the continuum as the interaction strength  is increased. We find the spectral profile for two particles initially located at nearest neighbor sites in a 1D lattice. Figure \ref{1D-dos-sw} shows the changes to the profile when the interaction strength is increased. At zero interaction strength, the profile is symmetric and has a sharp peak at $E=0$ with linear drop to the end of the band edge. For $V=2$, the profile has a sharp rise at $E=2$ and a sharp drop at the band edge. For stronger interactions, the profile matches with the distribution of bound states. These spectral weights also appear in the context of many-particle systems with different filling fractions \cite{rausch}.

\begin{figure}[H]
\centering
\includegraphics[width = 0.32\textwidth]{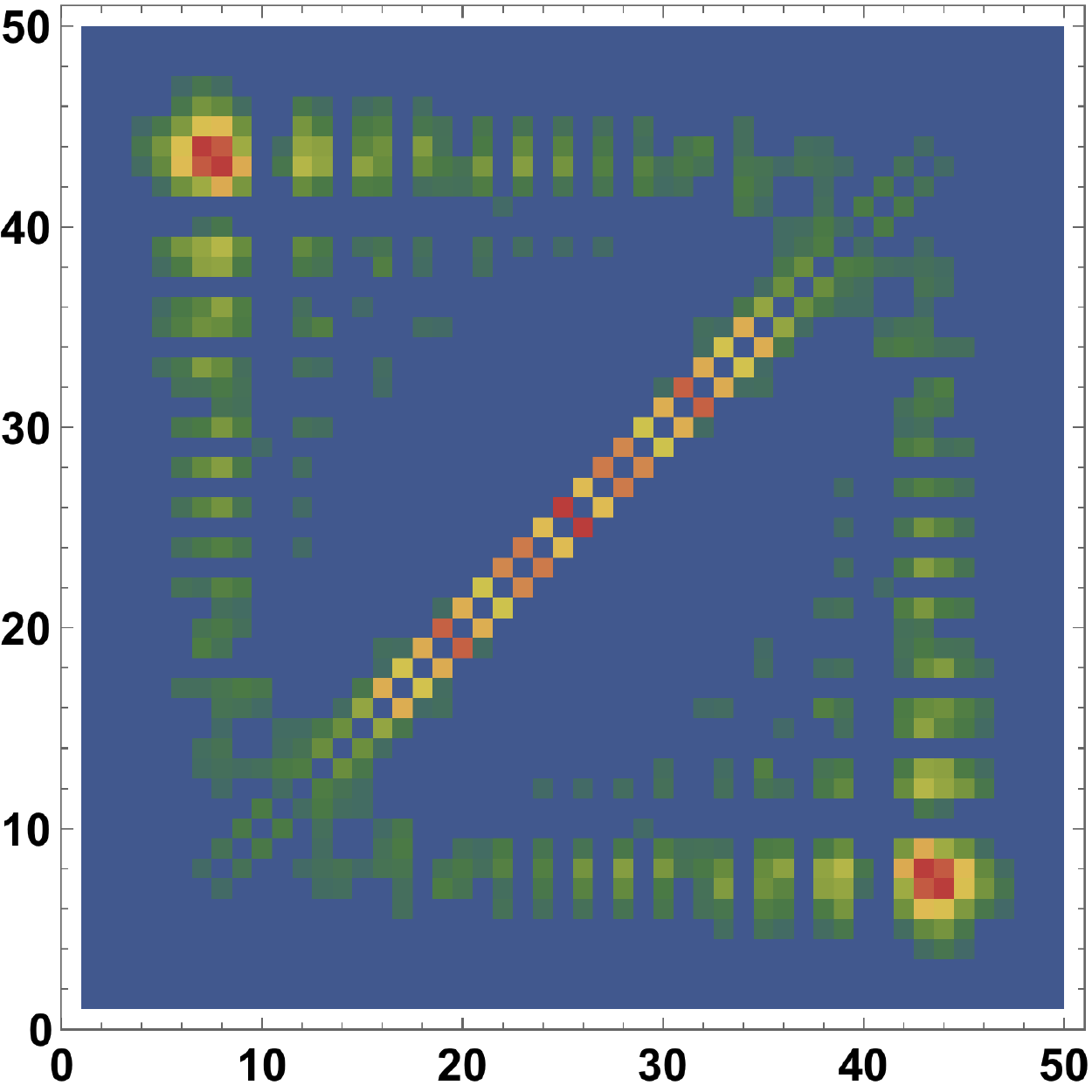}
\includegraphics[width = 0.32\textwidth]{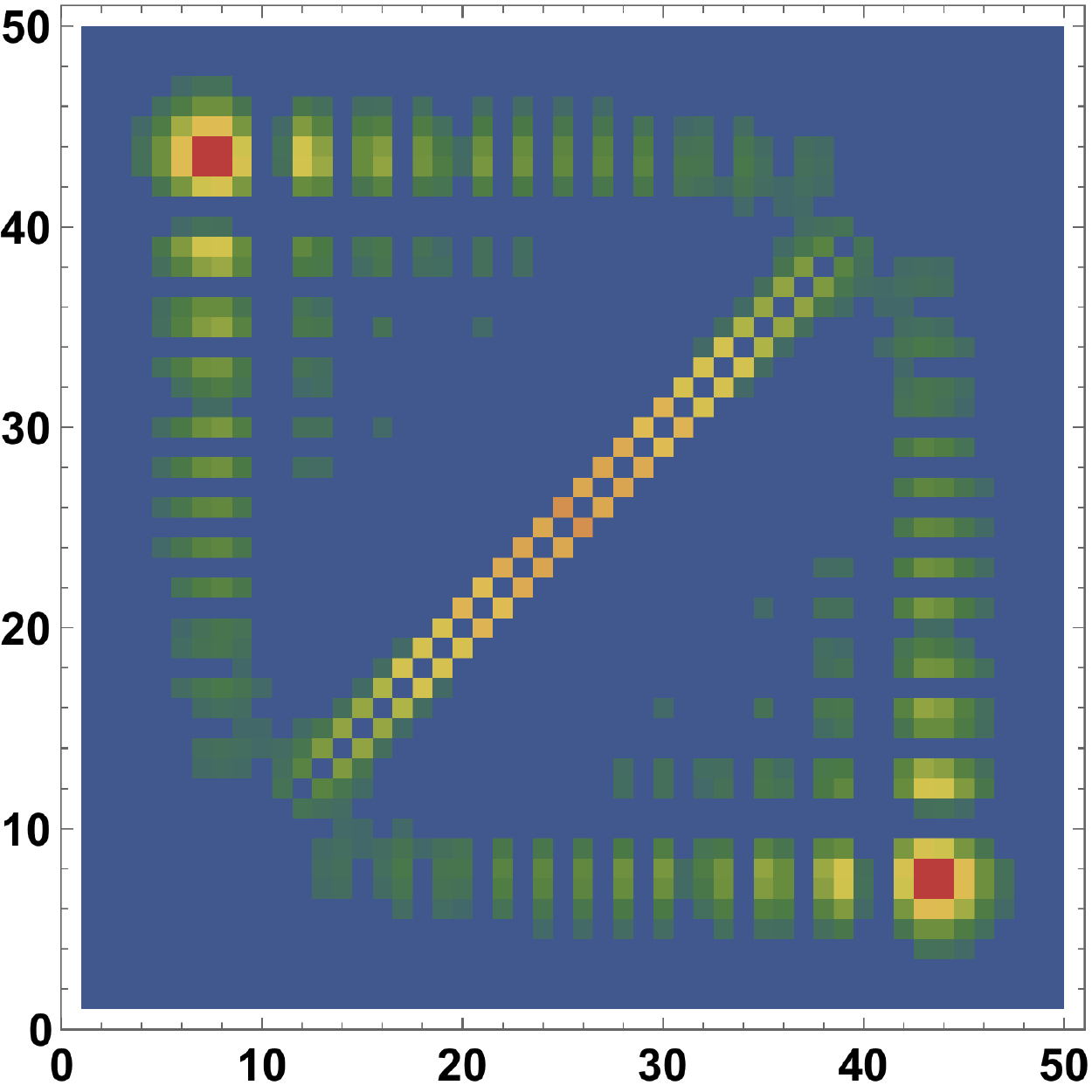}
\includegraphics[width = 0.32\textwidth]{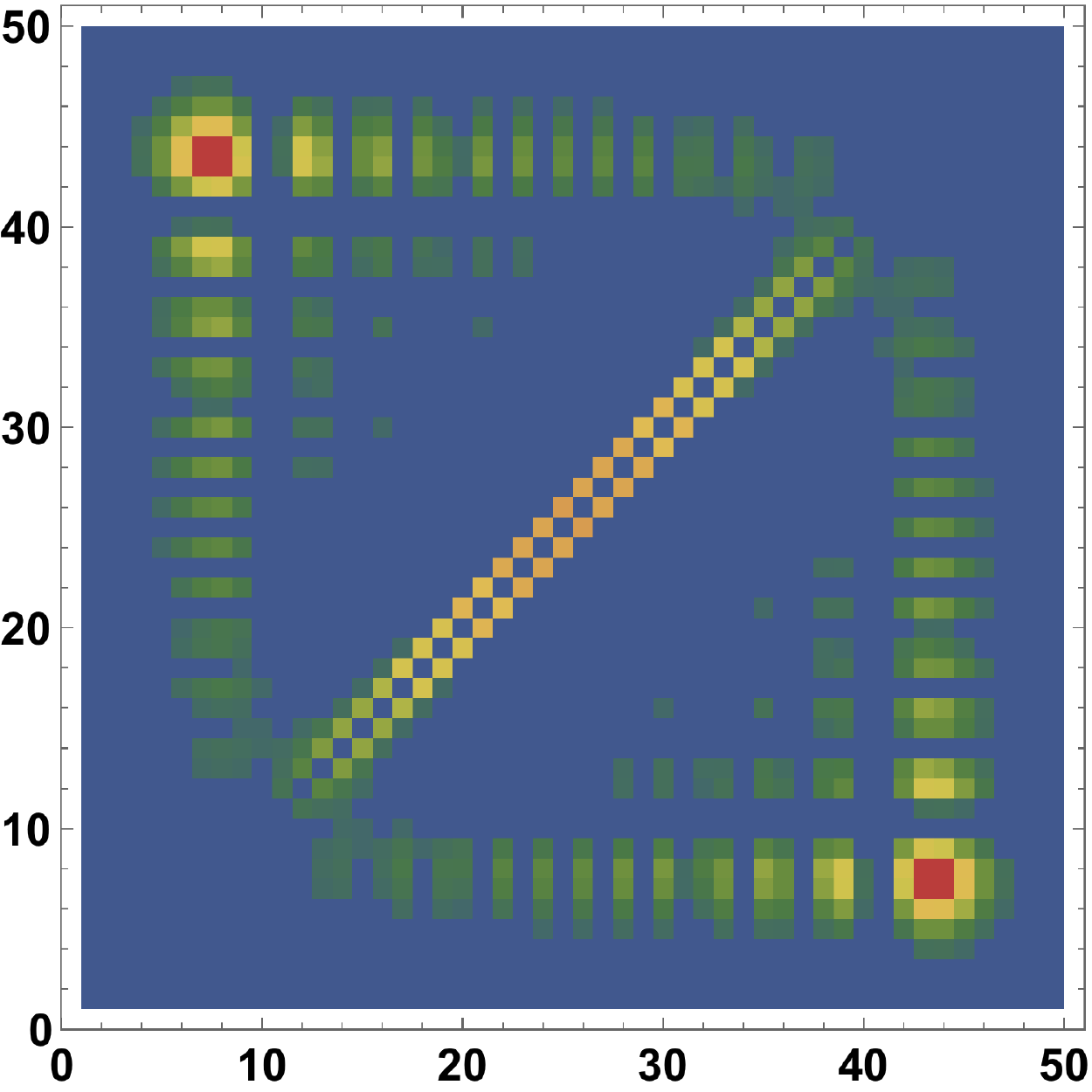}
\caption[Correlation dynamics calculated from Green's functions]{Comparison between recursive calculation using only 10(left) and 50(middle) search points within unit energy bandwidth with that of full diagonalization(right) for correlation dynamics of two particles in a 1D ideal lattice for $V=1$ and at $time=10$ (in the unit of hopping).}
\label{1Dyn}
\end{figure}

The correlation dynamics  also shows excellent agreement with that obtained from full diagonalization as shown in Fig. \ref{1Dyn}. The effectiveness of the method can be realized even by searching less than 500 points (for each $\omega$ needed in Eq. \ref{dyn}) within the full the energy band ($-4 \le E \le +4$). As shown in Fig. \ref{1Dyn}, searching less than 50 points per unit of the  energy width turns out to start producing errors in the range of a unit percentage in such calculations.  We prefer the approach of first finding the most important bandwidth from the calculated spectral weight which minimizes the number of search points and enhances the efficiency of the computation in such calculations.

\section{Two interacting particles in 2D}

In the two dimensional systems, the difficulty of doing full diagonalization for two particles grows approximately as $N^{12}$, where $N$ is the number of sites per dimension. As in the case of 2D lattices, one needs to consider a large number of sites to avoid significant finite size effects in the calculations of transport and localization properties, these numbers soon become not viable for doing any reasonable calculations. The recursive calculations, as described earlier, break the total calculation into multiple parts, which can be solved recursively as described before. 

\begin{figure}[H]
\centering
\includegraphics[width = 0.85\textwidth]{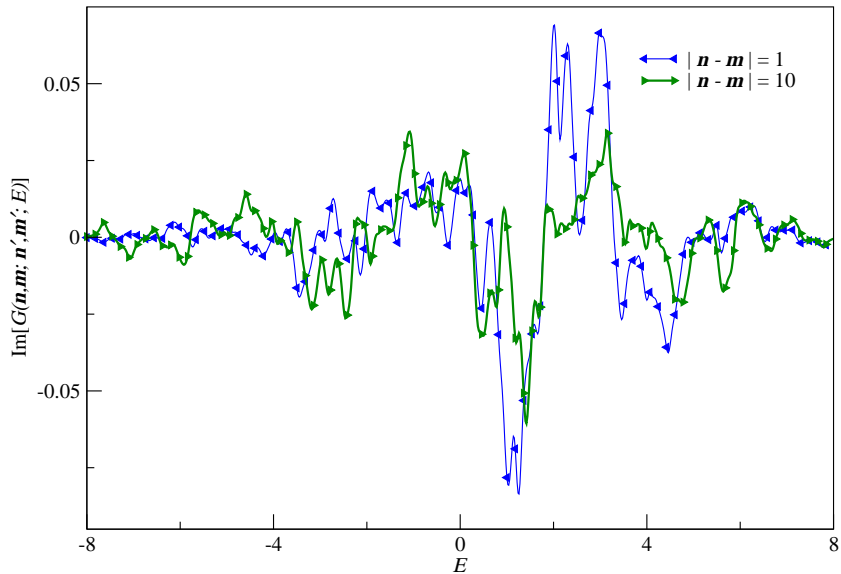}
\caption[Recursive calculations of 2D Green's functions]{Comparison between recursive calculation using only with that of full diagonalization for two particles in a 2D disordered lattice for $W =1, V=1$ (in the unit of hopping) as in Eq. \ref{ham} with $\epsilon$ chosen randomly from box distribution $\left[-\frac{W}{2}, \frac{W}{2}\right]$. $ r = |\textbf{\em{n}} - \textbf{\em{m}}|$ is the distance between two particles in number of minimum steps in a lattice. Full symbols are for full diagonalization, smaller filled symbols inside are for recursive calculations.}
\label{2D-greens}
\end{figure}

The recursive calculations are exact when the recursion includes the boundaries. The recursions  can also be done locally limited to only a small part of the lattice.  Figure \ref{2D-greens} shows the imaginary part of a few randomly selected Green's functions for a fixed initial state and final occupations at different  distance (minimum number of steps between two particles) in a 2D lattice. As it shows, the recursive calculations match exactly with that of full diagonalization irrespective of the distance between the two particles. However, there can be some numerical errors depending on the implementation of the algorithm.

For a 2D system as large as consisting of more than two thousand sites, this recursive method also becomes very difficult to implement. In the case of disordered cases, where one needs to average over many realizations of disorder to account for any robust results, implementation of the full recursive method becomes challenging. 
In these circumstances, one finds it necessary to employ some approximations which helps in reducing the size of the calculations significantly while producing accurate results. We propose and test such approximations. These approximations are very useful in the cases of the disordered systems.

We had found in our previous study \cite{tirtha} that the presence of the disorder enhances correlations for smaller distances between particles, effectively enhancing cowalking and binding. This effect allows us to make some approximation on the maximum distance between two particles. Green's functions with larger distances can be approximated to make no contribution to the calculations and hence neglected. This selection of Green's elements depending on the distance between two particles can also be done dynamically. One can select elements of importance differently at different times of propagation.

\begin{figure}[H]
\centering
\includegraphics[width = 0.35\textwidth]{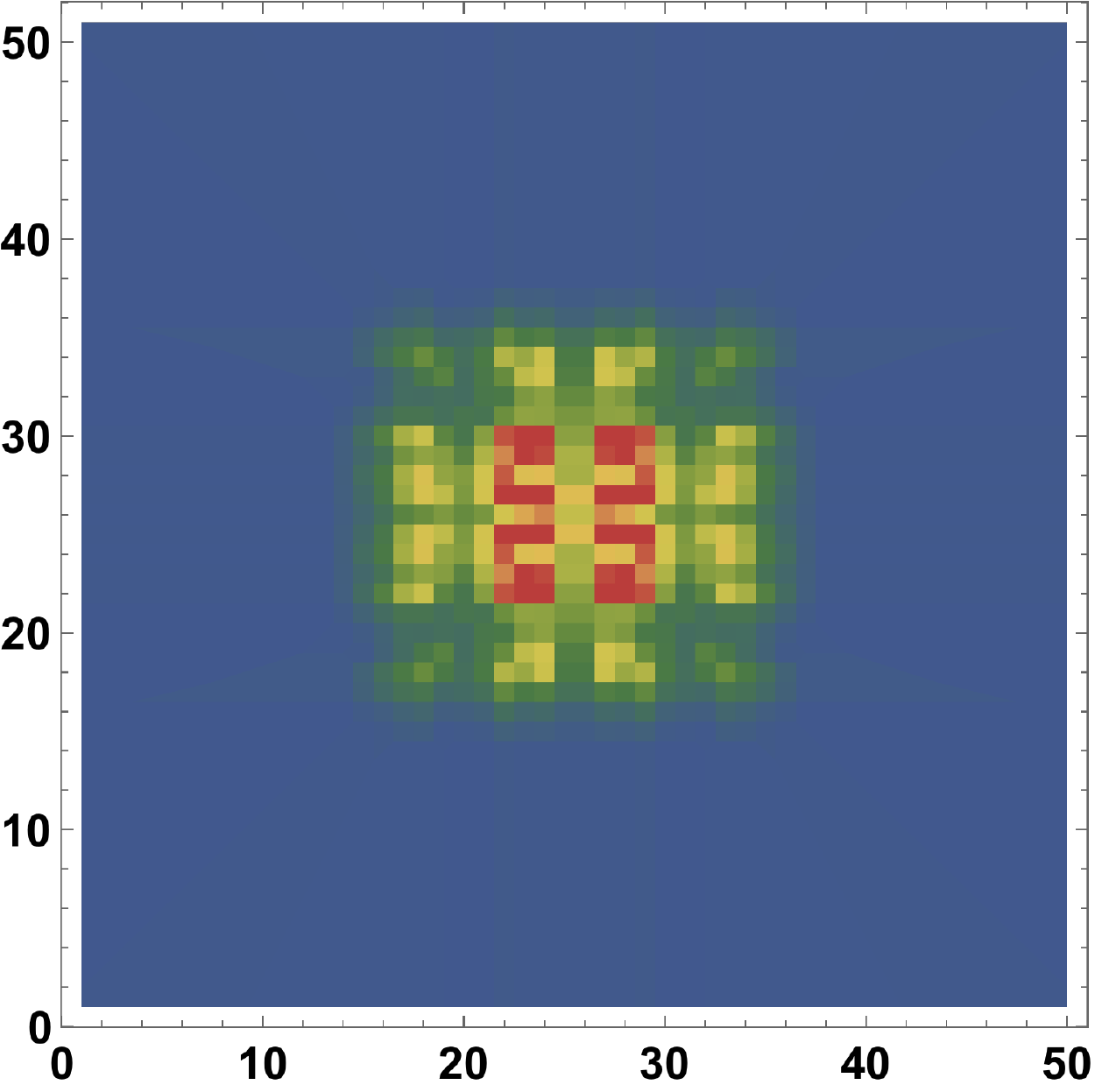}
\includegraphics[width = 0.35\textwidth]{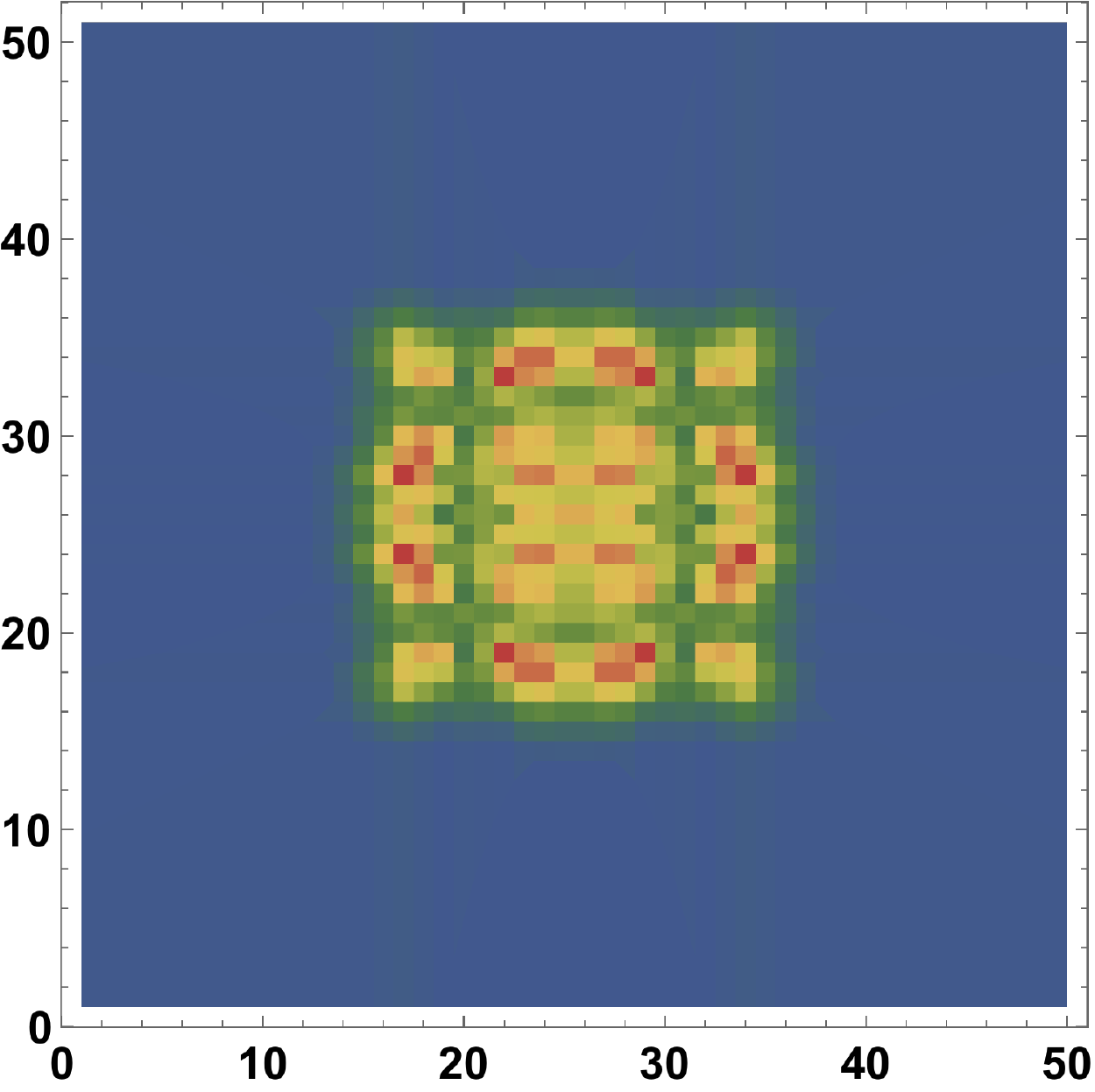}
\caption[Two particle dynamics in 2D]{Two particle dynamics in 2D ideal lattice (Green's probability, Eq. \ref{dyn}), calculated recursively using approximation of maximum allowed distance between two particles $r=5$ (left) and $r=10$ (right) at time$=5$ for $V=1$. Color scheme- red, yellow, green, blue show lower probability in that order.}
\label{2D-density-ideal}
\end{figure}

In Fig. \ref{2D-density-ideal}, the approximation was applied for the case of a 2D ideal lattice with 20 sites in each dimension. While the overall spread shows similarity for two limiting distance approximations, the exact distributions are different as expected. This shows that in the case of ideal lattices applying this approximation will not be accurate at large times.

\begin{figure}[H]
\centering
\includegraphics[width = 0.495\textwidth]{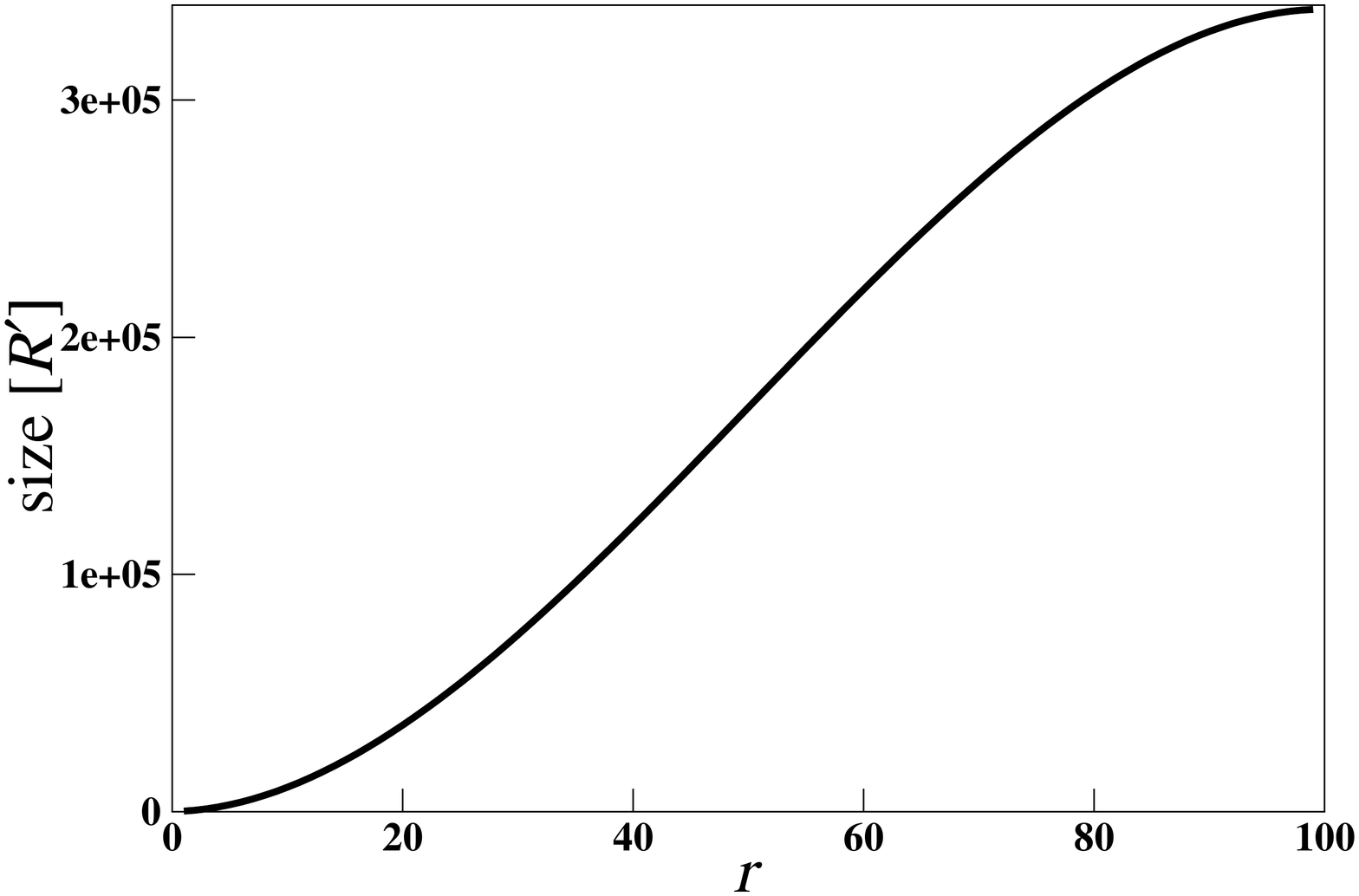}
\includegraphics[width = 0.495\textwidth]{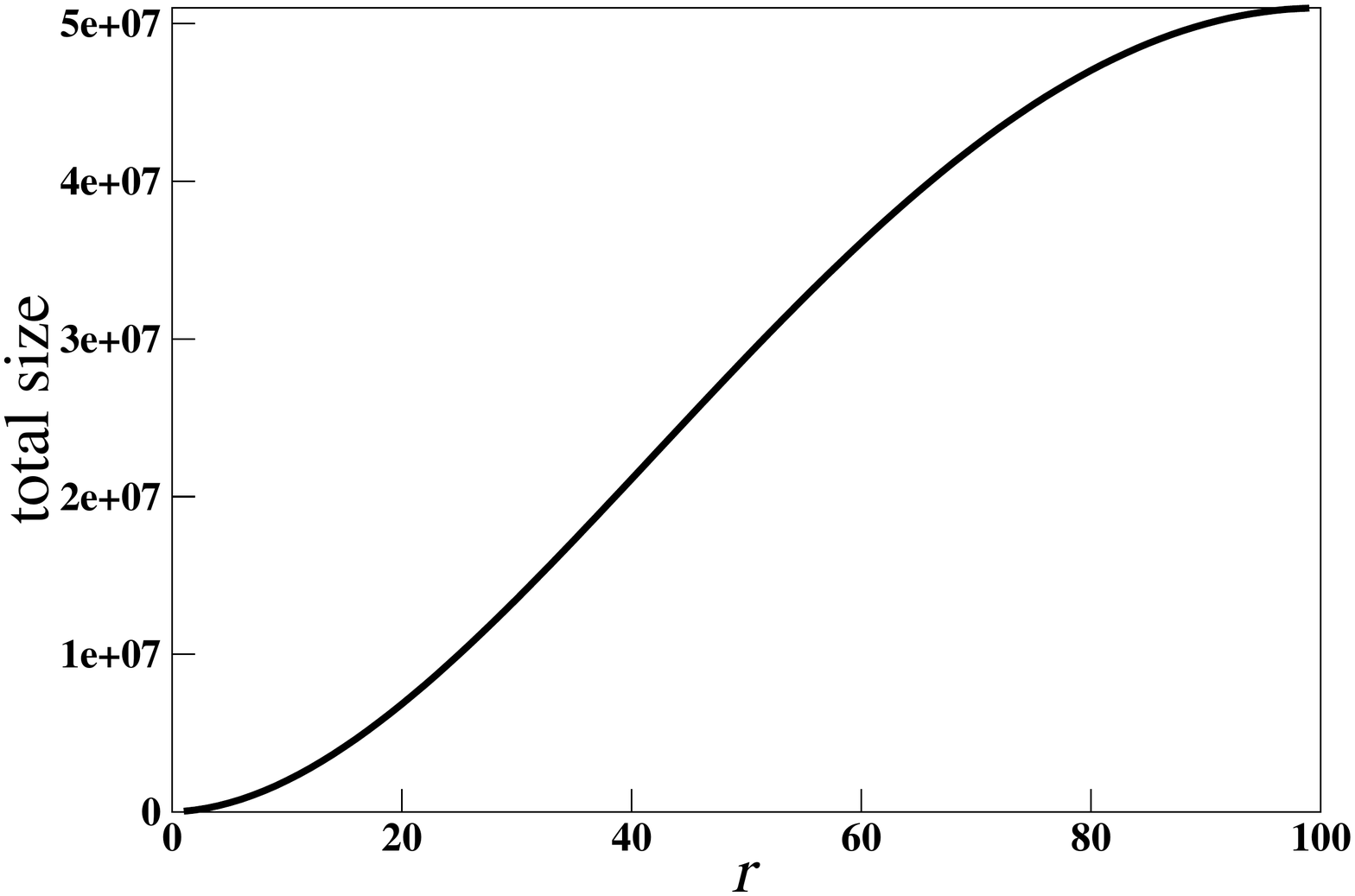}
\caption[Scaling of size of calculation with approximation in 2D]{Size of largest vector (on the left) and total number of elements involved in recursive calculations for given maximum allowed relative minimum step distance $r$ (Hamming distance) between two particles in a 2D lattice of 100x100 sites.}
\label{2D-recursion}
\end{figure}

For large systems, employment of such approximations are inevitable. As shown in Fig. \ref{2D-recursion}, the total number of elements for a full calculation in 2D becomes close to tens of millions for size with 100 sites per dimension. The largest vector involved in the recursive calculation also includes close to a few million Green's elements without approximation. One simple approximation to make is to neglect the propagators with larger relative distance between  the two particles and set a maximum allowed relative distance

\begin{equation}\label{limiting-distance}
G(\mathbf{n},\mathbf{m}; \mathbf{n'},\mathbf{m'}; \omega)  \simeq 0     ~~~~~~~~~\mbox{for} ~~~~~ |\mathbf{n-m}| > r
\end{equation}
where $r$ is a limiting (Hamming) distance in the number of minimum steps between two particles.

\begin{figure}[H]
\centering
\includegraphics[width = 0.68\textwidth]{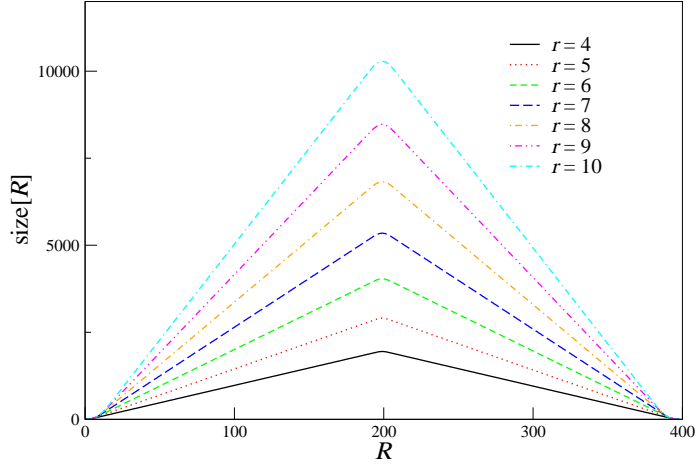}
\caption[Size of the vectors with approximation in 2D]{Size of vectors (Eq. \ref{gv}) in recursive calculations for maximum allowed  $r$ (Hamming distance) in a 2D lattice of 100x100 sites. The difficulty of calculations are determined by the size of the vectors with maximum number of elements.  }
\label{2D-ksize-approx}
\end{figure}

\begin{figure}[H]
\centering
\includegraphics[width = 0.68\textwidth]{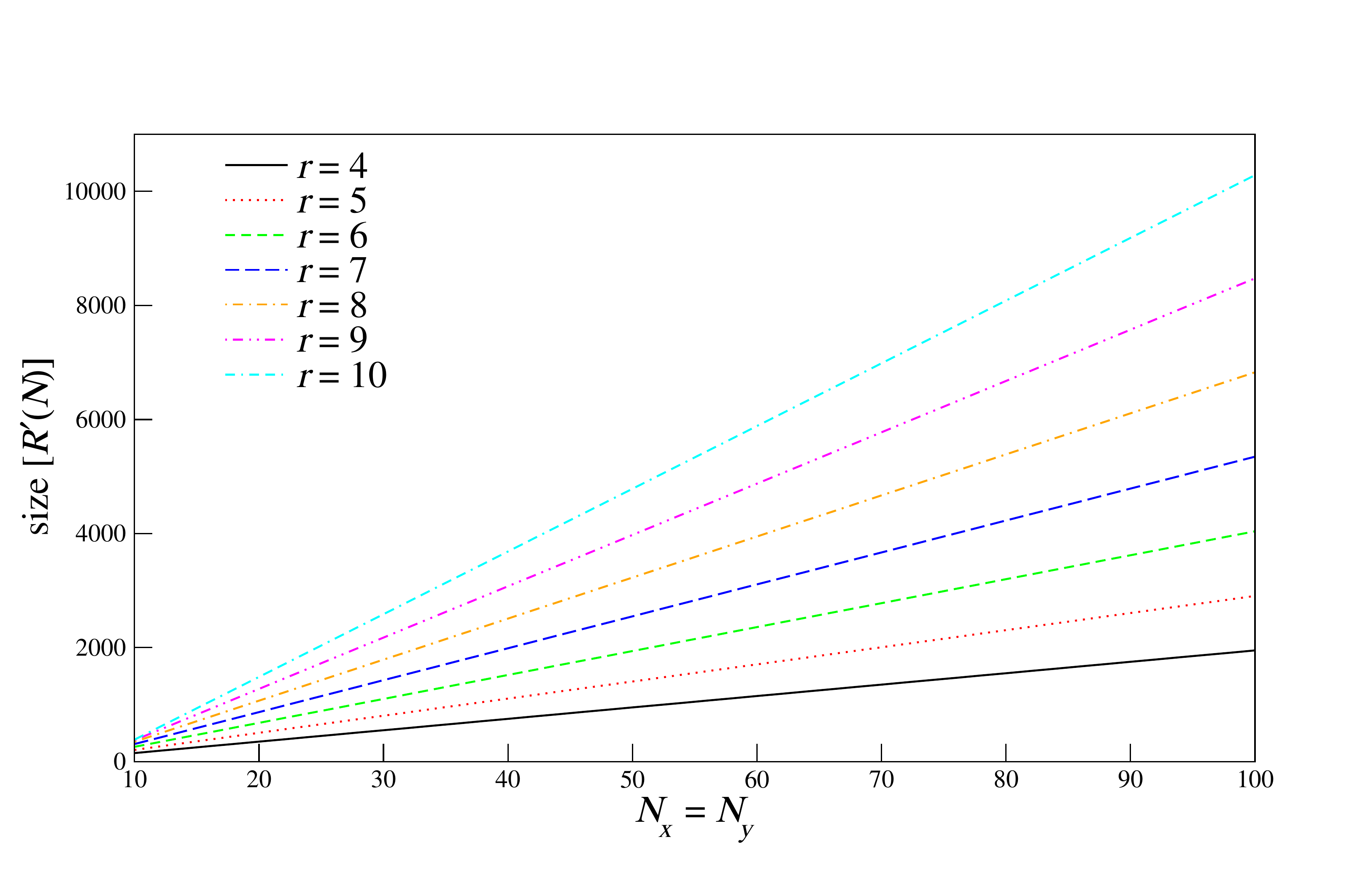}
\caption[Size of largest vector with approximation in 2D]{Size of largest vector (most often the $R'$ in Eq. \ref{gv})  in recursive calculation for maximum allowed  $r$  in a 2D lattice of $N_x$,$N_y$ sites. For larger $r$, the size  increases faster with the size of the lattices. $N = N_x + N_y$.}
\label{2D-Nsize-approx}
\end{figure}

This approximation  allows one to perform calculations considering the whole lattice but with much fewer number of  elements compared to the full recursion. This approximation does not constrain the particles over the lattice but neglects the elements that can contribute to larger distances between the particles than a maximum chosen distance. As can be seen from Figs. \ref{2D-ksize-approx} and \ref{2D-Nsize-approx}, doing calculations for a lattice of 100 sites per dimension with the maximum relative distance $> 10$ will be very difficult.

\newpage

Using this method, we calculate  all Green's elements for a given onsite energy disorder, chosen randomly from a uniform distribution of width $W$  ($\left[ -\frac{W}{2}, \frac{W}{2}\right]$), for particles initially at adjacent sites at very large times ($\tau  = 1000$)

   \begin{equation}\label{dyn}
     G(m,n,\tau)  = \sum_{\omega} e^{-\imath \omega \tau}  G(m, n, \omega).
    \end{equation}

\begin{figure}[H]
\centering
\includegraphics[width=0.8\textwidth]{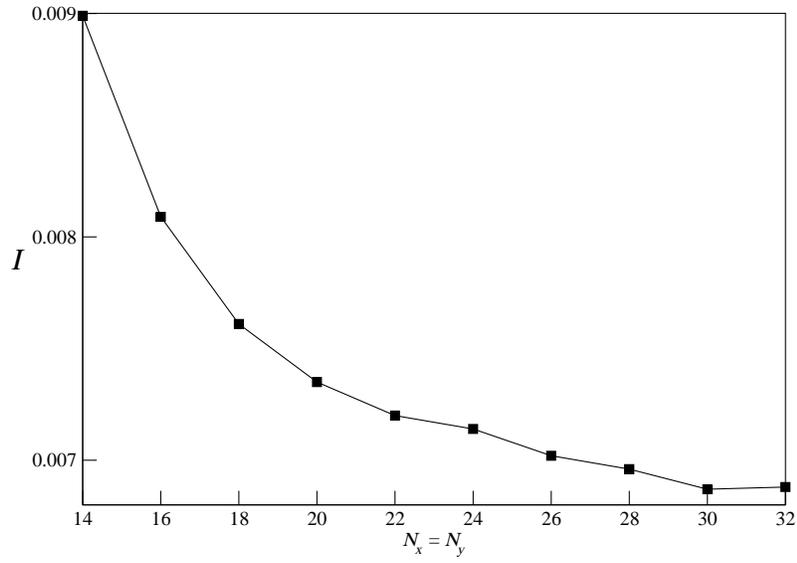}
\caption[Scaling of IPR with lattice sizes in 2D]{ IPR for increasing lattice sizes in 2D for the case of $W=1, V=4$ averaged over 50 realizations of disorder. }
\label{2D-IPR-scaling}
\end{figure}

Once we find all such Green's elements, that is for every pair of site indices ($i, j$), populated by two particles, we calculate the joint density distribution ($\varrho$), density distribution ($\rho$) and inverse participation ratio ($\mathcal{I}$) for each realization of disorder:

   \begin{equation}\label{density}
     \varrho (m,n,\tau)  =  |G(m, n, \tau)|^2,
    \end{equation}

   \begin{equation}\label{density}
     \rho (m,\tau)  =  \frac{1}{2} \sum_{n\neq m}  \varrho (m,n,\tau),
    \end{equation}

   \begin{equation}\label{ipr}
       \mathcal{I} =  \frac{\sum_{m}\rho (m,\tau)^2}{\sum_{m}\rho (m,\tau)}.
    \end{equation}
We average them over many realizations. From the scaling of IPR calculated for 2D disordered systems in the range of $W=4, V=4$, as shown in Fig. \ref{2D-IPR-scaling}, we find that a minimum lattice size of 30x30 sites should be considered for results that would be close to results in larger system sizes. This scaling even for the case under consideration, where the most delocalized behaviour is expected, hints at localization for two weakly interacting particles in disordered 2D lattices. The curve appears to a constancy for larger system sizes which is a signature of localized state rather than an exponential decay, characteristic of delocalization. 

\begin{figure}[H]
\centering
\includegraphics[width=0.49\textwidth]{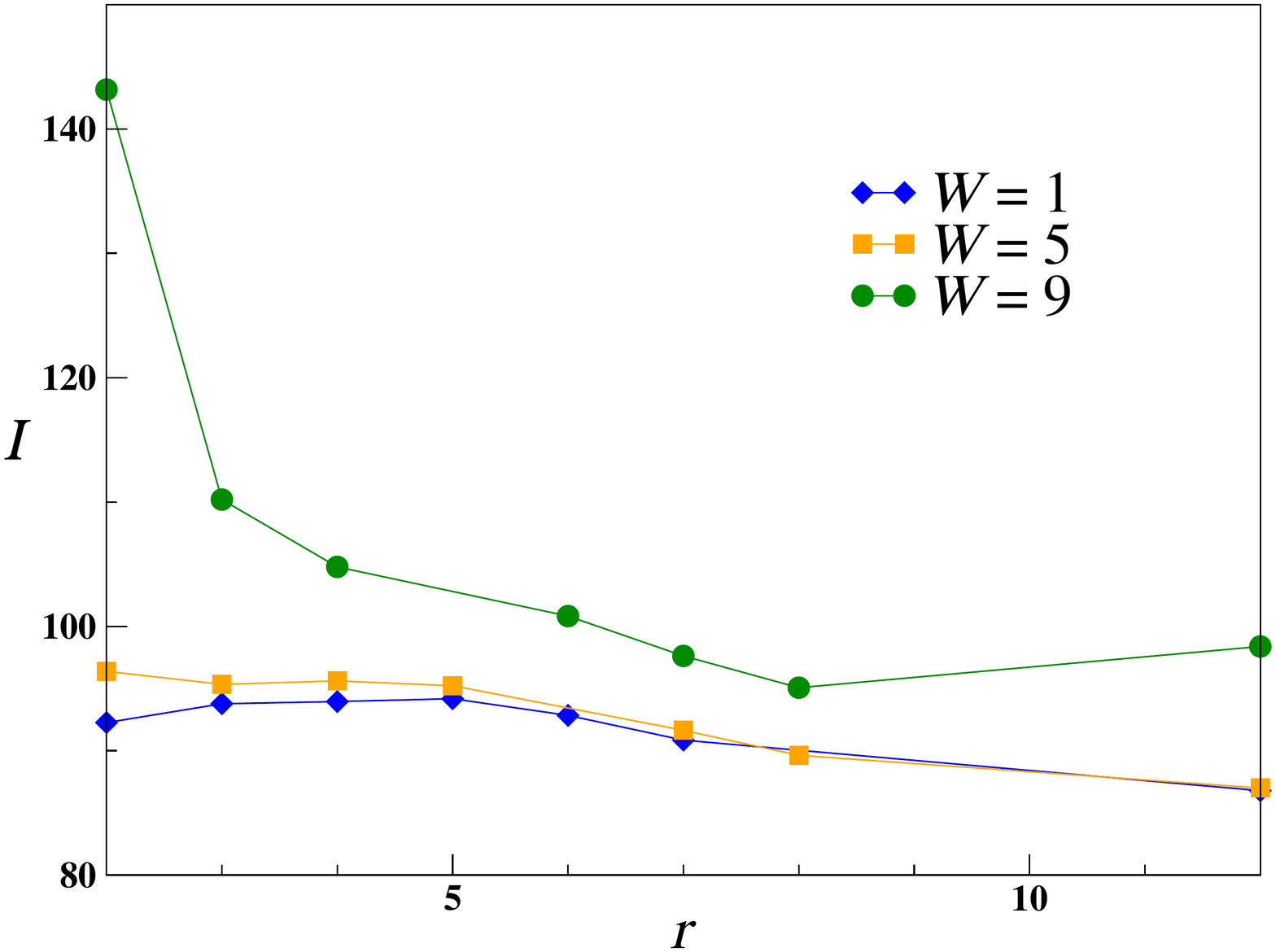}
\includegraphics[width=0.49\textwidth]{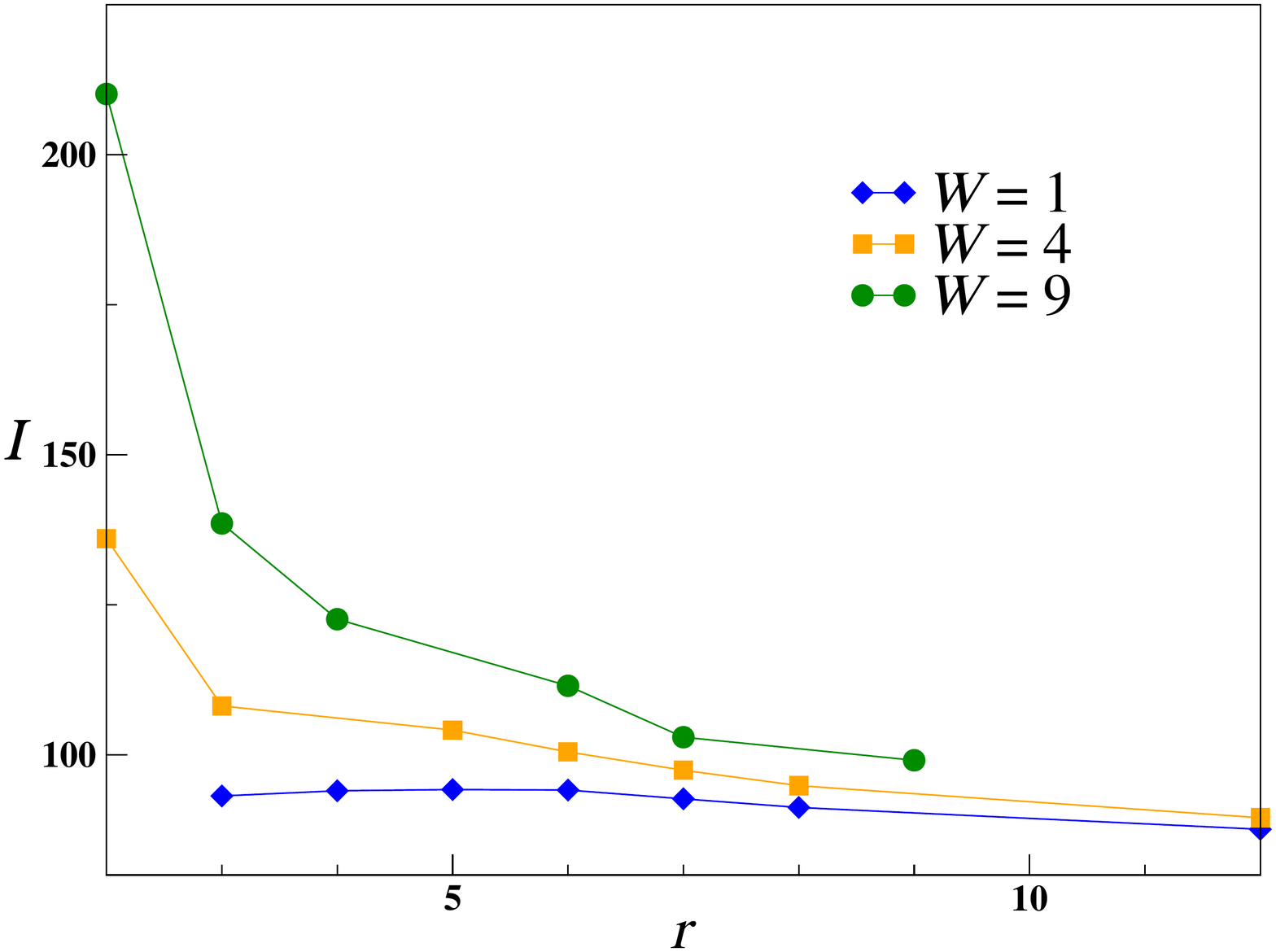}
\caption[Calculation of IPR with approximations]{Relative IPR with maximum allowed relative distance $r$ for a small 2D lattice of 12x13 sites. $V=0$ (left), $V=4$ (right). The IPR is not normalized as $\eta$ makes the calculated densities not normalized. Averaged over 50 realizations of disorder.}
\label{2D-IPR-approx}
\end{figure}

In disordered systems, the calculations of macroscopic properties such as the inverse participation ratio calculated from density distributions averaged at time much larger than that required to hit the boundaries, do not produce large errors even when the maximum allowed distance is kept as small as $r=5$. As shown in Fig. \ref{2D-IPR-approx}, with  $r=5$, the results are within the range of $10\%$ errors. However for specific Green's elements, these errors might be large. Specifically the elements describing transport from center to boundaries are expected to have large finite size effects for small or medium sized systems. Localization lengths calculated from such elements can have errors that are not negligible.

\begin{figure}
\centering
\includegraphics[width=0.5\textwidth]{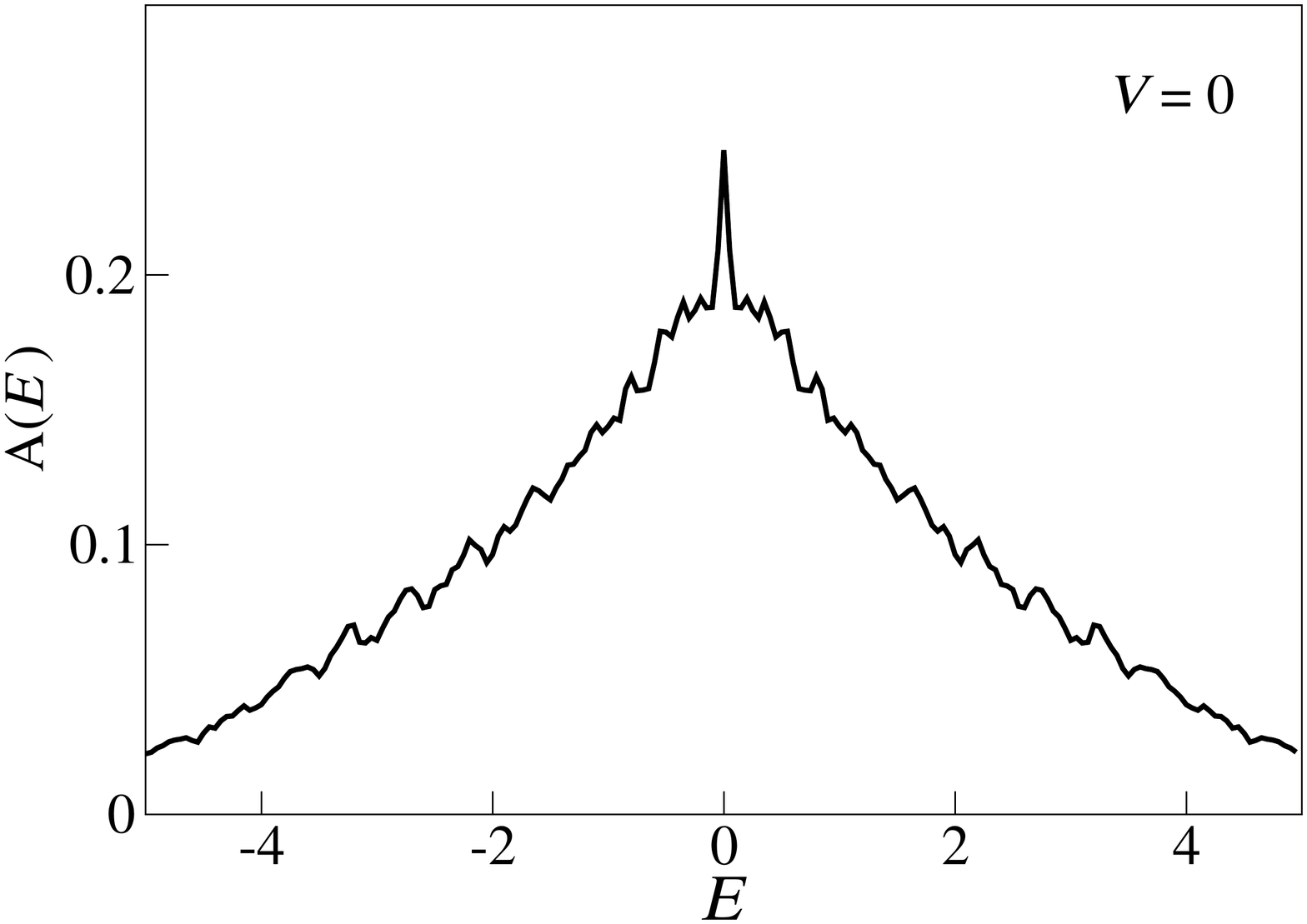}
\includegraphics[width=0.5\textwidth]{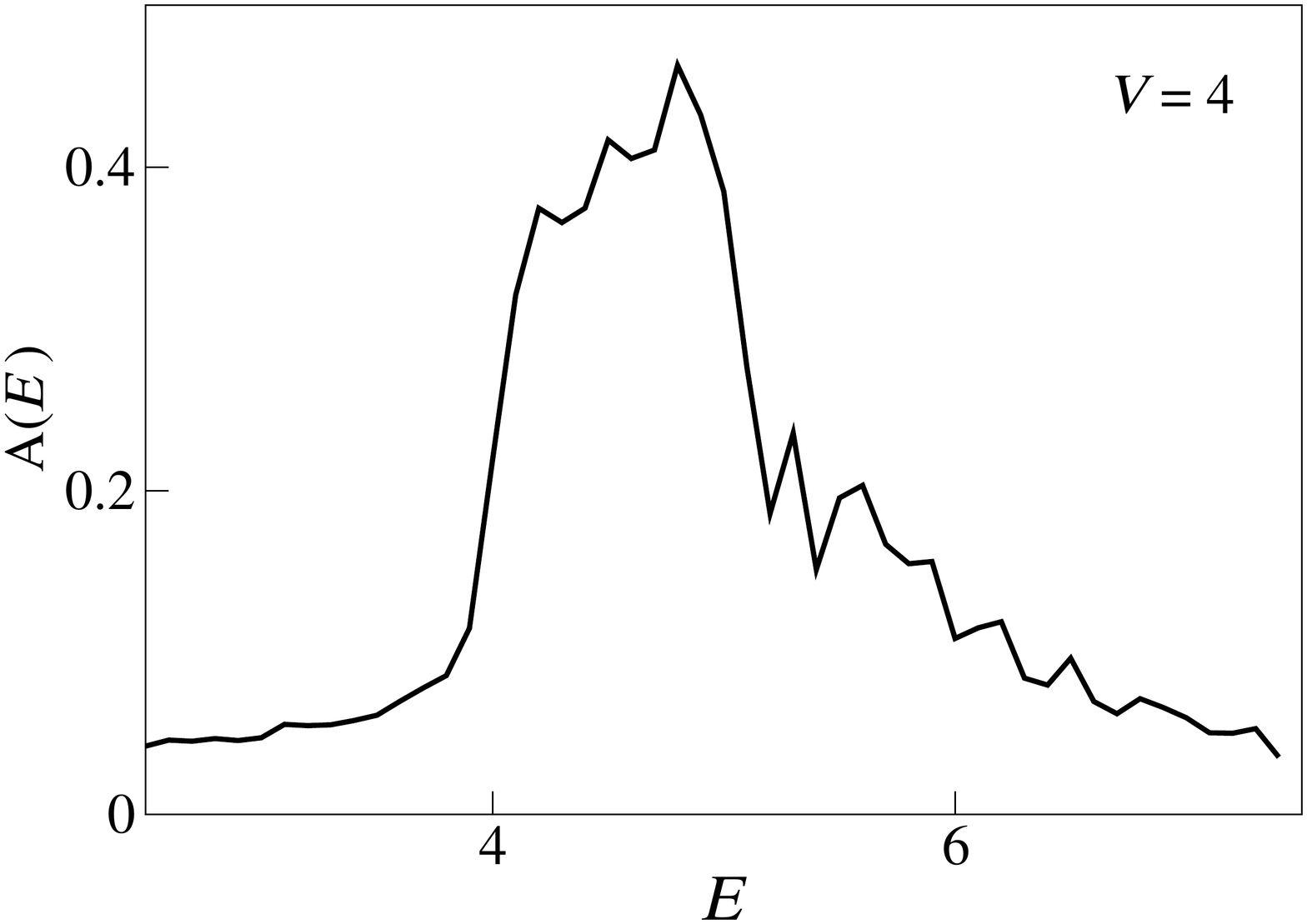}
\includegraphics[width=0.5\textwidth]{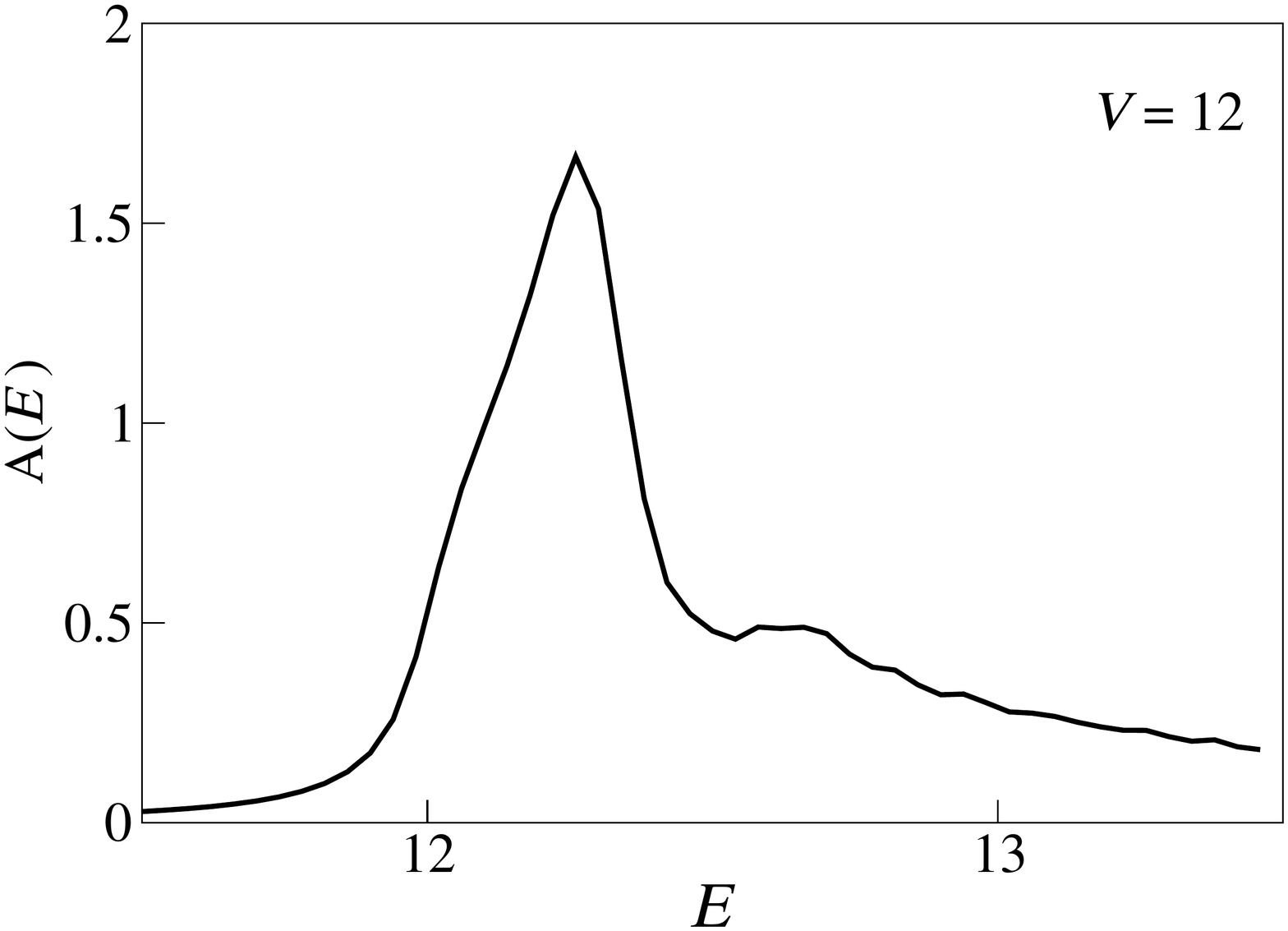}
\caption[Spectral weight of bound state in 2D]{The spectral weights (\ref{spec}) for two particles located side by side in a 2D lattice of 20 sites in each direction. For the non-interacting case, the spectral weight show a peak at $E = 0$ with broad wings on both sides. For the interacting case, the spectra becomes narrow with a long tail as the interaction strength is increased. }
\label{2D-spectral-weight}
\end{figure}

The spectral weight (Eq. \ref{spec}) for the two particles located at adjacent sites on a 2D lattice with nearest neighbor interaction can be calculated from the imaginary part of a single Green's propagator. Figure \ref{2D-spectral-weight} shows the spectral weight for such two particles, that is the bound state for different interaction strength. The spectral weight shows a sharp peak at $E=0$ for the non-interacting case with an exponential drop to the band edges. The spectral weight for the interacting case shows a sharp rise following a linear drop to the band edges.

\subsection*{Spectral weights of doublon in Hofstadter model}

In  recent years, there has been a lot of interest in implemeting the Hofstadter model \cite{hofstadter} in optical lattices for neutral atoms \cite{goldman, spielman1, spielman2, struck}. The Hofstadter model takes account the effect of external magnetic fields on electrons in lattices by making the hopping amplitude complex. The model is mimicked for neutral atoms by periodic modulation of lattice potentials, which averages to zero force, but produces a complex phase factor on momentum dependent hopping or tunneling amplitudes of atoms in lattices \cite{struck, monika-thesis}. This opens the possibility of simulating integer and fractional quantum Hall  \cite{yoshioka} systems and topological insulators \cite{bernevig} in disordered 2D optical lattice systems \cite{perczel}.

 The model can be derived by Peierls substitution \cite{peierls} from the tight binding Hubbard model and accounts for a phase for hopping

\begin{equation}
\mathcal{H}_{} = \sum_{\langle i j \rangle} e^{-\imath \phi_{ij}} J_{ij} a_i^\dagger a_j + \sum_{i} U_{i}  a_i^\dagger a_j^\dagger a_j a_i.
\end{equation}

Experimentally, a periodically modulated potential after averaging over the full time period \cite{eckardt} can effectively  add the directional phases to the hopping terms and have the same dispersion as after performing the  Peierls substitution \cite{struck}.



\begin{figure}
\centering
\hspace{0.0cm}
\begin{tikzpicture}[scale=1.45]
\draw [fill] (0,0) circle [radius=0.03];
\draw [fill] (1,0) circle [radius=0.03];
\draw [fill] (2,0) circle [radius=0.03];
\draw [fill] (3,0) circle [radius=0.03];
\draw [fill] (4,0) circle [radius=0.03];
\draw [fill] (5,0) circle [radius=0.03];
\draw [fill] (0,1) circle [radius=0.03];
\draw [fill] (1,1) circle [radius=0.03];
\draw [fill] (2,1) circle [radius=0.03];
\draw [fill] (3,1) circle [radius=0.03];
\draw [fill] (4,1) circle [radius=0.03];
\draw [fill] (5,1) circle [radius=0.03];
\draw (0,1) --(5,1);
\draw (1,0) --(1,4);
\draw [fill] (0,2) circle [radius=0.03];
\draw [fill] (1,2) circle [radius=0.03];
\draw [fill] (4,2) circle [radius=0.03];
\draw [fill] (5,2) circle [radius=0.03];
\draw (0,2) --(5,2);
\draw (2,0) --(2,4);
\draw [fill] (0,3) circle [radius=0.03];
\draw [fill] (1,3) circle [radius=0.03];
\draw [fill] (2,3) circle [radius=0.03];
\draw [fill] (3,3) circle [radius=0.03];
\draw [fill] (4,3) circle [radius=0.03];
\draw [fill] (5,3) circle [radius=0.03];
\draw (0,3) --(5,3);
\draw (3,0) --(3,4);
\draw [fill] (0,4) circle [radius=0.03];
\draw [fill] (1,4) circle [radius=0.03];
\draw [fill] (2,4) circle [radius=0.03];
\draw [fill] (3,4) circle [radius=0.03];
\draw [fill] (4,4) circle [radius=0.03];
\draw [fill] (5,4) circle [radius=0.03];
\draw (0,4) --(5,4);
\draw (4,0) --(4,4);
\draw (5,0) --(5,4);
\draw (0,0) -- (5,0) -- (5,4) -- (0,4) -- (0,0);
\draw [blue, <->, ultra thick]  (2.2,2) -- (2.8,2);
\node [blue] at (2.5,2.15) {\large $V$};
\draw [red, ->, ultra thick]  (3,2) -- (4,2);
\draw [green, -> , ultra thick]  (2,2) -- (1,2);
\draw [ ->, ultra thick]  (3,2) -- (3,3);
\draw [ -> , ultra thick]  (3,2) -- (3,1);
\draw [ ->, ultra thick]  (2,2) -- (2,1);
\draw [ -> , ultra thick]  (2,2) -- (2,3);
\draw [blue, fill=blue] (2,2) circle [radius=0.13];
\draw [blue, fill=blue] (3,2) circle [radius=0.13];
\node at (1.85,1.5) {\large $J$};
\node at (3.15,1.5) {\large $J$};
\node at (1.85,2.5) {\large $J$};
\node at (3.15,2.5) {\large $J$};
\node [red] at (3.5,2.15) {\large $J e^{i\phi}$};
\node [green] at (1.5,1.85) {\large $J e^{-i\phi'}$};
\draw [ ->]  (5,0) -- (5.5,0);
\node at (5.65,0) {\large $X$};
\draw [ ->]  (0,4) -- (0,4.5);
\node at (0,4.65) {\large $Y$};
\end{tikzpicture}
\vspace{0.5cm}
\caption[Terms in Hofstadter model]{The hopping and interaction terms in 2D Hofstadter model for hardcore bosons. The phases $\phi$, $\phi'$ for hopping terms on $X$ axis depends on lattice site indices of $Y$ axis.}
\label{Hoffig}
\end{figure}
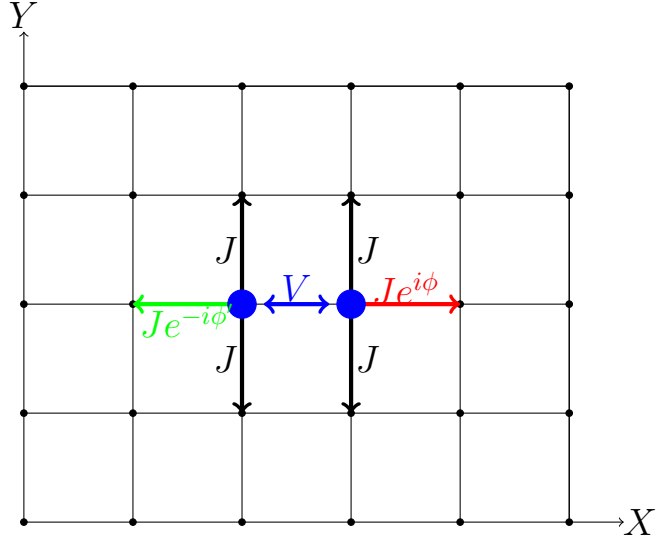

In the 2D lattices, we implement the same hamiltonian for two interacting  hard-core bosons

\begin{equation}
\mathcal{H}_{pq} = \sum_{\langle i j \rangle}\left[ e^{-\imath 2\pi \frac{p}{q} i_y} J_{i_x, i_{x+1}} a_{i_x}^\dagger a_{i_x +1} + J_{i_y, i_{y+1}} a_{i_y}^\dagger a_{i_y +1}  + h.c.  \right]+ \sum_{\langle i j \rangle} V  a_i^\dagger a_j^\dagger a_j a_i,
\end{equation}
where $i, j$ are the site indices of two particles and the axes dependency is removed from the interaction term for simplicity. The terms are elaborated in Fig. \ref{Hoffig}.

\begin{figure}[H]
\begin{subfigure}{0.325\textwidth}
\includegraphics[width = 1.15\textwidth]{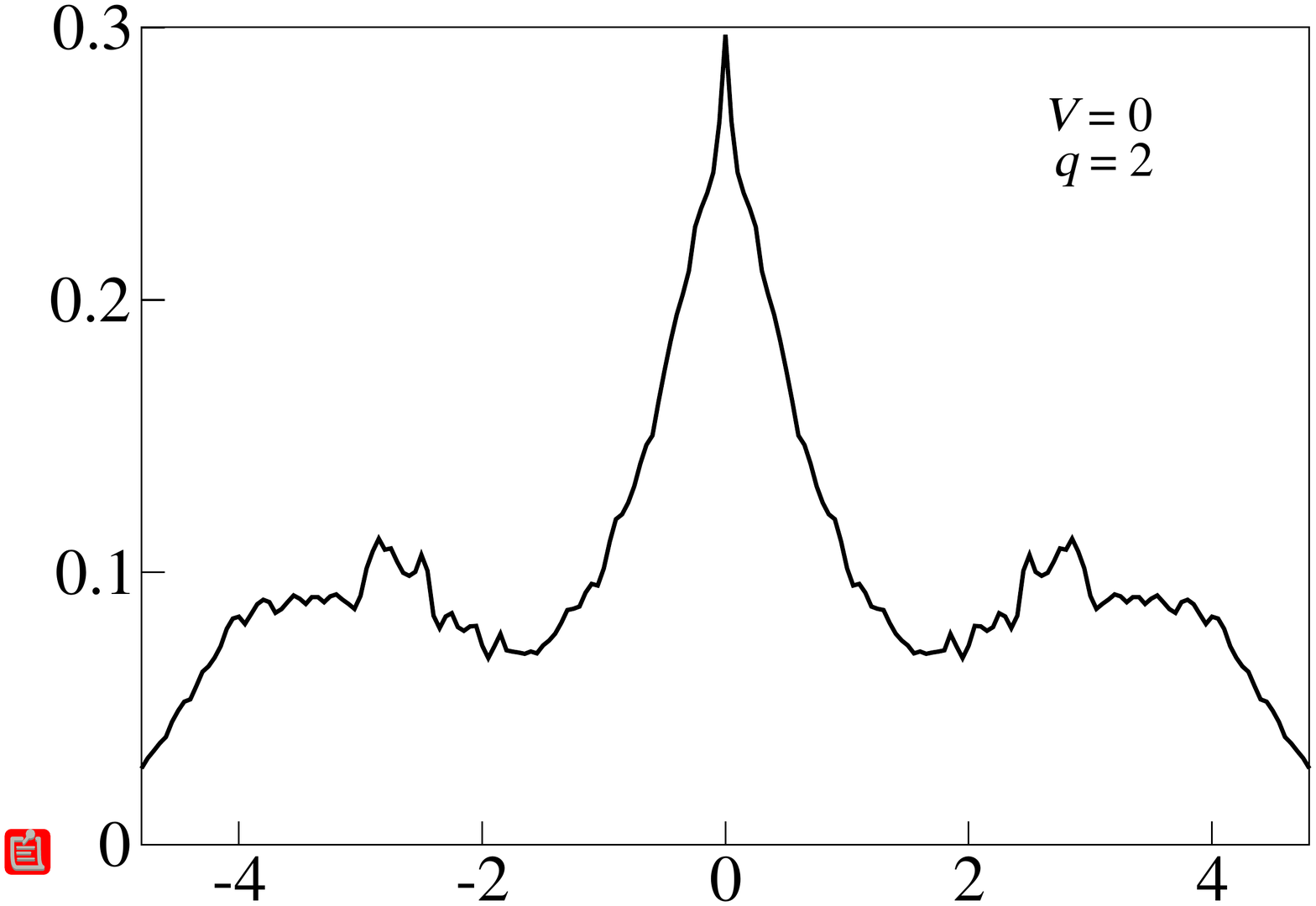}
\end{subfigure}
\begin{subfigure}{0.325\textwidth}
\includegraphics[width = 1.15\textwidth]{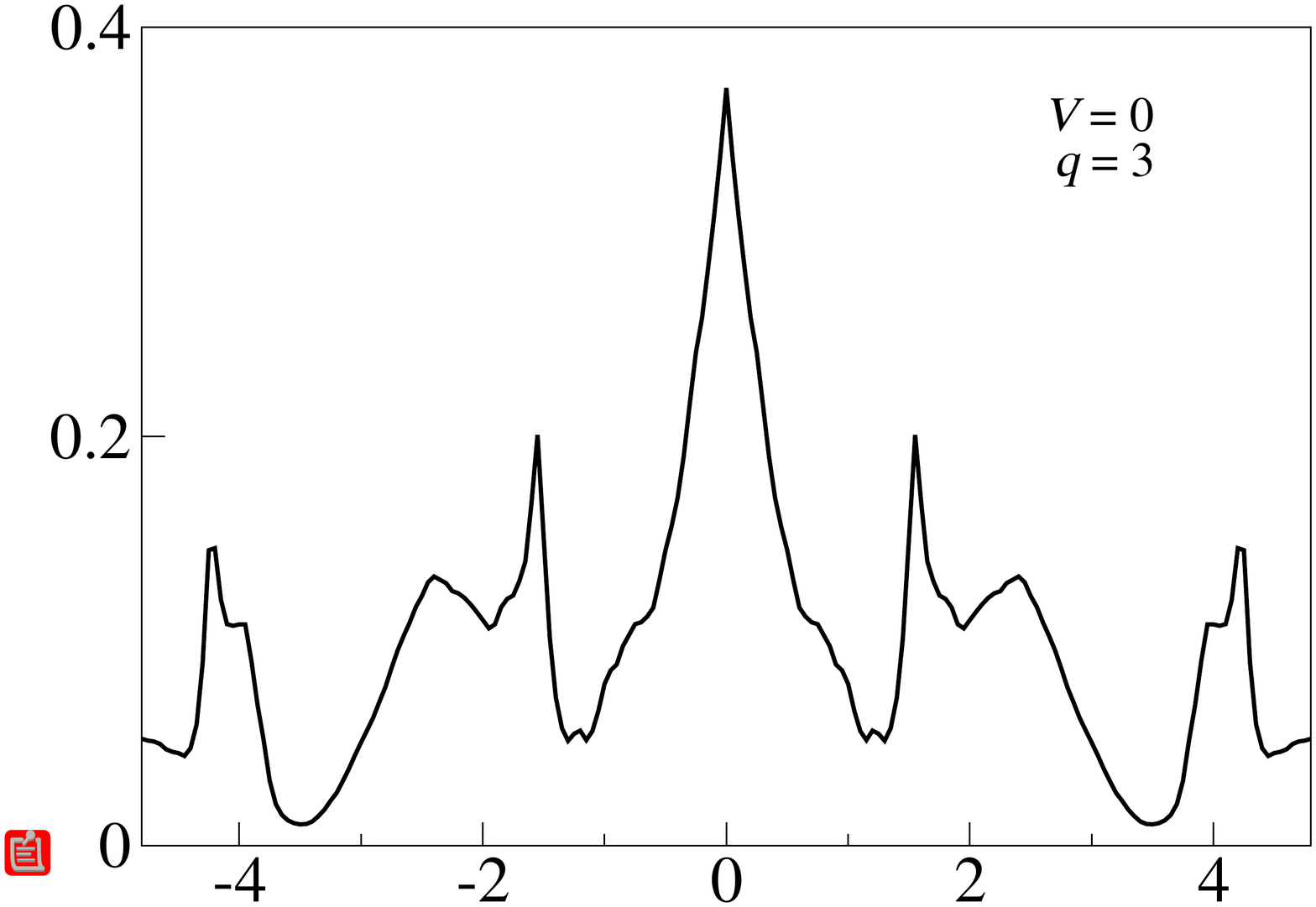}
\end{subfigure}
\begin{subfigure}{0.325\textwidth}
\includegraphics[width = 1.15\textwidth]{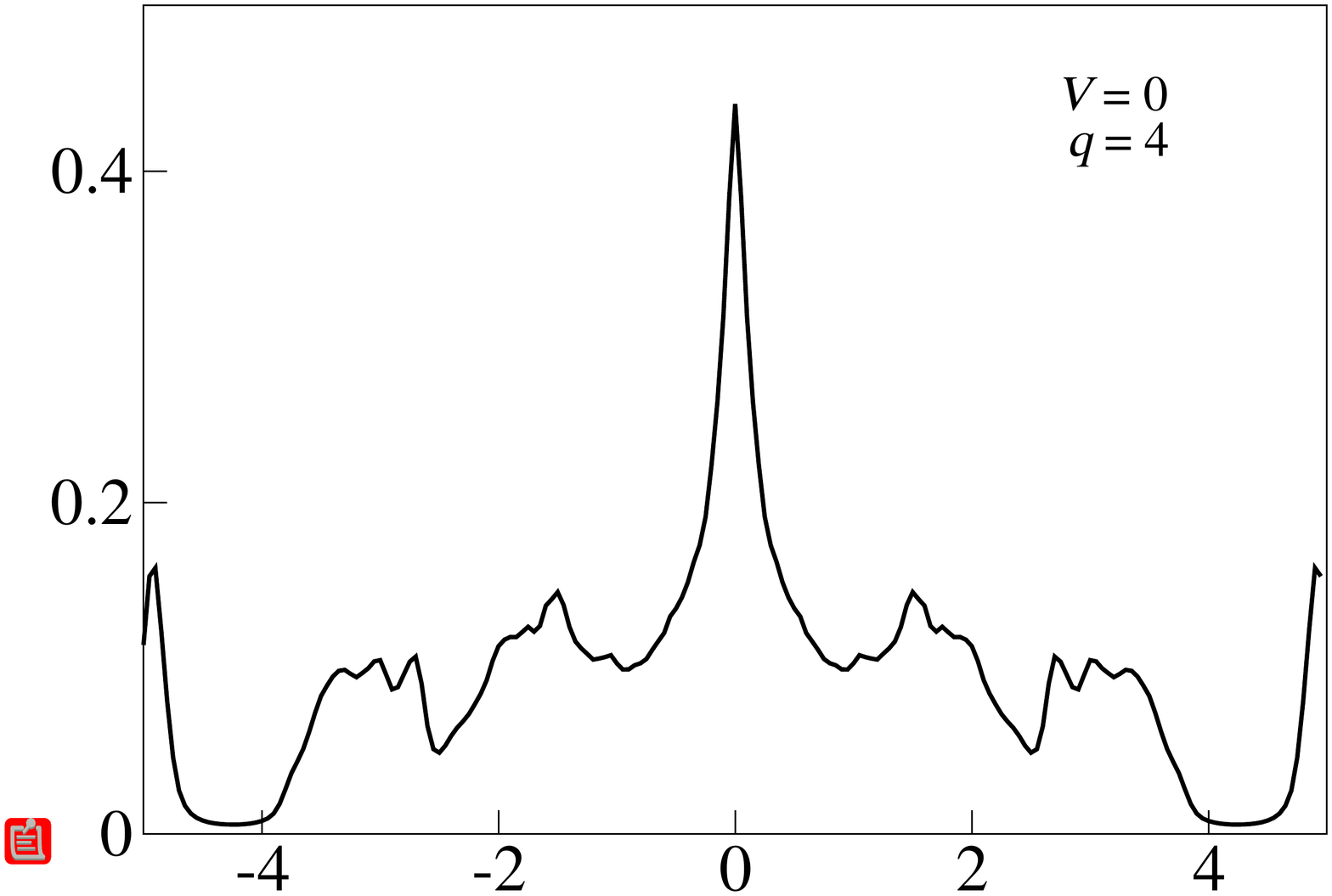}
\end{subfigure}
\vspace{-0.1cm}
\begin{subfigure}{0.325\textwidth}
\includegraphics[width = 1.15\textwidth]{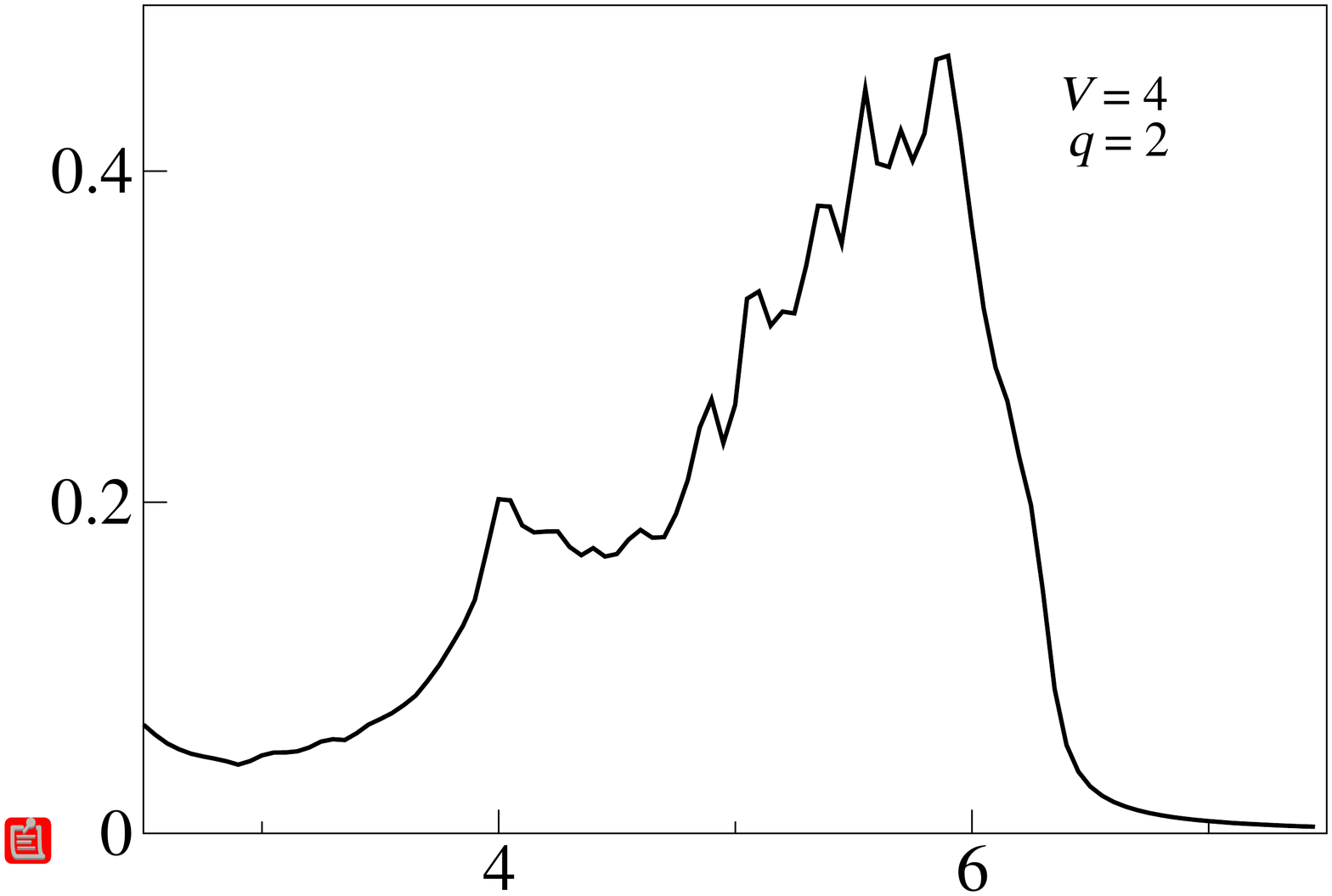}
\end{subfigure}
\begin{subfigure}{0.325\textwidth}
\includegraphics[width = 1.15\textwidth]{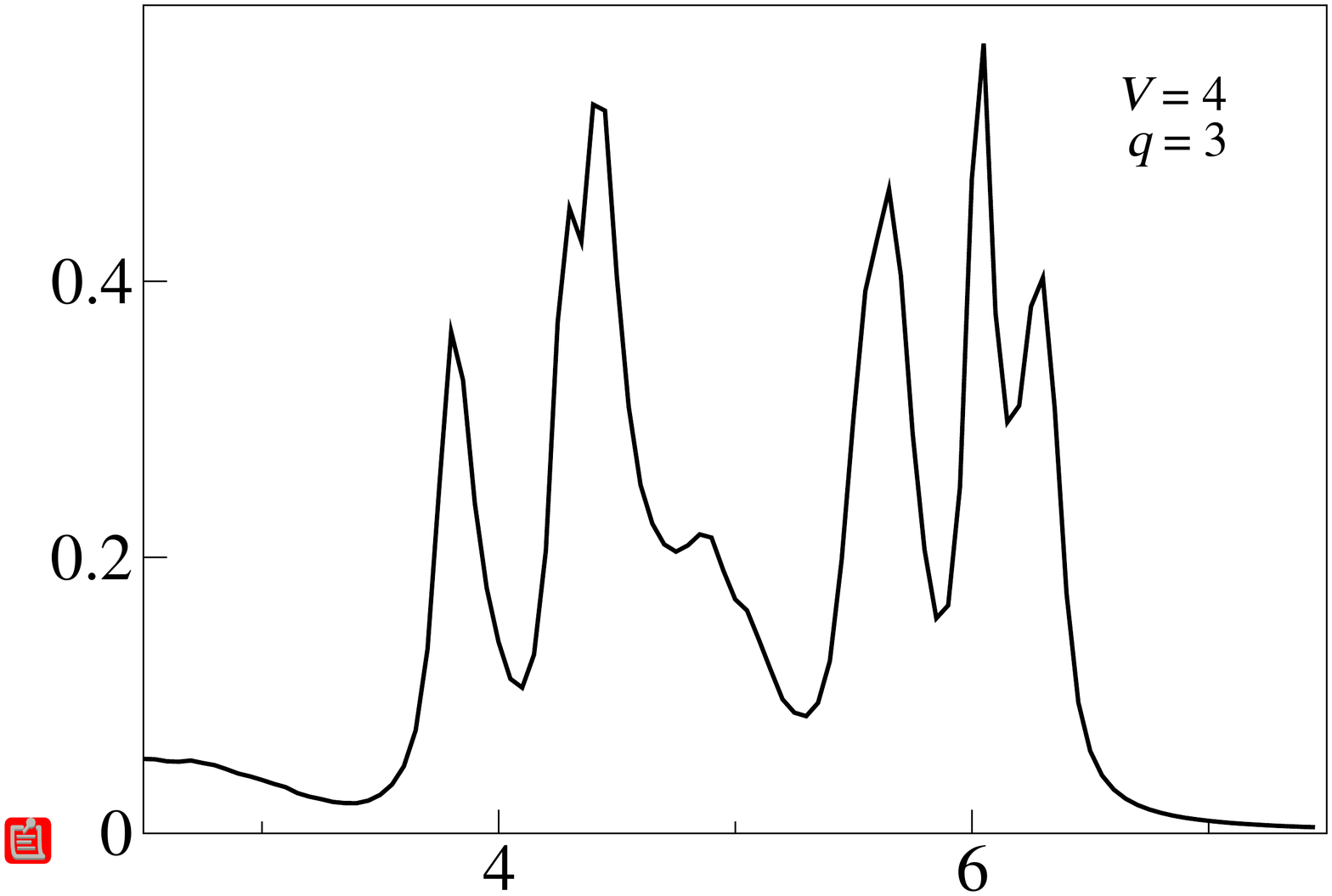}
\end{subfigure}
\begin{subfigure}{0.325\textwidth}
\includegraphics[width = 1.15\textwidth]{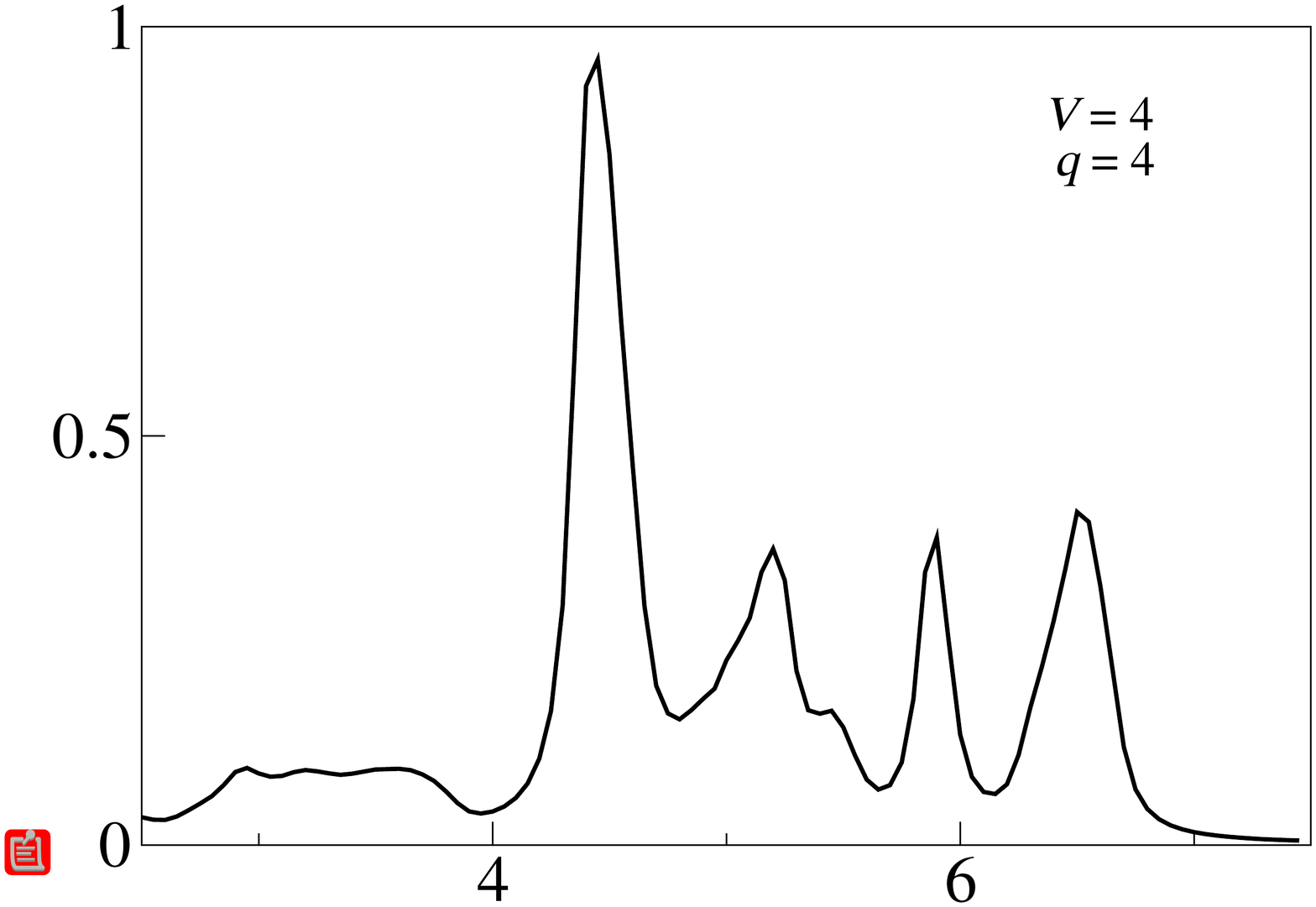}
\end{subfigure}
\begin{subfigure}{0.325\textwidth}
\includegraphics[width = 1.15\textwidth]{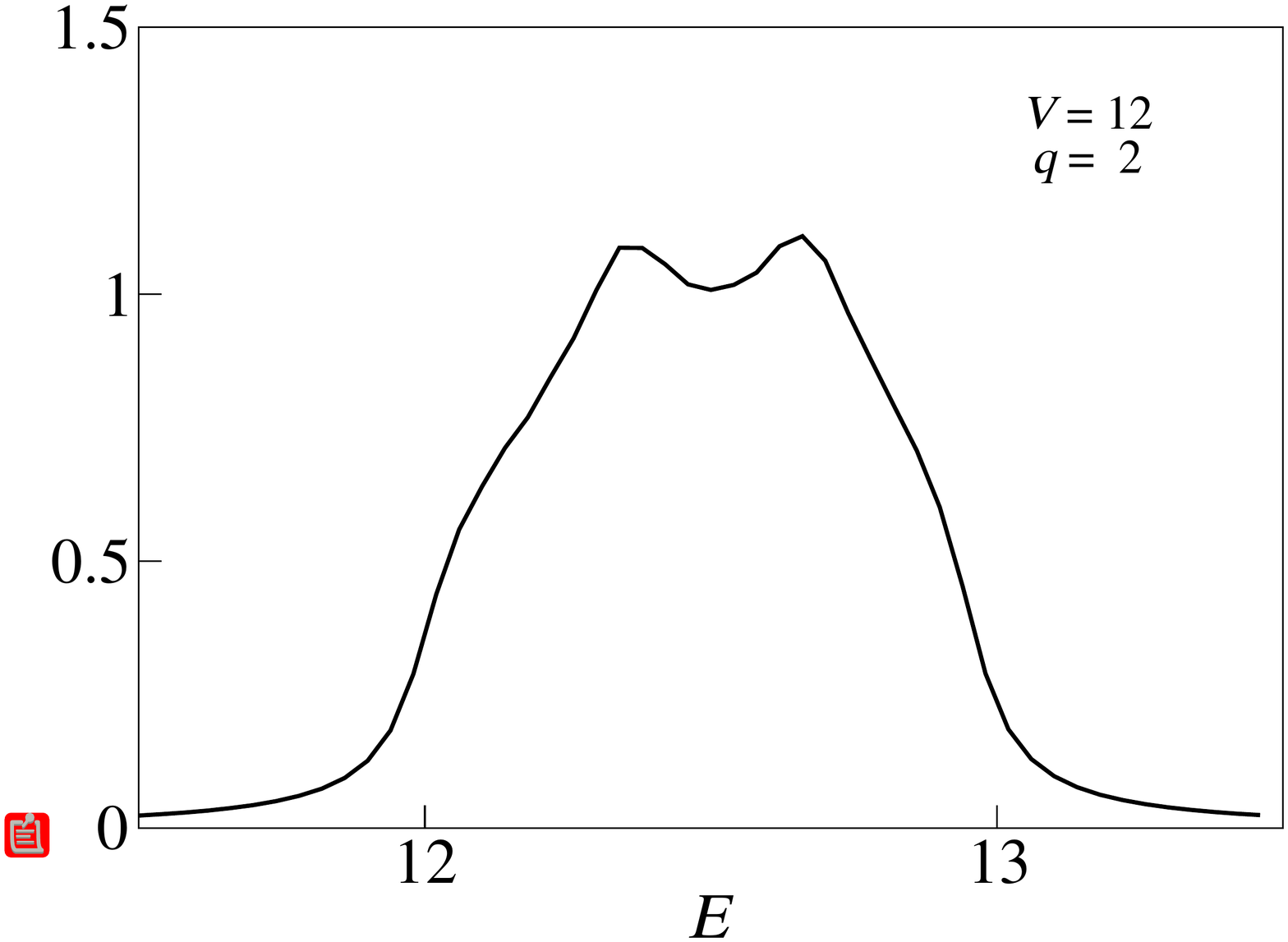}
\end{subfigure}
\begin{subfigure}{0.325\textwidth}
\includegraphics[width = 1.15\textwidth]{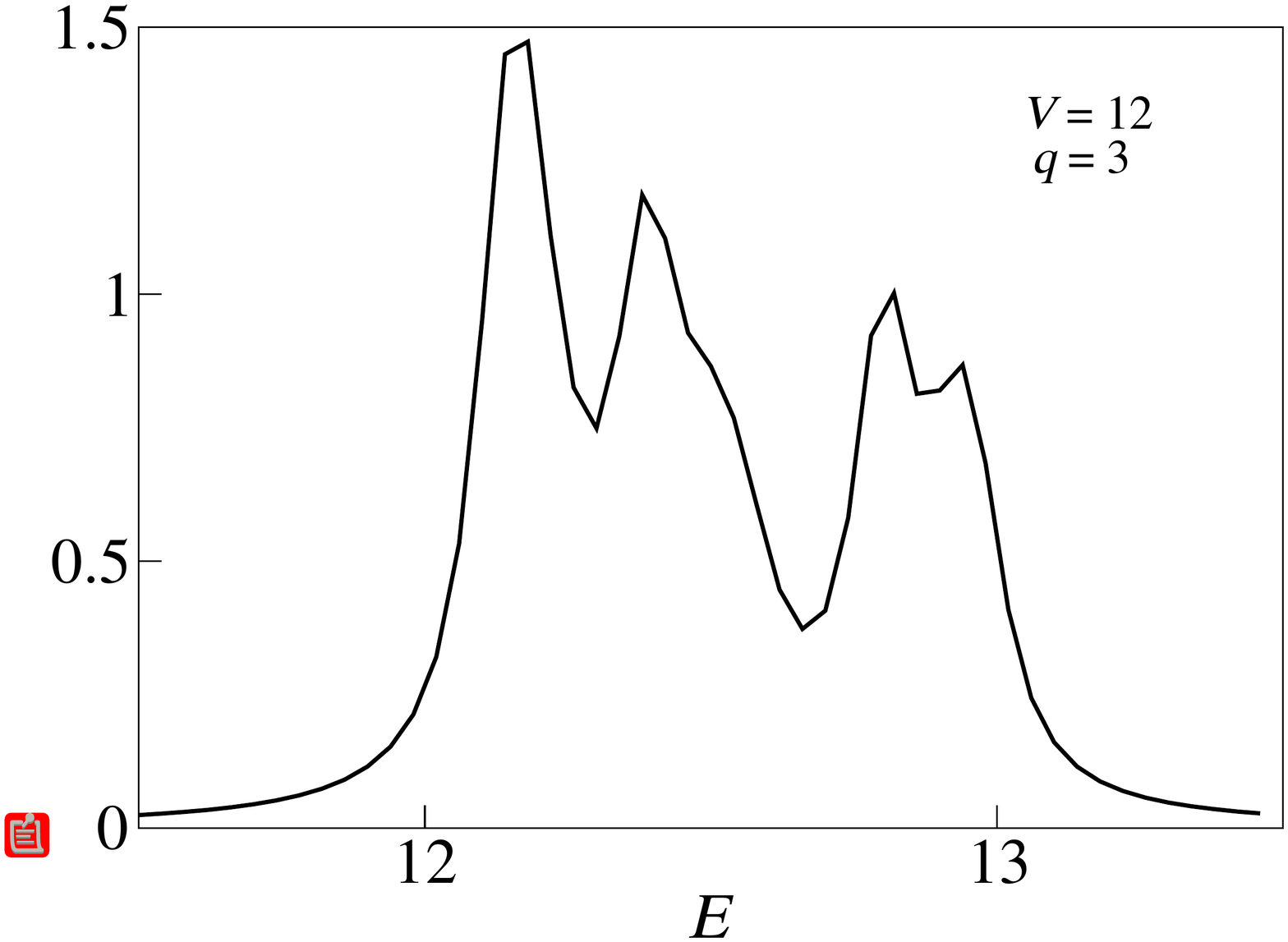}
\end{subfigure}
\begin{subfigure}{0.325\textwidth}
\includegraphics[width = 1.15\textwidth]{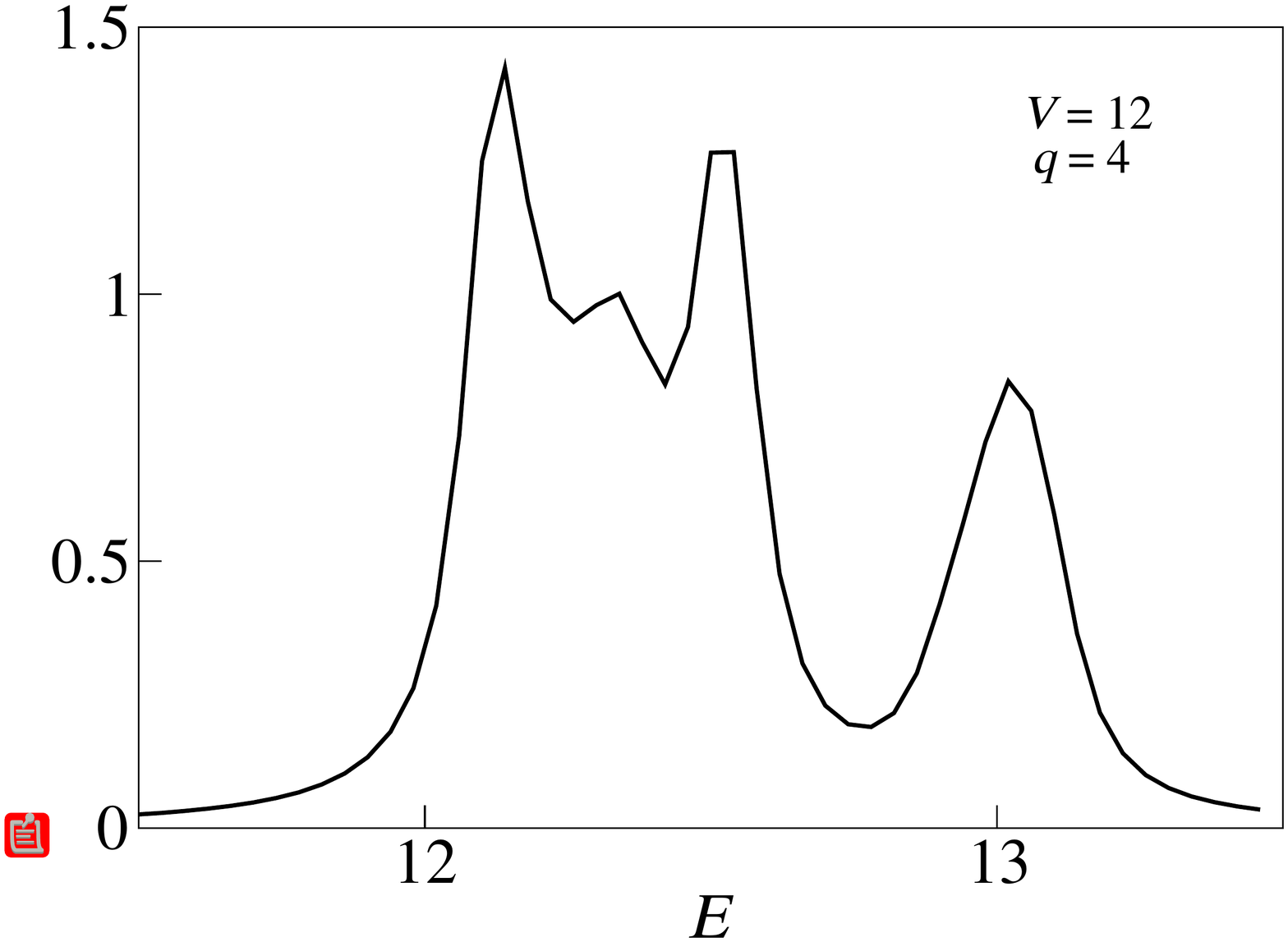}
\end{subfigure}
\caption[Spectral weights of doublon in Hofstadter model]{Spectral weight of two interacting particles in 2D Hofstadter model. The number of broad peaks show clear dependence of $q$ while stronger interaction seems to be merging these peaks.  $p = 1$.}
\label{2D-phase-sw}
\end{figure}

For this hamiltonian we calculate two-particle Green's functions and  spectral weight for two particles located at adjacent sites. We find each spectrum splits into several bands depending on the value of $q$ as shown in Fig. \ref{2D-phase-sw}. The non-interacting particles show a sharp peak at $E = 0$ with the $q-1$ number of broad peaks on both sides. Each of these broad peaks has more peaks inside them.  For increasing interactions, these broads peaks seem to be merging with each other while the overall shape for $q \neq \infty$ appearing totally different from the $q = \infty$ (\ref{2D-spectral-weight}) case. One observation can be made from the calculated results, which is, the weight of the spectra shifts toward lower side of the energy bandwidth for higher interaction strength ($V$) and higher  ratio of $\frac{q}{p}$ until $\frac{q}{p} < \frac{1}{2}$. For $\frac{q}{p} > \frac{1}{2}$, this trend is expected to reverse as $\frac{q}{p}$ and $\frac{q-p}{p}$ has same spectra.

\newpage
\section{Two interacting particles in binary tree}

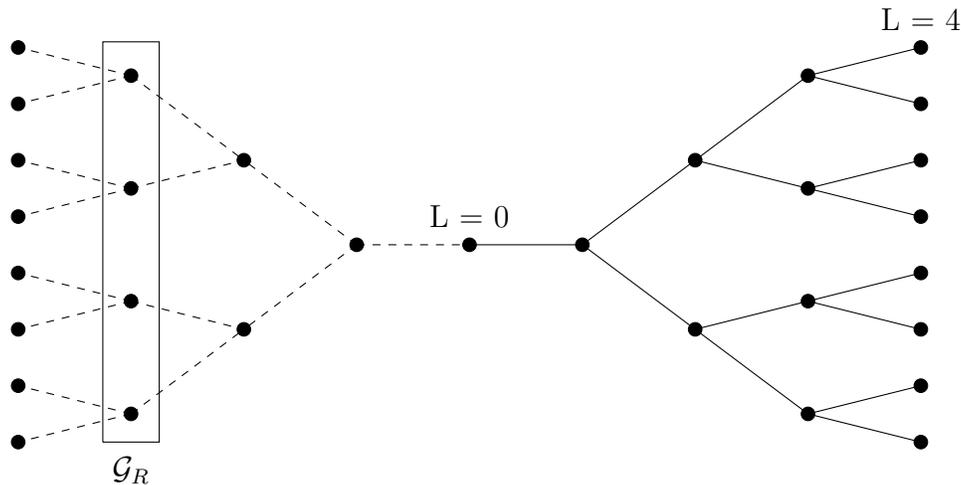
\begin{figure}[H]
\centering
\hspace{0.0cm}
\begin{tikzpicture}[scale=1.5]
\draw [fill] (0,0) circle [radius=0.06];
\draw [fill] (0,0.5) circle [radius=0.06];
\draw [fill] (0,1) circle [radius=0.06];
\draw [fill] (0,1.5) circle [radius=0.06];
\draw [fill] (0,2) circle [radius=0.06];
\draw [fill] (0,2.5) circle [radius=0.06];
\draw [fill] (0,3) circle [radius=0.06];
\draw [fill] (0,3.5) circle [radius=0.06];
\draw [fill] (1,0.25) circle [radius=0.06];
\draw [fill] (1,1.25) circle [radius=0.06];
\draw [fill] (1,2.25) circle [radius=0.06];
\draw [fill] (1,3.25) circle [radius=0.06];
\draw [fill] (2,1) circle [radius=0.06];
\draw [fill] (2,2.5) circle [radius=0.06];
\draw [fill] (3,1.75) circle [radius=0.06];
\draw [fill] (4,1.75) circle [radius=0.06];
\draw [fill] (5,1.75) circle [radius=0.06];
\draw [fill] (6,1) circle [radius=0.06];
\draw [fill] (6,2.5) circle [radius=0.06];
\draw [fill] (7,0.25) circle [radius=0.06];
\draw [fill] (7,1.25) circle [radius=0.06];
\draw [fill] (7,2.25) circle [radius=0.06];
\draw [fill] (7,3.25) circle [radius=0.06];
\draw [fill] (8,0) circle [radius=0.06];
\draw [fill] (8,0.5) circle [radius=0.06];
\draw [fill] (8,1) circle [radius=0.06];
\draw [fill] (8,1.5) circle [radius=0.06];
\draw [fill] (8,2) circle [radius=0.06];
\draw [fill] (8,2.5) circle [radius=0.06];
\draw [fill] (8,3) circle [radius=0.06];
\draw [fill] (8,3.5) circle [radius=0.06];
\draw (4,1.75) --(5,1.75);
\draw (6,1)  --(5,1.75);
\draw (6,2.5)  --(5,1.75);
\draw (6,1)  --(7,0.25);
\draw (6,1)  --(7,1.25);
\draw (6,2.5)  --(7,2.25);
\draw (6,2.5)  --(7,3.25);
\draw (8,0)  --(7,0.25);
\draw (8,1)  --(7,1.25);
\draw (8,2)  --(7,2.25) ;
\draw (8,3)  --(7,3.25) ;
\draw (8,0.5)  --(7,0.25);
\draw (8,1.5)  --(7,1.25);
\draw (8,2.5)  --(7,2.25);
\draw (8,3.5)  --(7,3.25);
\draw [dashed] (3,1.75) --(4,1.75);
\draw [dashed] (3,1.75) --(2,2.5);
\draw [dashed] (3,1.75) --(2,1);
\draw [dashed] (1,2.25) --(2,2.5);
\draw [dashed] (1,0.25) --(2,1);
\draw [dashed] (1,3.25) --(2,2.5);
\draw [dashed] (1,1.25) --(2,1);
\draw [dashed] (1,2.25) --(0,2);
\draw [dashed] (1,0.25) --(0,0);
\draw [dashed] (1,3.25) --(0,3);
\draw [dashed] (1,1.25) --(0,1);
\draw [dashed] (1,2.25) --(0,2.5);
\draw [dashed] (1,0.25) --(0,0.5);
\draw [dashed] (1,3.25) --(0,3.5);
\draw [dashed] (1,1.25) --(0,1.5);
\draw (0.75,0) -- (1.25,0) -- (1.25,3.55) -- (0.75,3.55) -- (0.75,0);
\node at (4,2) {L = 0};
\node at (8,3.75) {L = 4};
\node at (1,-0.25) {$\mathcal{G}_R$};
\end{tikzpicture}
\vspace{0.5cm}
\caption[Recursion scheme for binary tree]{Binary tree of 4 levels. Each level separated between left and right branches. Each level within each branch can be considered as a vector involved in recursive calculation as they are not connected to each other by hamiltonian. }
\label{binarytree}
\end{figure}

The structure of the recursive calculations maps directly to  binary trees when each level of the branches of the tree is taken as a full vector involved in recursive calculations. These tree structures are also known as the Bethe lattice (with a boundary). The root node ($L = 0$) of the tree splits into two branches of same level ($L = 1$). Each node on these branches splits into two different nodes.  Any node within the vectors does not connect to each other by the hamiltonian and each such vector is connected to nearest neighbor vectors only.
The two boundary conditions necessary for the computation of the Green's functions   correspond to the vectors at the highest level on the left and right branches as shown in Fig. \ref{binarytree}. Systems such as binary trees not only act as a model system interesting for its mathematical form but  similar forms can be found in biological systems where transport of excitations may prove to be relevant.

\begin{figure}[H]
\centering
\includegraphics[width=0.8\textwidth]{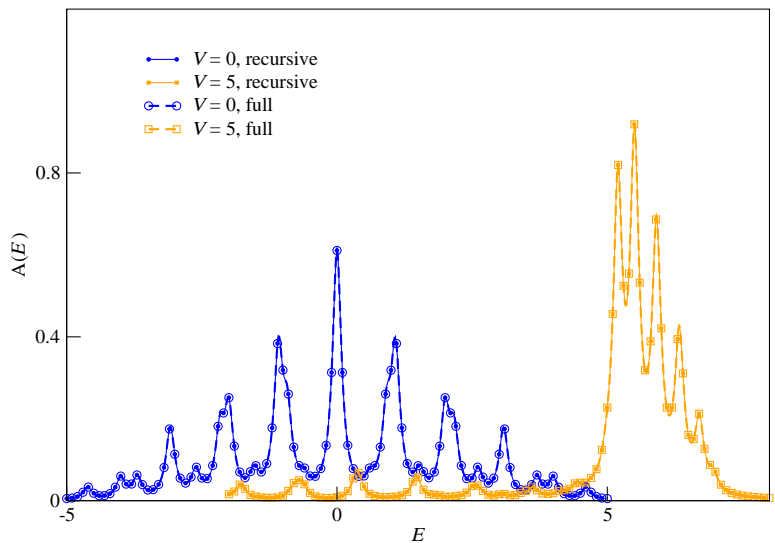}
\caption[Spectral weight of bound state in binary tree]{Spectral weight for two particles located on same site calculated for a binary tree of L=8. For non-interacting particles, the spectrum has multiple peaks with the highest peak at $E = 0$. For strongly interacting particles, the individual peaks are compacted into one peak.}
\label{binary-tree-spectra}
\end{figure}

The spectral weight of interacting particles on a binary tree can be calculated from Eq. \ref{spec}. For two particles placed on the same site of $L = 0$ on this graph (with maximum $L = 8$), Fig. \ref{binary-tree-spectra} describes the spectra. The spectrum shows discontinuous peaks as opposed to continuous spectra in 1D and 2D lattices. With stronger interactions, these peaks tend to merge together and a single continuous spectrum enveloping multiple peaks seems to be emerging. 

\section{Conclusion}

In this chapter two-particle  Green's functions have been  calculated efficiently using a recursive algorithm. These calculations provide insights into the problem of interacting particles. Possible extensions for calculations of response properties from two-particle correlations can be  avenues of further research. In the next chapter we attempt to understand the behaviour of two particles and their correlations in disordered one- and two-dimensional systems.

\newpage

\chapter{Quantum Localization of Interacting Particles}

     After a few experimental observations from 1990s  \cite{ambegaokar, chandrasekhar, levy, kravchenko1, kravchenko2}, there has been renewed interest in understanding the effect of interactions on the localization of particles in 1D and 2D systems. These experiments had reported observations of persistent currents in 1D wires  \cite{ambegaokar, chandrasekhar, levy} and a localization-delocalization transition in 2D lattices \cite{kravchenko1, kravchenko2}. This is of high interest as the scaling theory  \cite{scaling} predicts an absence of such transition in 1D and 2D systems. Since then there has been a plethora of studies. Investigations on whether the inter-particle interaction is responsible for such phenomena were started immediately. To understand the effect of interparticle interaction on localization, understanding the case of two particles was necessary. However, while some studies \cite{shepelyansky, imry, ponomarev} predicted the effect of interaction in delocalizing the particles in disordered lattices, some numerically found that the interaction-induced delocalization effect is limited to weak interaction cases \cite{flach, oppen} and for strong interactions the two particles become more localized \cite{pichard, schreiberm1}. The differences arose from the calculations using random matrix theory \cite{guhr}. Some studies have also noted a universal sub-diffusive behaviour after transient localization induced by the interaction \cite{flach2}. The localization-delocalization transition was also supported by some numerical studies in 1D \cite{schreiberm2, schreiberm3, schreiberm4} and in 2D \cite{ortuno}, although the latter were based on significant approximations.

In this chapter we perform numerical calculations to understand not only the effect of interaction on localization in 1D systems, but also the effect of the range of both tunnelling and interaction. In the case of 2D, where calculations are very difficult to perform, we apply the recursive method described in Section 3.1 and find localization parameters for the short-range tunnelling and interaction case. 

\section{Scattering with single impurities in 1D}

\begin{figure}[H]
\centering
\includegraphics[width=1.0\textwidth]{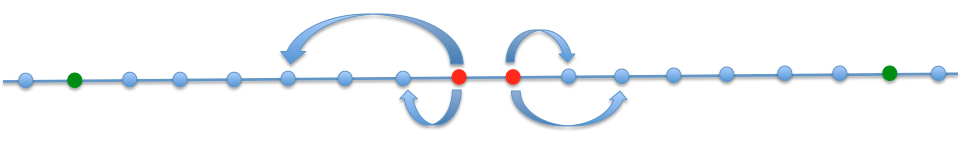}
\caption[Two particle scattering in 1D with impurity]{Initial preparation of two particles (red)  with two impurities placed towards the edges (green) of the 1D lattice.}
\label{1d-initial}
\end{figure}

Here we study the case of two interacting hardcore bosons in 1D systems by exact diagonalization. We first study the   particles interacting with impurities. In particular the  case of two particles initially placed in adjacent sites, in the middle of a 1D lattice with two impurities, each placed towards the edges of the lattice as in Fig. \ref{1d-initial}. We let the wavepackets tunnel out of the impurities for a certain time and examine the dependency of that dynamics on inter-particle interactions.  We note that weak interaction increases the tunnelling of the particles through the impurities while strong interactions reduce  the tunnelling. The long-range nature of tunnelling permits tunnelling through the impurities.  We observe  that the particles with strong interactions get bound hence heavier as their dispersion also becomes flatter, resulting in the slow tunnelling through the impurities. The impurities were modelled by $\delta$-function potentials and particles outside the impurities were assumed to not scatter back inside the impurities again. 

The hamiltonian is as given in Eq. \ref{H2p1d},

\begin{equation}\label{disorder_Ham}
\mathcal{H} = \sum_m \varepsilon_m a_m^\dagger a_m + \sum_{mn} t_{mn} a_m^\dagger a_n + \sum_{mn} V_{mn} a_m^\dagger a_n^\dagger a_n a_m
\end{equation}
where 

\begin{equation}\label{lattice_impurities}
 \varepsilon_m = 0 ~~~~ \forall ~~~ m \neq m_1, m_2 ~~~~\mbox{and} ~~~~ \varepsilon_m = \infty ~~~ \mbox{for}~~~ i = m_1, m_2
\end{equation}
and

\begin{equation}\label{lr-hoping}
t_{mn} = \frac{t}{|m-n|^\alpha} ~ \mbox{,} ~~~~~ V_{mn} = \frac{V}{|m-n|^\beta} ~~~~~ \mbox{for}~~~  \alpha, \beta \in \{1, 3, \infty\}
\end{equation}



\begin{figure}[H]
\centering
\includegraphics[width=0.8\textwidth]{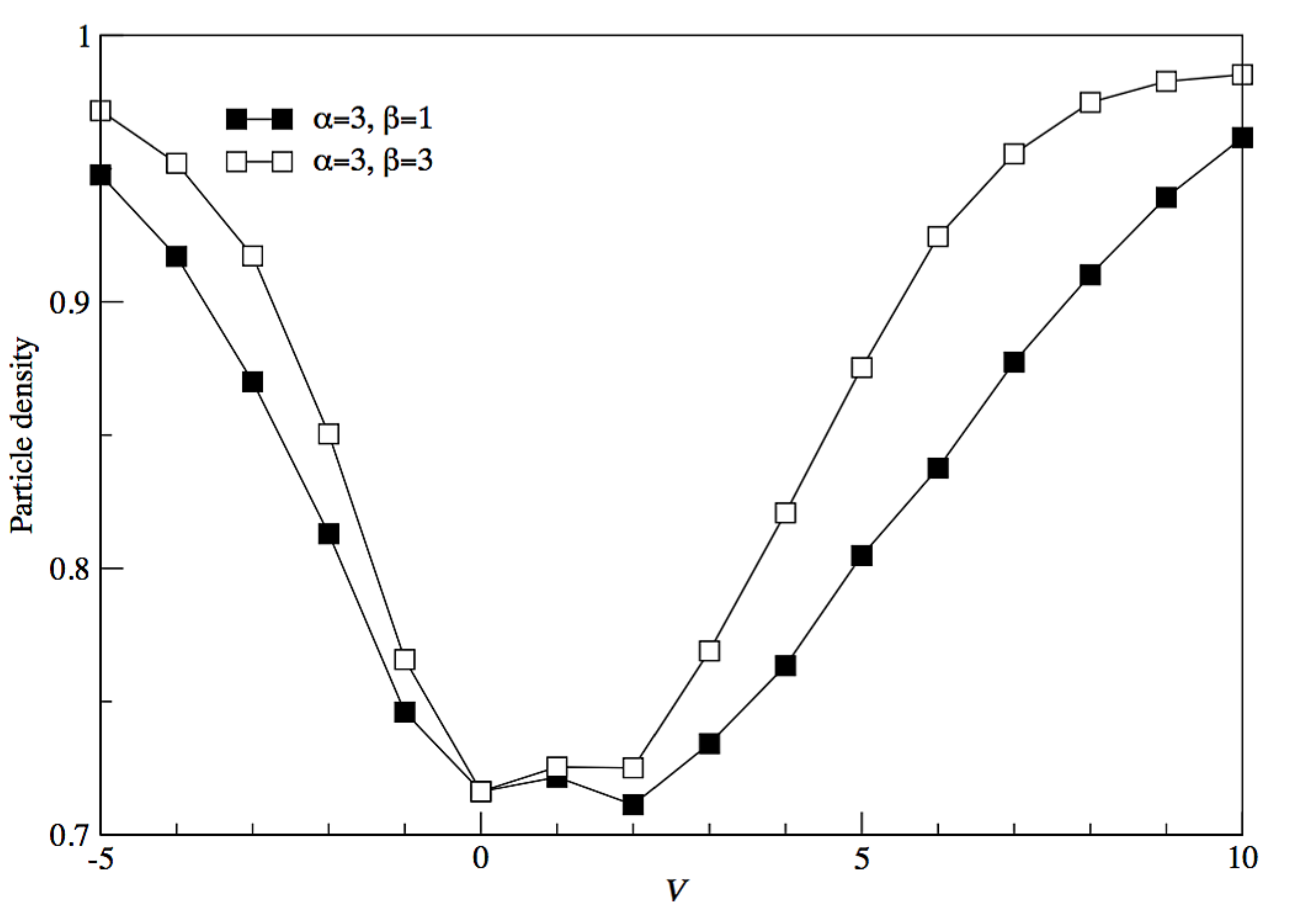}
\includegraphics[width=0.8\textwidth]{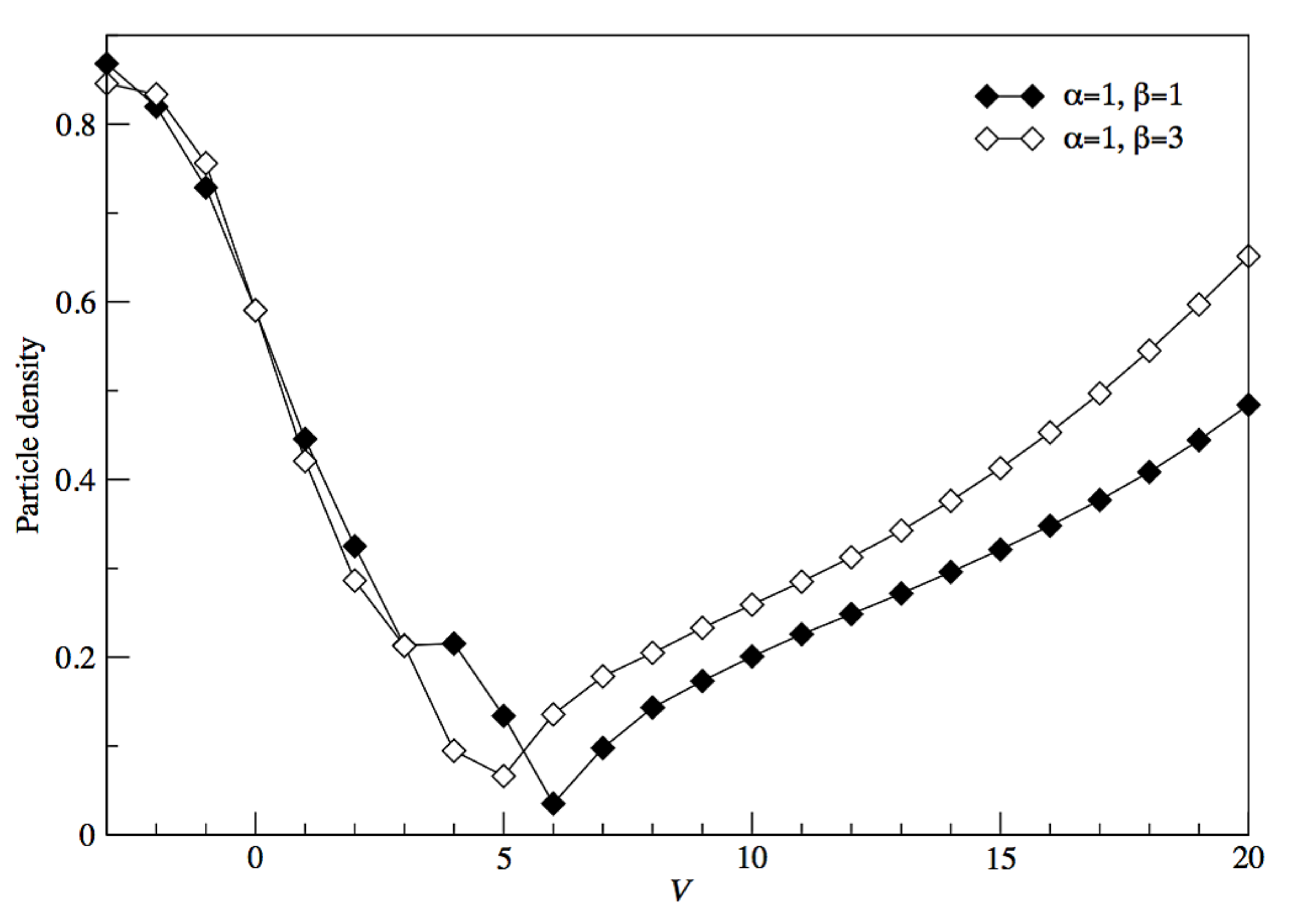}
\caption[Tunneling of composite particles through impurity]{Tunneling out of interacting particles through impurities in case of long-range hopping and interaction Eq. \ref{lr-hoping}. The particle density is the density remaining after time $20/t$ within a length of 10 sites between two impurities.}
\label{1D-2-impurity}
\end{figure}

Figure \ref{1D-2-impurity} shows the wavepacket density remaining inside the impurities after a certain time allowing multiple scattering with the impurities. The effect of the interaction increases the tunnelling through the impurities in the weak interaction cases and increases trapping of the particles in the strong interaction cases. For long-range hopping, the particles tunnel out faster as expected, however, the long-range nature of interaction has  very minimal effect on controlling the scattering through the impurities compared to the effect of long-range nature of tunnelling. As Fig. \ref{1D-2-impurity} illustrates, the tunnelling probability goes through a maximum for an interaction strength $V/t > 0$  for both dipolar  and Coulombic isotropic hopping. The asymmetry in tunnelling with respect to the sign of the interaction in the case of long-range hopping can also be noted. In these calculations the particles that tunnel through the impurities were dynamically removed from the calculations, with very short time steps in the unit of the inverse of the hopping parameter $t$ (typically $1000/t$). The length between the two impurities is chosen to be 10 sites to allow a few scattering events to take place. However, the particles with strong attractive interaction and very strong repulsive interaction behave as very slow particles. For very strongly bound particles the number of scattering events is less compared to weakly attractive particles which exhibit faster dynamics. At larger times the interference between the scattered part of the wavepackets within the impurities makes the character of the propagating wavepacket different from that of purely bound wavepacket projected toward impurities, which further modifies the scattering with the impurities. A time of $20/t$ was chosen for the results plotted in Fig. \ref{1D-2-impurity}. For the weakly interacting particles, when a sufficient number of collisions with the impurities is allowed, it is found that the weak repulsively interacting particles tunnel more through the impurities than the non-interacting ones. 

Longer tunnelling range leads to more tunnelling through the impurities. For the long-range interactions, the effect is not very different from the short-range interactions between the particles. However, for weakly repulsive interactions, the short-range interaction leads to more tunnelling than in the long-range interaction cases. This can be seen prominently present in the case of long-range tunnelling in Fig. \ref{1D-2-impurity}. For strong repulsive interaction, the short-range interactions lead to less tunnelling compared to the long-range interaction cases.

\section{Localization in 1D}

After gaining some insight into the scattering with isolated impurities, a distribution of impurities is placed (Eq. \ref{lattice_impurities}) in the lattice to understand the localization properties of disordered 1D systems. The disorders implemented here are both onsite and offsite in nature. The sites that cannot be occupied become disconnected from the rest of the lattice. The long-range character of the hopping makes it possible for the particles to hop over the impurities. After the dynamics is frozen at a very long time compared to the hopping parameter ($\tau \gg t^{-1}$), the joint densities ($\varrho_{mn}$) and densities ($\rho_m$) were calculated from the exact eigenfunctions and eigenenergies in the two-particle basis. The density-density correlations ($C_{mn}$) are nothing but the joint densities calculated from the two-particle basis

   \begin{equation}\label{1Dfull}
       \varrho_{m,  n}(\tau)= |\langle m,  n | \sum_{\lambda} e^{-\imath E_{\lambda} \tau} |\lambda\rangle \langle \lambda| m',  n'\rangle|^2
    \end{equation}

   \begin{equation}\label{1Dfull}
      C_{mn} (\tau) =  \varrho_{m,  n}(\tau)
    \end{equation}
   \begin{equation}\label{density}
     \rho_m (\tau)  =  \frac{1}{2} \sum_{ n\neq  m}  \varrho ( m,  n, \tau).
    \end{equation}
Alternatively, one can calculate the localization length as suggested by Oppen et al \cite{oppen} from Green's functions and gain insight into the localization behaviour.

\begin{figure}[H]
\centering
\includegraphics[width=0.8\textwidth]{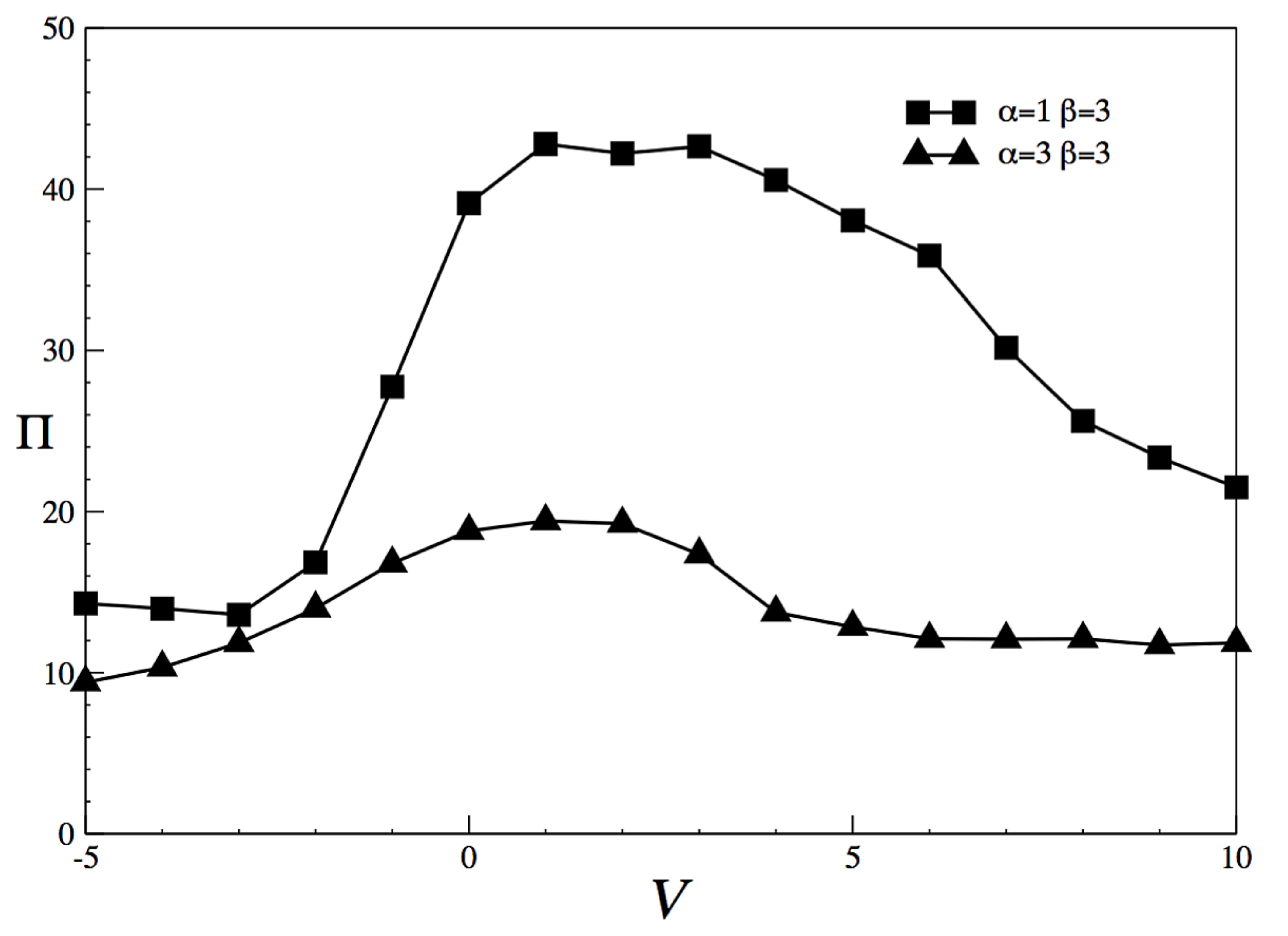}
\caption[Localization of interacting particles in 1D]{Localization parameter, participation ratio ($\Pi$)  calculated from dynamics of two interacting particles for interaction strength $V$ initially occupying two sites side by side in disordered 1D lattices for the cases of long and short-range hopping in presence of short-range interaction. Averaged over 5000 realizations.}
\label{1D-localization}
\end{figure}

The inverse participation ratio (IPR) of second rank is calculated from the density distribution as in the following equation
   \begin{equation}\label{ipr}
       \mathcal{I} =  \lim_{\tau \gg t}  \sum_{m}\rho_m (\tau)^2.
    \end{equation}
The participation ratio ($\Pi = \mathcal{I}^{-1}$) is the parameter which gives the number of sites participating in the distribution and hence is larger for the delocalized systems. On the other hand, a higher inverse participation ratio refers to more localized states. Figure \ref{1D-localization} presents the calculations performed using the method of full diagonalization for a lattice of 50 sites. The lattice was disordered by 10$\%$ of vacancies and the results were averaged over 5000 such disorders. It can be clearly observed from our calculations that the particles become more localized for the strong interaction cases compared to the non-interacting ones. The weak repulsive interaction, however, reduces the localization of the particles. 

\begin{figure}[H]
\centering
\includegraphics[width=1.0\textwidth]{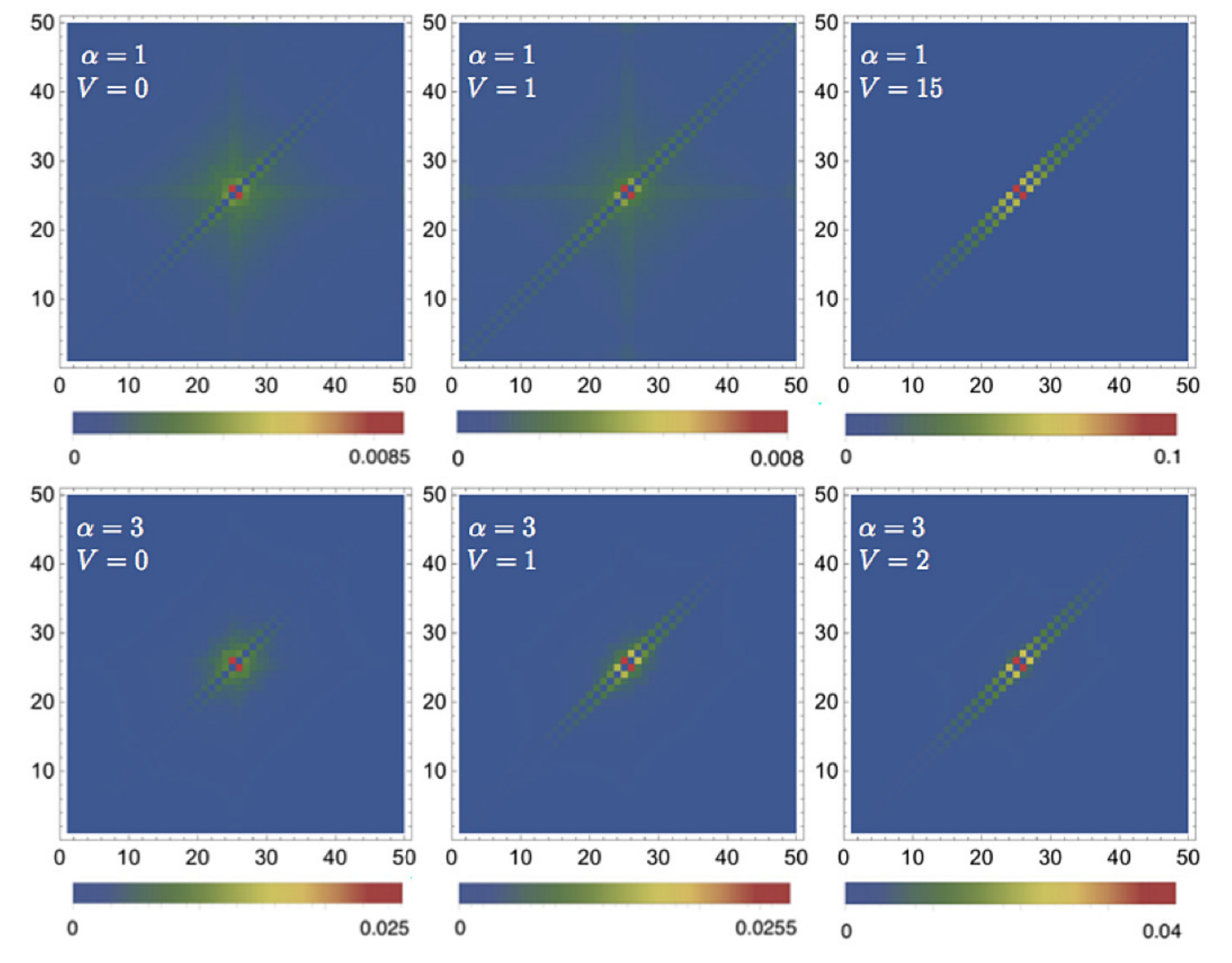}
\caption[ Correlations of localized interacting particles in 1D]{Correlations ($C_{mn}$, shown in the legends) calculated from dynamics of two interacting particles initially occupying adjacent sites in disordered 1D lattices for the cases of long and short-range hopping ($\alpha = 1, 3$) in presence of short-range interaction ($\beta = 3$) as in Eq. \ref {lr-hoping}. Averaged over 5000 realizations.}
\label{1D-correlation-in-localization}
\end{figure}

The correlations ($C_{mn}$)  between  particles in disordered lattices  show an enhancement of cowalking between the particles. Figure \ref{1D-correlation-in-localization} clearly illustrates disorder induced enhancement of cowalking correlations even for the weakly interacting particles. For non-interacting particles, emergence of correlations in between that of cowalking and antiwalking is observed. It can also be seen that the cowalking correlations extend toward the edges more prominently than any other correlations. It can be inferred that, if the correlations in disordered systems are measured, there will be a high probability of finding the particles close together. Figure 4.4 also shows, that in disordered cases, even in the weak interaction limits, there are very few correlations that are important. The particles might be spread over a large part of the lattice depending on the localization length but only a few correlations, mainly that of the cowalking type, should be taken into account in any such calculations for disordered systems.

\section{Localization in 2D}

To gain understanding on localization properties of 2D systems the same Hamiltonian as in Eq. \ref{disorder_Ham}
is simulated with onsite energies ($\varepsilon_m$) selected  randomly from a uniform distribution of fixed width ($W$)
\begin{equation}
\varepsilon_m \in \left[-\frac{W}{2}, \frac{W}{2}\right].
\end{equation}
The calculations of localization properties for two interacting particles in two dimensional disordered systems cannot be done by the method of full diagonalization as the basis size grows beyond what can be accounted for, even in the case of small 2D systems. To perform such calculations, we use the recursive Green's function method described in Section 3.1. The recursive method breaks down the full problem into multiple smaller size matrix-vector multiplications which make the calculations more efficient while maintaining accuracy. 

For a fairly large 2D lattice of 50 sites per dimension, a full diagonalization  for two particles would entail a total basis size of around three million Green's functions. This scale is impossible to fully diagonalize  even with the help of most sophisticated computers.

\begin{figure}[H]
\centering
\includegraphics[width=0.49\textwidth]{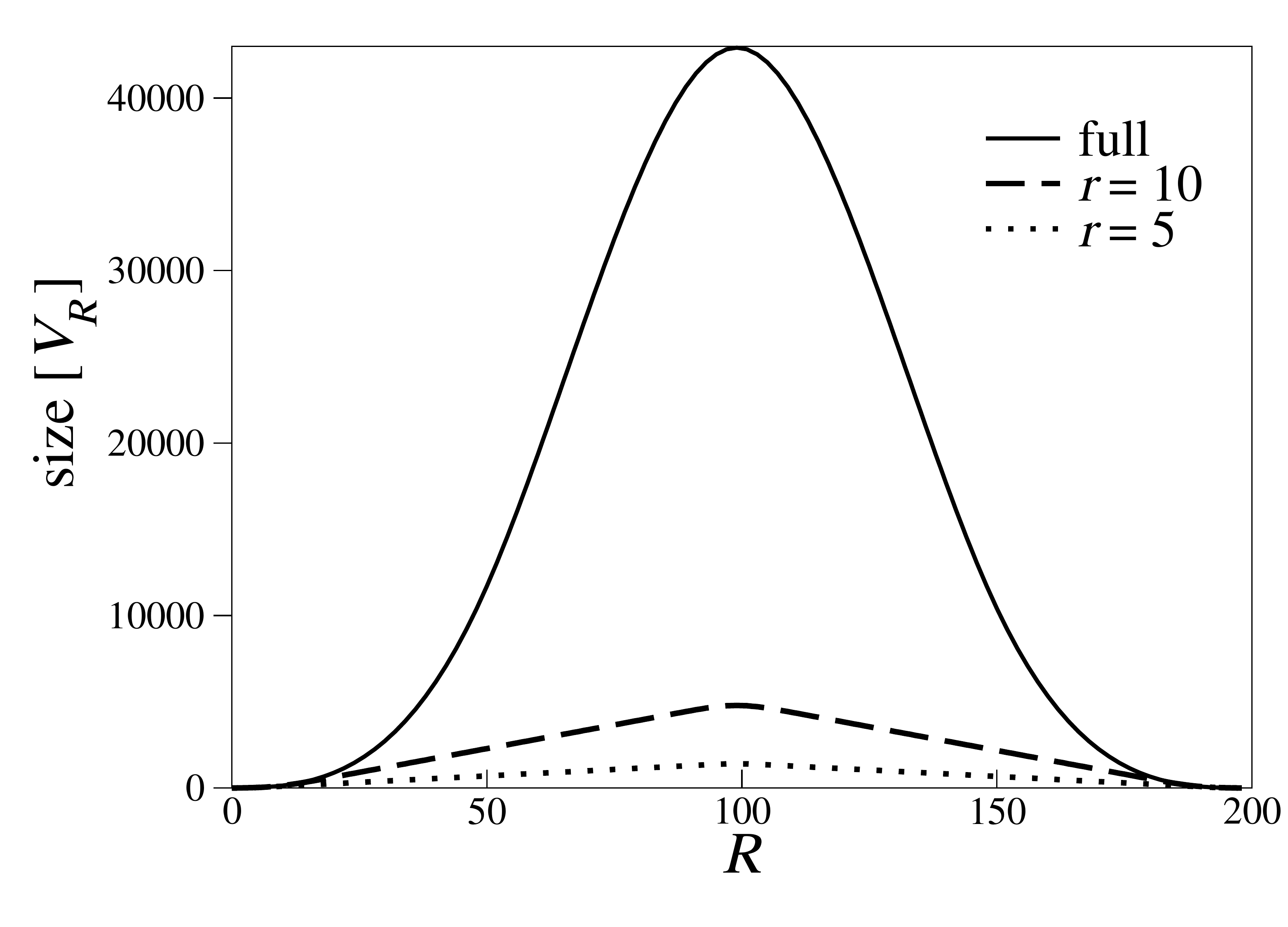}
\includegraphics[width=0.49\textwidth]{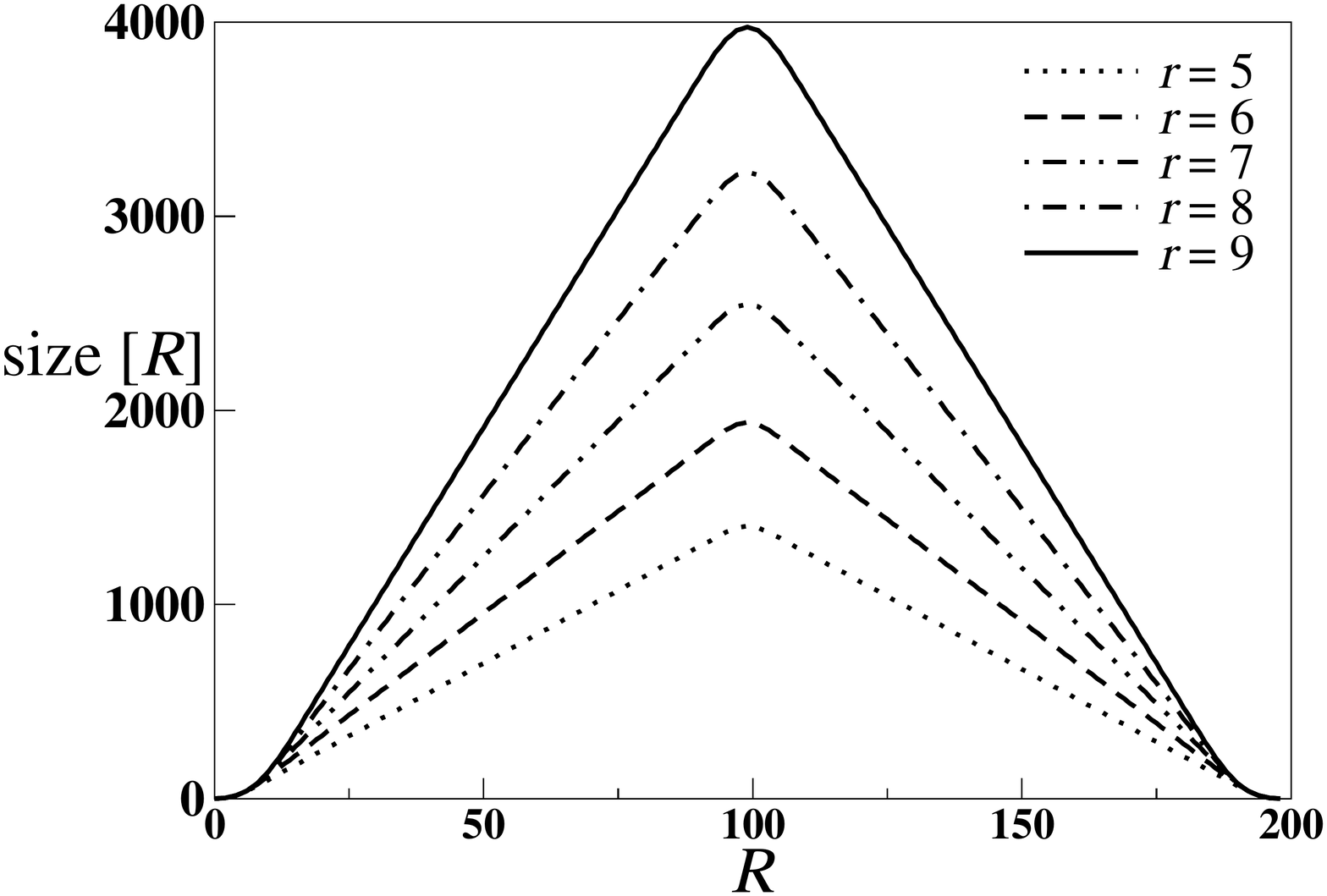}
\caption[Recursive calculation size for a large 2D lattice]{Sizes of vectors involved in recursive calculations for a 2D lattice with 50 sites per dimension. The Gaussian shaped full distribution reduces to a triangular distribution in presence of approximations.}
\label{2D-ksize}
\end{figure}

As shown in Fig. \ref{2D-ksize}, the total number of elements to be considered in the calculations even after applying the  approximations as referred in Eq. \ref{limiting-distance}, can become as large as a few hundred thousand (apply the triangle area law to get the total number from the figures). The recursive algorithm can break the calculation to those with vectors having a few thousand of Green's functions as shown in the figure. As the calculations involve inversion of matrices, this reduction makes the calculations significantly more efficient compared to full diagonalization. However, one now has to perform calculations over many search points effectively doing the same iterations many times to understand the dynamics and correlations of the interacting particles.   

The recursive method allows the exact calculations of these Green's functions by taking advantage of the sparsity of the whole matrix.  With the recursion method, the full calculation is split  into a Gaussian-shaped distribution of vectors as shown in Fig. \ref{2D-ksize}. The vectors are coupled as explained in Chapter 3 through Eq. \ref{vrec}. However as can be observed from Fig. \ref{2D-ksize}, a full calculation for a fairly large 2D lattice of 50 sites per dimension, even with the help of recursion, remains difficult as it involves tens of matrices with dimensions of the order of tens of  thousands, to be considered a few hundred times for every energy point within the band. This implies  enormous computational time and resource requirements. 

\begin{figure}
\centering
\includegraphics[width = 0.32\textwidth]{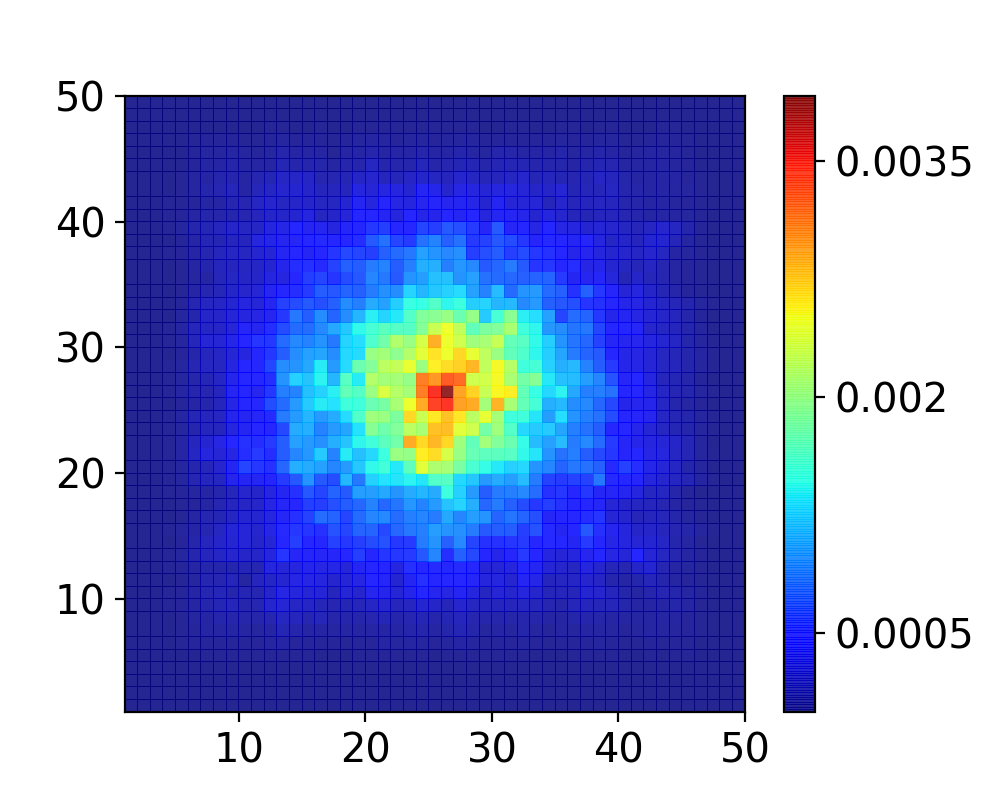}
\includegraphics[width = 0.32\textwidth]{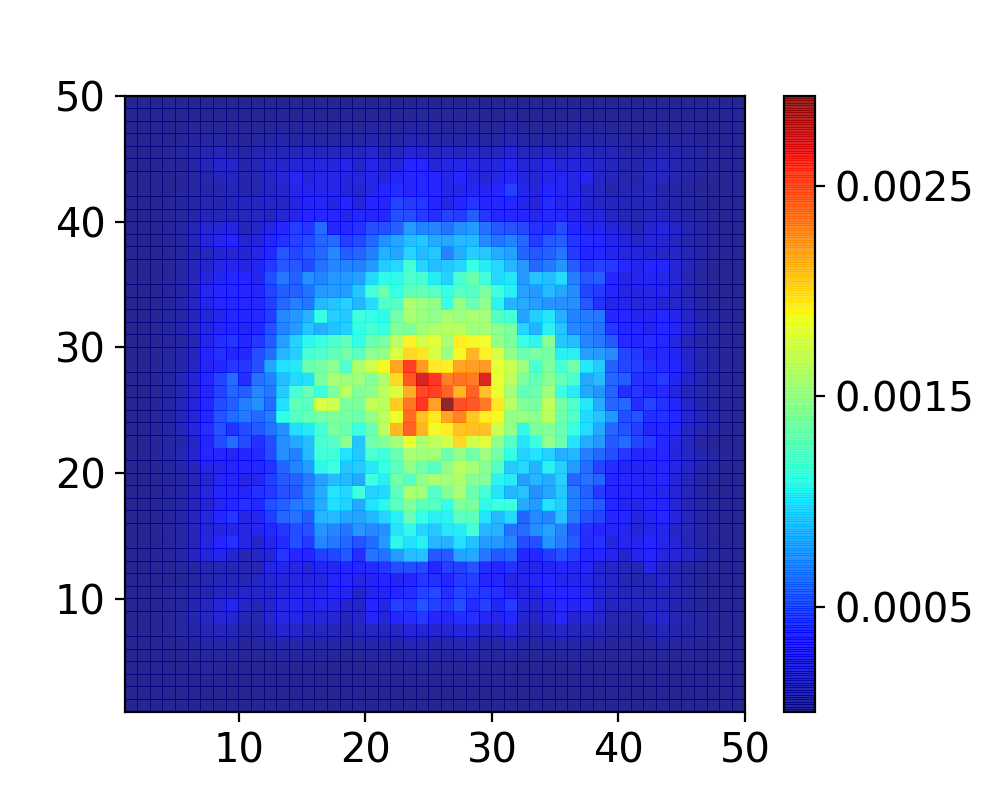}
\includegraphics[width = 0.32\textwidth]{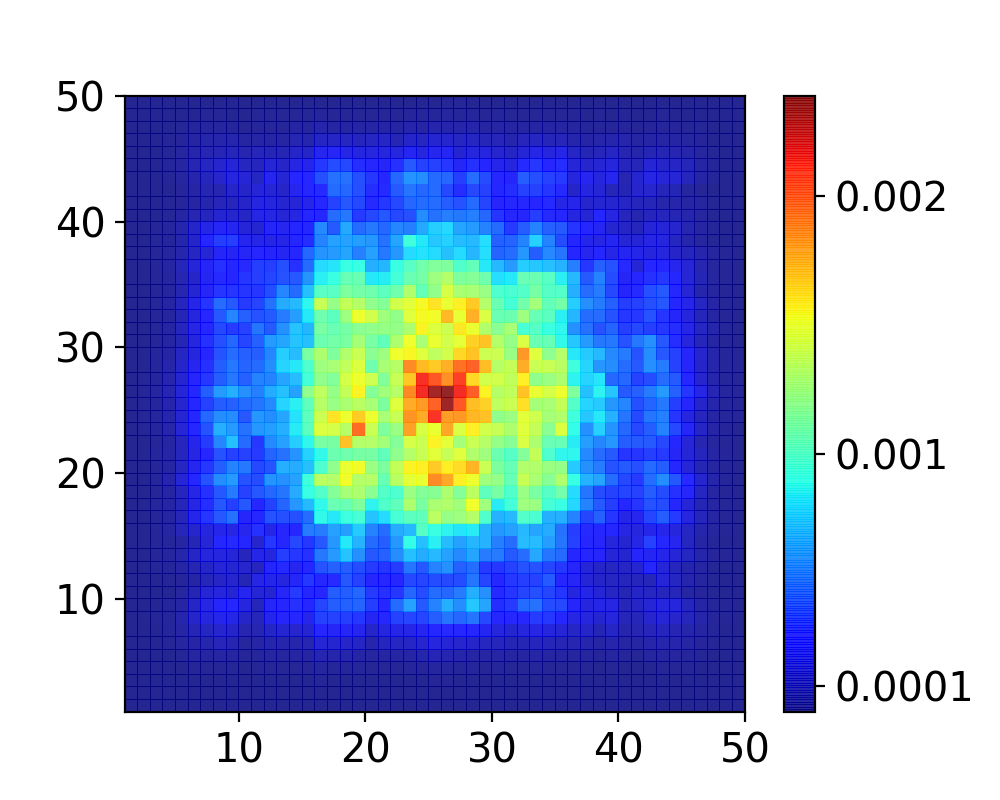}
\caption[Density distribution in disordered 2D lattice]{The density distribution for a single realization of same disorder in a 2D lattice with 50 sites per dimension calculated using approximation of Eq. \ref{limiting-distance}. Left panel show the density for $r = 5$, middle panel for $r = 7$ and right panel for $r = 9$.}
\label{2D-density}
\end{figure}

The approximation of the maximum relative distance that has been applied for the calculations of the time-dependent densities in Fig. \ref{2D-density}, for a disordered 2D lattice of 50 sites per dimension, makes the calculations significantly faster. The approximations can be used to reduce the total number of Green's functions involved in the calculation from a few millions to a few hundred thousands. These total number of elements, in the case of a small maximum relative distance to the total lattice size, has a linear distribution of the elements. 

\begin{figure}
\centering
\includegraphics[width = 0.32\textwidth]{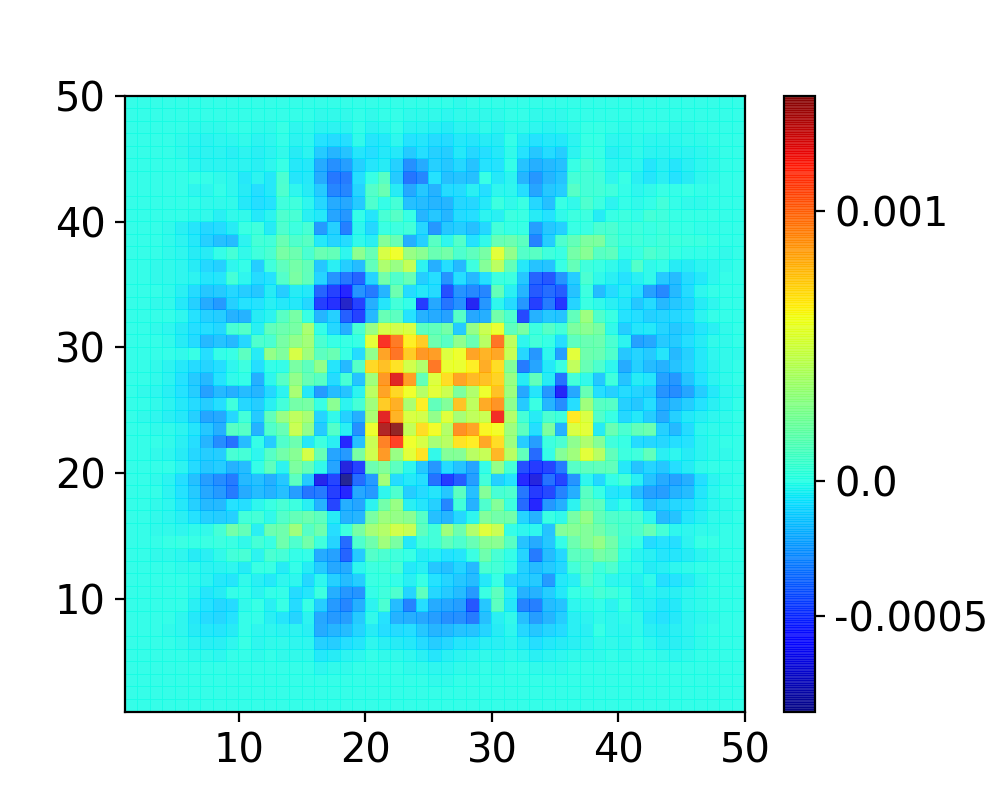}
\includegraphics[width = 0.32\textwidth]{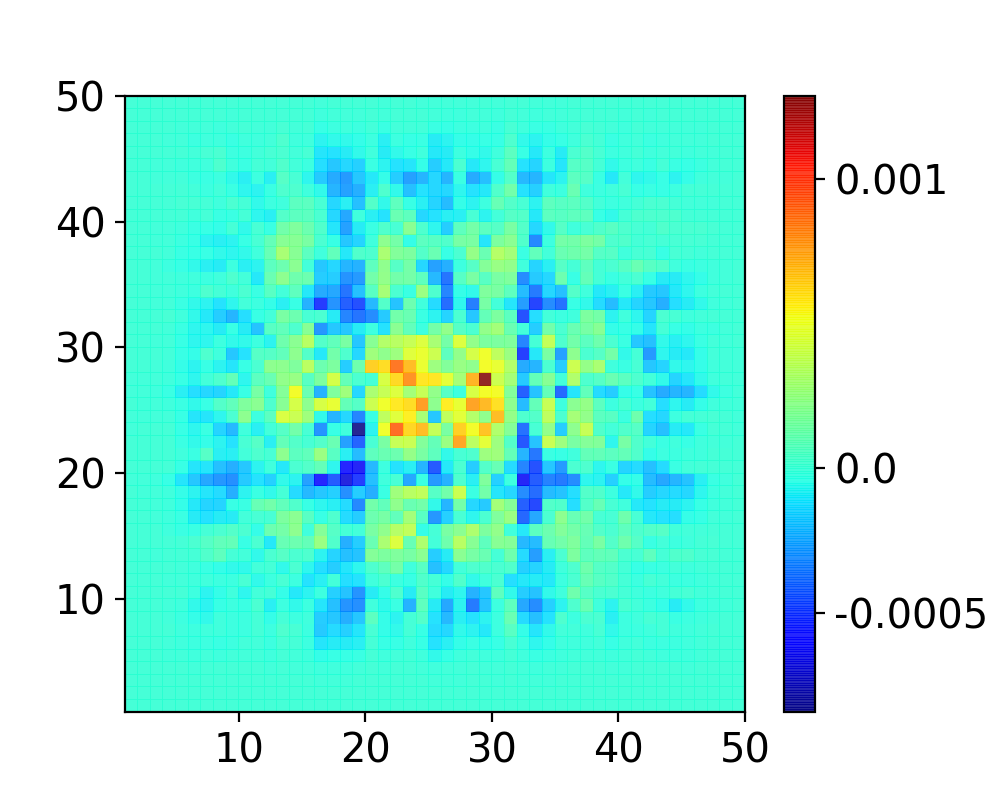}
\includegraphics[width = 0.32\textwidth]{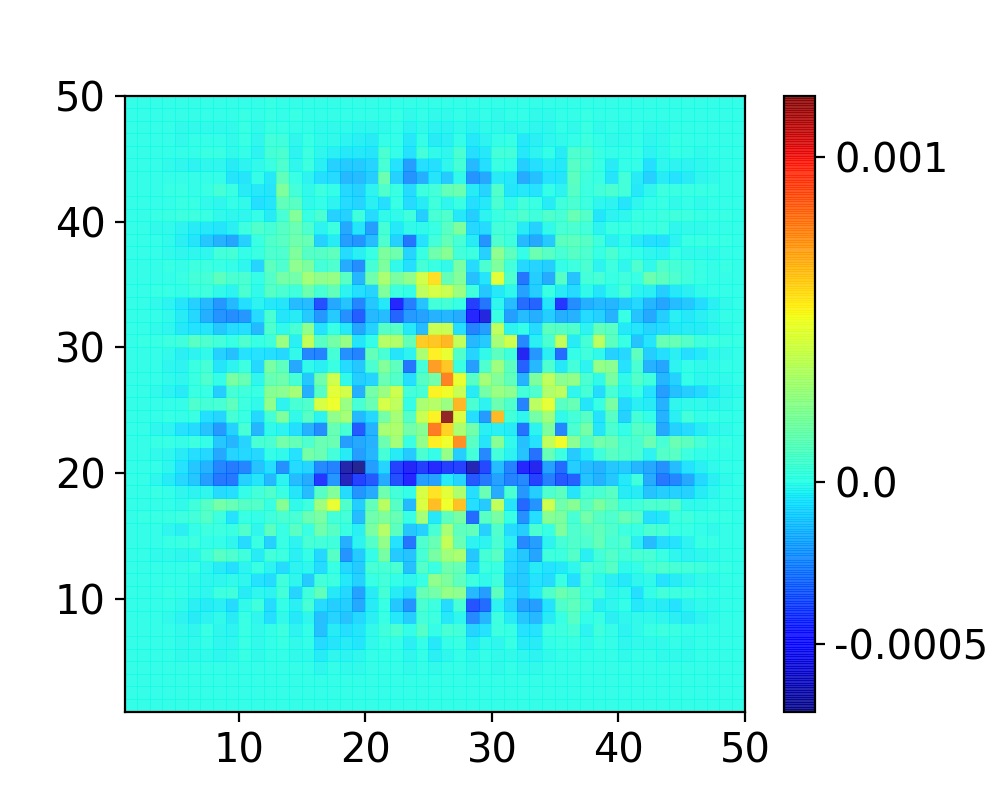}
\caption[Comparison between densities with approximations]{Difference in density distribution for a single realization of same disorder in a 2D lattice with 50 sites per dimension calculated using approximation of Eq. \ref{limiting-distance}. Left panel show difference between $r = 6$ and $r = 9$, middle panel between $r = 7$ and $r = 9$, and right panel  between $r = 8$ and $r = 9$.}
\label{2D-density-error}
\end{figure}

The method lets us perform the calculations which would be impossible  by full diagonalization, and is highly accurate and efficient as described in the previous chapter. With the method, we can proceed to take on the challenge of calculating the dynamics of a few interacting particles in disordered 2D lattices. We can calculate the localization parameters such as the inverse participation ratio (IPR) or any Green's function of interest from such calculations. As shown in Fig. \ref{2D-density}, the density distribution of two weakly interacting particles in a weakly disordered 2D lattice  appears to be localized. Comparisons between different degrees of approximations (increasing \textit{r}) with same disorder show that the density distributions are converging, as can be seen from Fig. \ref{2D-density-error}.  The density distributions are calculated for a single realization of fixed disorder. The differences between the approximations are not significantly large, even in the absence of averaging over many realizations of disorders, which indicates that such approximations, limiting the relative distance, can be used to calculate the properties of disordered lattices.

Alternatively, only a few Green's functions of interest are needed to gain insight into the localization properties, as suggested by von Oppen et al \cite{oppen}. However, a medium-size lattice that can be considered for the calculations by the recursion method will produce significant finite size effect, and render the calculations of localization lengths from Green's functions involving edges of the lattice highly inaccurate. Thus, the macroscopic properties such as the IPR were employed to understand the localization behaviours.

As described in the previous chapter, for calculations of the localizations properties, one requires  averaging over many realizations of disorder. The averaging minimizes differences in results between different realizations of disorders and takes account of the different degrees of randomness in each different realization of disorder. The averaging also produces a density distribution that can be expected of any realization of disorder. 

As explained before, even after approximations, a fairly large lattice size would be difficult to consider for the computation of the localization parameters. These calculations have to not only take into account the number of times the recursion has to be performed for each point of energy within the bandwidth, but also the number of times the same calculations have to be performed for averaging for each realization of disorder. However, from Fig. \ref{2D-IPR-scaling}, it can be observed that even in the ranges of weakly disordered and weakly interacting cases, a lattice of medium size, such as containing 20 sites per dimension, won't produce significant errors. These errors are found to be in the range of 10-20$\%$. Thus, the limitations that we confront, force us to make a choice of doing the calculations for a medium sized lattice for the localization calculations.

We try to find the overall pattern from the localization parameters over a vast range of disorder in the lattices and the interactions between particles. From these patterns, we attempt to infer if any localization-delocalization transition  exists  for a 2D disordered system of interacting particles. We also calculate the correlations between the particles over these vast ranges of disorder and interaction strenth, which helps us to understand the effect of disorder on correlations of the interacting particles that cannot be obtained from density distributions.

More importantly, these calculations for medium-size lattice indicate the length scales involved in localization of interacting particles in regular 2D lattices and how the interactions and disorder affect these length scales. The exact numbers that we find from our calculations might vary because of inherent errors due to the size of the lattices and the approximation that has been applied. However, a broad initial understanding can be achieved from these calculations.

As explained in Eq. \ref{joint-density}, these calculations directly compute the correlations between particles even in the disordered systems. These correlations reveal the underlying structures of localized particles in disordered systems. Measurements of total correlations for different relative distances reveal how the most probable relative distance between the particles changes when disorder and interaction strengths are varied.  

\begin{figure}
\begin{tikzpicture}
     \node[anchor=south west, inner sep=0] at (0,0) {\includegraphics[width=1.0\textwidth]{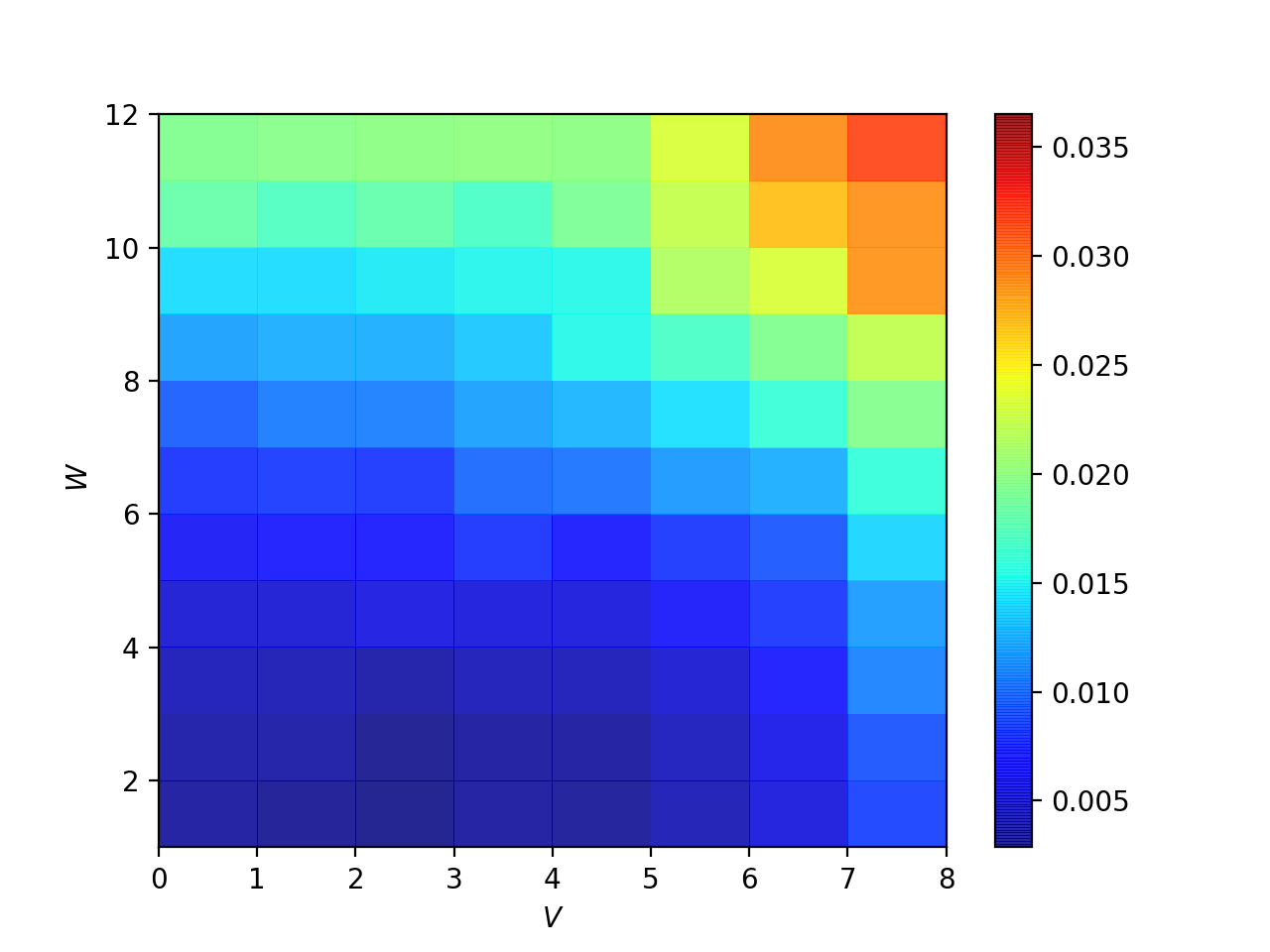}};   
     \draw[black,ultra thick,rounded corners] (3.1,9.2) rectangle (4.25,10.0);
     \draw[black,ultra thick,rounded corners] (4.25,7.6) rectangle (5.4,8.4);
    \draw[black,ultra thick,rounded corners] (5.4,6.0) rectangle (6.55,6.8);
     \draw[black,ultra thick,rounded corners] (6.6,5.2) rectangle (7.75,6.0);
     \draw[black,ultra thick,rounded corners] (7.75,5.2) rectangle (8.9,6.0);
     \draw[black,ultra thick,rounded corners] (8.9,4.4) rectangle (10.05,5.2);
     \draw[black,ultra thick,rounded corners] (10.05,2.0) rectangle (11.25,2.8);
\end{tikzpicture}
\caption[Disorder-interaction diagram of IPR]{Two dimensional disorder-interaction diagram with black squares showing the regions where the nature of correlations changes. The inverse participation ratios plotted are averaged over 320 realizations of disorder, shown in the color legend.}
\label{2D-IPR-2D}
\end{figure}
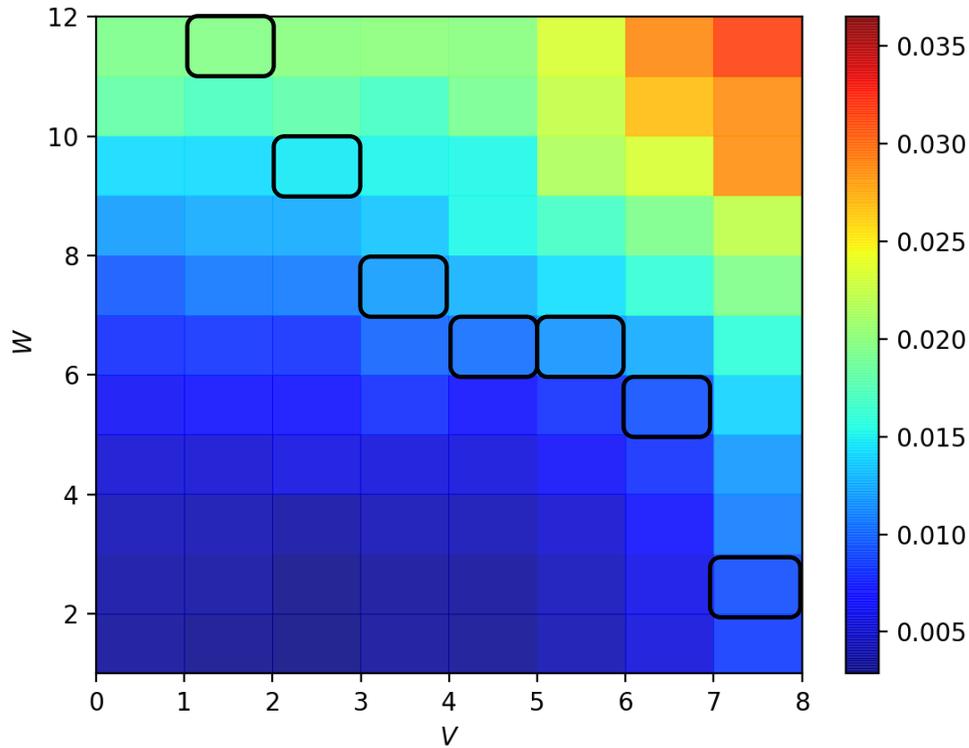

\begin{figure}[H]
\centering
\includegraphics[width = 1.18\textwidth]{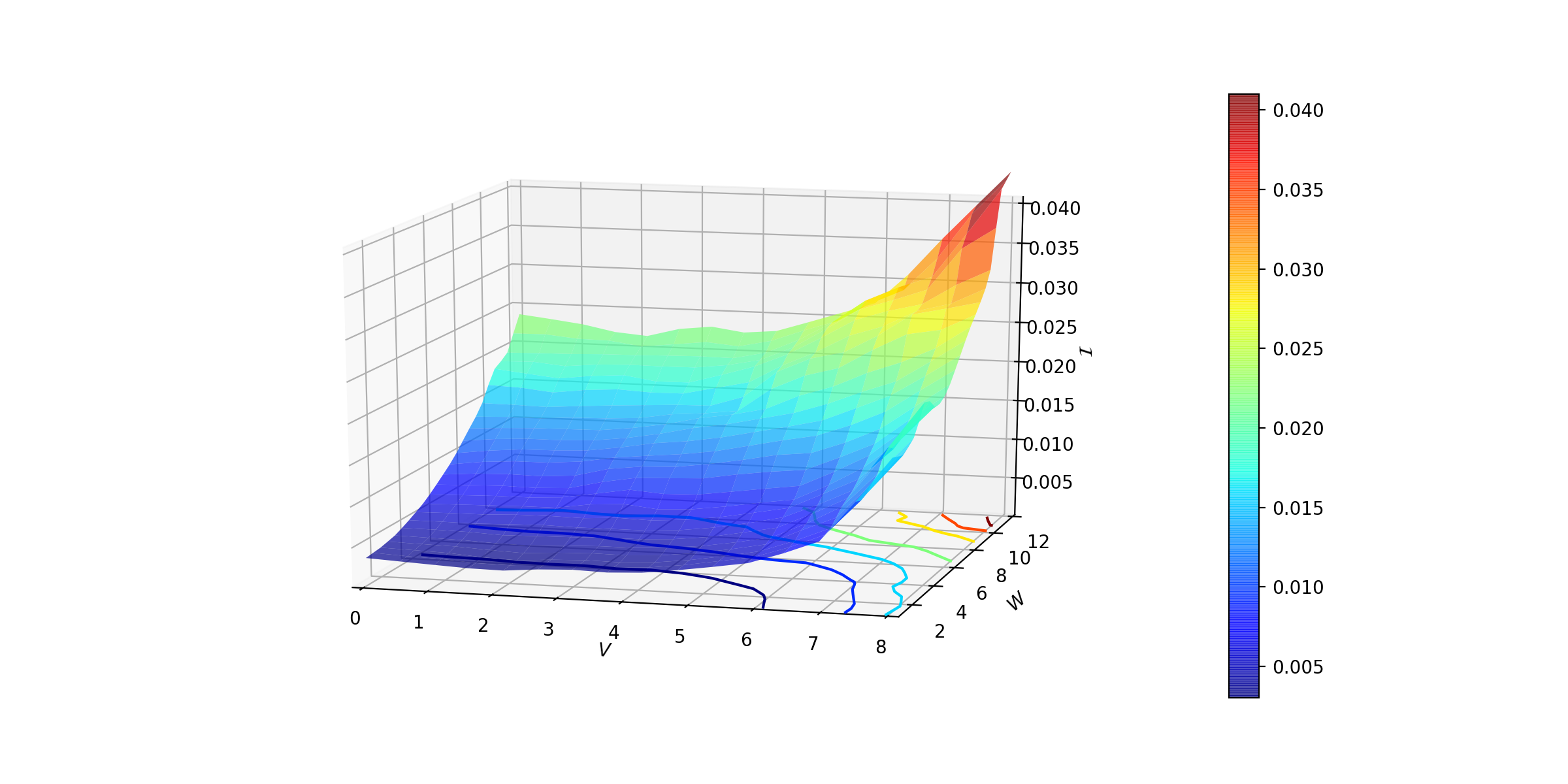}
\caption[3D disorder-interaction diagram of IPR]{3D disorder-interaction diagram of IPR computed for a broad range of interactions ($V = 0$ to $V = 8$) between particles and disorder strengths ($W = 1$ to $W = 12$) in 2D lattices. Averaged over 320 realizations of disorders.}
\label{2D-IPR-3D}
\end{figure}

In Fig. \ref{2D-IPR-2D}, the IPR is plotted for a broad range of the interaction and disorder. The same values are shown in Fig. \ref{2D-IPR-3D}  where a three-dimensional plot with the contours reveals the difference between the weak interaction - weak disorder limits and the strong interaction - strong disorder limits.  For the case of 2D lattice systems, these limits can be equated to half of the full bandwidth ($V = 4$, $W = 4$) , where the systems are found to be least localized.  The disorder strength is varied from $W =1$ to $W=12$. The interaction strength is varied from $V = 0$ to $V=8$. The work involves every combination of the  integer points for both the disorder and the interaction strength within this broad range.

\begin{figure}[H]
\centering
\includegraphics[width = 0.75\textwidth]{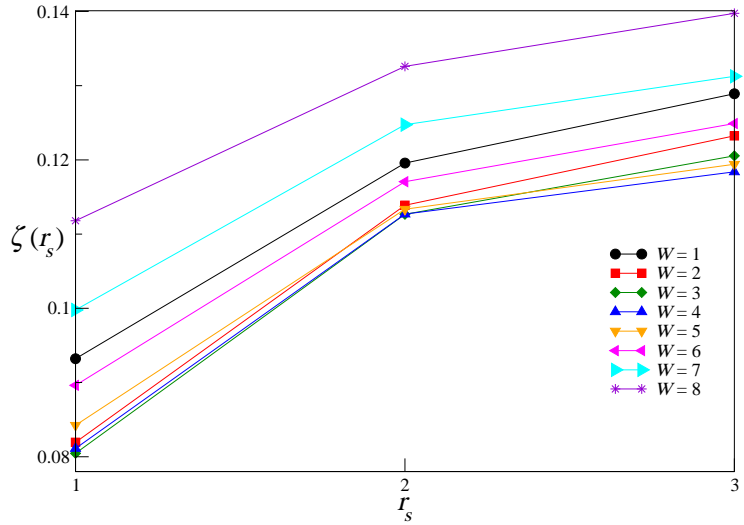}
\includegraphics[width = 0.75\textwidth]{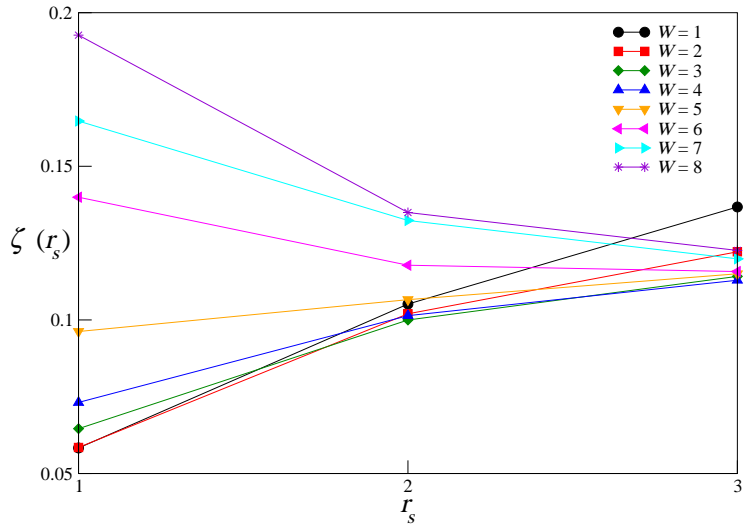}
\caption[Correlations in disordered 2D lattics]{Total correlations in disordered 2D lattics between for two interacting particles with increasing relative distances. Top and bottom panels show the correlations between particles for $V = 0$ and $V=4$ respectively for disorder strengths ranging from $W=1$ to $W=8$. Averaged over 320 realizations of disorders.}
\label{2D-disordered-correlation}
\end{figure}

In Fig. \ref{2D-IPR-3D}, the difference can be noted between the two contours colored as light blue and light green. It is an indication of some physical difference between the two regions. In Fig. \ref{2D-IPR-2D},  the squares highlighted by black rectangles signify a change in the prominent character of the correlations. On these marked squares, the nature of correlations changes from that of predominantly nearest neighbor ones to that of next nearest neighbor ones, between the localized particles. Figure \ref{2D-disordered-correlation} illustrates how the  nearest neighbor correlations become more prominent than the next nearest neighbor correlations in the range of strong interaction and strong disorder. These changes indicate some change in the character of the localized particles that cannot be understood from IPR calculations or from density distributions alone. We quantify these changes by defining a total correlation parameter $\zeta$ which depends on the minimum step distance between the two particles.

\begin{equation}
\zeta (r_s) = \sum_{\textbf{\em{mn}}}  C_{\textbf{\em{mn}}}\big|_{|\textbf{\em{m}} - \textbf{\em{n}}|=r_s}
\end{equation}

\begin{figure}[H]
\centering
\includegraphics[width = 1.0\textwidth]{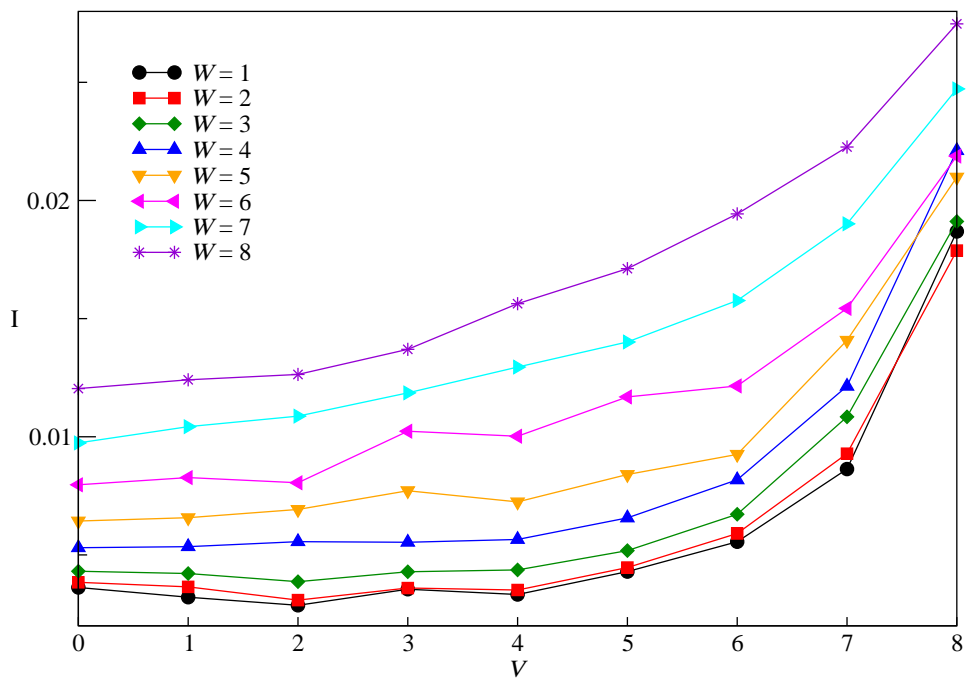}
\caption[IPR vs interaction for fixed disorders]{IPR vs interaction for  disorder strengths ranging from $W=1$ to $W=8$. Averaged over 320 realizations of disorders.}
\label{2D-IPR-W}
\end{figure}

\begin{figure}[H]
\centering
\includegraphics[width = 1.23\textwidth]{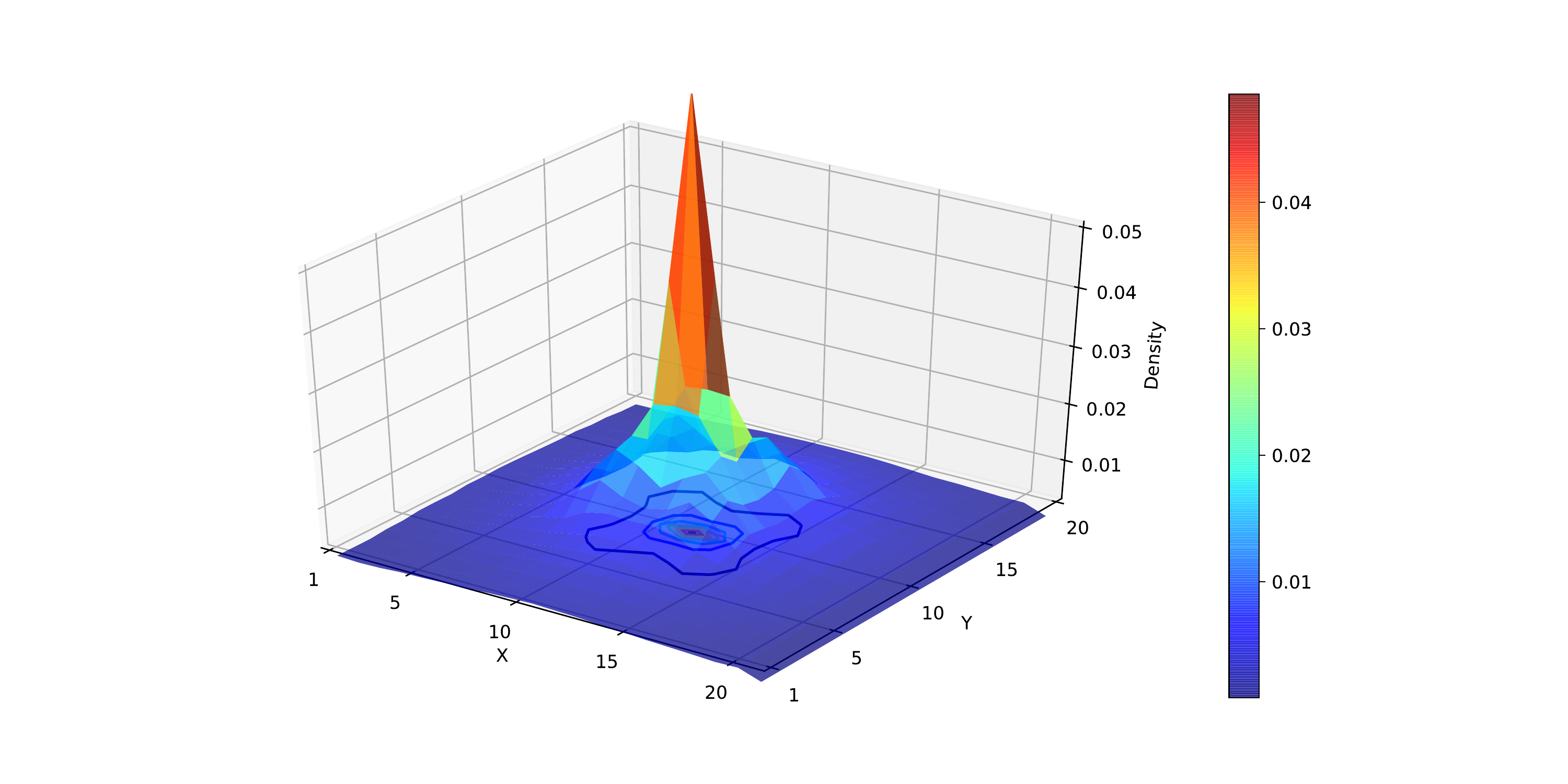}
\caption[Density distribution of interacting particles in disordered 2D lattice]{Density distribution of interacting particles in disordered 2D lattice for $W=1$ and $V=4$. Averaged over 320 realizations of disorders.}
\label{2D-density-W1V4}
\end{figure}

To understand the trend in the computed parameters, we plot the iso-disorder surfaces from Fig. \ref{2D-IPR-3D} in Fig. \ref{2D-IPR-W}. The calculations show very similar behaviour of localization within a broad range of the weak interaction and the weak disorder ($0 < W \le 4$ and $0 \le V \le 4$) strengths.  Beyond this region, particles start to localize strongly. The density distribution for one of the most delocalized points ($W=1$ and $V=4$) from the weakly localized region is shown in Fig. \ref{2D-density-W1V4}.

As can be observed, no significant difference in the localization parameters in the range of strong disorder - strong interaction limits are observed compared to the weak disorder - weak interaction regime. The transition between the regions is smooth. This can be further inferred as an absence of delocalization for interacting particles in disordered 2D lattices. However, this conclusion remains to be confirmed with other methods. As observed before from the contours  in Fig. \ref{2D-IPR-3D} and the change of behaviour in correlations in Fig. \ref{2D-disordered-correlation}, there appear to be some physical changes that should be further studied for any conclusion on the localization-delocalization transitions. Besides, a detailed scaling analysis on the whole range of the parameters from the lattice of size with 30 sites per dimension to a much larger number of sites per dimension is needed to draw any final conclusion on the  localization-delocalization transition of interacting particles in two dimensional disordered lattices. Such scale of calculations can only be considered after further improvements in the computational algorithms and the facilities.

\section{Conclusion}

In this chapter a basic understanding on the length scales of interacting particles in disordered one- and two-dimensional systems has been obtained by numerical methods. Enhancement of cowalking correlations has been observed in 1D disordered systems. The length scales  and correlations calculated for a vast range of parameters between interacting particles in disordered 2D systems provide an understanding of the localization of interacting particles in two dimensional disordered systems.

\newpage

\chapter{Conclusion}

\vspace{9mm}
The thesis has attempted to achieve the following main goals:

\vspace{5mm}
1. Develop an understanding of the effect of interaction between particles on their correlations in lattices. The correlations were calculated between two hard core bosons in 1D ideal and disordered lattices. 

\vspace{5mm}
2. Develop an understanding of the effect of the range of tunnelling on correlations between particles in lattices. The range of tunnelling was modelled as decaying isotropically with some power of distance. The nature of dynamical correlations for both dipolar and Coulombic hopping were found to be different from nearest neighbor hopping models. 

\vspace{5mm}
3.  Extend the method of recursive computation of Green's functions to interacting particles in disordered 2D lattices and introduce approximations to make it more accurate and efficient. The method is shown to be helpful in calculating dynamics and correlations of interacting particles in lattices of larger size than which can be computed  by full diagonalization method. 

\vspace{5mm}
4. Extend the recursive method for computation of Green's functions of interacting particles in some arbitrary graphs. The cases of binary trees were specifically taken as an example of such graphs. 

\vspace{5mm}
5. Develop an understanding of correlations and localization properties of interacting particles in disordered 2D lattices. The localization parameters of interacting particles were calculated for a broad range of disorder and interaction strengths.

\vspace{9mm}

More specifically, we have found for 1D lattice systems that two interacting particles can distinguish between the nature of the interactions (whether repulsive or attractive) when the tunnelling of the particles in lattices becomes long-range. The particles become bound with lesser strength of interaction for the attractive case compared to the repulsive case when tunnelling is long-range. The difference in binding for the different kind of interactions (attractive vs repulsive) makes the dynamics significantly different in the case of long-range hopping. These features may be used to control the dynamics or transport of particles in lattices by tuning the sign and strength of the interaction between the particles. In contrast to the effect of long-range tunnelling, the effects of long-range interaction were found to be insignificant. Although such effects are studied only since few years before \cite{felix, rausch2, laurent, covey, eckstein, platero}, it seems many more observations are yet to be observed.

The method of recursion has been developed to calculate Green's functions of interacting particles in  one and two dimensional disordered lattices and in  binary trees, which was used to extend the size of the calculations significantly compared to lattice systems which can be fully diagonalized. The spectral weights for the interacting particles have also been calculated for various systems of interest, which provides significant insights into those systems. The calculated Green's functions of interacting particles in real space can be Fourier transformed to momentum space. Using the recursion method, exact Green's functions were  calculated for disordered 2D systems, which provided insights into the behaviour of interacting particles in disordered systems. The approximations which have been introduced, make the calculations significantly more efficient while maintaining accuracy. The insights into correlations of interacting particles obtained here can also be incorporated into electronic structure calculations. For example, the spectral weights obtained from two-particle basis (Fig. \ref{1D-dos-sw}) provides the most significant parts of spectral weights calculated from full many-body calculations using methods as DMRG \cite{rausch}.  This indicates, going from single density basis to two density basis can significantly improve predictions on properties of interacting electrons in various material systems.

For two particles on binary trees, the spectral weight for non-interacting  particles  is  found to be discontinuous. For stronger interactions, the individual peaks of the spectra tend to merge into one profile, which  becomes continuous. The calculations of Green's functions of interacting particles on such trees is shown to be exact when compared to full diagonalization.

Some preliminary calculations to model the effects of uniform magnetic fields on interacting particles show the  splitting of spectra as expected. However, more calculations remain to be performed to gain further insights into the underlying physics of the model systems that have been considered. Calculations of two-particle correlations in the presence of inhomogeneous magnetic fields can  be considered in future research.

Localization properties of one and two dimensional finite disordered systems have been calculated for a vast range of parameters which provides important insights about the length scales of the spread of interacting particles in disordered systems. The effect of long range tunnelling on localization of interacting particles in disordered 1D systems has been calculated. Long-range tunnelling changes the localization parameters significantly. However, the prominent disorder-induced correlations  were found to be similar for different ranges of tunnelling. While our results indicate the absence of a localization-delocalization transition in disordered 2D systems, further scaling analysis is required to reach a non-ambiguous conclusion. The calculations for the scaling analysis  need to be performed for larger lattice sizes,  which in the case of 2D lattices with two correlated particles, will require huge computing resources.  The  effects of interactions on dynamical localization or delocalization must also be understood.  The role of dissipation on the dynamics and localization are also yet to be understood.

The calculations of two-particle Green's functions have  shown several interesting features. The bound state  has been found to play an important role in controlling the dynamics of the particles. This bound state only becomes effective after a critical interaction strength that depends on the band-width of the continuous states. In disordered systems, the particles were found to be correlated differently in comparison to the case of ordered systems. However, for strong interactions, disorder was found to enhance the co-walking correlations. These calculations can be used as the basis to understand the behaviour of a larger number of particles in disordered lattices. The recursive method is expected to be useful for calculations of higher order many-body terms  such as three-particle Green's functions. However, in two and three dimensions, for the calculations of Green's functions for a larger number of particles, one might require more efficient methods. The large basis size in higher dimensions for more  particles still remains a hindrance for understanding the physics through numerical approaches. However, for interacting particles in disordered systems, this approach seems to be the only one that can provide meaningful results.

\cleardoublepage
\addcontentsline{toc}{chapter}{Bibliography}
\bibliographystyle{unsrt}
\bibliography{18_bibliography}

\begin{thebibliography}{100}

\bibitem{hbt}
R.~Hanbury Brown and R.~Q. Twiss.
\newblock Correlation between photons in two coherent beams of light.
\newblock {\em Nature}, 177:27, 1956.

\bibitem{hom}
C.~K. Hong{,} Z.~Y. Ou and L.~Mandel.
\newblock Measurement of subpicosecond time intervals between two photons by
  interference.
\newblock {\em Phys. Rev. Lett.}, 59:2044, 1987.

\bibitem{chu}
S.~Chu{,} J. E. Bjorkholm{,}~A. Ashkin and A.~Cable.
\newblock Experimental observation of optically trapped atoms.
\newblock {\em Phys. Rev. Lett.}, 57:314, 1986.

\bibitem{aharonov}
Y.~Aharonov{,}~L. Davidovich and N.~Zagury.
\newblock Quantum random walks.
\newblock {\em Phys. Rev. A}, 48:1687, 1993.

\bibitem{aharonovd}
D.~Aharonov􏰁{,} A. Ambainis{,}~J. Kempe and U.~Vazirani.
\newblock Quantum walks on graphs.
\newblock In {\em STOC'01 33rd ACM symposium on Theory of computing}, page~50,
  2001.

\bibitem{ambainis}
A.~Ambainis{,} E. Bach{,} A. Nayak{,}~A. Vishwanath and J.~Watrous.
\newblock One-dimensional quantum walks.
\newblock In {\em STOC'01 33rd ACM symposium on Theory of computing}, page~37,
  2001.

\bibitem{karski}
M.~Karski{,} L. Forster{,} J. Choi{,} A. Steffen{,} W. Alt{,}~D. Meschede and
  A.~Widera.
\newblock Quantum walk in position space with single optically trapped atoms.
\newblock {\em Science}, 325:174, 2009.

\bibitem{schreibera}
A.~Schreiber{,} K. N. Cassemiro{,} V. Potocek{,} A. Gabris{,}~I. Jex{,} and Ch.
  Silberhorn.
\newblock Decoherence and disorder in quantum walks: From ballistic spread to
  localization.
\newblock {\em Phys. Rev. Lett.}, 106(180403), 2011.

\bibitem{schreibera1}
A.~Schreiber{,} A. G{\'a}bris{,} P. P. Rohde{,} K. Laiho{,} M. {\v S}tefa{\v
  n}{\'a}k{,} V. Poto{\v c}ek{,} C. Hamilton{,}~I. Jex and Ch. Silberhorn.
\newblock A 2d quantum walk simulation of two-particle dynamics.
\newblock {\em Science}, 336:55, 2012.

\bibitem{peruzzo}
A.~Peruzzo{,} M. Lobino{,} J. C. F. Matthews{,} N. Matsuda{,} A. Politi{,} K.
  Poulios{,} X. Zhou{,} Y. Lahini{,} N. Ismail{,} K. W{\"o}rhoff{,} Y.
  Bromberg{,} Y. Silberberg{,} M.~G. Thompson and J.~L. OBrie.
\newblock Quantum walks of correlated photons.
\newblock {\em Science}, 329:1500, 2010.

\bibitem{schreibera2}
A.~Schreiber{,} K. N. Cassemiro{,} V. Poto{\v c}ek{,} A. G{\'a}bris{,} P. J.
  Mosley{,} E. Andersson{,}~I. Jex and Ch. Silberhorn.
\newblock Photons walking the line: A quantum walk with adjustable coin
  operations.
\newblock {\em Phys. Rev. Lett.}, 104(050502), 2010.

\bibitem{zahringer}
F.~Z{\"a}hringer{,} G. Kirchmair{,} R. Gerritsma{,} E. Solano{,}~R. Blatt and
  C.~F. Roos.
\newblock Realization of a quantum walk with one and two trapped ions.
\newblock {\em Phys. Rev. Lett.}, 104(100503), 2010.

\bibitem{ashwin}
F.~Maginez{,} A. Nayak{,}~J. Roland and M.~Santha.
\newblock Search via quantum walk.
\newblock In {\em STOC'07 39th ACM symposium on theory of computing}, page 575,
  2007.

\bibitem{childs1}
A.~M. Childs and J.~Goldstone.
\newblock Spatial search by quantum walk.
\newblock {\em Phys. Rev. A}, 70(022314), 2004.

\bibitem{shenvi}
N.~Shenvi{,}~J. Kempe{,} and K.~B. Whaley.
\newblock Quantum random-walk search algorithm.
\newblock {\em Phys. Rev. A}, 67(052307), 2003.

\bibitem{childs2}
A.~M. Childs{,}~D. Gosset and Z.~Web.
\newblock Universal computation by multiparticle quantum walk.
\newblock {\em Science}, 339:791, 2013.

\bibitem{coppersmith1}
K.~Rudinger{,} J. K. Gamble{,} M. Wellons{,} E. Bach{,} M. Friesen{,}~R. Joynt
  and S.~N. Coppersmith.
\newblock Noninteracting multiparticle quantum random walks applied to the
  graph isomorphism problem for strongly regular graphs.
\newblock {\em Phys. Rev. A}, 86(022334), 2012.

\bibitem{coppersmith2}
J.~K. Gamble{,} M. Friesen{,} D. Zhou{,}~R. Joynt and S.~N. Coppersmith.
\newblock Two-particle quantum walks applied to the graph isomorphism problem.
\newblock {\em Phys. Rev. A}, 81(052313), 2010.

\bibitem{wang}
S.~D. Berry and J.~B. Wang.
\newblock Two-particle quantum walks: Entanglement and graph isomorphism
  testing.
\newblock {\em Phys. Rev. A}, 83:042317, 2011.

\bibitem{sougoto}
E.~Compagno{,} L. Banchi{,}~C. Gross and S.~Bose.
\newblock Noon states via a quantum walk of bound particles.
\newblock {\em Phys. Rev. A}, 95(012307), 2017.

\bibitem{mohan}
M.~Sarovar{,} A. Ishizaki{,} G.~R. Fleming and K.~B. Whaley.
\newblock Quantum entanglement in photosynthetic light-harvesting complexes.
\newblock {\em Nat. Phys.}, 6:462, 2010.

\bibitem{engel}
G.~S. Engel{,} T. R. Calhoun{,} E. L. Read T-K Ahn{,} T. Mancal{,} Y-C Cheng{,}
  R.~E. Blankenship and G.~R. Fleming.
\newblock Evidence for wavelike energy transfer through quantum coherence in
  photosynthetic systems.
\newblock {\em Nature}, 446:782, 2007.

\bibitem{qin}
X.~Qin{,} Y. Ke{,} X. Guan{,} Z. Li{,}~N. Andrei and C.~Lee.
\newblock Statistics-dependent quantum co-walking of two particles in
  one-dimensional lattices with nearest-neighbor interactions.
\newblock {\em Phys. Rev. A}, 90(062301), 2014.

\bibitem{bordone}
I.~Siloi{,} C. Benedetti{,} E. Piccinini{,} J. Piilo{,} S. Maniscalco{,} M.
  G.~A. Paris{,} and P.~Bordone.
\newblock Noisy quantum walks of two indistinguishable interacting particles.
\newblock {\em Phys. Rev. A}, 95(022106), 2017.

\bibitem{bakr}
W.~S. Bakr{,} A. Peng{,} M. E. Tai{,} R. Ma{,} J. Simon{,} J. I. Gillen{,} S.
  F{\"o}iling{,}~L. Pollet and M.~Greiner.
\newblock Probing the superfluid to {Mott} insulator transition at the
  single-atom level.
\newblock {\em Science}, 329:547, 2010.

\bibitem{sherson}
J.~F. Sherson{,} C. Weitenberg{,} M. Endres{,} M. Cheneau{,}~I. Bloch and
  S.~Kuhr.
\newblock Single-atom-resolved fluorescence imaging of an atomic {Mott}
  insulator.
\newblock {\em Nature}, 467:68, 2010.

\bibitem{folling}
S.~F{\"o}lling{,} F. Gerbier{,} A. Widera{,} O. Mandel{,}~T. Gericke and
  I.~Bloch.
\newblock Spatial quantum noise interferometry in expanding ultracold atom
  clouds.
\newblock {\em Nature}, 434:481, 2005.

\bibitem{greiner1}
M.~Greiner{,} C. A. Regal{,} J.~T. Stewart and D.~S. Jin.
\newblock Probing pair-correlated fermionic atoms through correlations in atom
  shot noise.
\newblock {\em Phys. Rev. Lett.}, 94(110401), 2005.

\bibitem{fukuhara}
T.~Fukuhara{,} P. Schau{\ss}{,} M. Endres{,} S. Hild{,} M. Cheneau{,}~I. Bloch
  and C.~Gross.
\newblock Microscopic observation of magnon bound states and their dynamics.
\newblock {\em Nature}, 502:76, 2013.

\bibitem{preiss}
P.~M. Preiss{,} R. Ma{,} M. E. Tai{,} A. Lukin{,} M. Rispoli{,} P. Zupancic{,}
  Y. Lahini{,}~R. Islam and M.~Greiner.
\newblock Strongly correlated quantum walks in optical lattices.
\newblock {\em Science}, 347:1229, 2015.

\bibitem{lahini1}
Y.~Lahini{,} M. Verbin{,} S. D. Huber{,} Y. Bromberg{,}~R. Pugatch and
  Y.~Silberberg.
\newblock Quantum walk of two interacting bosons.
\newblock {\em Phys. Rev. A}, 86(011603), 2012.

\bibitem{lahini2}
Y.~Lahini{,} Y. Bromberg{,} D.~N. Christodoulides and Y.~Silberberg.
\newblock Quantum correlations in two-particle {Anderson} localization.
\newblock {\em Phys. Rev. Lett.}, 105(163905), 2010.

\bibitem{anderson}
P.~W. Anderson.
\newblock Absence of diffusion in certain random lattices.
\newblock {\em Phys. Rev.}, 109:1492, 1958.

\bibitem{billy}
D.~Cl{\'e}ment{,} L. Sanchez-Palencia{,} P.~Bouyer J.~Billy{,} V. Josse{,} Z.
  Zuo{,} A. Bernard{,} B. Hambrecht{,} P.~Lugan and A.~Aspect.
\newblock Direct observation of {Anderson} localization of matter waves in a
  controlled disorder.
\newblock {\em Nature}, 453:891, 2008.

\bibitem{shepelyansky}
D.~L. Shepelyansky.
\newblock Coherent propagation of two interacting particles in a random
  potential.
\newblock {\em Phys. Rev. Lett.}, 73:2607, 1994.

\bibitem{imry}
Y.~Imry.
\newblock Coherent propagation of two interacting particles in a random
  potential.
\newblock {\em Europhys. Lett.}, 30:405, 1995.

\bibitem{pichard}
S.~De Toro Arias{,}~X. Waintal and J.-L. Pichard.
\newblock Two interacting particles in a disordered chain {III}: Dynamical
  aspects of the interplay disorder-interaction.
\newblock {\em Eur. Phys. J B}, 10:149, 1999.

\bibitem{schreiberm1}
R.~A. Romer and M.~Schreiber.
\newblock No enhancement of the localization length for two interacting
  particles in a random potential.
\newblock {\em Phys. Rev. Lett.}, 78:515, 1997.

\bibitem{ortuno}
M.~Ortuno and E.~Cuevas.
\newblock Localized to extended states transition for two interacting particles
  in a two-dimensional random potential.
\newblock {\em Europhys. Lett.}, 46:224, 1999.

\bibitem{flach}
D.~O. Krimer{,}~R. Khomeriki and S.~Flach.
\newblock Two interacting particles in a random potential.
\newblock {\em JETP Lett.}, 94:406, 2011.

\bibitem{guidoni}
L~Guidoni and P~Verkerk.
\newblock Optical lattices: cold atoms ordered by light.
\newblock {\em J. Opt. B: Quantum Semiclass. Opt.}, 1:R23, 1999.

\bibitem{ping}
P.~Xiang{,}~M. Litinskaya and R.~V. Krems.
\newblock Tunable exciton interactions in optical lattices with polar
  molecules.
\newblock {\em Phys. Rev. A}, 85(061401), 2011.

\bibitem{deborah}
A.~Chotia{,} B. Neyenhuis{,} S. A. Moses{,} B. Yan{,} J. P. Covey{,} M.
  Foss-Feig{,} A. M. Rey{,} D.~S. Jin and J.~Ye.
\newblock Long-lived dipolar molecules and {F}eshbach molecules in a {3D}
  optical lattice.
\newblock {\em Phys. Rev. Lett.}, 108(080405), 2012.

\bibitem{jun}
B.~Yan{,} S. A. Moses{,} B. Gadway{,} J. P. Covey{,} K. R. A. Hazzard{,} A. M.
  Rey{,} D.~S. Jin and J.~Ye.
\newblock Observation of dipolar spin-exchange interactions with
  lattice-confined polar molecules.
\newblock {\em Nature}, 501:521, 2013.

\bibitem{volz}
T.~Volz{,} N. Syassen{,} D. M. Bauer{,} E. Hansis{,}~S. Durr and G.~Rempe.
\newblock Preparation of a quantum state with one molecule at each site of an
  optical lattice.
\newblock {\em Nat. Phys.}, 2:692, 2006.

\bibitem{ospelkaus}
C.~Ospelkaus{,} S. Ospelkaus{,} L. Humbert{,} P. Ernst{,}~K. Sengstock and
  K.~Bongs.
\newblock Ultracold heteronuclear molecules in a {3D} optical lattice.
\newblock {\em Phys. Rev. Lett.}, 97(120402), 2006.

\bibitem{sowinski}
T.~Sowinski{,} O. Dutta{,} P. Hauke{,}~L. Tagliacozzo and M.~Lewenstein.
\newblock Dipolar molecules in optical lattices.
\newblock {\em Phys. Rev. Lett.}, 108(115301), 2012.

\bibitem{cote}
R.~Cote.
\newblock Quantum random walk with {Rydberg} atoms in an optical lattice.
\newblock {\em New J. Phys.}, 8(156), 2006.

\bibitem{hubbard}
J.~Hubbard.
\newblock Electron correlations in narrow energy bands.
\newblock {\em Proc. R. Soc. Lond. A Math. Phys. Sc.}, 276:238, 1963.

\bibitem{spielman1}
Y.-J. Lin{,} R. L. Compton{,} K. Jimenez-Garcia{,} J.~V. Porto and I.~B.
  Spielman.
\newblock Synthetic magnetic fields for ultracold neutral atoms.
\newblock {\em Nature}, 462:628, 2009.

\bibitem{spielman2}
N.~Goldman{,} G. Juzeliunas{,}~P. Ohberg and I.~B. Spielman.
\newblock Light-induced gauge fields for ultracold atoms.
\newblock {\em Rep. Prog. Phys.}, 77(126401), 2014.

\bibitem{ping-thesis}
P.~Xiang.
\newblock {\em Quantum control of dynamics of quasiparticles in periodic and
  disordered lattice potentials.}
\newblock PhD thesis, Univ. of British Columbia, 2014.

\bibitem{greiner-thesis}
M.~Greiner.
\newblock {\em Ultracold quantum gases in three-dimensional optical lattice
  potentials}.
\newblock PhD thesis, Ludwig-Maximilians-Universit{\"a}t M{\"u}nchen, 2003.

\bibitem{rajibul-thesis}
K.~R. Islam.
\newblock {\em Quantum simulation of interacting spin models with trapped
  ions}.
\newblock PhD thesis, University of Maryland, 2012.

\bibitem{feshbach}
H.~Feshbach.
\newblock Unified theory of nuclear reactions.
\newblock {\em Ann. Phys.}, 5:357, 1958.

\bibitem{chin}
C~Chin{,} R. Grimm{,}~P. Ju;ienne and E.~Tiesinga.
\newblock Feshbach resonances in ultracold gases.
\newblock {\em Rev. Mod. Phys.}, 82:1225, 2010.

\bibitem{volz1}
T.~Volz{,} S. Durr{,} S. Ernst{,}~A. Marte and G.~Rempe.
\newblock Characterization of elastic scattering near a feshbach resonance in
  87{Rb}.
\newblock {\em Phys. Rev. A}, 68(010702), 2003.

\bibitem{molmer-sorensen}
K.~Molmer and A.~Sorensen.
\newblock Multiparticle entanglement of hot trapped ions.
\newblock {\em Phys. Rev. Lett.}, 82:1835, 1999.

\bibitem{jurcevic}
P.~Jurcevic{,} B. P. Lanyon{,} P. Hauke{,} C. Hempel{,} P. Zoller{,}~R. Blatt
  and C.~F. Roos.
\newblock Quasiparticle engineering and entanglement propagation in a quantum
  many-body system.
\newblock {\em Nature}, 511:202, 2014.

\bibitem{griffith}
D.~F.~Schroeter D.~J.~Griffiths.
\newblock {\em Introduction to quantum mechanics}.
\newblock Cambridge University Press, 3 edition, 2018.

\bibitem{winkler}
K.~Winkler{,} G. Thalhammer{,} F. Lang{,} R. Grimm{,} J. Hecker Denschlag{,} A.
  J. Daley{,} A. Kantian{,} H.~P. Buchler{,} and P.~Zoller.
\newblock Repulsively bound atom pairs in an optical lattice.
\newblock {\em Nature}, 441:853, 2006.

\bibitem{levitov}
L.~S. Levitov.
\newblock Delocalization of vibrational modes caused by electric dipole
  interaction.
\newblock {\em Phys. Rev. Lett.}, 64:547, 1990.

\bibitem{malyshev}
A.~Rodriguez{,} V. A. Malyshev{,} G. Sierra{,} M. A. Martin-Delgado{,}~J.
  Rodriguez-Laguna and F.~Dominguez-Adame.
\newblock Anderson transition in low-dimensional disordered systems driven by
  long-range nonrandom hopping.
\newblock {\em Phys. Rev. Lett.}, 90(027404), 2003.

\bibitem{malysheva}
F.~A. B.~F. de~Moura{,} A. V. Malyshev{,} M. L. Lyra{,} V. A.~Malyshev and
  F.~Dominguez-Adame.
\newblock Localization properties of a one-dimensional tight-binding model with
  nonrandom long-range intersite interactions.
\newblock {\em Phys. Rev. B}, 71(174203), 2005.

\bibitem{pathak}
P.~K. Pathak and G.~S. Agarwal.
\newblock Quantum random walk of two photons in separable and entangled states.
\newblock {\em Phys. Rev. A}, 75(032351), 2007.

\bibitem{valiente1}
M.~Valiente and D.~Petrosyan.
\newblock Two-particle states in the {Hubbard} model.
\newblock {\em J. Phys. B: Mol. Opt. Phys.}, 41(161002), 2008.

\bibitem{hecker}
J.~Hecker Denschlag and A.~J. Daley.
\newblock Exotic atom pairs: Repulsively bound states in an optical lattice.
\newblock {\em arXiv:cond-mat}, (0610393), 2006.

\bibitem{molmer1}
R.~Piil and K.~Molmer.
\newblock Tunneling couplings in discrete lattices, single-particle band
  structure, and eigenstates of interacting atom pairs.
\newblock {\em Phys. Rev. A}, 76(023607), 2007.

\bibitem{vektaris}
G.~Vektaris.
\newblock A new approach to the molecular biexciton theory.
\newblock {\em J. Chem. Phys.}, 101:3031, 1994.

\bibitem{fumika-thesis}
F.~Suzuki.
\newblock {\em Quantum mechanics of composite objects with internal
  entanglement}.
\newblock PhD thesis, The University of British Columbia, 2017.

\bibitem{valiente3}
M.~Valiente and D.~Petrosyan.
\newblock Scattering resonances and two-particle bound states of the extended
  {Hubbard} model.
\newblock {\em J. Phys. B: Mol. Opt. Phys.}, 42(121001), 2009.

\bibitem{lieb-wu}
E.~H. Lieb and F.~Y. Wu.
\newblock Absence of {Mott} transition in an exact solution of the short-range,
  one-band model in one dimension.
\newblock {\em Phys. Rev. Lett.}, 20:1445, 1968.

\bibitem{christodoulides}
D.~N. Christodoulides{,}~F. Lederer and Y.~Silberberg.
\newblock Discretizing light behaviour in linear and nonlinear waveguide
  lattices.
\newblock {\em Nature}, 424:817, 2003.

\bibitem{sebabrata}
S.~Mukherjee{,} M. Valiente{,} N. Goldman{,} A. Spracklen{,} E. Andersson{,}~P.
  Ohberg and R.~R. Thomson.
\newblock Observation of pair tunneling and coherent destruction of tunneling
  in arrays of optical waveguides.
\newblock {\em Phys. Rev. A}, 94(055853), 2016.

\bibitem{hartmann}
M.~J. Hartmann.
\newblock Quantum simulation with interacting photons.
\newblock {\em J. Opt}, 18(104005), 2016.

\bibitem{firstenberg}
O.~Firstenberg{,} T. Peyronel{,} Q-Y Liang{,} A. V. Gorshkov{,} M.~D. Lukin and
  V.~Vuletic.
\newblock Attractive photons in a quantum nonlinear medium.
\newblock {\em Nature}, 502:71, 2013.

\bibitem{gorshkov}
A.~V. Gorshkov{,} J. Otterbach{,} M. Fleischhauer{,}~T. Pohl{,} and M.~D.
  Lukin.
\newblock Photon-photon interactions via {Rydberg} blockade.
\newblock {\em Phys. Rev. Lett.}, 107(133602), 2011.

\bibitem{tirtha}
T.~Chattaraj and R.~V. Krems.
\newblock Effects of long range hopping and interactions on quantum walk in
  ordered and disordered lattices.
\newblock {\em Phys. Rev. A}, 94(023601), 2016.

\bibitem{bloch}
I.~Bloch{,}~J. Dalibard and W.~Zwerger.
\newblock Many-body physics with ultracold gases.
\newblock {\em Rev. Mod. Phys.}, 80:885, 2008.

\bibitem{berciu1}
M.~Berciu.
\newblock On computing the square lattice {Green's} function without any
  integrations.
\newblock {\em J. Phys. A: Math. Theor.}, 42(395207), 2009.

\bibitem{berciu2}
M.~Berciu.
\newblock Few-particle {Green's} functions for strongly correlated systems on
  infinite lattices.
\newblock {\em Phys. Rev. Lett.}, 107(246403), 2011.

\bibitem{berciu3}
M.~Berciu and A.~M. Cook.
\newblock Efficient computation of lattice {Green's} functions for models with
  nearest-neighbour hopping.
\newblock {\em Europhys. Lett.}, 92(40003), 2010.

\bibitem{haydock}
V.~Heine R.~Haydock and M.~J. Kelly.
\newblock Electronic structure based on the local atomic environment for
  tight-binding bands: {II}.
\newblock {\em J. Phys. C: Solid State Phys.}, 8:2591, 1975.

\bibitem{morita}
T.~Morita.
\newblock Use of a recurrence formula in computing the lattice {Green}
  function.
\newblock {\em J. Phys. A: Math. Gen.}, 8:478, 1975.

\bibitem{economou}
E.~N. Economou.
\newblock {\em Green's Functions in Quantum Physics}.
\newblock Springer-Verlag, 3rd edition, 2010.

\bibitem{licciardello}
D.~C. Licciardello and E.~N. Economou.
\newblock Study of localization in {Anderson's} model for random lattices.
\newblock {\em Phys. Rev. B}, 11:3697, 1975.

\bibitem{thouless}
D.~J. Thouless and S.~Kirkpatrick.
\newblock Conductivity of the disordered linear chain.
\newblock {\em J. Phys. C: Solid State Phys.}, 14:235, 1981.

\bibitem{sawatzky}
G.~A. Sawatzky.
\newblock Quasiatomic {Auger} spectra in narrow-band metals.
\newblock {\em Phys. Rev. Lett.}, 39:504, 1977.

\bibitem{gutzwiller}
M.~C. Gutzwiller.
\newblock Effect of correlation on the ferromagnetism of transition metals.
\newblock {\em Phys. Rev. Lett.}, 10:159, 1963.

\bibitem{cini}
M.~Cini.
\newblock Theory of {Auger} lineshapes of solids.
\newblock {\em J. Phys.: Condens. Matter}, 1:SB55, 1989.

\bibitem{kanamori}
J.~Kanamori.
\newblock Electron correlation and ferromagnetism of transition metals.
\newblock {\em Prog. Theo. Phys.}, 30:275, 1963.

\bibitem{nolting}
W.~Nolting.
\newblock Influence of electron correlations on {Auger} electron- and
  appearance potential - spectra of solids.
\newblock {\em Z. Phys. B - Condensed Matter}, 80:73, 1990.

\bibitem{rausch}
R.~Rausch and M.~Potthoff.
\newblock Multiplons in the two-hole excitation spectra of the one-dimensional
  {Hubbard} model.
\newblock {\em New J. Phys.}, 18(023033), 2016.

\bibitem{hofstadter}
D.~R. Hofstadter.
\newblock Energy levels and wave functions of {Bloch} electrons in rational and
  irrational magnetic fields.
\newblock {\em Phys. Rev. B}, 14:2239, 1976.

\bibitem{goldman}
C.~E. Creffield{,} G. Pieplow{,}~F. Sols and N.~Goldman.
\newblock Realization of uniform synthetic magnetic fields by periodically
  shaking an optical square lattice.
\newblock {\em New J. Phys.}, 18(093013), 2016.

\bibitem{struck}
J.~Struck{,} C. Olschlager{,} M. Weinberg{,} P. Hauke{,} J. Simonet{,} A.
  Eckardt{,} M. Lewenstein{,}~K. Sengstock{,} and P.~Windpassinger.
\newblock Tunable gauge potential for neutral and spinless particles in driven
  optical lattices.
\newblock {\em Phys. Rev. Lett.}, 108(225304), 2012.

\bibitem{monika-thesis}
M.~Aidelsburger.
\newblock {\em Artificial gauge fields with ultracold atoms in optical
  lattices}.
\newblock PhD thesis, Ludwig-Maximilians-Universitat Munchen, 2014.

\bibitem{yoshioka}
D.~Yoshioka.
\newblock {\em The quantum {Hall} effect}.
\newblock Springer, 2010.

\bibitem{bernevig}
B.~A. Bernevig and T.~L. Hughes.
\newblock {\em Topological insulators and topological superconductors}.
\newblock Princeton Univ. Press, 2013.

\bibitem{perczel}
J.~Perczel{,} J. Borregaard{,} D. E. Chang{,} H. Pichler{,} S. F. Yelin{,}~P.
  Zoller{,} and M.~D. Lukin.
\newblock Topological quantum optics in two-dimensional atomic arrays.
\newblock {\em Phys. Rev. Lett.}, 119(023603), 2017.

\bibitem{peierls}
R.~Peierls.
\newblock On the theory of diamagnetism of conduction electrons.
\newblock {\em Z. Phys.}, 80:763, 1933.

\bibitem{eckardt}
A.~Eckardt{,} P. Hauke{,} P. Soltan-Panahi{,} C. Becker{,}~K. Sengstock and
  M.~Lewenstein.
\newblock Frustrated quantum antiferromagnetism with ultracold bosons in a
  triangular lattice.
\newblock {\em Europhys. Lett.}, 89(10010), 2010.

\bibitem{ambegaokar}
V.~Ambegaokar and U.~Eckern.
\newblock Coherence and persistent currents in mesoscopic rings.
\newblock {\em Phys. Rev. Lett.}, 65:381, 1990.

\bibitem{chandrasekhar}
V.~Chandrasekhar{,} R. A. Webb{,} M. J. Brady{,} M. B. Ketchen{,} W.~J.
  Gallagher and A.~Kleinsasser.
\newblock Magnetic response of a single, isolated gold loop.
\newblock {\em Phys. Rev. Lett.}, 67:3578, 1991.

\bibitem{levy}
L.~P. Levy{,} G. Dolan{,}~J. Dunsmuir and H.~Bouchiat.
\newblock Magnetization of mesoscopic copper rings: Evidence for persistent
  currents.
\newblock {\em Phys. Rev. Lett.}, 64:2074, 1990.

\bibitem{kravchenko1}
S.~V. Kravchenko{,} G. V. Kravchenko{,} J. E. Furneaux{,} V.~M. Pudalov and
  M.~DIorio.
\newblock Possible metal-insulator transition at {B = 0} in two dimensions.
\newblock {\em Phys. Rev. B}, 50:8039, 1994.

\bibitem{kravchenko2}
S.~V. Kravchenko{,} D. Simonian{,} M. P. Sarachik{,}~W. Mason and J.~E.
  Furneaux.
\newblock Electric field scaling at a {B = 0} metal-insulator transition in two
  dimensions.
\newblock {\em Phys. Rev. Lett.}, 77:4938, 1996.

\bibitem{scaling}
E.~Abrahams{,} P. W. Anderson{,} D.~C. Licciardello and T.~V. Ramakrishnan.
\newblock Scaling theory of localization: Absence of quantum diffusion in two
  dimensions.
\newblock {\em Phys. Rev. Lett.}, 42:673, 1978.

\bibitem{ponomarev}
I.~V. Ponomarev and P.~G. Silvestrov.
\newblock Coherent propagation of interacting particles in a random potential:
  the mechanism of enhancement.
\newblock {\em Phys. Rev. B}, 56:3742, 1997.

\bibitem{oppen}
P.~H. Song and F.~von Oppen.
\newblock General localization lengths for two interacting particles in a
  disordered chain.
\newblock {\em Phys. Rev. B}, 59:46, 1999.

\bibitem{guhr}
T.~Guhr{,}~A. Muller-Groeling and H.~A. Weidenmuller.
\newblock Random-matrix theories in quantum physics: common concepts.
\newblock {\em Phys. Rep.}, 299:189, 1998.

\bibitem{flach2}
S.~Flach{,} D.~O. Krimer and C.~Skokos.
\newblock Universal spreading of wave packets in disordered nonlinear systems.
\newblock {\em Phys. Rev. Lett.}, 102(024101), 2009.

\bibitem{schreiberm2}
A.~Eilmes{,} U. Grimm{,} R.~A. Romer and M.~Schreiber.
\newblock Two interacting particles at a metal-insulator transition.
\newblock {\em Eur. Phys. J. B}, 8:547, 1999.

\bibitem{schreiberm3}
A.~Eilmes{,} R.~A. Romer and M.~Schreiber.
\newblock Localization properties of two interacting particles in a
  quasi-periodic potential with a metal-insulator transition.
\newblock {\em Eur. Phys. J. B}, 23:229, 2001.

\bibitem{schreiberm4}
R.~A. Romer{,}~M. Leadbeater and M.~Schreiber.
\newblock Numerical results for two interacting particles in a random
  environment.
\newblock {\em Ann. Phys.}, 8:675, 1999.

\bibitem{felix}
F.~Hofmann and M.~Potthoff.
\newblock Doublon dynamics in the extended {Fermi-Hubbard} model.
\newblock {\em Phys. Rev. B}, 85(205127), 2012.

\bibitem{rausch2}
R.~Rausch and M.~Potthoff.
\newblock Filling-dependent doublon dynamics in the one-dimensional {Hubbard}
  model.
\newblock {\em Phys. Rev. B}, 95(045152), 2017.

\bibitem{laurent}
L.~Cevolani{,} J. Despres{,} G. Carleo{,}~L. Tagliacozzo and
  L.~Sanchez-Palencia.
\newblock Universal scaling laws for correlation spreading in quantum systems
  with short- and long-range interactions.
\newblock {\em Phys. Rev. B}, 98(024302), 2018.

\bibitem{covey}
J.~P. Covey{,} S. A. Moses{,} M.~Ga ̈rttner{,} A.~Safavi-Naini{,} M. T.
  Miecnikowski{,} Z. Fu{,} J. Schachenmayer{,} P. S. Julienne{,} A. M. Rey{,}
  D. S.~Jin and J.~Ye.
\newblock Doublon dynamics and polar molecule production in an optical lattice.
\newblock {\em Nat. Commun}, 7(11279), 2016.

\bibitem{eckstein}
P.~Werner and M.~Eckstein.
\newblock Effective doublon and hole temperatures in the photo-doped dynamic
  hubbard model.
\newblock {\em Structural Dynamics}, 3(023603), 2016.

\bibitem{platero}
M.~Bello{,} C.~E. Creffield and G.~Platero.
\newblock Long-range doublon transfer in a dimer chain induced by topology and
  ac fields.
\newblock {\em Sci. Rep}, 6(22562), 2016.

\bibitem{jook}
J.~T.~M. Walraven.
\newblock Quantum gases.
\newblock \url{https://staff.fnwi.uva.nl/j.t.m.walraven/walraven/Lectures.htm},
  Lecture notes Les Houches 2017.

\bibitem{toyoda}
K.~Toyoda{,} R. Hiji{,}~A. Noguchi and S.~Urabe.
\newblock {Hong-Ou-Mandel} interference of two phonons in trapped ions.
\newblock {\em Nature}, 527:74, 2015.

\end{thebibliography}

\newpage

\begin{appendices}

\chapter{Few Operator Algebra}

\section*{Bose operators}

In this section we list few general uses of bosonic operators. The bosonic creation ($a^\dagger$) and annihilation ($a$) operators satisfy their commutation relation as in the following equation.

\begin{equation}\label{bosonic}
aa^\dagger - a^\dagger a = 1
\end{equation}

Transforming these operators by addition of constants don't change their commutation properties.
\begin{equation*}
\tilde{a}^\dagger = a^\dagger + \alpha    ~~~\mbox{and}~~~~~ \tilde{a} = a + \alpha^{*}
\end{equation*}

\begin{equation}
\tilde{a}\tilde{a}^\dagger - \tilde{a}^\dagger \tilde{a} = 1
\end{equation}

For the commutation relations with higher powers of bose operators one can easily find the following.

\begin{align}
a\left(a^\dagger\right)^n - \left(a^\dagger\right)^n a &= n \left(a^\dagger\right)^{n-1} = \frac{\partial\left(a^\dagger\right)^n}{\partial a^\dagger} \nonumber \\
a^\dagger\left(a\right)^n - \left(a\right)^n a^\dagger &= n \left(a\right)^{n-1} = - \frac{\partial \left(a\right)^n}{\partial a}
\end{align}

This can be generalized further for functions that can be expanded in power series of bose operators. 

\begin{align}
a f\left(a^\dagger\right) - f\left(a^\dagger\right) a & = \frac{\partial f\left(a^\dagger\right)}{\partial a^\dagger} \nonumber \\
a^\dagger f\left(a\right) - f\left(a\right) a^\dagger &=  - \frac{\partial f\left(a\right)}{\partial a}
\end{align}

One can define the exponential function of these operators as in the following equation.

\begin{equation}
e^{\alpha a^\dagger} = \sum_{n = 0}^{\infty} \frac{\alpha^n}{n!} \left(a^\dagger\right)^n
\end{equation}

From previous equations it is easy to show the following.
\begin{align}
a e^{\alpha a^\dagger}  - e^{\alpha a^\dagger} a & =  \alpha e^{\alpha a^\dagger}  \nonumber \\
a^\dagger e^{\alpha a}  - e^{\alpha a}  a^\dagger &=  - \alpha e^{\alpha a} 
\end{align}

Multiplying inverse of the respective exponentials in the previous equation the following relations can be derived.

\begin{align}
e^{- \alpha a^\dagger} a e^{\alpha a^\dagger}   & =  \alpha + a  \nonumber \\
e^{- \alpha a}  a^\dagger e^{\alpha a}  &=  - \alpha + a^\dagger
\end{align}

Similarly one can obtain the following relations.

\begin{align}
e^{- \alpha a^\dagger} e^{\beta a} e^{\alpha a^\dagger}   & =  e^{\beta \left(e^{- \alpha a^\dagger} a e^{\alpha a^\dagger}\right)}  \nonumber \\
&= e^{\beta (\alpha + a)}  \nonumber \\
&= e^{\beta\alpha} e^{\beta a}  \nonumber \\
e^{- \alpha a} e^{\beta a^\dagger} e^{\alpha a}   & =  e^{\beta \left(e^{- \alpha a} a^\dagger e^{\alpha a}\right)}  \nonumber \\
&= e^{\beta (-\alpha + a^\dagger)}  \nonumber \\
&= e^{-\beta\alpha} e^{\beta a^\dagger}
\end{align}

For a similar operator algebra where $a$ and $a^\dagger$ appear together, following relations can be derived.

\begin{align}
e^{ \alpha a^\dagger a} a e^{- \alpha a^\dagger a}   & =  e^{-\alpha} a  \nonumber \\
e^{ \alpha a^\dagger a} a^\dagger e^{- \alpha a^\dagger a}   & =  e^{\alpha} a^\dagger 
\end{align}

Note that 

\begin{align}
\frac{\partial}{\partial \alpha} e^{ \alpha a^\dagger a} a e^{- \alpha a^\dagger a}   & =  e^{ \alpha a^\dagger a} (a^\dagger a - a a^\dagger) a e^{- \alpha a^\dagger a} \nonumber \\
&= - e^{ \alpha a^\dagger a} a e^{- \alpha a^\dagger a}
\end{align}

For such functions $f'(\alpha) = - f(\alpha)$ has solutions $f(\alpha) = e^{-\alpha}\cdot C$ and the constant ($C$) can be found from initial condition $f(0) = a$.

\section*{Fermi operators}

 Operators of creation ($c^\dagger$) and annihilation ($c$) of fermions satisfy following commutation relations.

\begin{align}
cc^\dagger + c^\dagger c &= 1 \\
\left( c^\dagger\right)^2 &= 0 \nonumber \\
c^2 &= 0 \nonumber 
\end{align}

From which it is easy to show that

\begin{align}
e^{- \alpha c^\dagger} c e^{\alpha c^\dagger}   & =   \left(1 - \alpha c^\dagger\right)c\left(1 + \alpha c^\dagger\right) \nonumber \\
&= c - \alpha^2 c^\dagger + \alpha \left(cc^\dagger - c^\dagger c\right)\nonumber \\
e^{- \alpha c}  c^\dagger e^{\alpha c}  &=  \left(1 - \alpha c\right) c^\dagger \left(1 + \alpha c\right) \nonumber \\
&= c^\dagger - \alpha^2 c - \alpha\left(cc^\dagger - c^\dagger c \right)
\end{align}

and

\begin{align}
e^{ \alpha c^\dagger c} c e^{- \alpha c^\dagger c}   & =  e^{-\alpha} c  \nonumber \\
e^{ \alpha c^\dagger c} c^\dagger e^{- \alpha c^\dagger c}   & =  e^{\alpha} c^\dagger 
\end{align}.

\chapter{s-wave scattering}
In this section we discuss the scattering of two  quantum particles  interacting through most simple potentials \cite{jook}.  This discussion helps in understanding interaction of quantum particles in most simple forms which can also be read from any introductory quantum mechanics book. The particles in a many body environment 'feel' each other when their wavelength is in the range of the interaction potential. This wavelength is of the same order as the thermal de Broglie wavelength

\begin{equation}
\Lambda \simeq [2 \pi /(m k_B T)]^{1/2}
\end{equation}
where $m, k_B, T$ are, respectively, the mass, the Boltzmann constant and the temperature. The interaction becomes important when $\Lambda \simeq n ^ {-1/3}$ ($n$ is the density) that is when the de Broglie wavelength reaches the limit of the interatomic distance.

Assuming a certain form of the interaction potential, the Schr{\"o}dinger equation  needs to be solved to understand the behaviour of the interacting particles. Assuming a centro-symmetric potential, the Schr{\"o}dinger equation for the radial part is (with $\hbar = 1$)

\begin{equation}\label{radial}
 \left[ \frac{1}{2m_r}  \left( - \frac{d^2}{dr^2} - \frac{2}{r} \frac{d}{dr} + \frac{l(l+1)}{r^2}  \right) + \mathcal{V}(r) \right] R_l (r) = E R_l (r)
\end{equation}
Here $l$ is the quantum number for angular momentum of the two particles. Multiplying by $2m_r$ on both sides and using the substitutions $\varepsilon = 2m_r E$ and $U(r) = 2m_r \mathcal{V}(r)$, Eq. \ref{radial} reduces to

\begin{equation}
R_l ^{''} + \frac{2}{r} R_l ^{'} +  \left[ \varepsilon - U(r) - \frac{l(l+1)}{r^2}  \right] R_l (r) = 0.
\end{equation}
With one more substitution $\chi_l (r) = r R_l (r)$, it becomes the following simple 1D Schr{\"o}dinger equation with $\varepsilon = k^2$ and $U_{eff}(r)$ taking account of the rotational energy

\begin{equation}
 \chi_l ^{''} +   \left[ k^2 - U_{eff} (r)  \right] \chi_l = 0.
\end{equation}

The rotational energy produces an effective barrier  in the potential and a classical turning point $r_{cl} =\frac{ \sqrt{l(l+1)}}{k}$. If the typical range of the potential  $r_0$ is less than the de Broglie wavelength $\Lambda$, then interaction is not important. Associating  a typical wavenumber $k$ (where $k = \frac{1}{\Lambda}$) for the relative motion of atoms, the condition emerges as $kr_0 \ll 1$ when interaction is negligible. Equating it with the previous equation, we find that the interaction is important when the range of the potential becomes closer to that of $r_{cl}$.

\begin{equation*}
kr_0 = \sqrt{l(l+1)} r_0 / r_{cl} \ll 1 \Longleftrightarrow r_{cl} \gg r_0 ~\text{for} ~ l \neq 0.
\end{equation*}

In the low energy limit, when $l=0$, this condition is not sufficient. Scattering can take place when the barrier is absent. In this \textit{s-wave regime}, the previous inequalities are valid at low collision energies.

Looking at the case of  a free particle for $l=0$  we can gain primary understanding of  the scattering event

\begin{equation}
 \chi_0 ^{''} +    k^2 \chi_0(r) = 0,
\end{equation}
 with the general solution $\chi_0 (k, r) = c_0 \sin (kr + \eta_0)$. The radial wave function is then

\begin{equation}
 R_0 (k, r)  = \frac{c_0}{kr} \sin (kr + \eta_0)
\end{equation}
where $\eta_0 = 0$ ensures non-singularity at the origin.

The general solution for arbitrary $l$ is found in terms of spherical Bessel functions $j_l $ and spherical Neumann functions $n_l $

\begin{equation}
 R_l (kr)  = c_l \left[ \cos\eta_l j_l (kr) + \sin\eta_l n_l (kr) \right].
\end{equation}

For $l=0$ this produces $R_0 (k,r) = c_0 j_0 (kr)$. This solution is also valid when $r \gg r_0$, where the free particle assumption is valid for short range potentials. At large distances $\left(r \gg \frac{1}{k} = \Lambda  \right)$ for short range potentials, the spherical Bessel and Neumann functions take their asymptotic form and we get the following formula with the asymptotic phase shift $\eta_l (k)$ even when the scattering is elastic and conserves asymptotic momentum $k$. 

\begin{align}
 R_l (kr)  & \underset{r \rightarrow \infty}{\simeq} \frac{c_l}{kr} \left[ \cos\eta_l \sin (kr - \frac{l\pi}{2}) + \sin\eta_l \cos (kr -  \frac{l\pi}{2}) \right]\nonumber \\ 
 & \underset{r \rightarrow \infty}{\simeq} \frac{c_l}{kr}  \sin (kr +\eta_l - \frac{l\pi}{2})  \\
& \underset{r \rightarrow \infty}{\simeq} \frac{c_l}{2k} \imath \left[ e^{-\imath \eta_l} \frac{e^{-\imath(kr - \frac{l\pi}{2})}}{r} - e^{\imath \eta_l} \frac{e^{\imath(kr - \frac{l\pi}{2})}}{r} \right] \nonumber \\
& \underset{r \rightarrow \infty}{\simeq} \frac{c_l ^{'}}{2k} \left[ \frac{e^{-\imath kr}}{r} - e^{\imath l\pi} e^{2 \imath \eta_l} \frac{e^{\imath kr}}{r} \right]
\end{align}

The final form shows that the outgoing wavefunction has a phase factor $S_l = e^{2\imath\eta_l}$  compared to the incoming wave  other than the sign. The length parameter $a$ ($a = -\frac{\eta_l}{k}$)  
 possesses valuable information about the interaction potential. One can infer whether the interaction is attractive ($a < 0$) or repulsive ($a > 0$) from it. It also provides information about the range of interaction.

In the presence of an interaction potential as simple as a spherical well with flat bottom as in the following equation, one can find resonant scattering energies where this interaction changes from attractive to repulsive or vice versa

\begin{equation}
U(r)=
\begin{cases}
-\kappa_0^2 ~ ~ \text{  for  } r \le r_0\\
0  ~~~~~~\text{  for  } r > r_0\\
\end{cases}.
\end{equation}

Let's define some parameters for this  potential. For energy in continuum range, $(\epsilon = k_{c}^2 > 0)~ K_{+} = \sqrt{\kappa_0^2 + k_{c}^2}$ and in the well depth $(\epsilon = k_{b}^2 < 0)~ K_{-} = \sqrt{\kappa_0^2 - k_{b}^2}$.

The 1D s-wave Schr{\"o}dinger equation is then

\begin{equation}
 \chi_0 ^{''} +    \left[k_{c}^2 - U(r) \right] \chi_0(r) = 0 ~~ ~~~ \mbox{for  ~} l=0 ~~ \mbox{and ~} \epsilon > 0
\end{equation}
which has the following solutions with a phase shift outside the well

\begin{equation}
\chi_0 (k, r)=
\begin{cases}
C_{<} \sin (K_{+}r) ~~~~~~\hspace{0.225cm}  \mbox{  for  } r \le r_0\\
C_{>} \sin (k_{c}r + \eta_0) ~~~  \mbox{  for  } r > r_0\\
\end{cases}.
\end{equation}

Applying the boundary condition to both of these solutions at $r = r_0$, the phase shift can be found.

\begin{equation}
 \chi_0 ^{'}/\chi_0 \vert_{r=r_0} =   K_{+} \cot K_{+} r_0 = k_{c} \cot (k_{c}r_0 + \eta_0)
\end{equation}

Now, define the  \textit{scattering length} $a$ in the limit $k_c \rightarrow 0$ (hence independent of $k_c$).

\begin{equation}
a = \lim_{k_c \rightarrow 0} a(k_c) = - \lim_{k_c \rightarrow 0} \frac{\eta_0 (k_c)}{k_c}
\end{equation}

In this limit, $K_{+} \rightarrow \kappa_0$ and the previous boundary condition becomes

\begin{equation}
\kappa_0 \cot \kappa_0 r_0 = \frac{1}{r_0 - a}.
\end{equation}

Following this we find
\begin{equation}
\frac{a}{r_0} = 1 - \frac{\tan \kappa_0 r_0}{\kappa_0 r_0}.
\end{equation}

\begin{figure}[h!]
\includegraphics[width=1.0\textwidth]{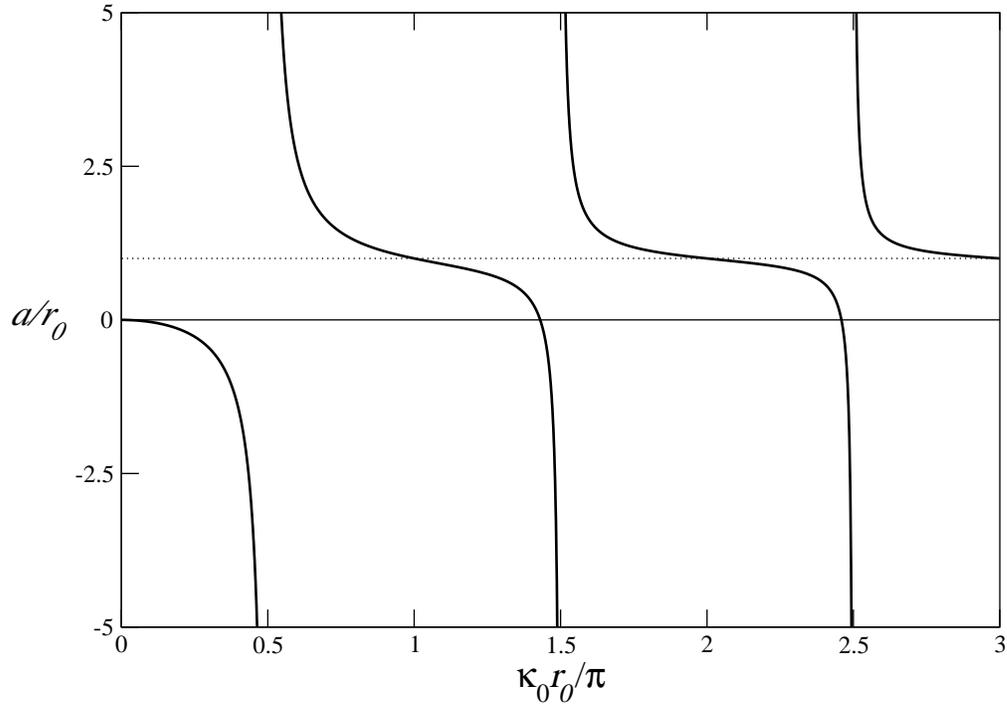}
\caption[s-wave scattering]{The  s-wave scattering length $a$ normalized to the well width as a function of the depth of the well for the case of very low energy scattering. The scattering length is one unit of the width of the well except in the resonant regions where it diverges for certain well depth.}
\label{swave}
\end{figure}

The behaviour of the scattering length as a function of the interaction strength is shown in  Figure \ref{swave}.  The Figure shows that  it diverges at half of odd integer values of $\kappa_0 r_0$.

For energies within the range of the well depth, the 1D Schr{\"o}dinger equation takes the following form.

\begin{equation}
 \chi_0 ^{''} +    \left[ - k_{b}^2 - U(r) \right] \chi_0(r) = 0 ~~ ~~~ \mbox{for  ~} l=0 ~~ \mbox{and ~} \epsilon < 0,
\end{equation}
which has the following solutions without any  phase shift

\begin{equation}
\chi_0 (k, r)=
\begin{cases}
C_{<} \sin (K_{-}r) ~~~~~~  \mbox{  for  } r \le r_0\\
C_{>} e^{- k_{b}r} ~~~\hspace{1.225cm}  \mbox{  for  } r > r_0 \mbox{  and  } k_b > 0.  \\
\end{cases}
\end{equation}

Applying the boundary condition to both of these solutions at $r = r_0$, one obtains the following relations

\begin{equation}
 \chi_0 ^{'}/\chi_0 \vert_{r=r_0} =   K_{-} \cot K_{-} r_0 = - k_{b}.
\end{equation}

In the limit $k_b \rightarrow 0$ hence $K_{-} \rightarrow \kappa_0$,  we find

\begin{equation}
\kappa_0 \cot \kappa_0 r_0 = 0.
\end{equation}

It has the solutions for

\begin{equation}
\kappa_0 r_0 =  \left(\upsilon + \frac{1}{2} \right) \pi
\end{equation}
where $\upsilon = 0, 1 , 2, ... \upsilon_{max}$ is the vibrational quantum number of the well. For existence of at least one bound state, it is required that $\kappa_0 r_0 \vert_{min} = \frac{\pi}{2}$. However, this solution is derived from a 3D spherically symmetric potential. In 2D, it turns out that a bound state exists for any shallow potential.

Understanding the basics of two particle scattering even at low energies show that the interaction between the two particles can be controlled if one can find the resonant scattering energies. Not only that, the interaction can also be made attractive or repulsive around the resonances with very small change in scattering energy. In the next section we learn how this is achieved experimentally.

\newpage
\chapter{Feshbach resonance}

In Appendix B, scattering of two particles  on one single potential was briefly described. Every atom or molecule has many  states that can couple while the scattering takes place. Here we will take a look at the most simple of such situation with only two states involved. One of them will be an \textit{open channel}, where particles come and leave without any binding. The other we will take as a \textit{closed channel} where particles can form a composite bound state in the process. The coupling between two such channels will allow  control of the scattering length and hence the effective interaction strength between the particles. In an experiment, the coupling can be controlled both optically or using external magnetic fields \cite{feshbach, chin}. For the case of states coupled via magnetic field, we will take one of the states to be a triplet and the other state as a singlet. Coupling between these kind of states can be achieved via spin flip of one of the particles. 

In the presence of a magnetic field the electron spins exhibit the Zeeman interaction

\begin{equation}
\mathcal{H}_Z =  \gamma_e \mathbf{s}_1 \cdot \mathbf{B} +  \gamma_e \mathbf{s}_2 \cdot \mathbf{B} =  \gamma_e \mathbf{S} \cdot \mathbf{B}
\end{equation}
where $\gamma_e$ is the gyromagnetic constant of the electron. As can be seen this Hamiltonian conserves total spin and does not couple singlet and triplet channels by any spin flip. There is also nuclear Zeeman interaction describing the coupling of nuclear spins  to magnetic field

\begin{equation}
\mathcal{H}_Z =  \gamma_n \mathbf{i}_1 \cdot \mathbf{B} +  \gamma_n \mathbf{i}_2 \cdot \mathbf{B} =  \gamma_n \mathbf{i} \cdot \mathbf{B}.
\end{equation}

This term  does not couple the singlet or triplet channels either. The hyperfine effect couples the electron spins with nuclear spins

\begin{equation}
\mathcal{H}_{hf} =  \alpha_1 \mathbf{i}_1 \cdot \mathbf{s_1} +  \alpha_2 \mathbf{i}_2 \cdot \mathbf{s_2},
\end{equation}
where $\alpha_{1,2}$ are the hyperfine interaction constants. This equation can be written in total spin conserving ($\mathcal{H}_{hf}^+$) and non-conserving terms ($\mathcal{H}_{hf}^-$). The non-conserving term can couple singlet and triplet channels via change in total spin

\begin{equation}
\mathcal{H}_{hf} ^{\pm}=  \frac{\alpha_1}{2} \left( \mathbf{s}_1 \pm \mathbf{s}_2 \right)\cdot \mathbf{i_1} \pm \frac{\alpha_2}{2} \left( \mathbf{s}_1 \pm \mathbf{s}_2 \right)\cdot \mathbf{i_2}.
\end{equation}

\begin{equation}
\mathcal{H}_{hf} =  \mathcal{H}_{hf}^+ + \mathcal{H}_{hf}^-
\end{equation} 

When two particles are identical ($\alpha_1 = \alpha_2$), these terms can be simplified further.

\begin{equation}
\mathcal{H}_{hf} ^{\pm}=  \frac{\alpha}{2} \left( \mathbf{s}_1 \pm \mathbf{s}_2 \right)\cdot  \left( \mathbf{i}_1 \pm \mathbf{i}_2 \right)
\end{equation}

Now solving the radial part similar to the Eq. \ref{radial} without any magnetic field applied for some Hamiltonian $H_{r}$ yields the following equation with spin indexes added

\begin{equation}
R_{l, S} ^{''} + \frac{2}{r} R_{l, S} ^{'} +  \left[ \varepsilon - U_{l, S}(r) - \frac{l(l+1)}{r^2}  \right] R_{l, S} (r) = 0.
\label{radial2}
\end{equation}
where $U_{l, S}(r)$ contains the spin-spin interaction part beside the singlet or triplet channel potential which also conserve total spin. In a basis like $\{\vert \nu, l, m_l\rangle \vert s_1, s_2, S, M_s \rangle\}$, the energies of the open channel ($\varepsilon > 0$) for wavefunctions  $R_{l, S}(k, r)$ belongs in continuum

\begin{equation}
\varepsilon_k = k_c^2
\end{equation} 
and for the closed channel ($\varepsilon < 0$) for wavefunctions  $R_{l, S}(v, r)$ is that of bound vibrational states with ro-vibrational energy

\begin{equation}
\varepsilon_v^{S, l} = -\kappa_{v, S}^2 + l(l+1) \mathcal{R}_{v, S}^l.
\end{equation} 
where $\mathcal{R}_{v, S}^l = \langle R_{l, S}(v, r)\vert r^{-2} \vert R_{l, S}(v,r)\rangle$.

In the presence of a magnetic field, as both channels get coupled, one needs to solve the following secular equation

\begin{equation}
\vert\langle i_1^\prime, i_2^\prime, m_1^\prime, m_2^\prime \vert \langle s_1^\prime, s_2^\prime, S^\prime, M_s^\prime\vert \langle \nu^\prime, l^\prime, m_l^\prime\vert \mathcal{H} - E \vert \nu, l, m_l \rangle \vert s_1, s_2, S, M_s\rangle \vert i_1, i_2, m_1, m_2\rangle\vert = 0
\end{equation} 
where the total Hamiltonian now contains both the Zeeman and hyperfine parts

\begin{equation}
\mathcal{H} = H_r + \mathcal{H}_Z + \mathcal{H}_{hf}^+ + \mathcal{H} _{hf}^-.
\end{equation} 

Although this Hamiltonian couples different electronic spin states, it doesn't change the total spin projection of the nuclear and electronic spin states ($M_F = m_1 + m_2 + M_s$). One can see that from rewriting the hyperfine terms with ($s_{zj}, s_{xj}, s_{yj}$)  and ($i_{zj}, i_{xj}, i_{yj}$) operators where $j\in \{1, 2\}$.

\begin{equation}
\mathcal{H}_{hf}^+ =  \sum_{j=1}^2 \frac{\alpha}{2} [ S_z i_{jz} + \frac{1}{2}  \{ S_+ i_{j-} + S_+ i_{j+} \} ]
\end{equation} 

\begin{equation}
\mathcal{H}_{hf}^- =  \sum_{j=1}^2   (-)^{j-1} \frac{\alpha}{2} [ (s_{1z} - s_{2z}) i_{jz} + \frac{1}{2}  \{ (s_{1+} -s_{2+}) i_{j-} + (s_{1-} - s_{2-}) i_{j+} \} ]
\end{equation} 

The secular equation now reads as follows.

\begin{equation}
|  ( \varepsilon_{v}^{l, S} + E_B^\sigma - E ) \delta_{\sigma, \sigma^\prime} + \langle v^\prime, l^\prime, S^\prime\vert v, l, S\rangle \langle \sigma^\prime \vert \mathcal{H}_{hf}^+ \vert \sigma\rangle +  \langle v^\prime, l^\prime, S^\prime\vert v, l, S\rangle \langle \sigma^\prime \vert \mathcal{H}_{hf}^- \vert \sigma\rangle   | = 0
\end{equation} 
where $\sigma (^\prime)$ has all the electronic and nuclear spin indices. In most cases the singlet-triplet coupling via the third term in the equation doesn't dominate as the radial wavefunctions are very different in small distances because the potentials are generally very different and contribute negligible overlap which suppresses the last two terms. However, a large overlap at long distances for both the singlet and triplet states, if any, makes this term significant. For cases when the energies of both  the asymptotic singlet and triplet  states are resonant at long distances, the radial overlap  can be taken approximately as full overlap and a spin flip can happen.

\begin{equation}
| ( \varepsilon_{v}^{l, S} + E_B^\sigma - E ) \delta_{\sigma, \sigma^\prime} + \langle \sigma^\prime \vert \mathcal{H}_{hf}^+ \vert \sigma\rangle +  \langle \sigma^\prime \vert \mathcal{H}_{hf}^- \vert \sigma\rangle | = 0
\end{equation} 

So, we look for a way to make any two of such potentials resonant with each other.  
Let's look at how the energies of the potentials change in response to an applied perturbation $U$ in general. For s-wave scattering the energy of the continuum states is given by $k^2$ and those of unperturbed bound states by $\varepsilon_\upsilon$. This bound state in the presence of a perturbation mixes with all the other bound and continuum states,

\begin{equation}
\mathcal{H} = H + U
\end{equation} 

\begin{equation}
\vert \upsilon \rangle \overset{U}{\rightarrow}  \vert \phi \rangle ~~~ \mbox{and} ~~  \varepsilon_\upsilon \overset{U}{\rightarrow} - \kappa^2
\end{equation}

\begin{equation}
 \vert \phi \rangle = \sum_{\upsilon^\prime} \vert \upsilon^\prime \rangle \langle \upsilon^\prime \vert \phi \rangle + \int d\mathbf{k} \vert \mathbf{k} \rangle \langle \mathbf{k} \vert \phi \rangle
\end{equation} 
which leads to the very general following form,
\begin{multline}
 \langle \upsilon \vert \mathcal{H} \vert \phi \rangle = \sum_{\upsilon^\prime}  \langle \upsilon \vert H \vert \upsilon^\prime \rangle \langle \upsilon^\prime \vert \phi \rangle+ \sum_{\upsilon^\prime}  \langle \upsilon \vert U \vert \upsilon^\prime \rangle \langle \upsilon^\prime \vert \phi \rangle \\ +  \int d\mathbf{k}  \langle \upsilon \vert H \vert \mathbf{k} \rangle \langle \mathbf{k} \vert \phi \rangle + \int d\mathbf{k}  \langle \upsilon \vert U \vert \mathbf{k} \rangle \langle \mathbf{k} \vert \phi \rangle.
\end{multline} 
Applying orthogonality of states results in a simpler form

\begin{equation} \label{pert1}
- ( \kappa^2 + \varepsilon_\upsilon )  \langle \upsilon \vert \phi \rangle  = \int d\mathbf{k}  \langle \upsilon \vert U \vert \mathbf{k} \rangle \langle \mathbf{k} \vert \phi \rangle.
\end{equation}

From coupling of the continuum states with that of perturbed states we find the following:

\begin{multline}
 \langle \mathbf{k} \vert \mathcal{H} \vert \phi \rangle = \sum_{\upsilon^\prime}  \langle\mathbf{k}\vert H \vert \upsilon^\prime \rangle \langle \upsilon^\prime \vert \phi \rangle+ \sum_{\upsilon^\prime}  \langle \mathbf{k} \vert U \vert \upsilon^\prime \rangle \langle \upsilon^\prime \vert \phi \rangle \\ +  \int d\mathbf{k}  \langle \mathbf{k} \vert H \vert \mathbf{k} \rangle \langle \mathbf{k} \vert \phi \rangle + \int d\mathbf{k}  \langle\mathbf{k} \vert U \vert \mathbf{k} \rangle \langle \mathbf{k} \vert \phi \rangle.
\end{multline} 
This equation can be simplified by assuming $ \langle \mathbf{k} \vert \mathcal{H} \vert \phi \rangle = -\kappa^2 \langle \mathbf{k} \vert \phi \rangle$ to lowest order of perturbation on left hand side and applying the orthogonalities

\begin{equation}
- ( \kappa^2 + k^2 )  \langle \mathbf{k}  \vert \phi \rangle  = \sum_{\upsilon^\prime}  \langle \mathbf{k} \vert U \vert \upsilon^\prime \rangle \langle \upsilon^\prime \vert \phi \rangle
\end{equation} 
Substituting this equation into Eq. \ref{pert1} we find 
\begin{equation} \label{pert2}
( \kappa^2 + \varepsilon_\upsilon )   = \int d\mathbf{k}  \frac{\vert\langle \upsilon \vert U \vert \mathbf{k} \rangle\vert^2 }{\kappa^2 + k^2}.
\end{equation} 

For the case of a general form for the potential $U$, which is short range and centro-symmetric, this equation can be approximated as
 
\begin{equation} 
\langle \upsilon \vert U \vert \mathbf{k} \rangle = \int d\mathbf{r} \langle \upsilon\vert r \rangle U(r)  e^{\imath \mathbf{k} \cdot \mathbf{r}} = \int d\mathbf{r} \langle \upsilon\vert r \rangle U(r) \underset{r<r_0}{\Big\vert} = u_0
\end{equation}

At long range this can be taken as zero for a short range perturbation potential. Eq. \ref{pert2}  now simplifies to a simple integral 

\begin{multline} 
( \kappa^2 + \varepsilon_\upsilon )   = u_0^2 \int_0^{k_{max}} dk  \frac{4\pi k^2 }{\kappa^2 + k^2}  = 4\pi u_0^2 \left[k_{max} - \kappa \tan^{-1}\left(\frac{k_{max}}{\kappa}\right)\right] \\ = \varepsilon_0 -   \kappa ~ 4\pi u_0^2 \tan^{-1}\left(\frac{k_{max}}{\kappa}\right) 
\end{multline} 
where  $\varepsilon_0 = 4\pi u_0^2 k_{max} $ and 
$ \varepsilon_0 \Big\vert_{\kappa \rightarrow 0} = \varepsilon_\upsilon$ which is the threshold for the perturbed energy for unperturbed asymptotic bound state of very small energy. The detuning from threshold is given by $\varepsilon_{res} = \varepsilon_{\upsilon} - \varepsilon_0$ which can be written as

\begin{equation} 
\varepsilon_{res} = -\kappa^2 -   \kappa ~ 4\pi u_0^2 \tan^{-1}\left(\frac{k_{max}}{\kappa}\right).
\end{equation}

In the range for $\kappa$, where $\tan^{-1}\left(\frac{k_{max}}{\kappa}\right)=\frac{\pi}{2}$ and independent of $k_{max}$, a characteristic parameter ($R^*$) can be introduced
\begin{equation} 
R^* = \frac{1}{2\pi^2 u_0^2} ~~~ \text{and} ~~~ \varepsilon_{res}R^* = -R^*\kappa^2 - \kappa
\end{equation} 
which gives us the relation of the perturbed energy to that of the unperturbed energy

\begin{equation} 
-\kappa^2 = - \frac{1}{4R^{*2}} \left[ -1 + \sqrt{1-4R^{*2}( \varepsilon_\upsilon - \varepsilon_0) }\right].
\end{equation} 

To generalize this form further, one can define a characteristic scattering length parameter $a_{res}$

\begin{equation} 
a_{res} = \frac{1}{\kappa + R^* \kappa^2} = - \frac{1}{R^*\varepsilon_{res}}.
\end{equation}

The open channel also contributes  to the scattering length which can be termed as the background contribution $a_{bg}$. The general total scattering length then takes  the following form:

\begin{equation} \label{res}
a = a_{res} + a_{bg} = a_{bg} - \frac{1}{R^*\varepsilon_{res}}.
\end{equation}

To bring two potentials at asymptote on resonance, the Zeeman interaction is very useful when both have different  spin projections. Suppose they reach a resonance at field $B_0$ at asymptotic energy $\varepsilon_0$, then the differential change of energy is linearly proportional to that of the  difference in field strength  close to a resonance 

\begin{equation} 
\varepsilon_\upsilon (B) - \varepsilon_0 = 2\mu \delta \mu_M (B - B_0)
\end{equation}

This can be substituted into Eq. \ref{res} to find the following general form.

\begin{equation} \label{res}
a = a_{bg} - \frac{1}{2\mu R^*\delta\mu_M(B - B_0)}
\end{equation}

\begin{equation} \label{res}
a = a_{bg}\left(1 - \frac{\Delta_B}{B - B_0}\right)
\end{equation}

where $\Delta_B = \frac{1}{a_{bg}R^* 2\mu \delta\mu_M }$.

\begin{figure}[h!]
\includegraphics[width=1.0\textwidth]{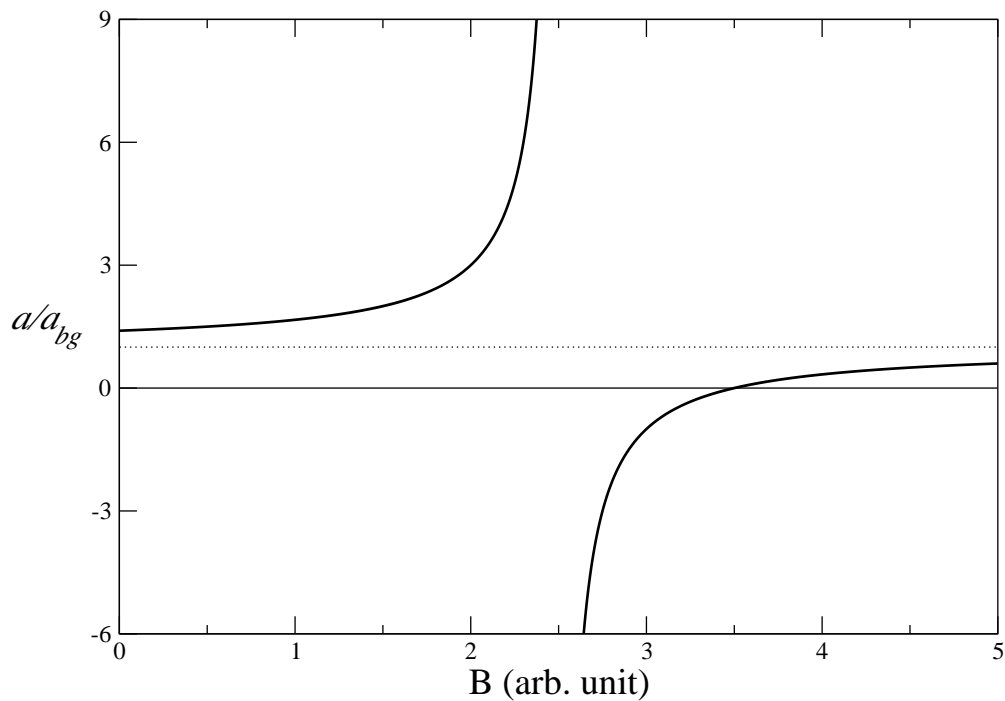}
\caption[Feshbach resonance]{s-wave scattering length in Feshbach resonance.}
\label{feshbach}
\end{figure}

In almost all cases $a_{bg} \neq 0$. This possibility of tuning resonant scattering by magnetic fields   has  applications ranging from formation of molecules (polar and non-polar) to control of interatomic interaction (both attractive and repulsive) in optical lattices. 

In this thesis, a broad range of interaction parameter (both attractive and repulsive) is considered to explore the physics between two interacting particles in both ordered and disordered lattices. The experimental method, described here only in brief, justifies for undertaking of such consideration where a broad range of interaction strength can be realized.

\newpage
\chapter{Interaction via phonons}

In lattices, there are always lattice motional states present as phonons. These phonons can also play an important role in controlling inter-particle interactions. 

The general Hamiltonian in a 1D lattice where the motions of lattice particles are coupled to nearest neighbors only and are harmonic, one can write the collective motional modes in terms of creation (annihilation) operators giving rise to the  bosonic quasiparticles well known as phonons

\begin{equation}
\mathcal{H} = \sum_l \frac{p_l^2}{2m} + V(y_1, y_2, .... )
\end{equation}

\begin{equation}
V = \sum_l \frac{1}{2} K (y_l - y_{l+1})^2  ~~~~ \text{where} ~~~ V_{l, l^\prime} = 
\begin{cases}
~2K~~~~~~~~\hspace{0.12cm}  \text{  if  } l=l^\prime\\
-K ~~~~~~~~~  \text{  if  } l = l^\prime \pm 1\\
~0 ~~~~~~~~~~~~~ \text{else}
\end{cases}
 \end{equation}

\begin{align}
V_q &= \sum_{|l-l^\prime| = 0, \pm 1} e^{\imath q (l - l^\prime)} V_{l,l^\prime}\nonumber \\
&= 2K - K(e^{\imath q} + e^{-\imath q}) \nonumber\\
& = 4K \sin^2\left(\frac{q}{2}\right)\\
& = M \omega_q^2 \nonumber\\
\omega_q &= 2 \sqrt{\frac{K}{M}} \sin \left( \frac{q}{2} \right)
\end{align}

The individual site ($l, l'$) displacement coordinates and momenta can be Fourier transformed to collective lattice modes

\begin{equation}
y_q = \frac{1}{\sqrt{N}} \sum_l e^{-\imath q l} y_l  ~~~ \text{and} ~~~ p_q = \frac{1}{\sqrt{N}} \sum_l e^{\imath q l} p_l
\end{equation}
where the periodic boundary condition $y_{l+N} = y_l$ would make it quantized within first Brillouin zone ($q = \frac{2\pi n}{N}, -\frac{N}{2}< n < \frac{N}{2} + 1, n\in \mathbb{Z} $).

\begin{equation}
\mathcal{H} = \sum_q \frac{1}{2M} p_q^\dagger p_q + \frac{1}{2} V_q y_q^\dagger y_q
\end{equation}

Writing the collective modes in terms of the creation (annihilation) operators gives rise to  quasiparticles well known as phonons
\begin{equation}
y_q = \sqrt{\frac{1}{2M\omega_q}} \left(  a_q^\dagger + a_{-q}\right) ~~\text{and}~~ p_q = \imath\sqrt{\frac{M\omega_q}{2}} \left(  a_q^\dagger - a_{-q}\right)
\end{equation}

\begin{equation}
\mathcal{H} = \omega_q \left( a_q^\dagger a_q + \frac{1}{2}\right)
\end{equation}

In general, when the range of coupling is taken beyond nearest neighbors and the potential is not so simple one will require to diagonalize full Hamiltonian in order to find the uncoupled normal modes and their energies.

In 3D lattices one needs to uncouple  the coordinate displacements along three separate axes, that are coupled via the potential term, into three orthogonal directions of \textit{polarizations}(\textbf{s}).

\begin{equation}
a_{\bf{qs}} = \frac{1}{\sqrt{2M\omega_{\bf{q,s}}}} \left( M\omega_{\bf{q,s}} \bf{y_q} + \imath \bf{p}^\dagger_{\bf q}\right)\cdot \bf{s_q}   
\end{equation}

\begin{equation}
 a_{\bf{qs}}^\dagger = \frac{1}{\sqrt{2M\omega_{\bf{q,s}}}} \left( M\omega_{\bf{q,s}} \bf{y_q} -\imath \bf{p}^\dagger_{\bf q}\right)\cdot \bf{s_q}
\end{equation}

\begin{equation}
\mathcal{H} = \omega_{\bf q,s} \left( a_{\bf q, s}^\dagger a_{\bf q, s} + \frac{1}{2}\right)   ~~~ \text{where} ~~~ \omega_{\bf q,s} = \sqrt{\frac{V_{\bf q}^{\bf s}}{M}}
\end{equation}

In an ideal lattice, without any electron-phonon coupling the unperturbed Hamiltonian is the sum of the individual electron and phonon energies

\begin{equation}
\mathcal{H}_0 = \sum_{\bf k} \varepsilon_{\bf k} c_{\bf k}^\dagger c_{\bf k} + \sum_{\bf q, s} \omega_{ \bf q,s} a_{\bf q, s}^\dagger a_{\bf q, s}.
\end{equation}

The interaction of phonons with electrons can be seen as  scattering different electronic states via the potential  due to the displacement of lattice particles from their equilibrium positions

\begin{align}
\mathcal{H_I} &= \sum_{\mathbf{k, k^\prime, l}} \langle \mathbf{k} \vert V(\mathbf{ r - l - y_l})\vert \mathbf{k^\prime}\rangle c_{\mathbf k}^\dagger c_{\mathbf k^\prime} \nonumber\\
&= \sum_{\mathbf{k, k^\prime, l}}  e^{\imath (\mathbf{k^\prime - k}) \cdot (\mathbf{l + y_l})} V_{\mathbf{k - k^\prime}} c_{\mathbf k}^\dagger c_{\mathbf k^\prime} \nonumber\\
&\simeq \sum_{\mathbf{k, k^\prime, l}}  e^{\imath (\mathbf{k^\prime - k})\cdot \mathbf{l}} \left( 1 + \imath  (\mathbf{k - k^\prime})\cdot \mathbf{y_l} \right) V_{\mathbf{k - k^\prime}} c_{\mathbf k}^\dagger c_{\mathbf k^\prime} \nonumber\\
&= \sum_{\mathbf{k, k^\prime, l}}  e^{\imath (\mathbf{k^\prime - k})\cdot \mathbf{l}} \left( 1 + \imath \frac{1}{\sqrt{N}} (\mathbf{k - k^\prime})\cdot \sum_\mathbf{q} e^{\imath \mathbf{q \cdot l}} \mathbf{y_q}  \right) V_{\mathbf{k - k^\prime}} c_{\mathbf k}^\dagger c_{\mathbf k^\prime} \nonumber\\
&= N V_0\sum_{\mathbf{k}}  c_{\mathbf k}^\dagger c_{\mathbf k} + \imath \sqrt{N}  \sum_{\mathbf{k, k^\prime}}   \left( \mathbf{k - k^\prime} \right) \cdot y_{\mathbf{k - k^\prime}} V_{\mathbf{k - k^\prime}} c_{\mathbf k}^\dagger c_{\mathbf k^\prime} \nonumber\\
&= N V_0\sum_{\mathbf{k}}  c_{\mathbf k}^\dagger c_{\mathbf k} + \imath \sqrt{\frac{N}{2M \omega_{\mathbf{q,s}}}}  \sum_{\mathbf{k, k^\prime, s}}   \left( \mathbf{k - k^\prime} \right) \cdot \mathbf{s} \left( a_{\mathbf{q,s}}^\dagger+ a_{\mathbf{-q,s}}\right)  V_{\mathbf{k - k^\prime}} c_{\mathbf k}^\dagger c_{\mathbf k^\prime}
\end{align}
where $\mathbf{q = k - k^\prime}$. After some readjustment we are led to the simple  form  which now has the electron-phonon interaction

\begin{equation}
\mathcal{H_F} =\mathcal{H}_0 + \mathcal{H}_{e-p} =  \sum_{\bf k} \varepsilon_{\bf k} c_{\bf k}^\dagger c_{\bf k} + \sum_{\bf q} \omega_{ \bf q} a_{\bf q}^\dagger a_{\bf q} +\sum_{\mathbf{k, k^\prime}} M_{\mathbf{kk^\prime}}  \left( a_{\mathbf{q}}^\dagger+ a_{\mathbf{-q}}\right)  c_{\mathbf k}^\dagger c_{\mathbf k^\prime}.
\end{equation}

We can now look for how this electron-phonon interaction can effectively mediate electron-electron interaction, both attractive and repulsive.

The energy of the electrons can be calculated by using the $\mathcal{H}_{e-p}$ as a perturbation. To zeroth order the electron and phonon energies contribute separately. The first order term vanishes as there are number non-conserving phonon operators in  $\mathcal{H}_{e-p}$. The effect of interaction can be found in the second order term

\begin{equation}
\mathcal{E}_2 = \langle \Phi \vert  \mathcal{H}_{e-p} \frac{1}{\mathcal{E}_0 - \mathcal{H}_0} \mathcal{H}_{e-p} \vert \Phi\rangle,
\end{equation}
where $ \vert \Phi\rangle =  \vert n_{\mathbf{k}}, n_{\mathbf{q}}\rangle$ has $n_{\mathbf{q}}$ number of phonons in mode $\mathbf{q}$ and $n_{\mathbf{k}}$ number of electrons in state $\mathbf{k}$.

With the full form of $\mathcal{H}_{e-p}$ inserted in the previous equation we find the following:

\begin{align}
\mathcal{E}_2 &= \langle \Phi \vert   \sum_{\mathbf{k, k^\prime}} M_{\mathbf{kk^\prime}}  \left( a_{\mathbf{q}}^\dagger+ a_{\mathbf{-q}}\right)  c_{\mathbf k}^\dagger c_{\mathbf k^\prime}  \frac{1}{\mathcal{E}_0 - \mathcal{H}_0}    \sum_{\mathbf{k^\dprime, k^\tprime}} M_{\mathbf{k^\dprime k^\tprime}}  \left( a_{\mathbf{q^\prime}}^\dagger+ a_{\mathbf{-q^\prime}}\right)  c_{\mathbf k^\dprime}^\dagger c_{\mathbf k^\tprime}  \vert \Phi\rangle \nonumber \\
 &=  \langle \Phi \vert   \sum_{\mathbf{k, k^\prime}} \vert M_{\mathbf{kk^\prime}} \vert^2  \{ a_{\mathbf{-q}}^\dagger c_{\mathbf k}^\dagger c_{\mathbf k^\prime}  \frac{1}{\mathcal{E}_0 - \mathcal{H}_0}    a_{\mathbf{-q}} c_{\mathbf k^\prime}^\dagger c_{\mathbf k}  + a_{\mathbf{q}}^\dagger  c_{\mathbf k}^\dagger c_{\mathbf k^\prime}  \frac{1}{\mathcal{E}_0 - \mathcal{H}_0} a_{\mathbf{q}}^\dagger c_{\mathbf k^\prime}^\dagger c_{\mathbf k} 
 \}\vert \Phi\rangle  \nonumber \\
&= \sum_{\mathbf{k, k^\prime}} \vert M_{\mathbf{kk^\prime}} \vert^2 \langle n_{\mathbf{k}} (1 - n_{\mathbf{k^\prime}}) \rangle  \left( \frac{ \langle n_{\mathbf{-q}} \rangle}{\mathcal{E}_{\mathbf{k}} - \mathcal{E}_{\mathbf{k^\prime}} + \omega_{\mathbf{-q}}} +  \frac{ \langle n_{\mathbf{q}} + 1 \rangle}{\mathcal{E}_{\mathbf{k}} - \mathcal{E}_{\mathbf{k^\prime}} - \omega_{\mathbf{q}}} \right)
\end{align}

With  $\omega_{\mathbf{q}}  = \omega_{\mathbf{-q}}$ (in systems with time reversal symmetry), this equation simplifies by letting $n_{\mathbf{q}} = n_{\mathbf{-q}}$

\begin{equation}\label{via-phonon}
\mathcal{E}_2 = \sum_{\mathbf{k, k^\prime}} \vert M_{\mathbf{kk^\prime}} \vert^2 \langle n_{\mathbf{k}}  \rangle  \left( \frac{ 2 (\mathcal{E}_{\mathbf{k}} - \mathcal{E}_{\mathbf{k^\prime}} ) \langle n_{\mathbf{q}} \rangle}{(\mathcal{E}_{\mathbf{k}} - \mathcal{E}_{\mathbf{k^\prime}})^2 - ( \omega_{\mathbf{q}})^2 } +  \frac{ \langle 1 - n_{\mathbf{k^\prime}}\rangle}{\mathcal{E}_{\mathbf{k}} - \mathcal{E}_{\mathbf{k^\prime}} - \omega_{\mathbf{q}}} \right).
\end{equation}

The first term of Eq. \ref{via-phonon} shows that the electron-phonon interaction is proportional to the phonon number.
The second term has phonon mediated electron-electron interaction term. As can be observed, this term can be both repulsive or attractive

\begin{equation}
\mathcal{E}_{pm} = \sum_{\mathbf{k, k^\prime}} \vert M_{\mathbf{kk^\prime}} \vert^2 \langle -  n_{\mathbf{k}}  n_{\mathbf{k^\prime}}\rangle   \frac{  \omega_{\mathbf{q}}}{(\mathcal{E}_{\mathbf{k}} - \mathcal{E}_{\mathbf{k^\prime}})^2 - ( \omega_{\mathbf{q}})^2 }.
\end{equation}

Another way of arriving at  this form of  electron-electron interaction  from the \textit{Frohlich} Hamiltonian, is by the powerful technique of canonical transformation. One needs to transform 
\begin{equation*}
\mathcal{H_F} =\mathcal{H}_0 + \mathcal{H}_{e-p} 
\end{equation*}
to
\begin{align}
\tilde{\mathcal{H}} &= e^{-s}~ \mathcal{H_F}~ e^s \\
&= (1 - s + \frac{s^2}{2} - \cdot \cdot \cdot )~ \mathcal{H_F}~   (1 + s + \frac{s^2}{2} + \cdot \cdot \cdot ) \nonumber \\
&=\mathcal{H}_0 + \mathcal{H}_{e-p}  + [\mathcal{H}_0, s] + [\mathcal{H}_{e-p}, s] + \cdot \cdot \cdot 
\end{align}

One can eliminate the phononic operators in $\tilde{\mathcal{H}}$ by a choice in $s$ up to certain orders of perturbation in $s$. Since $\mathcal{H}_0 $ does not include the phononic operators, one choice is to have phononic operators in $s$ in such a way that $\mathcal{H}_{e-p} $ gets cancelled by  $[\mathcal{H}_0, s] $ and the final form gives rise to the following form where only the electron-electron interaction terms will be retained. 

\begin{equation*}
\tilde{\mathcal{H}}  =\mathcal{H}_0 + [\mathcal{H}_{e-p}, s] 
\end{equation*}

Chosing $s$ in the following way 

\begin{equation}
s= \sum_{\mathbf{k, k^\prime}} M_{\mathbf{kk^\prime}}  \left(A a_{\mathbf{-q}}^\dagger+ Ba_{\mathbf{q}}\right)  c_{\mathbf k}^\dagger c_{\mathbf k^\prime}
\end{equation}
one can find $A =  -\frac{ 1}{\mathcal{E}_{\mathbf{k}} - \mathcal{E}_{\mathbf{k^\prime}} + \omega_{\mathbf{-q}}} $ and 
$B =  -\frac{ 1}{\mathcal{E}_{\mathbf{k}} - \mathcal{E}_{\mathbf{k^\prime}} - \omega_{\mathbf{q}}}$.

The transformed Hamiltonian ($\tilde{\mathcal{H}} $) then takes the following form where the electron-electron interaction term without any direct phonon term included can be observed. 

\begin{equation*}
\tilde{\mathcal{H}}  =\mathcal{H}_0 +  \sum_{\mathbf{k, k^\prime}} \vert M_{\mathbf{kk^\prime}} \vert^2   \frac{  \omega_{\mathbf{q}}}{(\mathcal{E}_{\mathbf{k}} - \mathcal{E}_{\mathbf{k^\prime}})^2 - ( \omega_{\mathbf{q}})^2 }  c_{\mathbf k^\prime + q}^\dagger c_{\mathbf k - q}^\dagger c_{\mathbf k}  c_{\mathbf k^\prime} + \cdot   \cdot  \cdot
\end{equation*}

In recent times, unprecedented control has been achieved to have a controlled number of phonons in a system. There are also experiments where one studies two phonon correlations\cite{toyoda}. Phonons can not only mediate interaction between particles in a lattice, but been used to make the particles hop to sites at long range in lattices as described in the introduction chapter.

\newpage
\chapter{Coupling two states coherently}

In this section we discuss how two states can be coherently coupled using optical tools of two detuned lasers. Preparation of states at lattice sites requires not only spatial control  but also temporal control. In this section we discuss the simplest case of effectively controlling the preparation of a site in a superposition of two states.  Let's consider the system of two closely lying energy levels ($\vert 1 \rangle$ and $\vert 2 \rangle$) coupled optically to a far lying excited state ($\vert e \rangle$) .

The Hamiltonian of the system is

\begin{equation}
\mathcal{H}_0 = \omega_1 \vert 1 \rangle + \omega_2 \vert 2 \rangle  + \omega_e \vert e \rangle.
\end{equation}

Two lasers with frequencies $\lambda_1 = (\omega_e - \Delta)  - \omega_1$ (which is proportional to the gap between  a detuned level below the excited state  from the first state) and $\lambda_2 = (\omega_e - \Delta) - (\omega_2 + \delta)$ (which corresponds to the gap between the detuned level slightly above the second state with the detuning $\Delta$ and the detuned level below the excited state with the detuning $\delta$) can be used to effectively couple two states to the excited state.

So we have (with $\omega_{e1}=\omega_e - \omega_1 $ and $\omega_{e2} = \omega_e - \omega_2$)

\begin{align}
\lambda_1 &= \omega_{e1} - \Delta \nonumber \\
\lambda_2 &= \omega_{e1} - \Delta - (\omega_{12} + \delta) \nonumber\\
\lambda_1 &= \omega_{e2} - \Delta + \omega_{12}  \nonumber\\
\lambda_2 &= \omega_{e2} - \Delta - \delta
\end{align}

In the presence of an external field the interaction Hamiltonian becomes the following:

\begin{align}
\mathcal{H_I} &= - \mu \cdot \mathbf{E}(t) \nonumber \\
& = - \left[ \mu_{e1} \vert 1 \rangle \langle e \vert  +\mu_{1e} \vert e \rangle \langle 1 \vert+   \mu_{e2} \vert 2 \rangle 
 \langle e \vert    +  \mu_{2e} \vert e \rangle \langle 2 \vert\right] \nonumber \\
& \hspace{5.5cm}  E_0 [\cos (\lambda_1 t) + \cos (\lambda_2 t) ]
\end{align}
where the spatial variance of the field is ignored.

Solving the Schrodinger equation 

\begin{equation}
\imath \hbar \frac{\partial \psi(t)}{\partial t} =  \left( \mathcal{H}_0 + \mathcal{H_I} \right) \psi(t) 
\end{equation}

 the wavefunction 

\begin{equation}
\psi (t) = C_1 (t) \vert 1 \rangle +  C_2 (t) \vert 2 \rangle  + C_e (t) \vert e \rangle 
\end{equation}

one can find the folowing relations:

\begin{align}
\imath \dot{C_1} &= \omega_1 C_1 - g C_e \left[  \cos (\lambda_1 t) + \cos (\lambda_2 t +\phi)\right]  \nonumber \\
\imath \dot{C_2} &= \omega_2 C_2 - g C_e \left[  \cos (\lambda_1 t) + \cos (\lambda_2 t +\phi)\right]  \nonumber \\
\imath \dot{C_e} &= \omega_e C_e - g^* \left[C_1 + C_2 \right]\left[  \cos (\lambda_1 t) + \cos (\lambda_2 t + \phi)\right]  
\end{align}
where explicit time dependence on the coefficients has not been shown and we assume $g = \mu_{e1} E_0 = \mu_{e2} E_0$. One can write these coefficients in their own rotating frame to make the equations look simpler

\begin{equation}
\tilde{C_i} = C_i e^{\imath \omega_i t} ~~~~~\text{for}~~~~ i \in {1, 2, e} ~~~\text{and}~~ \imath = \sqrt{-1} 
\end{equation}
and transform the previous set of equation into the following form with cosines now expanded into exponentials

\begin{align}
\imath \dot{\tilde{C_1}} &=  \frac{g}{2} \tilde{C_e} e^{-\imath \omega_{e1} t}  \left[  e^{\imath \lambda_{1} t} + e^{-\imath \lambda_{1} t} +e^{\imath \lambda_{2} t}  + e^{-\imath \lambda_{2} t}\right]  \nonumber \\
\imath \dot{\tilde{C_2}} &=  \frac{g}{2} \tilde{C_e} e^{-\imath \omega_{e2} t}  \left[  e^{\imath \lambda_{1} t} + e^{-\imath \lambda_{1} t} +e^{\imath \lambda_{2} t}  + e^{-\imath \lambda_{2} t}\right]  \nonumber \\
\imath \dot{\tilde{C_e}}  &=  \frac{g^*}{2} \left[ \tilde{C_1} e^{\imath \omega_{e1} t} + \tilde{C_2} e^{\imath \omega_{e2} t}\right]   \left[  e^{\imath \lambda_{1} t} + e^{-\imath \lambda_{1} t} +e^{\imath \lambda_{2} t}  + e^{-\imath \lambda_{2} t}\right].
\end{align}

After change of the variables we find the following:

\begin{align}
\imath \dot{\tilde{C_1}} &=  \frac{g}{2} \tilde{C_e}  \left[  e^{- \imath \Delta t}  + e^{-\imath (\Delta +\delta +  \omega_{12} ) t}\right]  \nonumber \\
\imath \dot{\tilde{C_2}} &=  \frac{g}{2} \tilde{C_e}  \left[  e^{- \imath (\Delta -\omega_{12})t}  + e^{-\imath (\Delta +\delta) t}\right]  \nonumber \\
\imath \dot{\tilde{C_e}}  &=  \frac{g^*}{2} \left[ \tilde{C_1}  \left[  e^{\imath \Delta t}  + e^{\imath (\Delta +\delta +  \omega_{12} ) t}\right] + \tilde{C_2}  \left[  e^{\imath (\Delta -\omega_{12})t}  + e^{\imath (\Delta +\delta) t}\right] \right]   
\end{align}

Solving for the excited state we find the following after integration at time $\tau$

\begin{align}
 \tilde{C_e}(\tau) &=  - \frac{g^*}{2} \left[ \tilde{C_1}(\tau)  \left[  \frac{e^{ \imath \Delta \tau} - 1}{\Delta}  + \frac{e^{\imath (\Delta +\delta +  \omega_{12} ) \tau} -1}{\Delta +\delta +  \omega_{12}}\right] \right. \nonumber\\ 
&\hspace{5cm} + \left. \tilde{C_2}(\tau)  \left[  \frac{e^{\imath (\Delta -\omega_{12})\tau} -1}{\Delta -\omega_{12}}  + \frac{e^{\imath (\Delta +\delta) \tau} -1}{\Delta +\delta}\right] \right]   \nonumber \\ 
&\simeq  - \frac{g^*}{2\Delta} \left[ \tilde{C_1} \left[  e^{ \imath \Delta \tau}  + e^{\imath (\Delta +\delta +  \omega_{12} ) \tau}  -2 \right] + \tilde{C_2} \left[  e^{\imath (\Delta -\omega_{12})\tau}  + e^{\imath (\Delta +\delta) \tau}  -2\right] \right]
\end{align}
where in the denominators we have applied the assumption that $\Delta \gg \omega_{12} > \delta$.

Putting this equation back into the Schr{\"o}dinger equation for  $\tilde{C_1}$ and  $\tilde{C_2}$  one can find the dynamics between the two states:

\begin{align}
\dot{\tilde{C_1}} &= \imath \frac{\vert g\vert^2}{4\Delta} \left[ \tilde{C_1} \left[  e^{ \imath \Delta \tau}  + e^{\imath (\Delta +\delta +  \omega_{12} ) \tau}  -2 \right]  \right. \nonumber\\ 
&\hspace{2cm} \left. + \tilde{C_2} \left[  e^{\imath (\Delta -\omega_{12})\tau}  + e^{\imath (\Delta +\delta) \tau}  -2\right] \right] 
\left[  e^{- \imath \Delta t}  + e^{-\imath (\Delta +\delta +  \omega_{12} ) t}\right]  \nonumber \\
&\simeq \imath \frac{\vert g\vert^2}{4\Delta}   \left[ \tilde{C_1} \left[  2 + e^{\imath (\delta +  \omega_{12} ) \tau}  +  e^{- \imath (\delta +  \omega_{12} ) \tau} \right]   \right. \nonumber\\ 
&\hspace{2cm} \left. + \tilde{C_2} \left[  2e^{- \imath  \omega_{12} \tau} + e^{ \imath \delta \tau}  + e^{- \imath (\delta +  2\omega_{12} ) \tau} \right]   \right. \nonumber\\ 
&\simeq \imath \frac{\vert g\vert^2}{4\Delta}   \left[ 2\tilde{C_1} + \tilde{C_2}  e^{ \imath \delta \tau} \right]
\end{align}
where the \textit{rotating-wave approximation} has been applied to disregard exponentials which have large frequencies as $\Delta$ in first step and $\omega_{12}$ in the second step.  

Similarly

\begin{equation}
\dot{\tilde{C_2}} \simeq \imath \frac{\vert g\vert^2}{4\Delta}   \left[ 2\tilde{C_2} + \tilde{C_1}  e^{ -\imath \delta \tau} \right]
\end{equation}
which results in the following effective Hamiltonian where the two states are coupled with the Rabi frequency $\Omega =  \frac{\vert g\vert^2}{2\Delta} $

\begin{align}
\mathcal{H}_{eff} &= - \frac{\vert g\vert^2}{2\Delta}  \left[ \vert 1\rangle \langle 1 \vert  +  \vert 2\rangle \langle 2 \vert \right] - \frac{\vert g\vert^2}{4\Delta}  \left[ e^{ \imath \delta \tau} \vert 1\rangle \langle 2 \vert + e^{ -\imath \delta \tau} \vert 2\rangle \langle 1 \vert \right] \\
&\simeq  - \frac{\Omega}{2}  \left[ e^{ \imath \delta \tau} \sigma^- + e^{ -\imath \delta \tau} \sigma^+ \right]  
\end{align}
where $\sigma^{+}(^-)$ are the raising (lowering) operators.

The phase between the two laser beams can now be put back into the equation
\begin{align}
\mathcal{H}_{eff} &=  - \frac{\Omega}{2}  \left[ e^{ \imath (\delta \tau + \phi)} \sigma^- + e^{ -\imath (\delta \tau + \phi)} \sigma^+ \right]  
\end{align}
Fixing the detuning  to $\delta=0$, makes the dynamics between the states easily controllable using the phase between two photons. See details in reference \cite{rajibul-thesis}.

\begin{align}
\mathcal{H}_{eff} &=  - \frac{\Omega}{2}  \left[ e^{ \imath  \phi} \sigma^- + e^{ -\imath  \phi} \sigma^+ \right]  
\end{align}

In the introductory chapter, this approach has been used to effectively couple sites at long range,  whcih has been recently implemented experimentally \cite{jurcevic}.

\newpage
\chapter{Optical lattices}
    
Optical lattices are created as standing electric  field waves by  constructive interference of multiple laser beams. The lattice constant of these systems is typically in the order of the $\mu$m. In these lattices, the dynamics of the atoms loaded, can be easily controlled as the inter-site tunnelling parameter can be determined by fixing the intensity of the laser beams. The atoms \textit{feel} the lattice potential according to their AC-Stark shift. The geometry, shape and trap depth of these lattice systems can be modified experimentally. The temperature of the atoms loaded  in these lattices can also be very low ($< 10^{-9} K$).

The atoms interact with the laser field through the dipole moment induced by the electric field

\begin{equation}
\mathcal{H_I} = \mathbf{d\cdot E} ~~~~~\mbox{where} ~~~~~\mathbf{E} = E(\mathbf{x}) \cos (\omega t) 
\end{equation}

The change in ground state energy due to this perturbation   can be calculated  as

\begin{equation}
\mathcal{E}_g^{(2)} = \sum_{e \neq g} \frac{\vert \langle e\vert \mathbf{d}\vert g \rangle\vert^2}{4} \left[ \frac{1}{\mathcal{E}_e - \mathcal{E}_g - \omega} - \frac{1}{\mathcal{E}_e - \mathcal{E}_g + \omega}\right] E(\mathbf{x})^2
\end{equation}

The first order term cancels as the expectation value of the dipole operator for the ground state is zero.

The lifetime of the excited state $\left(\frac{1}{\Gamma_e}\right)$ can be accounted in the equation as modified excited state energy from uncertainty principle.

\begin{equation}
\mathcal{E}_g^{(2)} = \sum_{e \neq g} \frac{\vert \langle e\vert \mathbf{d}\vert g \rangle\vert^2}{4} \left[ \frac{1}{\mathcal{E}_e -  \frac{\imath\Gamma_e}{2}-\mathcal{E}_g - \omega} - \frac{1}{\mathcal{E}_e - \frac{\imath\Gamma_e}{2}- \mathcal{E}_g + \omega}\right] E(\mathbf{x})^2
\end{equation}

Assuming a two level system,  after applying the rotating wave approximation, one obtains the spatial potential felt by the particles locally

\begin{equation}
V_g (\mathbf{x}) = \frac{\Omega^2 \delta}{ 4\delta^2 + \Gamma_e^2} \simeq \frac{\Omega^2}{ 4\delta^2} \delta
\end{equation}
where $\Omega(\mathbf{x}) = \langle e\vert \mathbf{d}\vert g \rangle E(\mathbf{x})$ is the Rabi frequency.

For the red detuned photons ($\delta <0$), this potential draws the particles to the intensity maximum while for the blue detuned ($\delta >0$) photons  to the intensity minimum. Particles in the excited state however will feel the opposite effect.

The imaginary part can be understood as the photon absorption rate related to the lifetime of the ground state and represents the rate of loss of the ground state. 

\begin{equation}
\Gamma (\mathbf{x}) =\frac{1}{2} \frac{\Omega^2 \Gamma_e}{ 4\delta^2 + \Gamma_e^2} \simeq \frac{\Omega^2}{ 8\delta^2} \Gamma_e
\end{equation}

The geometry of 1D, 2D and 3D optical lattices can be ideally represented by the following equations for standing waves

\begin{equation}
V_x^{1D} = V_0 \cos^2 (kx), ~~~~~~~~~~ k=\frac{2\pi}{\lambda}, a = \frac{\lambda}{2}
\end{equation}
where $a$ is the lattice constant. In  two dimensions four beams can be used to create a rectangular optical lattice with many desired types of  potentials

\begin{equation}
V_{x, y}^{2D} = V_0 \left[ \cos^2 (kx) + \cos^2(ky) + 2 \epsilon_1\cdot \epsilon_2 \cos(\phi) \cos(kx)\cos(ky) \right] 
\end{equation}
while using three beams one can prepare hexagonal optical lattices. 

For the case of a 3D optical lattice, one can use three pairs of beams

\begin{equation}
V_{x, y, z}^{3D} = V_0 \left[ \cos^2 (kx) + \cos^2(ky) + \cos^2(kz) + \frac{m}{2}\left(\omega_x^2 + \omega_y^2 + \omega_z^2 \right) \right] 
\end{equation}


The states of particles generally spread over a few lattice sites and can be approximated by Wannier orbitals 
(see Appendix G). In many cases of interest of such systems, it would suffice to consider the overlap of the wavefunctions localized on each lattice site with that of nearest neighbors sites giving rise to hopping of particles limited to nearest neighbors.

\newpage

\chapter{Excitons}

Excitons are composite quasiparticles of interacting electrons and holes. Its energy depends on electron-electron interactions, hole-hole interactions and electron-hole interactions. The theoretical models that were analyzed in this thesis also hold relevant for few cases of exciton physics, specifically to physics of interacting Frenkel excitons. In this Appendix we discuss the basic understanding of excitonic Hamiltonians.

\section*{electron-electron interaction}

Electrons in lattices are understood through the following Hamiltonian which consists of kinetic energy of the electrons, electron-electron Coulombic interaction and interaction with the lattice potential. In absence of any magnetic interaction, the Schr{\"o}dinger equation takes the following form.

\begin{align}
\left[ \int d^3x \psi^\dagger (x) \left(-\frac{\hbar^2\Delta}{2m} + V_L (x)\right)\psi(x) \right. &\nonumber \\
+ \frac{1}{2} \int d^3x d^3x' \psi^\dagger (x) \psi^\dagger (x')&   \left. \frac{e}{|x-x'|} \psi (x)\psi (x') \right]\vert\Phi\rangle = E\vert\Phi\rangle
\end{align}
where the interaction is approximated to two body form only and the wavefunctions are written in terms of fermionic operators $\psi^\dagger(x)$ and $\psi(x)$. These operators can be expanded in terms of eigenfunctions $\phi_j(x)$ and $\phi_{j}^*(x)$.

\begin{align}
\psi(x) &= \sum_j c_j \phi_j(x) \nonumber \\
\psi^\dagger(x) &= \sum_j c_{j}^\dagger \phi_{j}^*(x)
\end{align}

The full wavefunction can be taken as that of electrons occupying the eigenstates.

\begin{equation}
\vert\Phi\rangle = c_{j1}^\dagger c_{j2}^\dagger ..... c_{jN}^\dagger \vert0\rangle
\end{equation}

The Hamiltonian now transforms into the following form.

\begin{align}
H =  \sum_{mn} c_m^\dagger c_n \int d^3x \phi^* (x) \left(-\frac{\hbar^2\Delta}{2m} + V_L (x)\right)\phi(x) &\nonumber \\
+ \frac{1}{2} \sum_{mnm'n'} c_m^\dagger c_n^\dagger c_{n'} c_{m'} \int d^3x d^3x' \phi^* (x) \phi^* (x')& \frac{e}{|x-x'|} \phi (x)\phi (x') 
\end{align}

From the kinetic energy part one can find the following.

\begin{align}
\langle\Phi \vert c_m^\dagger c_n \vert\Phi\rangle &= \langle 0\vert c_{j1} c_{j2} ..... c_{jN} ~ c_m^\dagger c_n ~ c_{j1}^\dagger c_{j2}^\dagger ..c_{j'}^\dagger.. c_{jN}^\dagger \vert0\rangle \nonumber \\
&=   \langle 0\vert c_{j1} c_{j2} ..... c_{jN} ~ c_m^\dagger  ~~~~ c_{j1}^\dagger c_{j2}^\dagger ..~~~.. c_{jN}^\dagger \vert0\rangle(-1)^{exchange} \nonumber \\
&= \langle 0\vert c_{j1} c_{j2} ..~.. c_{jN} ~~~~~~~~ c_{j1}^\dagger c_{j2}^\dagger ..~~~.. c_{jN}^\dagger \vert0\rangle \nonumber \\
&= \delta_{mn} ~\mbox{for} ~~ m \in {j_1,j_2, .... , j_N}
\end{align}
where the number of exchanges are same for application of both $c_m^\dagger$ and $c_n$ hence no minus sign. Similarly in the interaction term the following terms can be found.

\begin{align}
\langle\Phi \vert c_m^\dagger c_n^\dagger  c_{n'} c_{m'} \vert\Phi\rangle &= \delta_{nn'}\delta_{mm'} -  \delta_{nm'}\delta_{mn'} ~~\mbox{for}~~ m,n \in {j_1,j_2, ... ,j_N}~\mbox{and} ~ m\neq n, m'\neq n'
\end{align}
where in the first term of right hand side, the number of exchanges of fermionic operators stay even as one can observe from the previous equation first applying $c_m^\dagger$ and $c_{m'}$ then $c_n^\dagger$ and $c_{n'}$. The second term will require one extra exchange between $c_m^\dagger$ and $c_n^\dagger$ hence having a minus sign. For bosonic operators, however, no minus sign will be there for both Coulombic interaction and exchange terms. 

Thus we arrive at the Hartree-Fock expression of total energy where the interaction term with no minus sign is described as that of Coulombic interaction and the term with minus sign is described as that of exchange interaction. 

\begin{align}
\langle \Phi\vert H\vert\Phi\rangle =  \sum_{j_1}\int d^3x \phi_{j_1}^* (x) \left(-\frac{\hbar^2\Delta}{2m} + V_L (x)\right)\phi_{j_1}(x) &\nonumber \\
+ \frac{1}{2} \sum_{j_1, j_2}  \int d^3x d^3x' \phi_{j_1}^* (x) \phi_{j_2}^* (x')& \frac{e}{|x-x'|} \phi_{j_2} (x')\phi_{j_1} (x)  \nonumber \\
- \frac{1}{2} \sum_{j_1, j_2}  \int d^3x d^3x' \phi_{j_1}^* (x) \phi_{j_2}^* (x')& \frac{e}{|x-x'|} \phi_{j_1} (x')\phi_{j_2} (x) 
\nonumber \\
 E = E_0 + E_C + E_{ex} \hspace{4.8cm} &
\end{align}

\section*{electron-hole interaction}

Empty states within a full valence band ($V$) can be represented by particles called hole for simplification.  

\begin{align}
c_{j,V} &= d_j^\dagger \nonumber \\
c_{j,V}^\dagger &= d_j
\end{align}

The fermionic operators $d$ here represents holes. The Hamiltonian now becomes

\begin{align}
H = E_V - \sum_j d_j^\dagger d_j E_{j,V} + \frac{1}{2} \sum_{j_1,j_2,j_3,j_4} d_{j_1}^\dagger d_{j_2}^\dagger d_{j_3} d_{j_4} \mathcal{V} (j_1 j_2 \vert j_3 j_4) 
\end{align}
where $E_{j,V}$ now contains both interaction and exchange terms of electrons and $\mathcal{V}$ contains explicit hole-hole interaction term as can be seen from following equations.

\begin{align}
c_m^\dagger c_n^\dagger c_{n'} c_{m'} &= d_m d_n d_{n'}^\dagger d_{m'}^\dagger \nonumber \\
&= \delta_{nn'} \delta_{mm'}- \delta_{nm'} \delta_{nm'} \nonumber \\
& ~~~ - \delta_{nn'} d_{m'}^\dagger d_m + \delta_{mn'} d_{m'}^\dagger d_n - \delta_{mm'} d_{n'}^\dagger d_n + \delta_{nm'} d_{n'}^\dagger d_m \nonumber \\
&~~~ + d_{n'}^\dagger d_{m'}^\dagger d_m d_n
\end{align}

\begin{align}
\mathcal{V} (j_1 j_2 \vert j_3 j_4) = \int d^3x d^3x' \phi_{j_1}^*(x)  \phi_{j_2}^*(x') \frac{e^2}{|x-x'|} \phi_{j_3}(x')  \phi_{j_4}(x)  
\end{align}

To deal with holes in valence band and electrons in conduction band we use the fermionic operators with extra notation.

\begin{align}
c_{j_1,J_1} c_{j_2, J_2}^\dagger + c_{j_2, J_2}^\dagger c_{j_1,J_1} = \delta_{j_1,j_2} \delta_{J_1,J_2} 
\end{align}
where $J$ symbols are band indexes.

Now the total interaction can be written as following.

\begin{align}
H' = \frac{1}{2}  \sum_{j_1,j_2,j_3,j_4;J_1,J_2,J_3,J_4}& c_{j_1,J_1}^\dagger c_{j_2,J_2}^\dagger c_{j_3,J_3} c_{j_4,J_4} \nonumber\\
&\cdot \int d^3xd^3x' \phi_{j_1,J_1}^*(x) \phi_{j_2,J_2}^*(x') \frac{e^2}{|x-x'|} \phi_{j_3,J_3}(x') \phi_{j_4,J_4}(x)   
\end{align}

Using the hole operators in valence band and electron operators in conduction band one can remove the band indexes and write the interaction Hamiltonian in terms of electron-electron interaction in conduction band, hole-hole interaction in valence band and electron-hole interaction between conduction and valence band.

\begin{align}
H' = H_{e-e} + H_{h-h} + H_{e-h}
\end{align}

\begin{align}
 H_{e-e} = \frac{1}{2} \sum_{j_1,j_2,j_3,j_4} c_{j_1}^\dagger c_{j_2}^\dagger c_{j_3} c_{j_4}^\dagger \mathcal{V} (j_1 j_2 | j_3 j_4)
\end{align}

\begin{align}
 H_{h-h} = \frac{1}{2} \sum_{j_1,j_2,j_3,j_4}  d_{j_1} d_{j_2} d_{j_3}^\dagger d_{j_4}^\dagger \mathcal{V} (j_1 j_2 | j_3 j_4)
\end{align}

The electron-hole interaction is combination of many parts.

\begin{align}
 H_{e-h} =  \sum_{j_1,j_2,j_3,j_4} \left[ c_{j_1}^\dagger d_{j_2} d_{j_3}^\dagger c_{j_4} + d_{j_1} c_{j_2}^\dagger c_{j_3} d_{j_4}^\dagger + c_{j_1}^\dagger d_{j_2} c_{j_3} d_{j_4}^\dagger + d_{j_1} c_{j_2}^\dagger d_{j_3}^\dagger c_{j_4} \right] \mathcal{V} (j_1 j_2 | j_3 j_4)
\end{align}

From each terms in electron-hole interaction Hamiltonian induced electron energy terms can be found. For example

\begin{align}
 c_{j_1}^\dagger d_{j_2} d_{j_3}^\dagger c_{j_4} =  c_{j_1}^\dagger c_{j_4} d_{j_2} d_{j_3}^\dagger = c_{j_1}^\dagger c_{j_4} \left[\delta_{j_2 j_3} - d_{j_3}^\dagger d_{j_2}  \right]
\end{align}

\section*{Wannier excitons}

For electrons and holes  with relative motion delocalized over whole lattice, one can describe these electrons  according to their occupation at valence band maximum and holes to  their occupation at conduction band minimum where the dispersion is parabolic. For such a single electron-hole pair in 3D lattices, they behave essentially as Hydrogen atom and similar spectra is expected beside the effect of lattice confinement. In such cases the Hamiltonian takes the following form.

\begin{align}
H =  \sum_{k}\left(E_{0C} \right.&+\left.  \frac{\hbar^2k^2}{2m_C}\right) + \sum_{k}\left(- E_{0V} +  \frac{\hbar^2k^2}{2m_V}\right) \nonumber\\
&+ \frac{1}{2}\sum_{k_1,k_2, k_3, k_4} c_{k_1}^\dagger c_{k_2}^\dagger c_{k_3} c_{k_4} \mathcal{V}(k_1 k_2\vert k_3 k_4) \nonumber \\
 &+ \frac{1}{2}\sum_{k_1,k_2, k_3, k_4} d_{k_1}^\dagger d_{k_2}^\dagger d_{k_3} d_{k_4} \mathcal{V}(k_3 k_4\vert k_1 k_2) \nonumber \\
&+ \sum_{k_1,k_2, k_3, k_4} c_{k_1}^\dagger c_{k_2} d_{k_3}^\dagger d_{k_4} \mathcal{V}(k_1 k_4\vert k_2 k_3) + \cdot \cdot\cdot
\end{align}
where $E_{0C}$ is the minimum of conduction band and $E_{0V}$ is the maximum of valence band.

Writing a single combined quasiparticle operator that can be called \textit{exciton} replacing separate operators for electrons and holes, one can show that these quasiparticles effectively are boson like.

\begin{align}
q_{j}^\dagger =  c_{k_e}^\dagger d_{k_h}^\dagger
\end{align}

\begin{align}
q_{j_1}^\dagger q_{j_2}^\dagger &= q_{j_2}^\dagger q_{j_1}^\dagger \nonumber \\ 
\end{align}

\begin{align}
 q_{j_1} q_{j_2}^\dagger &= d_{k_h'}c_{k_e'}c_{k_e}^\dagger d_{k_h}^\dagger \nonumber \\ 
&=  d_{k_h'} \left( \delta_{k_e',k_e} - c_{k_e}^\dagger c_{k_e'}  \right)  d_{k_h}^\dagger   \nonumber \\ 
&=  \delta_{k_e',k_e}  \left( \delta_{k_h',k_h}  - d_{k_h}^\dagger d_{k_h'}  \right)  - c_{k_e}^\dagger  \left( \delta_{k_h',k_h}  - d_{k_h}^\dagger d_{k_h'}  \right)c_{k_e'}   
\end{align}

\begin{align}
\left[q_{j_1}^\dagger , q_{j_2}^\dagger \right]_- &=\left[q_{j_1} , q_{j_2} \right]_-= 0 \nonumber \\
\left[q_{j_1} , q_{j_2}^\dagger \right]_- &= \delta_{j_1,j_2} - D_{j_1,j_2}
\end{align}
where $D_{j_1,j_2}$ is the operator describing deviation from bosonic commutation rule essentially consisting of scattering components between electron and hole.

\begin{align}
 D_{j_1,j_2} = \delta_{k_e',k_e}d_{k_h}^\dagger d_{k_h'} + \delta_{k_h',k_h} c_{k_e}^\dagger c_{k_e'}
\end{align}

\section*{Frenkel excitons}

For electron hole pairs in lattices with relative motion limited to single sites only are another extreme of exciton physics. These excitons are named after Frenkel. Their states can be better expanded in terms of atomic states rather than Bloch waves or in terms of Wannier functions (e.g. atoms in optical lattices) or in terms of delta functions for further simplification. In such cases the lattice site index replaces the state index for electrons and holes.

\begin{align}
H = \sum_{mn} E_e^{eff} & c_m^\dagger c_n +  \sum_{mn} E_h^{eff} d_m^\dagger d_n \nonumber\\
&- \sum_{mn} c_m^\dagger c_m d_n^\dagger d_n \mathcal{V}(mn|mn) +  \sum_{mn} c_m^\dagger c_n d_m^\dagger d_n \mathcal{V}(mn|mn) + \cdot \cdot \cdot
\end{align}
where $E_e^{eff}$ and $E_h^{eff}$ consists of not only that of kinetic energies of electrons and holes respectively but also of electron and hole scattering terms.




\newpage

\chapter{On Numerical Computations}

The numerics involves many intricacies which spans through basis selection, basis transformation, algorithm implementation and sorting of computed results. A few LAPACK and BLAS routines (dgemm, zgemm, zgemv, dsyev, zsyev) were extensively used combined with random number generator (Mersenne Twister) and parallelization method (OpenMP). The results are sensitive to use of these libraries at proper sequences within each numerical calculations. The calculations were also dynamically optimized to produce numbers within tested accuracy while using minimum disk space of both running and static type. Each of these steps involves further complications. Each of the results was benchmarked against either previously existing reports or some alternatively calculated result. The most computational difficulty involves handling of available running memory on computing nodes for optimization of efficiency and accuracy.  While each calculation had to pass rigorous testing, the scale of the calculations were not easy for implementation either, specifically for large system sizes of higher dimension. The calculations for disordered systems involves averaging over many disorders till the results converges, which is time consuming. The recursive calculations add significant multiplicity for each calculation with a separate point of energy within full bandwidth, which although allows for calculations of larger system sizes, limits the size of the calculations from consideration of consumption of time. For large system sizes, the available running memory on computing nodes limits the use of parallel computation method. Splitting of each calculation over several computing nodes is often the case. Sorting of elements within dynamic multidimensional arrays is also important for the efficiency of the calculations. While providing prototypes for each code is not very useful here, as use of coding platforms and styles changes from person to person, this appendix only intends to give the reader very brief sense on the detail of numerical implementations which lies underneath the presented results in this thesis.

\end{appendices}

\end{document}